\documentstyle[preprint,aps,epsfig]{revtex}

\def\trademark{{\ooalign{\hfil\raise.07ex\hbox{\scriptsize R}\hfil\crcr\mathhexbox20D}}}
\def\registered{{\ooalign {\hfil\raise .05ex\hbox{\scriptsize
R}\hfil\crcr\mathhexbox20D}}}
\begin{document}
\draft
\tighten

\title{Correlations of errors
in measurements of CP violation at neutrino factories}

\author{John Pinney{$^1$}
\footnote{Email: jpinney@phys.metro-u.ac.jp}
and Osamu Yasuda{$^2$}
\footnote{Email: yasuda@phys.metro-u.ac.jp}
}
\address{
\qquad\\{$^1$}The Daiwa Anglo-Japanese Foundation\\
Daiwa Securities Kabuto-cho Bldg 9F\\
1-1-9 Nihonbashi Kayabacho, Chuo-ku, Tokyo 103-0025, Japan\\
\qquad \\ {$^2$}Department of Physics,
Tokyo Metropolitan University \\
Minami-Osawa, Hachioji, Tokyo 192-0397, Japan
}

\date{May, 2001}
\preprint{
\parbox{5cm}{
hep-ph/0105087\\
}}

\maketitle

\begin{abstract}

Using $\Delta\chi^2$ which is defined by the difference
of the number of events with the CP phase $\delta$ and
the hypothetical one with $\delta=0$,
we discuss correlations of
errors of the CP phase and other oscillation parameters as well as
the matter effect in measurements at neutrino
factories.  By varying the oscillation parameters and the
normalization of the matter effect, we evaluated the
data size required to reject a hypothesis with $\delta=0$
at 3$\sigma$CL.
The optimum muon energy and the baseline depends on
the magnitude of $\theta_{13}$, the background fraction,
the uncertainty of the normalization of the matter effect,
but in general lie in the ranges
20GeV $\lesssim E_\mu\lesssim$ 50GeV,
1000km$\lesssim L\lesssim$3000km.
If we assume that the uncertainty of the matter effect is as large
as 20\% then the optimum values
may be modified to $E_\mu\lesssim$ 10GeV, $L\lesssim$1000km
due to the strong correlation of $\delta$ and the matter effect.
We show analytically that sensitivity to CP violation
is lost for $E_\mu\ll$ 10GeV or for $E_\mu\gg$ 50GeV.
We also discuss the possibility of measuring
CP violation at the upgraded JHF experiment by
taking all the error correlations into account, and
show that it is possible to demonstrate $\delta\ne0$ at 3$\sigma$CL
for $\theta_{13}\gtrsim 3^\circ$.

\end{abstract}
\vskip 0.1cm

\section{Introduction}
There have been several experiments
\cite{homestake,Kamsol,SKsol,sage,gallex,Kamatm,IMB,SKatm,SKup,soudan2,lsnd}
which suggest neutrino oscillations.
It has been shown in the two flavor framework that the solar neutrino deficit
can be explained by neutrino oscillation with the set of parameters
$(\Delta m^2_\odot,\sin^22\theta_\odot)\simeq$ $({\cal
O}(10^{-5}{\rm eV}^2),{\cal O}(10^{-2}))$ (small angle MSW solution),
$({\cal O}(10^{-5}{\rm eV}^2),{\cal O}(1))$ (large angle MSW
solution), or $({\cal O}(10^{-10}{\rm eV}^2),{\cal O}(1))$ (vacuum
oscillation solution), and the atmospheric neutrino anomaly
can be accounted for by $(\Delta m_{\mbox{\rm
atm}}^2,~\sin^22\theta_{\mbox{\rm atm}})\simeq (10^{-2.5}{\rm
eV}^2,1.0)$.

In the three flavor framework there are two independent mass squared
differences and it is usually assumed that these two mass differences
correspond to $\Delta m^2_\odot$ and $\Delta m_{\mbox{\rm atm}}^2$.
Throughout this paper we will assume three neutrino species which
can account for only the solar neutrino deficit and the atmospheric
neutrino anomaly \footnote{To explain the LSND
anomaly \cite{lsnd} one needs at least four neutrino species.}.
Without loss of generality we assume
$|\Delta m_{21}^2|<|\Delta m_{32}^2|<|\Delta m_{31}^2|$
where $\Delta m^2_{ij}\equiv m^2_i-m^2_j$.
The flavor eigenstates are related to the mass eigenstates by
$U_{\alpha j}$ ($\alpha=e,\mu,\tau$),
where $U_{\alpha j}$ are the elements
of the MNS mixing matrix U \cite{mns}:
\begin{eqnarray}
&{\ }&\left( \begin{array}{c} \nu_e  \\ \nu_{\mu} \\ 
\nu_{\tau} \end{array} \right)
=U\left( \begin{array}{c} \nu_1  \\ \nu_2 \\ 
\nu_3 \end{array} \right),\nonumber\\
U&\equiv&\left(
\begin{array}{ccc}
U_{e1} & U_{e2} &  U_{e3}\\
U_{\mu 1} & U_{\mu 2} & U_{\mu 3} \\
U_{\tau 1} & U_{\tau 2} & U_{\tau 3}
\end{array}\right)
%\nonumber\\
=\left(
\begin{array}{lll}
c_{12}c_{13} & s_{12}c_{13} &  s_{13}e^{-i\delta}\nonumber\\
-s_{12}c_{23}-c_{12}s_{23}s_{13}e^{i\delta} & 
c_{12}c_{23}-s_{12}s_{23}s_{13}e^{i\delta} & s_{23}c_{13}\nonumber\\
s_{12}s_{23}-c_{12}c_{23}s_{13}e^{i\delta} & 
-c_{12}s_{23}-s_{12}c_{23}s_{13}e^{i\delta} & c_{23}c_{13}\nonumber\\
\end{array}\right).
\label{eqn:mns}
\end{eqnarray}
With the mass hierarchy $|\Delta m_{21}^2|\ll|\Delta m_{32}^2|$
there are two possible
mass patterns which are depicted in Fig. 1a and 1b,
depending on whether $\Delta m_{32}^2$ is positive or negative.
If the matter effect is relevant then  the sign of
$\Delta m_{32}^2$ can be determined by distinguishing neutrinos and
anti-neutrinos.
The Superkamiokande experiment uses water Cherenkov detectors
and events of neutrinos and anti-neutrinos are unfortunately
indistinguishable, so the sign of $\Delta m_{32}^2$ is unknown
to date.

It has been shown in the three flavor framework
\cite{yasuda1,gmpv,flmmp}
of the CHOOZ reactor data \cite{chooz}
and the atmospheric neutrino data implies very small
$\theta_{13}$, i.e., $\sin^22\theta_{13} < 0.1$ which is
essentially the result of the CHOOZ data.  When $|\theta_{13}|$ is
small, the MNS matrix looks like
\begin{eqnarray}
U\simeq\left(
\begin{array}{lll}
{\ }~c_\odot  & {\ }~s_\odot  &  \epsilon\nonumber\\
-s_\odot c_{\mbox{\rm atm}} & 
{\ }~c_\odot c_{\mbox{\rm atm}} & s_{\mbox{\rm atm}}\nonumber\\
{\ }~s_\odot s_{\mbox{\rm atm}} & 
-c_\odot s_{\mbox{\rm atm}} & c_{\mbox{\rm atm}}\nonumber\\
\end{array}\right),\nonumber
\end{eqnarray}
where $\theta_{12}$, $\theta_{23}$ have been replaced by $\theta_\odot$
and $\theta_{\mbox{\rm atm}}$, respectively.

Measurements of $\theta_{13}$ and even CP phase $\delta$ may be
possible in the near future projects \cite{minos,jhf}.  Neutrino
factories \cite{Geer,nf} have been proposed to measure $\theta_{13}$, the
sign of $\Delta m_{32}^2$ and CP phase $\delta$, and a lot of people
have worked out many aspects in detail
\cite{nf,BGW,Golden,BGRW,BGRW2,GH,DFLR,FLPR,NuFact,BCR,KS,Yasuda0,sato,mtky}.

Since measurement of CP violation is the main motivation for neutrino
factories, it is important to examine errors of the CP phase $\delta$
taking into consideration all possible sources of uncertainties in the
oscillation parameters as well as matter effects.  In this paper we
discuss correlations of errors of $\delta$ and the matter
effect, as well as $\theta_{k\ell}$ and $\Delta m^2_{k\ell}$.
\footnote{Some of the correlations of errors of $\delta$ and
other oscillation parameters have been discussed
in the past \cite{Golden,BCR}, but they used relatively
optimistic sets of the oscillation parameters, and
the behaviors we obtain with the most up to date best fit values
for the oscillation parameters
are sometimes different from theirs.}
We find that the correlation of $\delta$ and the matter effect
is strong for ($E_\mu$=50 GeV, $L\simeq$3000km),
which has been advocated as the best choice,
and if the uncertainty of the matter effect is larger than
5 \% then baselines shorter than 3000km are advantageous.

There have been works \cite{sato,mn,richter}
which advocated the advantage of conventional low energy neutrino beams
over neutrino factories with high energy, and
there does not seem to be consensus among the community of
neutrino physicists about the optimum neutrino energy
and the baseline.
One of the purposes of this paper is to show
that neutrino factories with high or medium muon energy
(20GeV$\lesssim E_\mu\lesssim$50GeV) are more advantageous
over experiments with low energy ($E_\mu\ll$10GeV).
We evaluate the statistical significance of
possible CP violation in neutrino factory type
experiments for a range of the muon energy
0.5GeV$\le E_\mu\le$50GeV and the baseline 10km $\le L\le$ 10$^4$km
and we show that
the case $E_\mu\lesssim$ a few GeV is always
inferior to the option with higher energy.
On the other hand, Lipari \cite{lipari}
has shown that long baseline experiments
lose the sensitivity for CP violation
for extremely high energy.
We will show analytically that
the two kinds of $\Delta\chi^2$ that we will
use, one of which is defined through
the difference of $\delta\ne0$
and $\delta=0$ and the other
of which through T violation, decrease
for large muon energy $E_\mu\gg$50GeV.

We also analyze the possibility to measure CP violation
at the upgraded JHF experiment with 4MW power and
a one mega ton detector.  We show that measurement of
CP violation is possible for a relatively large value
of $\theta_{13}$.  It is possible mainly because the detector
size is so huge.

In section 2 we analyze numerically the effect of the CP phase
$\delta$ in neutrino factories
by introducing $\Delta\chi^2$ which is defined as the square
of the difference of the number of events with $\delta\ne0$
and that with $\delta=0$, divided by errors.
We show that the optimal muon energy and baseline
vary depending on the magnitude of $\theta_{13}$,
the background fraction and the uncertainty of the matter effect.
In section 3 we discuss T violation assuming ideal polarization.
In section 4 we present analytical treatments
of the two kinds of $\Delta\chi^2$ introduced
in sections 2 and 3, and show that these
$\Delta\chi^2$ decrease as the muon energy becomes
very small or very large.
In section 5 we discuss the possibility of measurement of
CP violation at the upgraded JHF experiment.
In section 6 we summarize our results and
we briefly comment on the uncertainty of the matter effects
from the viewpoint of geophysics.

\section{Analysis of CP Violation}

\subsection{Definition of $\Delta\chi^2$}

Our strategy here is to examine whether a hypothesis with
a vanishing CP phase is rejected or not by taking into
consideration all channels $\nu_e\rightarrow\nu_\mu$,
${\bar\nu}_e\rightarrow{\bar\nu}_\mu$,
$\nu_\mu\rightarrow\nu_\mu$ and
${\bar\nu}_\mu\rightarrow{\bar\nu}_\mu$.
For this purpose we define $\Delta\chi^2$ to test a hypothesis that a
CP phase is given by $\bar{\delta}$ in the case where the true value is
$\delta$:
\begin{eqnarray}
&{\ }&\Delta\chi^2(\theta_{k\ell},\Delta m_{k\ell}^2,
\delta,C;\bar{\theta}_{k\ell},\overline{\Delta m_{k\ell}^2},
\bar{\delta},\bar{C})\nonumber\\
&\equiv&
\sum_j
{\left[N_j^{\mbox{\rm wrong}}(\mu^-;\theta_{k\ell},\Delta m_{k\ell}^2,\delta,C)
-N_j^{\mbox{\rm wrong}}(\mu^-;
\bar{\theta}_{k\ell},\overline{\Delta m_{k\ell}^2},\bar{\delta},\bar{C}
)\right]^2
\over N_j^{\mbox{\rm wrong}}(\mu^-;\theta_{k\ell},\Delta m_{k\ell}^2,\delta,C)}
\nonumber\\
&+&
%+
\sum_j
{\left[N_j^{\mbox{\rm wrong}}(\mu^+;\theta_{k\ell},\Delta m_{k\ell}^2,\delta,C)
-N_j^{\mbox{\rm wrong}}(\mu^+;
\bar{\theta}_{k\ell},\overline{\Delta m_{k\ell}^2},\bar{\delta},\bar{C}
)\right]^2
\over N_j^{\mbox{\rm wrong}}(\mu^+;\theta_{k\ell},\Delta m_{k\ell}^2,\delta,C)}
\nonumber\\
&+&\sum_j
{\left[N_j^{\mbox{\rm right}}(\mu^-;\theta_{k\ell},\Delta m_{k\ell}^2,\delta,C)
-N_j^{\mbox{\rm right}}(\mu^-;
\bar{\theta}_{k\ell},\overline{\Delta m_{k\ell}^2},\bar{\delta},\bar{C}
)\right]^2
\over N_j^{\mbox{\rm right}}(\mu^-;\theta_{k\ell},\Delta m_{k\ell}^2,\delta,C)}
\nonumber\\
&+&
%+
\sum_j
{\left[N_j^{\mbox{\rm right}}(\mu^+;\theta_{k\ell},\Delta m_{k\ell}^2,
\delta,C)
-N_j^{\mbox{\rm right}}(\mu^+;
\bar{\theta}_{k\ell},\overline{\Delta m_{k\ell}^2},\bar{\delta},\bar{C}
)\right]^2
\over N_j^{\mbox{\rm right}}(\mu^+;\theta_{k\ell},\Delta m_{k\ell}^2,
\delta,C)},
\label{eqn:delchi2}
\end{eqnarray}
where $j$ runs over energy bins and the numbers of events are given by
\begin{eqnarray}
&{\ }&N_j^{\mbox{\rm wrong}}(\mu^-;\theta_{k\ell},\Delta m_{k\ell}^2,\delta,C)
\nonumber\\&=&
{12N_0E_\mu\over \pi L^2 m_\mu^2}
\int_{E_j}^{E_{j+1}} dE_\nu
\left({E_\nu \over E_\mu}\right)^2
\left(1-{E_\nu \over E_\mu}\right)
\sigma_{\nu N}(E_\nu)P(\nu_e\rightarrow\nu_\mu
;\theta_{k\ell},\Delta m_{k\ell}^2,\delta,C)\nonumber\\
&{\ }&N_j^{\mbox{\rm wrong}}(\mu^+;\theta_{k\ell},\Delta m_{k\ell}^2,\delta,C)
\nonumber\\&=&
{12N_0E_\mu\over \pi L^2 m_\mu^2}
\int_{E_j}^{E_{j+1}} dE_{\bar\nu}
\left({E_{\bar\nu} \over E_\mu}\right)^2
\left(1-{E_{\bar\nu} \over E_\mu}\right)
\sigma_{{\bar\nu} N}(E_{\bar\nu})P({\bar\nu}_e\rightarrow{\bar\nu}_\mu
;\theta_{k\ell},\Delta m_{k\ell}^2,\delta,C)
\nonumber\\
&{\ }&N_j^{\mbox{\rm right}}(\mu^-;\theta_{k\ell},\Delta m_{k\ell}^2,\delta,C)
\nonumber\\&=&
{2N_0E_\mu\over \pi L^2 m_\mu^2}
\int_{E_j}^{E_{j+1}} dE_\nu
\left({E_\nu \over E_\mu}\right)^2
\left(3-2{E_\nu \over E_\mu}\right)
\sigma_{\nu N}(E_\nu)P(\nu_\mu\rightarrow\nu_\mu
;\theta_{k\ell},\Delta m_{k\ell}^2,\delta,C)\nonumber\\
&{\ }&N_j^{\mbox{\rm right}}(\mu^+;\theta_{k\ell},\Delta m_{k\ell}^2,\delta,C)
\nonumber\\&=&
{2N_0E_\mu\over \pi L^2 m_\mu^2}
\int_{E_j}^{E_{j+1}} dE_{\bar\nu}
\left({E_{\bar\nu} \over E_\mu}\right)^2
\left(3-2{E_{\bar\nu} \over E_\mu}\right)
\sigma_{{\bar\nu} N}(E_{\bar\nu})P({\bar\nu}_\mu\rightarrow{\bar\nu}_\mu
;\theta_{k\ell},\Delta m_{k\ell}^2,\delta,C),
\nonumber
\end{eqnarray}
where $E_\mu$ is the muon energy, $L$ is the length of the neutrino path,
$N_0$ is the number of the target nucleons times the number of
useful decays of muons,
$\sigma_{\nu N}(E_\nu)$ and
$\sigma_{{\bar\nu} N}(E_{\bar\nu})$ are
the (anti-)neutrino nucleon cross sections.
We adopt
the cross section which is the sum of
those \cite{nakahata} of the quasi elastic scattering,
one pion production, and inelastic scattering,
where double counting of the latter two is suitably
subtracted \cite{lls}.
Throughout this paper the threshold energy is assumed to be
0.1GeV which is close to what has been assumed for
liquid argon detectors \cite{BCR} and which
may be realized in possible mega ton water Cherenkov
detectors \cite{nakamura,casper}.

The number of the free parameters in the present case is six ($\delta$,
$\theta_{12}$, $\theta_{13}$, $\theta_{23}$, $\Delta m^2_{21}$,
$\Delta m^2_{21}$), but the density $N_e(x)=Y_e(x) \rho(x)$ of electrons is
not known exactly ($Y_e(x)$ is the ratio of the number of electrons to
that of protons and neutrons, and $\rho(x)$ is the density of the
Earth at a distance $x$ from the beam production point), so we have to
vary $N_e(x)$ also.  Here for simplicity we assume the PREM
(Preliminary Reference Earth Model) \cite{stacey} and vary the overall
normalization of the PREM:
\begin{eqnarray}
A(x)=CA_0(x)
=\sqrt{2} C G_F Y_e(x) N_e(x),\nonumber
\end{eqnarray}
where $C=1$ corresponds to the PREM.
We have to consider correlations of errors
of the CP phase and six other quantities and taking into account all
these errors we obtain the probability of rejecting a hypothesis
$\bar{\delta}=0$.  To do that we look for the minimum value of
$\Delta\chi^2(\theta_{k\ell},\Delta m_{k\ell}^2,
\delta,C;\bar{\theta}_{k\ell},\overline{\Delta m_{k\ell}^2},
\bar{\delta},\bar{C})$ by varying the six parameters
($\bar{\theta}_{12}$, $\bar{\theta}_{13}$,
$\bar{\theta}_{23}$,
$\overline{\Delta m^2_{21}}$,
$\overline{\Delta m^2_{32}}$, $\bar{C}$):

\begin{eqnarray}
\Delta\chi^2_{\mbox{\rm min}}\equiv
\min_{{\ }_{\bar{\theta}_{k\ell},
\overline{\Delta m^2_{k\ell}},\bar{C}}}\Delta\chi^2
(\theta_{k\ell},\Delta m_{k\ell}^2,\delta,C;
\bar{\theta}_{k\ell},\overline{\Delta m_{k\ell}^2},\bar{\delta}=0,\bar{C}),
\nonumber
\end{eqnarray}
where $C$ stands for the overall normalization
of the electron density.

\subsection{Correlations of errors of $\delta$ and other parameters}

Let us first discuss correlations of two variables ($\bar{\delta}$,
$\bar{X}$)
where a parameter $X$ stands for $C$,
$\theta_{13}$, $\theta_{12}$, $\theta_{23}$, 
$\Delta m^2_{21}$ and $\Delta m^2_{32}$.

We have studied numerically
correlations of errors between $\delta$ and the other
oscillation parameters ($\theta_{k\ell}$, $\Delta m^2_{k\ell}$)
as well as the normalization $C$ of the matter effect for the case
where the central values for these parameters are those of
the best fit point, i.e., $\sin^22\theta_{12}=0.75$, $\Delta
m^2_{21}=3.2\times10^{-5}{\mbox{\rm eV}}^2$; $\sin^22\theta_{23}=1.0$,
$\Delta m^2_{32}=3.2\times10^{-3}{\mbox{\rm eV}}^2$, C=1.0
and we have used a reference value $8^\circ$.
The values of
$\Delta\chi^2(\theta_{12},\theta_{13},\theta_{23},
\Delta m_{21}^2,\Delta m_{32}^2,\delta,C;
\theta_{12},\theta_{13},\theta_{23},
\Delta m_{21}^2,\Delta m_{32}^2,\bar{\delta},\bar{C})$,
$\Delta\chi^2(\theta_{12},\theta_{13},\theta_{23},
\Delta m_{21}^2,\Delta m_{32}^2,\delta,C;
\theta_{12},\bar{\theta}_{13},\theta_{23},
\Delta m_{21}^2,\Delta m_{32}^2,\bar{\delta},C)$,
$\Delta\chi^2(\theta_{12},\theta_{13},\theta_{23},
\Delta m_{21}^2,\Delta m_{32}^2,\delta,C;
\bar{\theta}_{12},\theta_{13},\theta_{23},
\Delta m_{21}^2,\Delta m_{32}^2,\bar{\delta},C)$,
$\Delta\chi^2(\theta_{12},\theta_{13},\theta_{23},
\Delta m_{21}^2,\Delta m_{32}^2,\delta,C;
\theta_{12},\theta_{13},\bar{\theta}_{23},
\Delta m_{21}^2,\Delta m_{32}^2,\bar{\delta},C)$,
$\Delta\chi^2(\theta_{12},\theta_{13},\theta_{23},
\Delta m_{21}^2,\Delta m_{32}^2,\delta,C;
\theta_{12},\theta_{13},\theta_{23},
\overline{\Delta m_{21}^2},\Delta m_{32}^2,\bar{\delta},C)$,
$\Delta\chi^2(\theta_{12},\theta_{13},\theta_{23},
\Delta m_{21}^2,\Delta m_{32}^2,\delta,C;
\theta_{12},\theta_{13},\theta_{23},
\Delta m_{21}^2,\overline{\Delta m_{32}^2},\bar{\delta},C)$
are plotted in Fig. 1 -- 6 in the case of $\delta=\pi/2$
for $E_\mu$=3, 20, 50 GeV,
$L$=100km, 1000km, 2500km, 6300km,
where the data size 10$^{21}\mu\cdot$10kt is used as a reference
value and no backgrounds are assumed.  Since the number of
degrees of freedom is 2,
$\Delta \chi^2$=0.18, 0.34, 0.73
correspond to 1$\sigma$, 90\%, 99\% confidence level
to reject a hypothesis with $\bar{\delta}=0$.

As can be seen in Fig. 1,
the correlation ($\bar{\delta}$, $\bar{C}$) for $L\sim$3000km
is strong for $\theta_{13}= 8^\circ$.
The correlation ($\bar{\delta}$, $\bar{C}$)
turns out to be small for larger values of $\Delta m^2_{21}$
or for smaller value of $\theta_{13}$ (i.e.,
$\theta_{13}\lesssim 3^\circ$; See Fig. 1b), as the gradient of the ellipse
in the ($\bar{\delta}$, $\bar{C}$) plane becomes smaller for larger values of
$\Delta m^2_{21}$.
This is why strong correlations were not found in \cite{BCR}
where the set of parameters ($\sin^22\theta_{12}=1.0$, $\Delta
m^2_{21}=1.0\times10^{-4}{\mbox{\rm eV}}^2$;
$\sin^22\theta_{23}=1.0$,
$\Delta m^2_{32}=3.5 (5, 7) \times10^{-3}{\mbox{\rm eV}}^2$) and
$\sin^22\theta_{13}=0.05$, $E_\mu=30$GeV were used.
If we assume that
the uncertainty in the overall normalization $C$ is at most 5\%,
then the correlation ($\bar{\delta}$, $\bar{C}$) is not so serious, but
if we assume that the uncertainty is as large as 20 \% then
the set of the parameters ($E_\mu\sim$ 50GeV, $L\sim$ 3000km)
is not a good option.  We will discuss this issue later.

From Figs. 2 -- 6,
we see that the correlations of ($\bar{\delta}$, $\bar{\theta}_{k\ell}$)
and ($\bar{\delta}$, $\overline{\Delta m^2_{k\ell}}$) are not large for
$L\gtrsim$1000km, $E_\mu\gtrsim$20GeV.
As we will show analytically later, the value of
$\Delta\chi^2(\theta_{k\ell},\Delta m_{k\ell}^2,\delta,C;
\bar{\theta}_{k\ell},\overline{\Delta m_{k\ell}^2},\bar{\delta}=0,\bar{C})$
increases for $E_\mu\gg$ 50GeV and $L\ll$ 1000km
unless we minimize it with respect to $\theta_{k\ell}$
and $\Delta m^2_{k\ell}$, but because of strong correlations in
($\bar{\delta}$, $\bar{\theta}_{13}$), ($\bar{\delta}$, $\bar{\theta}_{23}$)
and ($\bar{\delta}$, $\overline{\Delta m^2_{32}}$), the value of
$\Delta \chi^2_{\mbox{\rm  min}}$,
which is minimized with respect to $\theta_{k\ell}$
and $\Delta m^2_{k\ell}$,
decreases for $E_\mu\gg$ 50GeV.

\subsection{Data size to reject a hypothesis with $\bar{\delta}=0$}

The quantity $\Delta\chi^2_{\mbox{\rm min}}$ can be regarded
as the deviation of $\chi^2$ from the best fit point (the best fit point in
eq. (\ref{eqn:delchi2}) is of course $\bar{\theta}_{k\ell}=\theta_{k\ell}$,
$\overline{\Delta m^2_{k\ell}}=\Delta m_{k\ell}^2$,
$\bar{\delta}=\delta$ and $\bar{C}=C$
for which we have $\Delta\chi^2_{\mbox{\rm min}}=0$) and for six
degrees of freedom the value of $\Delta\chi^2_{\mbox{\rm
min}}$ which corresponds to $3\sigma$ ($4\sigma$) is 20.1
(28.9).  From this we can estimate the necessary data size $D$ to
reject a hypothesis $\bar{\delta}=0$ at $3\sigma$ by dividing 20.1 by
$\Delta\chi^2_{\mbox{\rm min}}$ for each value of $\delta$.
On the other hand, it is important to include the effect of
the backgrounds in the analysis \cite{Golden,BCR,superbeam}.
Here we assume that the fraction $f_B$ of backgrounds to right sign
muon events
is given by $f_B=10^{-3}$ or $10^{-5}$ and that
the systematic error of backgrounds is $\sigma_B=0.1$ as in \cite{superbeam}
for simplicity.
We also assume the number of muons $10^{21} \mu\cdot 10$kt
as a reference value.
Thus $\Delta\chi^2$ is modified as
\begin{eqnarray}
&{\ }&\left.\Delta\chi^2(\theta_{k\ell},\Delta m_{k\ell}^2,
\delta,C;\bar{\theta}_{k\ell},\overline{\Delta m_{k\ell}^2},
\bar{\delta},\bar{C})\right|_{f_B}\nonumber\\
&\equiv&
\sum_j
{\left[N_j^{\mbox{\rm wrong}}(\mu^-)
-\bar{N}_j^{\mbox{\rm wrong}}(\mu^-)\right]^2
\over \left[\sqrt{N_j^{\mbox{\rm wrong}}(\mu^-)
+f_B N_j^{\mbox{\rm right}}(\mu^+)+1}+{11 \over 9}\right]^2
+\left[\sigma_Bf_B N_j^{\mbox{\rm right}}(\mu^+)\right]^2}
\nonumber\\
&+&
%+
\sum_j
{\left[N_j^{\mbox{\rm wrong}}(\mu^+)
-\bar{N}_j^{\mbox{\rm wrong}}(\mu^+)\right]^2
\over \left[\sqrt{N_j^{\mbox{\rm wrong}}(\mu^+)
+f_B N_j^{\mbox{\rm right}}(\mu^-)+1}+{11 \over 9}\right]^2
+\left[\sigma_Bf_B N_j^{\mbox{\rm right}}(\mu^-)\right]^2}
\nonumber\\
&+&\sum_j
{\left[N_j^{\mbox{\rm right}}(\mu^-)
-\bar{N}_j^{\mbox{\rm right}}(\mu^-)\right]^2
\over N_j^{\mbox{\rm right}}(\mu^-)}
%\nonumber\\
%&+&
+
\sum_j
{\left[N_j^{\mbox{\rm right}}(\mu^+)
-\bar{N}_j^{\mbox{\rm right}}(\mu^+)\right]^2
\over N_j^{\mbox{\rm right}}(\mu^+)},
\label{eqn:delchi2m}
\end{eqnarray}
where $\bar{N}_j^{\mbox{\rm wrong}}(\mu^\pm)$,
$\bar{N}_j^{\mbox{\rm right}}(\mu^\pm)$ stand for
$N_j^{\mbox{\rm wrong}}(\mu^\pm)$,
$N_j^{\mbox{\rm right}}(\mu^\pm)$ with arguments
$\bar{\theta}_{k\ell},\overline{\Delta m_{k\ell}^2},
\bar{\delta},\bar{C}$, respectively,
and the corrections in the statistical errors
are due to the Poisson statistical \cite{superbeam}.
Then we minimize $\Delta\chi^2$ with respect
$\bar{\theta}_{k\ell}$, $\overline{\Delta m^2_{k\ell}}$
and $\bar{C}$:
\begin{eqnarray}
\left.\Delta\chi^2_{\mbox{\rm min}}\right|_{f_B}\equiv
\left.\min_{{\ }_{\bar{\theta}_{k\ell},
\overline{\Delta m^2_{k\ell}},\bar{C}}}\Delta\chi^2
(\theta_{k\ell},\Delta m_{k\ell}^2,\delta,C;
\bar{\theta}_{k\ell},\overline{\Delta m_{k\ell}^2},\bar{\delta}=0,
\bar{C})\right|_{f_B},
\label{eqn:chicpv}
\end{eqnarray}
where the values of oscillation parameters we use
in (\ref{eqn:chicpv}) are the best fit values in the analyses of
the solar and atmospheric neutrinos \cite{ggpg}
as in Figs. 1 -- 6,
and we take $\theta_{13}$= $1^\circ$, $5^\circ$, $8^\circ$
and $\delta=\pi/2$ as a reference value.
In varying the overall normalization $C$ we assume
$0.95\le C \le 1.05$.  We will mention the results
for $|\Delta C|\le0.1$ and for $|\Delta C|\le0.2$ later.
For other oscillation parameters, we vary
($\theta_{12}$, $\Delta m^2_{21}$) and
($\theta_{23}$, $\Delta m^2_{32}$) within
the allowed region at 90\%CL of the solar and the atmospheric
neutrino data, i.e., $25^\circ\le\theta_{12}\le 41^\circ$,
$35^\circ\le\theta_{23}\le 55^\circ$,
$1.5\times10^{-5}{\mbox{\rm eV}}^2\le\Delta m^2_{21}
\le2.2\times10^{-4}{\mbox{\rm eV}}^2$,
$1.6\times10^{-3}{\mbox{\rm eV}}^2\le\Delta m^2_{32}
\le4\times10^{-3}{\mbox{\rm eV}}^2$.
It should be emphasized that
in minimizing $\Delta\chi^2$ in (\ref{eqn:chicpv})
all the six parameters are varied at the same time, unlike
in Figs. 1 -- 6
which are obtained by varying only one of $\bar{\theta}_{k\ell}$,
$\overline{\Delta m_{k\ell}^2}$, $\bar{C}$.

The result is given in Fig. 7
for a neutrino factory
with 0.5GeV $\le E_\mu\le$ 50GeV, 10km $\le L\le$ 10000km and for
three values of $\theta_{13}$= $1^\circ$, $5^\circ$, $8^\circ$
and two different values of
the background fraction $f_B=10^{-5}$, $10^{-3}$.
The behavior of
the figures change a little depending on the value of $\theta_{13}$.  For
$f_B=10^{-3}$,
the sensitivity to CP violation, i.e., the ability to reject
a hypothesis with $\bar{\delta}=0$ is {\it not}
optimized by the set of parameters $(E_\mu, L)\simeq$ (50GeV, 3000km),
which has been advocated as the best choice,
but rather by $(E_\mu, L)\simeq$ (20GeV, 2000km).
This is because with a nonnegligible fraction $f_B$
the contribution of the systematic uncertainty
$\sigma_Bf_B N_j^{\mbox{\rm right}}$ to the total error
becomes so large for high energy such as $E_\mu\sim$ 50GeV
and sensitivity to CP violation is lost.
For $f_B=10^{-5}$ and $\theta_{13}$= $1^\circ$, on the other hand,
the contribution of $\sigma_Bf_B N_j^{\mbox{\rm right}}$
is not so large and the sensitivity is optimized by
$(E_\mu, L)\simeq$ (50GeV, 3000km).
We note in passing that we have also optimized the sensitivity
with respect to the number of energy bins but the conclusion
does not depend very much on the number of energy bins.
This result disagrees with the claim in \cite{sato}.

We have also evaluated the data size assuming a larger
uncertainty of the matter effect, i.e., $|\Delta C|\le0.1$ and $|\Delta
C|\le0.2$.  The results for $\theta_{13}=8^\circ$ are shown in
Fig. 8.  If we have to assume an uncertainty of the
matter effect which is as large as 20\%, then the optimum baseline and
muon energy become even smaller than the results with $|\Delta
C|\le0.05$.  The situation is less serious for smaller value of
$\theta_{13}$, i.e., $\theta_{13}\lesssim 3^\circ$, for which the
correlation ($\bar{\delta}$, $\bar{C}$) is not so strong.  It should be noted that
we have assumed in our analysis that the detection efficiency does not
decrease down to the neutrino energy $E_\nu\sim$ a few GeV, so if this
assumption is not satisfied then the optimum muon energy may not be as
low as Fig. 8 indicates.

In Figs. 7 and 8 we have taken
$\delta=\pi/2$ as a reference value.  It is possible to do the same
analysis for a value of $\delta$ other than $\pi/2$.  The results for
$\theta_{13}$= $8^\circ$, $5^\circ$ and $1^\circ$ are given in
Fig. 9 for three sets of the parameters
($E_\mu$=50GeV, $L$=3000km), ($E_\mu$=20GeV, $L$=1000km)
and ($E_\mu$=20GeV, $L$=2000km).  We
observe that ($E_\mu$=50GeV, $L$=3000km) is better than
($E_\mu$=20GeV, $L$=1000km) for smaller values of $\theta_{13}$, but
for larger values of $\theta_{13}$ ($E_\mu$=20GeV, $L$=1000km) can be
more advantageous than the other.  It should be emphasized that in all
cases in Fig. 9 we can distinguish the case of
$\delta=\pi$ from that of $\delta=0$, since the necessary data size to
reject $\delta=0$ is finite even for $\delta=\pi$.  This is because
there are both contributions from $\sin\delta$ and $\cos\delta$
for the muon energy $E_\mu\lesssim$50 GeV.  As we will see in
section 4, for extremely high energy $E_\mu\gg$50 GeV we can show
analytically that our $\Delta\chi^2$(CPV) becomes proportional to
$\sin^2\delta$ and distinction between $\delta=\pi$ and $\delta=0$ is
no longer possible.

Sato et al. \cite{sato} have been emphasizing the importance
of the low energy option ($E_\mu\lesssim$ 1GeV),
since the shape of sine curves
in the CP violating probabilities becomes more conspicuous
for low energies.  As can be seen from Fig. 7,
the data set which is require to reject a hypothesis
$\bar{\delta}=0$ is much larger
than the value for the high energy case.
This is because the small boost factor $(E_\mu/m_\mu)^2$ means
that the number of events is statistically insufficient
even though the CP conserving probabilities become relatively
large in this case.  Furthermore,
in obtaining Fig. 7, we have assumed that
the detection efficiency does not change for low energy.
In reality, however, it becomes more and more difficult to
distinguish wrong sign muons from right sign ones at lower energy
so that the background fraction is expected to increase as
$E_\mu\rightarrow$ small.
Therefore, we conclude that the low energy
option offers no advantage over the high energy
region as long as we have the same detector size.

Sato et al. \cite{sato} gave another criticism that
neutrino factories at high energy are looking at
only terms which are proportional to $\cos\delta$.
However, as we will see in section 4, we can show
analytically in the high energy limit
that our $\Delta\chi^2$(CPV), when minimized with respect
to one of the oscillation parameters, is proportional
to the square $J^2$ of the Jarlskog factor and has the same behavior
$\propto 1/E_\mu$ as $\Delta\chi^2$ in the case of T violation.
Therefore a naive argument that the square of difference of
the quantity at $\delta$ and the one at $\delta=0$
must depend only on $\cos\delta$ does not necessarily apply.

\section{Analysis of T violation}

It is known that measurements of T violation at neutrino
factories are very difficult since electrons and positrons are
difficult to distinguish at high energy.  Here we discuss the
possibility of measurements of T violation assuming perfect polarized
muon beams for simplicity, i.e., $P_\mu\cos\theta$ in eqs. (3) and (4)
in \cite{Geer} is assumed to be $\pm 1$ so that either the flux of
$\nu_e$ or $\bar{\nu}_e$ vanishes.  If we assume $P_\mu\cos\theta=-1$,
then we have
\begin{eqnarray}
\left.{d^2N_{\bar{\nu}_e} \over dxd\Omega}\right\vert_{P_\mu\cos\theta=-1}
&=&0\nonumber\\
\left.{d^2N_{\nu_\mu} \over dxd\Omega}
\right\vert_{P_\mu\cos\theta=-1}&=&
{2x^2 \over 4\pi}\left[(3-2x)+(1-2x)\right]
%\nonumber\\
%&=&
={2 \over 3}
\left.{d^2N_{\bar{\nu}_e} \over dxd\Omega}
\right\vert_{P_\mu\cos\theta=0},
\end{eqnarray}
where $x\equiv E_\nu/E_\mu$,
so that we can compare the numbers of events
$(3/2)N_j(\nu_\mu\rightarrow\nu_e;P_\mu\cos\theta=-1)$ from $\nu_\mu$
in decays of polarized $\mu^-$ and
$N_j(\nu_e\rightarrow\nu_\mu;P_\mu\cos\theta=0)$
from $\nu_e$ in decays of unpolarized $\mu^+$.
$\Delta\chi^2$ in this case is thus given by
\begin{eqnarray}
%&{\ }&
\left.\Delta\chi^2({\mbox{\rm TV}})\right|_{f_B}
%\nonumber\\
%&\equiv&
\equiv\sum_j
{\left[N_j(\nu_e\rightarrow\nu_\mu;P_\mu\cos\theta=0)
-{3 \over 2}N_j(\nu_\mu\rightarrow\nu_e;P_\mu\cos\theta=-1)
\right]^2
\over \Delta N^2},
\label{eqn:chi2tv}
\end{eqnarray}
where
\begin{eqnarray}
&{\ }&N_j(\nu_\mu\rightarrow\nu_e;P_\mu\cos\theta=-1)
%&\equiv&
\equiv
{8N_0E_\mu\over \pi L^2 m_\mu^2}
\int_{E_j}^{E_{j+1}} dE_\nu
\left({E_\nu \over E_\mu}\right)^2
\left(1-{E_\nu \over E_\mu}\right)
\sigma_{\nu N}(E_\nu)P(\nu_\mu\rightarrow\nu_e)\nonumber\\
&{\ }&N_j(\nu_e\rightarrow\nu_\mu;P_\mu\cos\theta=0)
%&\equiv&
\equiv
{12N_0E_\mu\over \pi L^2 m_\mu^2}
\int_{E_j}^{E_{j+1}} dE_\nu
\left({E_\nu \over E_\mu}\right)^2
\left(1-{E_\nu \over E_\mu}\right)
\sigma_{\nu N}(E_\nu)P(\nu_e\rightarrow\nu_\mu)\nonumber\\
&{\ }&\Delta N^2\equiv\left[\sqrt{N_j
(\nu_e\rightarrow\nu_\mu;P_\mu\cos\theta=0)
+f_B N_j
(\bar{\nu}_\mu\rightarrow\bar{\nu}_\mu;P_\mu\cos\theta=0)+1}
+{11 \over 9}\right]^2
\nonumber\\
&{\ }&
%&+&
\qquad~+\left({3 \over 2}\right)^2
N_j(\nu_\mu\rightarrow\nu_e;P_\mu\cos\theta=-1)
+\left[\sigma_Bf_B N_j
(\bar{\nu}_\mu\rightarrow\bar{\nu}_\mu;P_\mu\cos\theta=0)\right]^2
\end{eqnarray}
and the systematic errors are taken into account only for the
$\mu^+$ beam.  In the definition (\ref{eqn:chi2tv})
we do not optimize $\Delta\chi^2({\mbox{\rm TV}})$
with respect to $\theta_{k\ell}$,
$\Delta m_{k\ell}^2$, $C$, since
the numerator is defined only by quantities
which are measured by experiments and a non zero value
of $\Delta\chi^2({\mbox{\rm TV}})$ immediately indicates
$\delta\ne0$.
The date size to reject a hypothesis
with $\bar{\delta}=0$ at 3$\sigma$ using $\Delta\chi^2({\mbox{\rm TV}})$
can be calculated as in the case of $\Delta\chi^2({\mbox{\rm CPV}})$
and is shown if Fig. 
10 for $\delta=\pi/2$ and
$\theta_{13}=8^\circ$ assuming the best fit values
for ($\theta_{12}$, $\Delta m^2_{21}$) and
($\theta_{23}$, $\Delta m^2_{32}$).
The number of degrees of freedom in this case
is one, i.e., the magnitude of the Jarlskog factor
$J\equiv{\tilde J}\sin\delta$, so that the data size
is in general smaller than the case of CP violation
for which we have six degrees of freedom.
In practice, however, obtaining $P_\mu\cos\theta\simeq\pm1$
is technically difficult and the possibility to use T violation
seems to be challenging.

\section{Low and high energy behaviors of $\Delta\chi^2$}
In this subsection we will show analytically that the sensitivity to
CP and T violation decreases as $E_\mu\rightarrow$ small ($E_\mu\ll$ 10GeV)
or $E_\mu\rightarrow$ large ($E_\mu\gg$ 50GeV).
Throughout this paper we assume
$\sin^22\theta_{13}\gtrsim 10^{-3}$ ($\theta_{13}\gtrsim 1^\circ$) so
that we are always in the atmospheric regime in the language of
\cite{bgghm}, i.e., $\sin^22\theta_{13}/\sin^22\theta_{12}\gg (\Delta
m^2_{21}/\Delta m^2_{31})^2$.  In this subsection we will ignore the
effects of backgrounds and systematic errors for simplicity.

To examine significance of CP/T violation analytically, we introduce
the following simplified quantities:

\begin{eqnarray}
\Delta\chi^2({\mbox{\rm CPV}})&\equiv&
\min_{{\ }_{\bar{\theta}_{k\ell},
\overline{\Delta m^2_{k\ell}},\bar{C}}}\Delta\chi^2
(\theta_{k\ell},\Delta m_{k\ell}^2,\delta,C;
\bar{\theta}_{k\ell},\overline{\Delta m_{k\ell}^2},\bar{\delta}=0,
\bar{C}),\nonumber\\
\Delta\chi^2({\mbox{\rm TV}})&\equiv&
{\left[\langle P(\nu_e\rightarrow\nu_\mu;\delta)\rangle -\langle
P(\nu_\mu\rightarrow\nu_e;\delta)\rangle\right]^2 \over \left\langle
P(\nu_e\rightarrow\nu_\mu;\delta)\right\rangle}, 
\end{eqnarray}
where
\begin{eqnarray}
&{\ }&\Delta\chi^2
(\theta_{k\ell},\Delta m_{k\ell}^2,\delta,C;
\bar{\theta}_{k\ell},\overline{\Delta m_{k\ell}^2},\bar{\delta},
\bar{C})\nonumber\\
&=& {\left[\langle
P(\nu_e\rightarrow\nu_\mu;\theta_{k\ell},\Delta m_{k\ell}^2,\delta,C)
\rangle -\langle
P(\nu_e\rightarrow\nu_\mu;
\bar{\theta}_{k\ell},\overline{\Delta m_{k\ell}^2},\bar{\delta},\bar{C})
\rangle\right]^2 \over \left\langle
P(\nu_e\rightarrow\nu_\mu;\delta)\right\rangle}
\end{eqnarray}
is defined as in (\ref{eqn:delchi2}),
\begin{eqnarray}
&{\ }&\langle P(\nu_\alpha\rightarrow\nu_\beta;\delta)\rangle\equiv
{12N_0E^2_\mu\over \pi L^2 m_\mu^2}
\int d\left({E_\nu \over E_\mu}\right)
\left({E_\nu \over E_\mu}\right)^2
\left(1-{E_\nu \over E_\mu}\right)
\sigma_{\nu N}(E_\nu)P(\nu_\alpha\rightarrow\nu_\beta;\delta)
\label{eqn:noevents}
\end{eqnarray}
are the number of events (($\alpha$, $\beta$) = ($e$, $\mu$) or
($\mu$, $e$); in the case of ($\alpha$, $\beta$) = ($\mu$, $e$)
we assume perfect polarization as in the previous section so that
the number of events is given by the same definition (\ref{eqn:noevents}))
and we have ignored effects of the backgrounds and
systematic errors
and correlations of errors for simplicity in this section.  Also
we will assume that the cross section is proportional to
the neutrino energy $E_\nu$ for any $E_\nu$, i.e.,
$\sigma=\sigma_0 E_\nu$.  Strictly speaking this assumption
is not accurate, but it is known \cite{lls} that
$0<\sigma<\sigma_0 E_\nu$ is satisfied
for low energy $E_\nu\ll$ 1GeV, so our approximation is sufficient to
give an upper bound on the value of $\Delta\chi^2$ for low energy.

Let us first look at the low energy limit ($E_\nu \ll$ 10GeV).  In
this case matter effects are negligible and the probability can be
replaced by that in vacuum.  Thus we have
\begin{eqnarray}
P(\nu_e\rightarrow\nu_\mu;\delta)&\simeq&
s_{23}^2 \sin^22\theta_{13}\sin\left({\Delta E_{31}L \over 2}\right)
+c_{23}^2 \sin^22\theta_{12}\sin\left({\Delta E_{21}L \over 2}\right)
\nonumber\\
&+&8{\tilde J}\sin\left({\Delta E_{21}L \over 2}\right)
\sin\left({\Delta E_{31}L \over 2}\right)
\cos\left(\delta+{\Delta E_{31}L \over 2}\right)
\end{eqnarray}
to the second order in ${\cal O}(\theta_{13})$ and
${\cal O}(\Delta E_{21}/\Delta E_{31})$,
where
\begin{eqnarray}
{\tilde J}\equiv {c_{13} \over 8}\sin2\theta_{12}\sin2\theta_{13}
\sin2\theta_{23},
\end{eqnarray}
and $\Delta E_{jk}\equiv\Delta m_{jk}^2/2E\equiv(m_j^2-m_k^2)/2E$.
The number of events are given by
\begin{eqnarray}
&{\ }&\langle P(\nu_e\rightarrow\nu_\mu;\delta)\rangle
-\langle P(\nu_e\rightarrow\nu_\mu;\delta=0)\rangle
\nonumber\\
&=&{96 N_0 E^3_\mu\sigma_0{\tilde J} \over \pi L^2m^2_\mu}
\int dx~x^3(1-x)\sin\left({\Delta m^2_{21}L \over 4xE_\mu}\right)
\nonumber\\
&\times&
\sin\left({\Delta m^2_{31}L \over 4xE_\mu}\right)\left[
\cos\left(\delta+{\Delta m^2_{31}L \over 4xE_\mu}\right)
-\cos\left({\Delta m^2_{31}L \over 4xE_\mu}\right)\right]
\label{eqn:cpvnum}\\
&{\ }&\langle P(\nu_e\rightarrow\nu_\mu;\delta)\rangle
-\langle P(\nu_\mu\rightarrow\nu_e;\delta)\rangle
\nonumber\\
&=&{192 N_0 E^3_\mu\sigma_0{\tilde J}\sin\delta \over \pi L^2m^2_\mu}
\int dx~x^3(1-x)\sin\left({\Delta m^2_{21}L \over 4xE_\mu}\right)
\sin\left({\Delta m^2_{31}L \over 4xE_\mu}\right)
\sin\left({\Delta m^2_{32}L \over 4xE_\mu}\right)
\label{eqn:tvnum}\\
&{\ }&\langle P(\nu_e\rightarrow\nu_\mu;\delta)\rangle
\nonumber\\
&\simeq&{12 s^2_{23}\sin^22\theta_{13}E^3_\mu\sigma_0
\over \pi L^2m^2_\mu}
\int dx~x^3(1-x)\sin^2\left({\Delta m^2_{31}L \over 4xE_\mu}\right),
\label{eqn:nemu}
\end{eqnarray}
where $x\equiv E_\nu/E_\mu$,
we have assumed conditions for the atmospheric regime
$\sin^22\theta_{13}/\sin^22\theta_{12}\gg
(\Delta m^2_{21}/\Delta m^2_{21})^2$, and
we have put $\bar{\theta}_{k\ell}=\theta_{k\ell}$,
$\overline{\Delta m_{k\ell}^2}=\Delta m_{k\ell}^2$,
$\bar{C}=C$ in (\ref{eqn:cpvnum}) instead of optimizing
$\langle P(\nu_e\rightarrow\nu_\mu;\delta)\rangle
-\langle P(\nu_e\rightarrow\nu_\mu;\delta=0)\rangle$
with respect to these variables, as that is sufficient
to demonstrate that $\Delta\chi^2({\mbox{\rm CPV}})$
decreases as $E_\mu\rightarrow0$.
If we keep $L/E_\mu$ fixed while $L, E_\mu\rightarrow$ small,
then all the quantities (\ref{eqn:cpvnum}), (\ref{eqn:tvnum})
and (\ref{eqn:nemu}) behave as ${\cal O}(E_\mu)$,
so $\Delta\chi^2({\mbox{\rm CPV}})\propto E_\mu$ and
$\Delta\chi^2({\mbox{\rm TV}})\propto E_\mu$
as $E_\mu\rightarrow0$ with $L/E_\mu$ fixed.  Thus
sensitivity to CP/T violation is asymptotically lost as
$E_\mu\rightarrow0$.  This is consistent with our
numerical results in previous sections.

Next let us discuss the behavior of $\Delta\chi^2$ in the high energy
limit ($E_\mu\gg$50GeV).  In this case we have to take into account
the matter effect and we use the probability which has been obtained
in \cite{Golden} to second order in ${\cal O}(\theta_{13})$,
${\cal O}(\Delta E_{21}/\Delta E_{31})$,
${\cal O}(\Delta E_{21}/A)$ and
${\cal O}(\Delta E_{21}L)$:
\begin{eqnarray}
P(\nu_e\rightarrow\nu_\mu;\delta)&\simeq&
s_{23}^2 \sin^22\theta_{13}\left({\Delta E_{31} \over B}\right)^2
\sin^2\left({BL \over 2}\right)
+c_{23}^2 \sin^22\theta_{12}\left({\Delta E_{21} \over A}\right)^2
\sin^2\left({AL \over 2}\right)
\nonumber\\
&+&8{\tilde J}{\Delta E_{21} \over A}{\Delta E_{31} \over B}
\sin\left({AL \over 2}\right)
\sin\left({BL \over 2}\right)
\cos\left(\delta+{\Delta E_{31}L \over 2}\right).
\label{eqn:pemu}
\end{eqnarray}
Since we assume $\sin^22\theta_{13}/\sin^22\theta_{12}\gg
(\Delta m^2_{21}/\Delta m^2_{21})^2=(\Delta E_{21}/\Delta E_{31})^2$
here, we can ignore the
second term in (\ref{eqn:pemu}).

It is straightforward to
get the following high energy limit of
$\Delta\chi^2({\mbox{\rm TV}})$.
Using (\ref{eqn:pemu}) we have
\begin{eqnarray}
P(\nu_e\rightarrow\nu_\mu;\delta)
-P(\nu_\mu\rightarrow\nu_e;\delta)
%\nonumber\\
&=&P(\nu_e\rightarrow\nu_\mu;\delta)
-P(\nu_e\rightarrow\nu_\mu;-\delta)
\nonumber\\
&\simeq&-2{\tilde J}
{\Delta m^2_{21}\left(\Delta m^2_{31}\right)^2 \over E_\nu^3}
{L \over A^2}\sin^2\left({AL \over 2}\right),
\end{eqnarray}
where we have expanded $\sin(\Delta E_{21}L/2)\simeq \Delta E_{21}L/2$
and have used the fact $B=[(\Delta E_{31}\cos2\theta_{13}
-A)^2+(\Delta E_{31}\sin2\theta_{13})^2]^{1/2}\simeq A$
as $E_\nu\rightarrow$ large.
Therefore the number of events is given by
\begin{eqnarray}
&{\ }&\langle P(\nu_e\rightarrow\nu_\mu;\delta)\rangle
-\langle P(\nu_\mu\rightarrow\nu_e;\delta)\rangle
\nonumber\\
&\simeq&{24 N_0 \sigma_0{\tilde J}\sin\delta
\Delta m^2_{21}\left(\Delta m^2_{31}\right)^2 \over \pi m^2_\mu}
{1 \over A^2L}\sin^2\left({AL \over 2}\right)
\int dx~(1-x)
\nonumber\\
&=&{12 N_0 \sigma_0{\tilde J}\sin\delta
\Delta m^2_{21}\left(\Delta m^2_{31}\right)^2 \over \pi m^2_\mu}
{1 \over A^2L}\sin^2\left({AL \over 2}\right),
\\
&{\ }&\langle P(\nu_e\rightarrow\nu_\mu;\delta)\rangle
\nonumber\\
&\simeq&{3 N_0\sigma_0s^2_{23}\sin^22\theta_{13}
\over \pi m^2_\mu}
{E_\mu \over A^2L^2}\sin^2\left({AL \over 2}\right)
\int dx~x(1-x)
\nonumber\\
&=&{N_0\sigma_0s^2_{23}\sin^22\theta_{13}
\over 2\pi m^2_\mu}
{E_\mu \over A^2L^2}\sin^2\left({AL \over 2}\right).
\end{eqnarray}
Hence we have the behaviors
\begin{eqnarray}
\Delta\chi^2({\mbox{\rm TV}})&\simeq&
{N_0\sigma_0 \over \pi m^2_\mu}
{288\sin^2\delta{\tilde J}^2
\left(\Delta m^2_{21}\right)^2\left(\Delta m^2_{31}\right)^4
 \over s^2_{23}\sin^22\theta_{13}}
{1 \over E_\mu A^2}\sin^2\left({AL \over 2}\right)
\label{eqn:tvh}\\
&{\ }&{\mbox{\rm as~}} E_\mu \rightarrow {\mbox{\rm large}}\nonumber.
\end{eqnarray}
(\ref{eqn:tvh}) indicates that the sensitivity to T violation
decreases as $E_\mu$ becomes very large.
Also for a fixed large $E_\mu$,
$\Delta\chi^2({\mbox{\rm TV}})$ is optimized for
$L\sim \pi/A\sim 3\times 2000$km/$(\rho/2.7$g$\cdot$cm$^{-3})\sim$ 5000km.
From numerical calculations we see that $\Delta\chi^2({\mbox{\rm TV}})$ is
optimized for $(L, E_\mu)\sim$ (3000km, 50GeV) (see Fig. 1), so
our analytic treatment is consistent with numerical calculations
qualitatively.

The behavior of $\Delta\chi^2({\mbox{\rm CPV}})$ is a little
more complicated, as we have to optimize $\Delta\chi^2$
with respect to $\bar{\theta}_{k\ell}$,
$\overline{\Delta m_{k\ell}^2}$, $\bar{C}$.
If we put $\bar{\theta}_{k\ell}=\theta_{k\ell}$,
$\overline{\Delta m_{k\ell}^2}=\Delta m_{k\ell}^2$,
$\bar{C}=C$ as we did in (\ref{eqn:cpvnum}),
we have
\begin{eqnarray}
%&{\ }&
P(\nu_e\rightarrow\nu_\mu;\delta)
- P(\nu_e\rightarrow\nu_\mu;\delta=0)
%\nonumber\\
&\simeq& 2{\tilde J}(\cos\delta-1)
{\Delta m^2_{21}\Delta m^2_{31} \over E_\nu^2}
{1 \over A^2}\sin^2\left({AL \over 2}\right),
\end{eqnarray}
\begin{eqnarray}
%&{\ }&
\langle P(\nu_e\rightarrow\nu_\mu;\delta)\rangle
-\langle P(\nu_e\rightarrow\nu_\mu;\delta=0)\rangle
%\nonumber\\
%&=&
={4 N_0 \sigma_0{\tilde J}(\cos\delta-1)
\Delta m^2_{21}\Delta m^2_{31}
\over \pi m^2_\mu}
{E_\mu \over A^2L^2}\sin^2\left({AL \over 2}\right),\nonumber\\
\end{eqnarray}
so that we naively have the following behavior
\begin{eqnarray}
\Delta\chi^2({\mbox{\rm naive CPV}})&\simeq&
{N_0\sigma_0 \over \pi m^2_\mu}
{32(\cos\delta-1)^2{\tilde J}^2
\left(\Delta m^2_{21}\right)^2\left(\Delta m^2_{31}\right)^2
 \over s^2_{23}\sin^22\theta_{13}}
{E_\mu \over A^2L^2}\sin^2\left({AL \over 2}\right).
\label{eqn:cpvh}
\end{eqnarray}

It turns out that it is sufficient to consider
the correlation of two variables ($\bar{\delta}$, $\bar{X}$),
where $X$ is $\theta_{k\ell}$, $\Delta m^2_{k\ell}$ or $C$,
to demonstrate $\Delta\chi^2({\mbox{\rm CPV}})\propto 1/E_\mu$.
Except for the correlations ($\bar{\delta}$, $\bar{\theta}_{12}$)
and ($\bar{\delta}$, $\overline{\Delta m^2_{21}}$), we can ignore
terms of order ${\cal O}((\Delta E_{21}/\Delta E_{31})^2)$.
From the assumption $\sin^22\theta_{13}/\sin^22\theta_{12}\gg
(\Delta m^2_{21}/\Delta m^2_{31})^2$, (\ref{eqn:pemu})
is approximately given by
\begin{eqnarray}
P(\nu_e\rightarrow\nu_\mu;\delta)&\simeq&
\left[
s_{23} \sin2\theta_{13}{\Delta E_{31} \over A}
\sin\left({AL \over 2}\right)\right.
\nonumber\\
&+&\left.{4{\tilde J} \over s_{23} \sin2\theta_{13}}{\Delta E_{21} \over A}
\sin\left({AL \over 2}\right)
\cos\left(\delta+{\Delta E_{31}L \over 2}\right)\right]^2,
\end{eqnarray}
where we have used $A-\Delta E_{31}\simeq A$ for $E_\nu\rightarrow$ large,
and we have ignored terms of order
${\cal O}((\Delta E_{21}/\Delta E_{31})^2)$.
In the case of the two variable correlation
($\bar{\delta}$, $\bar{\theta}_{13}$), to minimize the square of
\begin{eqnarray}
&{\ }&
P(\nu_e\rightarrow\nu_\mu;\theta_{13},\delta)-
P(\nu_e\rightarrow\nu_\mu;\bar{\theta}_{13},\bar{\delta})
\nonumber\\
&\simeq&\left[
s_{23} \sin2\theta_{13}{\Delta E_{31} \over A}
\sin\left({AL \over 2}\right)+
{4{\tilde J} \over s_{23} \sin2\theta_{13}}{\Delta E_{21} \over A}
\sin\left({AL \over 2}\right)
\cos\left(\delta+{\Delta E_{31}L \over 2}\right)\right]^2
\nonumber\\
&-&
\left[
s_{23} \sin2\bar{\theta}_{13}{\Delta E_{31} \over A}
\sin\left({AL \over 2}\right)+
{4{\tilde J} \over s_{23} \sin2\theta_{13}}{\Delta E_{21} \over A}
\sin\left({AL \over 2}\right)
\cos\left(\bar{\delta}+{\Delta E_{31}L \over 2}\right)\right]^2,
\label{eqn:pdiff13}
\end{eqnarray}
it is sufficient to take\footnote{
Here we do not discuss the other solution of
the quadratic equation which was discussed by \cite{bgghm}, since
we are mainly interested in rejecting
$\delta=0$ rather than determining the
precise value of $\delta$.}
\begin{eqnarray}
\sin2\bar{\theta}_{13}=\sin2\theta_{13}
-{4{\tilde J} \over s_{23}^2 \sin2\theta_{13}}
{\Delta m^2_{21} \over \Delta m^2_{31}}
\left(\cos\bar{\delta}-\cos\delta\right),
\label{eqn:diff13}
\end{eqnarray}
where we have used in (\ref{eqn:pdiff13}) and (\ref{eqn:diff13})
the fact
${\tilde J}/\sin2\theta_{13}=\cos\theta_{13}\times
(\mbox{\rm independent of~} \theta_{13})\simeq
\cos\bar{\theta}_{13}\times
(\mbox{\rm independent of~} \theta_{13})$ which
holds because $\sin^2\theta_{13}\ll1$.
Notice that the phase $\Delta E_{31}L / 2$ which appears together
with $\delta$ in cosine
in (\ref{eqn:pdiff13}) disappears as $E_\nu\rightarrow$ large.
Plugging (\ref{eqn:diff13}) in (\ref{eqn:pdiff13}), we find
\begin{eqnarray}
&{\ }&
P(\nu_e\rightarrow\nu_\mu;\theta_{13},\delta)-
P(\nu_e\rightarrow\nu_\mu;\bar{\theta}_{13},\bar{\delta})
\nonumber\\
\simeq&{\ }&
s_{23}^2 \left({\Delta E_{31} \over A}\right)^2
\sin^2\left({AL \over 2}\right)
\left(\sin^22\theta_{13}-\sin^22\bar{\theta}_{13}\right)
\nonumber\\
&+&8{\tilde J}
{\Delta E_{21}\Delta E_{31} \over A^2}
\sin^2\left({AL \over 2}\right)
%\nonumber\\
%&\times&
\left[
\cos\left(\delta+{\Delta E_{31}L \over 2}\right)-
\cos\left(\bar{\delta}+{\Delta E_{31}L \over 2}\right)\right],
\nonumber\\
\simeq&{\ }&
8{\tilde J}
{\Delta E_{21}\Delta E_{31} \over A^2}
\sin^2\left({AL \over 2}\right)\nonumber\\
&\times&\left[
\cos\left(\delta+{\Delta E_{31}L \over 2}\right)-
\cos\left(\bar{\delta}+{\Delta E_{31}L \over 2}\right)
-\cos\delta+\cos\bar{\delta}
\right],
\nonumber\\
\simeq&{\ }&
8{\tilde J}{\Delta E_{21}\Delta E_{31} \over A^2}
\sin^2\left({AL \over 2}\right)
\left(\sin\bar{\delta}-\sin\delta\right)
{\Delta E_{31}L \over 2},
\label{eqn:pdiff132}
\end{eqnarray}
where we have expanded $\sin(\Delta E_{31}L/2)\simeq\Delta E_{31}L/2$,
$\cos(\Delta E_{31}L/2)-1\simeq-(\Delta E_{31}L)^2/2\simeq0$
in the last step in (\ref{eqn:pdiff132}).
Hence we get
\begin{eqnarray}
&{\ }&\left\langle
P(\nu_e\rightarrow\nu_\mu;\theta_{13},\delta)-
P(\nu_e\rightarrow\nu_\mu;\bar{\theta}_{13},\bar{\delta}=0)
\right\rangle\nonumber\\
\simeq&{\ }&
-{6N_0\sigma_0 \over \pi m_\mu^2}
{\tilde J}\sin\delta{\Delta m^2_{21}\left(\Delta m^2_{31}\right)^2
\over A^2 L^2}
\sin^2\left({AL \over 2}\right)\int dx(1-x)\nonumber\\
=&{\ }&{3N_0\sigma_0 \over \pi m_\mu^2}
{\tilde J}\sin\delta\Delta m^2_{21}\left(\Delta m^2_{31}\right)^2
{\sin^2\left(AL/2\right) \over A^2 L^2}.
\label{eqn:pdiff133}
\end{eqnarray}
We see from (\ref{eqn:pdiff133}) that if we optimize
$\Delta\chi^2({\mbox{\rm CPV}})$ with respect only to
$\bar{\theta}_{13}$ then
$\Delta\chi^2({\mbox{\rm CPV}})$ behaves as
\begin{eqnarray}
\Delta\chi^2({\mbox{\rm CPV}};(\delta,\theta_{13}))
%=&{\ }&
\simeq{18N_0\sigma_0 \over \pi m_\mu^2}
{{\tilde J}^2\sin^2\delta\left(\Delta m^2_{21}\right)^2
\left(\Delta m^2_{31}\right)^2 \over
s_{23}^2\sin^22\theta_{13}}
{\sin^2\left(AL/2\right) \over E_\mu A^2}.
\label{eqn:chi2cpv13}
\end{eqnarray}
Note that the behavior of
$\Delta\chi^2({\mbox{\rm CPV}};(\delta,\theta_{13}))$
which is optimized with respect to $\bar{\theta}_{13}$
is quite different from that of $\Delta\chi^2({\mbox{\rm naive CPV}})$
in (\ref{eqn:cpvh}).  We observe that the dependence of
$\Delta\chi^2({\mbox{\rm CPV}};(\delta,\theta_{13}))$
on $E_\mu$ is the same as that of $\Delta\chi^2({\mbox{\rm TV}})$.
It should be also emphasized that
$\Delta\chi^2({\mbox{\rm CPV}};(\delta,\theta_{13}))$
is proportional to $\sin^2\delta$ and does not
depend on $\cos\delta$ unlike
$\Delta\chi^2({\mbox{\rm naive CPV}})$
in (\ref{eqn:cpvh}).

We can play the same game for $\theta_{23}$,
$\Delta m^2_{32}$ and $C$.
In the case of the two variable correlation
($\bar{\delta}$, $\bar{\theta}_{23}$),
\begin{eqnarray}
\sin\bar{\theta}_{23}=\sin\theta_{23}
-{4{\tilde J} \over s_{23} \sin^22\theta_{13}}
{\Delta m^2_{21} \over \Delta m^2_{31}}
\left(\cos\bar{\delta}-\cos\delta\right)
\label{eqn:diff23}
\end{eqnarray}
minimizes $\Delta\chi^2({\mbox{\rm CPV}};(\delta,\theta_{23}))$
and we have
\begin{eqnarray}
\Delta\chi^2({\mbox{\rm CPV}};(\delta,\theta_{23}))
%=&{\ }&
\simeq{18N_0\sigma_0 \over \pi m_\mu^2}
{{\tilde J}^2\sin^2\delta\left(\Delta m^2_{21}\right)^2
\left(\Delta m^2_{31}\right)^2 \over
s_{23}^2\sin^22\theta_{13}}
{\sin^2\left(AL/2\right) \over E_\mu A^2},
\end{eqnarray}
which is the same as
$\Delta\chi^2({\mbox{\rm CPV}};(\delta,\theta_{13}))$.
In the case of the two variable correlation
($\bar{\delta}$, $\overline{\Delta m^2_{32}}$), using
\begin{eqnarray}
P(\nu_e\rightarrow\nu_\mu;\delta)&\simeq&
\left[
s_{23} \sin2\theta_{13}{\Delta E_{31} \over A}
\sin\left({AL \over 2}\right)
%\right.
%\nonumber\\
%&+&
%+\left.
+{4{\tilde J} \over s_{23} \sin2\theta_{13}}{\Delta E_{21} \over A}
\sin\left({AL \over 2}\right)
\cos\delta\right]^2,
\end{eqnarray}
we find
\begin{eqnarray}
\overline{\Delta m^2_{31}}=\Delta m^2_{31}
-{4{\tilde J} \over s_{23}^2 \sin^22\theta_{13}}
\Delta m^2_{21}
\left(\cos\bar{\delta}-\cos\delta\right)
\label{eqn:diffm31}
\end{eqnarray}
minimizes $\Delta\chi^2({\mbox{\rm CPV}};(\delta,\Delta m^2_{31}))$.
We obtain
\begin{eqnarray}
\Delta\chi^2({\mbox{\rm CPV}};(\delta,\Delta m^2_{31}))
%=&{\ }&
\simeq{18N_0\sigma_0 \over \pi m_\mu^2}
{{\tilde J}^2\sin^2\delta\left(\Delta m^2_{21}\right)^2
\left(\Delta m^2_{31}\right)^2 \over
s_{23}^2\sin^22\theta_{13}}
{\sin^2\left(AL/2\right) \over E_\mu A^2},
\end{eqnarray}
which again is the same as
$\Delta\chi^2({\mbox{\rm CPV}};(\delta,\theta_{13}))$.
In the case of the two variable correlation
($\bar{\delta}$, $\bar{C}$),
\begin{eqnarray}
{\sin\left(\bar{A}L/2\right) \over \bar{A}}
={\sin\left(AL/2\right) \over A}
\left[1-{4{\tilde J} \over s_{23}^2 \sin^22\theta_{13}}
{\Delta m^2_{21} \over \Delta m^2_{31}}
\left(\cos\bar{\delta}-\cos\delta\right)\right]
\label{eqn:diffa}
\end{eqnarray}
minimizes $\Delta\chi^2({\mbox{\rm CPV}};(\delta,C))$
and we get
\begin{eqnarray}
\Delta\chi^2({\mbox{\rm CPV}};(\delta,C))
%=&{\ }&
\simeq{18N_0\sigma_0 \over \pi m_\mu^2}
{{\tilde J}^2\sin^2\delta\left(\Delta m^2_{21}\right)^2
\left(\Delta m^2_{31}\right)^2 \over
s_{23}^2\sin^22\theta_{13}}
{\sin^2\left(AL/2\right) \over E_\mu A^2},
\end{eqnarray}
which once again is the same as
$\Delta\chi^2({\mbox{\rm CPV}};(\delta,\theta_{13}))$.
The expressions (\ref{eqn:diff13}), (\ref{eqn:diff23}), (\ref{eqn:diffm31})
and (\ref{eqn:diffa}) for the optimal values for
$\bar{\theta}_{13}$, $\bar{\theta}_{23}$,
$\overline{\Delta m^2_{31}}$ and $\bar{C}$
explain why the correlation has a cosine curve
for large $E_\mu$ and small $L$ in Figs. 1,
2, 4 and 6.

In the case of the correlations ($\bar{\delta}$, $\bar{\theta}_{12}$),
and ($\bar{\delta}$, $\overline{\Delta m^2_{21}}$),
we have to take into account of terms of order
${\cal O}((\Delta E_{21}/\Delta E_{31})^2)$.
For ($\bar{\delta}$, $\bar{\theta}_{12}$), we have
\begin{eqnarray}
\sin2\bar{\theta}_{12}=-{4{\tilde J} \over c^2_{23}}
{\Delta m^2_{31} \over \Delta m^2_{21}}
\cos\bar{\delta}
+\sqrt{\left({4{\tilde J} \over c^2_{23}}
{\Delta m^2_{31} \over \Delta m^2_{21}}
\right)^2
\cos^2\bar{\delta}
+{8{\tilde J} \over c^2_{23}}
{\Delta m^2_{31} \over \Delta m^2_{21}}
\cos\delta\sin2\theta_{12}
+\sin^22\theta_{12}},
\label{eqn:diff12}
\end{eqnarray}
and this optimizes $\Delta\chi^2({\mbox{\rm CPV}})$.
We find
\begin{eqnarray}
&{\ }&
P(\nu_e\rightarrow\nu_\mu;\theta_{12},\delta)-
P(\nu_e\rightarrow\nu_\mu;\bar{\theta}_{12},\bar{\delta})
\nonumber\\
\simeq&{\ }&
{8{\tilde J} \over \sin2\theta_{12}}
{\Delta E_{21}\Delta E_{31} \over A^2}
\sin^2\left({AL \over 2}\right)
\left[\sin2\theta_{12}\cos\left(\delta+{\Delta E_{31}L \over 2}
\right)\right.
\nonumber\\
&-&
\left.\sin2\bar{\theta}_{12}\cos\left(\bar{\delta}
+{\Delta E_{31}L \over 2}\right)
-\sin2\theta_{12}\cos\delta
+\sin2\bar{\theta}_{12}\cos\bar{\delta}\right]
\nonumber\\
\simeq&{\ }&
{8{\tilde J} \over \sin2\theta_{12}}
{\Delta E_{21}\Delta E_{31} \over A^2}
\sin^2\left({AL \over 2}\right)
\left(\sin\bar{\delta}\sin2\bar{\theta}_{12}
-\sin\delta\sin2\theta_{12}\right){\Delta E_{31}L \over 2},
\end{eqnarray}
where we have expanded $\sin(\Delta E_{31}L/2)\simeq\Delta E_{31}L/2$.
By putting $\bar{\delta}=0$, we obtain
\begin{eqnarray}
\Delta\chi^2({\mbox{\rm CPV}};(\delta,\theta_{12}))
%=&{\ }&
\simeq{18N_0\sigma_0 \over \pi m_\mu^2}
{{\tilde J}^2\sin^2\delta\left(\Delta m^2_{21}\right)^2
\left(\Delta m^2_{31}\right)^2 \over
s_{23}^2\sin^22\theta_{13}}
{\sin^2\left(AL/2\right) \over E_\mu A^2},
\end{eqnarray}
which once again is the same as
$\Delta\chi^2({\mbox{\rm CPV}};(\delta,\theta_{13}))$.
For ($\bar{\delta}$, $\overline{\Delta m^2_{21}}$), we have
\begin{eqnarray}
\overline{\Delta m^2_{21}}&=&-{4{\tilde J} \over c^2_{23}
\sin^22\theta_{12}}\Delta m^2_{31}\cos\bar{\delta}\nonumber\\
&+&\sqrt{\left({4{\tilde J} \over c^2_{23}\sin^22\theta_{12}}\right)^2
\cos^2\bar{\delta}\left(\Delta m^2_{31}\right)^2
+{8{\tilde J} \over c^2_{23}\sin^22\theta_{12}}
\Delta m^2_{21}\Delta m^2_{31}\cos\delta
+\left(\Delta m^2_{21}\right)^2}
\label{eqn:diffm21}
\end{eqnarray}
which leads to
\begin{eqnarray}
&{\ }&
P(\nu_e\rightarrow\nu_\mu;\Delta m^2_{21},\delta)-
P(\nu_e\rightarrow\nu_\mu;\overline{\Delta m^2_{21}},\bar{\delta})
\nonumber\\
\simeq&{\ }&
8{\tilde J}
{\Delta E_{21}\Delta E_{31} \over A^2}
\sin^2\left({AL \over 2}\right)
\left(\sin\bar{\delta}
-\sin\delta\right){\Delta E_{31}L \over 2}.
\end{eqnarray}
Thus we get
\begin{eqnarray}
\Delta\chi^2({\mbox{\rm CPV}};(\delta,\Delta m^2_{21}))
%=&{\ }&
\simeq{18N_0\sigma_0 \over \pi m_\mu^2}
{{\tilde J}^2\sin^2\delta\left(\Delta m^2_{21}\right)^2
\left(\Delta m^2_{31}\right)^2 \over
s_{23}^2\sin^22\theta_{13}}
{\sin^2\left(AL/2\right) \over E_\mu A^2},
\end{eqnarray}
which once again is the same as
$\Delta\chi^2({\mbox{\rm CPV}};(\delta,\theta_{13}))$.
Unlike the cases for ($\bar{\delta}$, $\bar{\theta}_{13}$),
($\bar{\delta}$, $\bar{\theta}_{23}$),
($\bar{\delta}$, $\overline{\Delta m^2_{31}}$)
and ($\bar{\delta}$, $\bar{C}$), the optimal values (\ref{eqn:diff12})
and (\ref{eqn:diffm21}) have nontrivial behaviors even for
large $E_\mu$ and small $L$, as we can see from
Figs. 3 and 5.

We have seen analytically that two variable correlations give us
the behavior $\Delta\chi^2({\mbox{\rm CPV}})\propto \sin^2\delta/E_\mu$
and this behavior is the same as
$\Delta\chi^2({\mbox{\rm TV}})$.
Although it is difficult to discuss correlations of
more than two variables analytically, the discussions above
are sufficient to demonstrate that sensitivity to
CP violation decreases as $E_\mu$ becomes larger.
In fact we have verified numerically that
$\Delta\chi^2({\mbox{\rm CPV}})$ decreases as the
muon energy increases ($E_\mu \gtrsim$ 100GeV).
The conclusion in this subsection
is qualitatively consistent with the work \cite{lipari} by Lipari 
who claims that sensitivity to CP violation decreases as
$E_\mu$ becomes large.
However it may not be quantitatively consistent with \cite{lipari}
in which it was suggests that sensitivity starts getting lost
for $E_\nu\gtrsim$ a few GeV.
In our discussion here it was
necessary to have $|\Delta E_{31}L|\ll 1$ which may not be
attained for $L\sim$ 3000km and $E_\mu\lesssim$ 50GeV.
Our numerical calculations in the previous section
indicate that the sensitivity is optimized for
20GeV $\lesssim E_\mu\lesssim$ 50GeV
which is quantitatively consistent with the results in
\cite{BGW,Golden,BGRW,BGRW2,GH,DFLR,FLPR,NuFact,BCR,KS,Yasuda0}.
This interval for $E_\mu$ is the intermediate energy region
which cannot be treated analytically using our arguments
in this section.  In fact it seems difficult to explain
analytically the strong correlation
of ($\bar{\delta}$, $\bar{C}$) for $E_\mu\simeq$50 GeV and $L\simeq$ 3000km.
(cf. Fig. 1)

\section{JHF experiment}

The JHF project \cite{jhf} has been proposed to
perform precise measurements of the oscillation
parameters.
The possible extension of this project includes
the upgrade of the power to 4MW and the construction
of a mega--ton detector \cite{nakamura}.
The possibility to measure CP violation at
the JHF project has been discussed by
\cite{obayashi,superbeam,kobayashi}.
Here we briefly discuss the possibility of measurements
of CP violation at the JHF experiment with power 4MW and
a 1 Mton detector as a comparison with neutrino factories.
As in previous sections, we will take into consideration
the correlations of all the oscillation parameters.
In the case of the JHF experiment, which has the baseline
$L\simeq$ 300km, the matter effect is almost negligible
and it is possible to compare the numbers of events
for $\nu_\mu\rightarrow\nu_e$ and $\bar{\nu}_\mu\rightarrow\bar{\nu}_e$
directly by taking into account the difference of the cross sections
between $\sigma_{\nu N}$ and $\sigma_{\bar{\nu} N}$.
However, we use the same $\Delta\chi^2$ as in section 2,
as discussions with the same criterion gives more transparent
comparisons between neutrino factories and the superbeam
at JHF.

The correlations of two variables ($\delta$, $X$), where $X$ is
$\theta_{k\ell}$, $\Delta m^2_{k\ell}$ or $C$, are shown in
Fig. 11, where the central values for these
parameters are those of the best fit point, i.e.,
$\sin^22\theta_{12}=0.75$, $\Delta m^2_{21}=3.2\times10^{-5}{\mbox{\rm
eV}}^2$; $\sin^22\theta_{23}=1.0$, $\Delta
m^2_{32}=3.2\times10^{-3}{\mbox{\rm eV}}^2$, C=1.0 and we have used a
reference value $\theta_{13}=8^\circ$. In this calculation the narrow
band beam (NBB) (the flux referred to as LE2$\pi$ in \cite{jhf}) is
used, and it is assumed for simplicity that there are no backgrounds
and the detection efficiency is 70\% in Fig. 11.
Note that for the purpose of
measurements of CP violation NBB is more advantageous than
the wide band beam, as the former has better energy resolution.

As in the section 2, we have evaluated numerically the data size
required to reject a hypothesis with $\bar{\delta}=0$.  Of course the data
size depends on the true value $\delta$ and the results obtained by
varying the six variables ($\theta_{k\ell}$, $\Delta m^2_{k\ell}$,
$C$) are plotted in Fig. 12, where we have taken the
best fit values for ($\theta_{12}$, $\Delta m^2_{21}$),
($\theta_{23}$, $\Delta m^2_{32}$), $\theta_{13}=8^\circ, 5^\circ,
1^\circ (2^\circ)$, and the NBB is used.  The vertical axis
of Fig. 12 stands for the data size required
per kt$\times$($\nu_\mu$ 1 year + $\bar{\nu}_\mu$ 2 years).
We have used two ways of
$\nu_e$ selections, one is 1-ring e-like selection which has the
background fraction $f_B=1.8\times10^{-2}$, the detection efficiency
70.4\%, and the other one is $\pi^0$ cut selection which has the
background fraction $f_B=2\times10^{-3}$, the detection efficiency
50.4\% \cite{obayashi}.  In the case of the 1-ring e-like selection,
for $\theta_{13}=1^\circ$ the systematic error becomes so large that
the data size required to reject $\delta=0$ becomes infinite.  Also in
this case the number of events for $\delta=\pi$ becomes almost the
same as that for $\delta=0$ up to the systematic errors and there is
no way to distinguish the case of $\delta=\pi$ and that of $\delta=0$.
However, as long as the value of $\delta$ is not close to 0 or $\pi$
and $\theta_{13}\gtrsim 3^\circ$, the JHF with 4MW power and a 1 mega ton
detector will be able to demonstrate $\delta\ne0$ at 3$\sigma$CL.

\section{discussions}

In order to be more concrete, we need the knowledge
on the uncertainty of the matter effect $A$.
The error of $A=\sqrt{2}G_F Y_e\rho$ comes from
those of $Y_e$ and $\rho$.  The error of $Y_e$
has been discussed by \cite{mp} and it is about 2\%
and geophysicists \cite{anderson,tanimoto2} agree with it.
Without any uncertainty of the matter effect,
it has been claimed that a medium baseline experiment
($L\sim 3000$km, $E_\mu\sim$ 50GeV) is best for measurements of CP violation.
In that case the depth of the neutrino path is at most
200km and most of the neutrino path
is in the upper mantle.  It is known in geophysics \cite{tanimoto1}
that the crust has relatively large
latitude-longitude dependent fluctuations around
constant density.  On the other hand, in the case of the upper mantle,
some geophysicists claim
that fluctuations around constant density are
a few \% \cite{resovsky,anderson,tanimoto2}
while another \cite{geller} says that they may be as large as 5 \%.
However, such discussions are based on
normal mode studies in seismology which are confined to
long wavelength features, and it was pointed out \cite{tanimoto2,geller}
that the fluctuations in the density in the analysis of
neutrino factories may be larger than 5 \%, since
the width of the neutrino beam is much smaller than typical
wavelengths in seismological studies.
If that is the case, then
it follows from Fig. 8 that the case $L\simeq$1000km
is better than the case $L\simeq$3000km,
since the former is insensitive to the uncertainty of
the matter effect.
More detailed analysis with seismological discussions
seems to be necessary to determine the optimal baseline
and the muon energy in neutrino factories.

We have also seen that if there is a nontrivial fraction of backgrounds,
which is expected in the case of water Cherenkov or liquid argon
detectors,
then medium energy (10GeV $\lesssim E_\mu\lesssim$ 20GeV)
is more advantageous than high energy ($E_\mu\sim$ 50GeV),
as systematic errors become so large in the latter case.

The bottom line of the present paper is that
either the high or medium energy option is certainly
better than the low energy option which has been
advocated by some people \cite{sato,mn}.
We have arrived at this conclusion on the assumption that
the energy threshold is as low as 0.1 GeV,
and the detection efficiency is independent
of the neutrino energy.
In practice, it may be very difficult to have such a
low threshold and to keep such a good detection
efficiency down to 0.1 GeV, so it is expected that
the low energy option ($E_\nu\lesssim$ 10GeV)
becomes less and less advantageous.

If we can realize a nearly perfectly polarized muon beam,
it may be possible to measure T violation, which
is much better than CP violation, since we do not
have to worry about the uncertainty of the density of
the Earth or other oscillation parameters
and a nonzero value of T violation immediately
indicates $\delta\ne0$.  Efforts should
be made along these lines.

If $\theta_{13}\gtrsim 3^\circ$ and if the value of $\delta$ is not close
to 0 or $\pi$, then the JHF experiment with 4MW power and a 1 mega ton
detector will be able to demonstrate $\delta\ne0$ at 3$\sigma$CL.  On
the other hand, if $\theta_{13}\lesssim 3^\circ$, then neutrino
factories seem to be the only experiment which can demonstrate
$\delta\ne0$.  In that case, depending on the situation such as the
fraction of backgrounds, the uncertainty of the matter effect and the
magnitude of $\theta_{13}$, the option with ($E_\mu\simeq$50GeV,
$L\simeq$3000km) may be advantageous (or disadvantageous) over
($E_\mu\simeq$20GeV, $L\simeq$1000km).  In both
a neutrino factory with (20GeV$\lesssim E_\mu\lesssim$50GeV,
1000km$\lesssim L\lesssim$3000km) and the
JHF experiment, our $\Delta\chi^2$(CPV)
depends not only on $\sin\delta$ but also $\cos\delta$,
\footnote{If we evaluate
$\Delta\chi^2({\mbox{\rm CPV}};(\delta,\theta_{13}))$
in (\ref{eqn:chi2cpv13}) to the next leading order
in $\Delta m^2_{31}L/E_\mu$ then
${\tilde J}^2\sin^2\delta$ in (\ref{eqn:chi2cpv13})
is replaced by
${\tilde J}^2(\sin\delta+\mbox{\rm const.}
(\Delta m^2_{31}L/E_\mu)\cos\delta)^2$,
and $\Delta m^2_{31}L/E_\mu$ is not necessarily
negligible in either case.}
so that we can in principle distinguish
$\delta=\pi$ from $\delta=0$ as long as the statistical
significance overcomes the systematic errors.
This is not the case for a neutrino factory with
large systematic errors for small $\theta_{13}$,
i.e., for $E_\mu$=50GeV and $\theta_{13}=1^\circ$ (cf. Fig. 9),
and for the JHF experiment with less S/N ratio
i.e., when the 1-ring e-like selection is adopted,
or when $\theta_{13}=8^\circ$ and the $\pi^0$ cut
selection is adopted (cf. Fig. 12).

In this paper
we adopted simplified assumptions such as that the detection
efficiency is independent of the neutrino energy, that the threshold
energy can be taken as low as 0.1GeV, and that the uncertainty of the matter
effect is at most 5 \%,
but we need much more detailed experimental information
to obtain the optimum muon energy and the baseline.

\section*{Note Added}

Just before we hit a return key to submit our paper
to the preprint archive, we became aware of the work by
M. Freund et al. \cite{fhl}, where the similar
topics has been discussed from a slightly different viewpoint.

%\vskip 0.2in
\section*{Acknowledgments}
O. Y. would like to thank R. Geller, Y. Fukao and H. Yamazaki for
discussions on the density of the Earth, J. Resovsky,
M. Ishii, C. Kuo, D.L. Anderson, T. Tanimoto
and J. Tromp for useful communications
on the uncertainty of the density of the Earth, and
Y. Kuno and Y. Mori for discussions on neutrino factories.
We thank Kenji Ogawa for his contribution in the earlier stage
of this work, P. Lipari, H. Minakata and H. Nunokawa for discussions.
This
research was supported in part by a Grant-in-Aid for Scientific
Research of the Ministry of Education, Science and Culture,
\#12047222, \#13640295.
%\vglue 2cm

%\newpage

\newpage
\noindent
{\Large{\bf Figures}}
\begin{description}
\item[Fig.1a] 
Correlations of errors of $\bar{\delta}$ and the normalization
$\bar{C}$ for $L$=100km, 1000km, 2500km, 6300km and
for $E_\mu$=3GeV, 20GeV, 50GeV.
$\Delta\chi^2=0.18, 0.37, 0.73$ corresponds to
1$\sigma$CL, 90\%CL, 99\%CL, respectively
for two degrees of freedom.
The oscillation parameters are $\Delta m^2_{21}=1.8\times 10^{-5}$eV$^2$,
$\Delta m^2_{32}=3.5\times 10^{-5}$eV$^2$,
$\sin^22\theta_{12}=0.76$, $\sin^22\theta_{23}=1.0$,
$\theta_{13}=8^\circ$, $\delta=\pi/2$.  The number of useful muon
decays is $10^{21}\mu\cdot10$kt.  No backgrounds are taken into consideration
in Figs. 1--6.

\item[Fig.1b] 
The same correlation as Fig.1a for $\theta_{13}=5^\circ$, $1^\circ$.
The oscillation parameters and other reference values
are the same as in Fig. 1a.

\item[Fig.2] 
Correlations of errors of $\bar{\delta}$ and $\bar{\theta}_{13}$.
The oscillation parameters and other reference values
are the same as in Fig. 1a.

\item[Fig.3]
Correlations of errors of $\bar{\delta}$ and $\bar{\theta}_{12}$.
The oscillation parameters and other reference values
are the same as in Fig. 1a.

\item[Fig.4]
Correlations of errors of $\bar{\delta}$ and $\bar{\theta}_{23}$.
The oscillation parameters and other reference values
are the same as in Fig. 1a.

\item[Fig.5]
Correlations of errors of $\bar{\delta}$ and
$\overline{\Delta m^2_{21}}$.
The oscillation parameters and other reference values
are the same as in Fig. 1a.

\item[Fig.6]
Correlations of errors of $\bar{\delta}$ and
$\overline{\Delta m^2_{32}}$.
The oscillation parameters and other reference values
are the same as in Fig. 1a.

\item[Fig.7]
The contour plot of equi-number of data size required
(in the unit of kt) to
reject a hypothesis $\bar{\delta}=0$ at $3\sigma$
using $\Delta\chi^2(\mbox{\rm CPV})$ (\ref{eqn:chicpv})
in the case of a neutrino factory with
$10^{21}$ useful muon decays, the background fraction
$f_B=10^{-5}$ or $10^{-3}$, $\theta_{13}=8^\circ$, $5^\circ$, $1^\circ$.
The other oscillation parameters are the same as in Fig. 1.

\item[Fig.8]
The same as Fig.7
with $\theta_{13}=8^\circ$, except that the uncertainty
of the matter effect is assumed to be larger
$|\Delta C|\le 0.1$ or $|\Delta C|\le 0.2$.

\item[Fig.9]
The number of data size required to
reject a hypothesis $\bar{\delta}=0$ at $3\sigma$ for
a neutrino factory using $\Delta\chi^2(\mbox{\rm CPV})$ (\ref{eqn:chicpv})
as a function of the true value of $\delta$ for $f_B=10^{-3}$.
All the assumptions
except for $\delta$ are the same as in Fig. 7.  The
situation is improved for smaller $\theta_{13}$.

\item[Fig.10]
The contour plot of equi-number of data size required
(in the unit of kt) to
reject a hypothesis $\bar{\delta}=0$ at $3\sigma$
using T violation (\ref{eqn:chi2tv})
in the case of a neutrino factory with
$10^{21}$ useful muon decays, the background fraction
$f_B=10^{-5}$ or $10^{-3}$, $\theta_{13}=8^\circ$, $5^\circ$, $1^\circ$.
The other oscillation parameters are the same as in Fig. 1.
Crucial assumption in this case is that the polarization is
perfect, and under this assumption T violation is 3 times
better than the CP violation analysis in section 2.

\item[Fig.11]
The correlations of errors of ($\bar{\delta}$, $\bar{C}$),
($\delta$, $\bar{\theta}_{13}$),
($\delta$, $\bar{\theta}_{12}$), ($\delta$, $\bar{\theta}_{23}$),
($\delta$, $\overline{\Delta m^2_{21}}$),
($\delta$, $\overline{\Delta m^2_{32}}$)
in the case of the JHF experiment with 4MW power, a 1 mega ton
detector and NBB.  No backgrounds are taken into consideration
in these figures.
The oscillation parameters used are the same as in Fig. 1.

\item[Fig.12]
The number of data size required (in the unit of kt)
to reject a hypothesis $\bar{\delta}=0$ at $3\sigma$ for
the JHF experiment with 4MW power, a 1 mega ton
detector and NBB using $\Delta\chi^2(\mbox{\rm CPV})$ (\ref{eqn:chicpv})
as a function of the true value of $\delta$.  Unlike
in the case of Fig.11, the effects of backgrounds are
taken into account in this figure.
The oscillation parameters used are the same as in Fig. 1,
and two ways of cuts (1-ring e-like and $\pi^0$ cut) \cite{obayashi}
are used.  In the case of the 1-ring e-like selection,
$\theta_{13}=1^\circ$ does not have a solution
because the systematic errors become so large.

\end{description}
\pagestyle{empty}

\newpage
\vglue -2.5cm
\hglue -6.0cm 
\epsfig{file=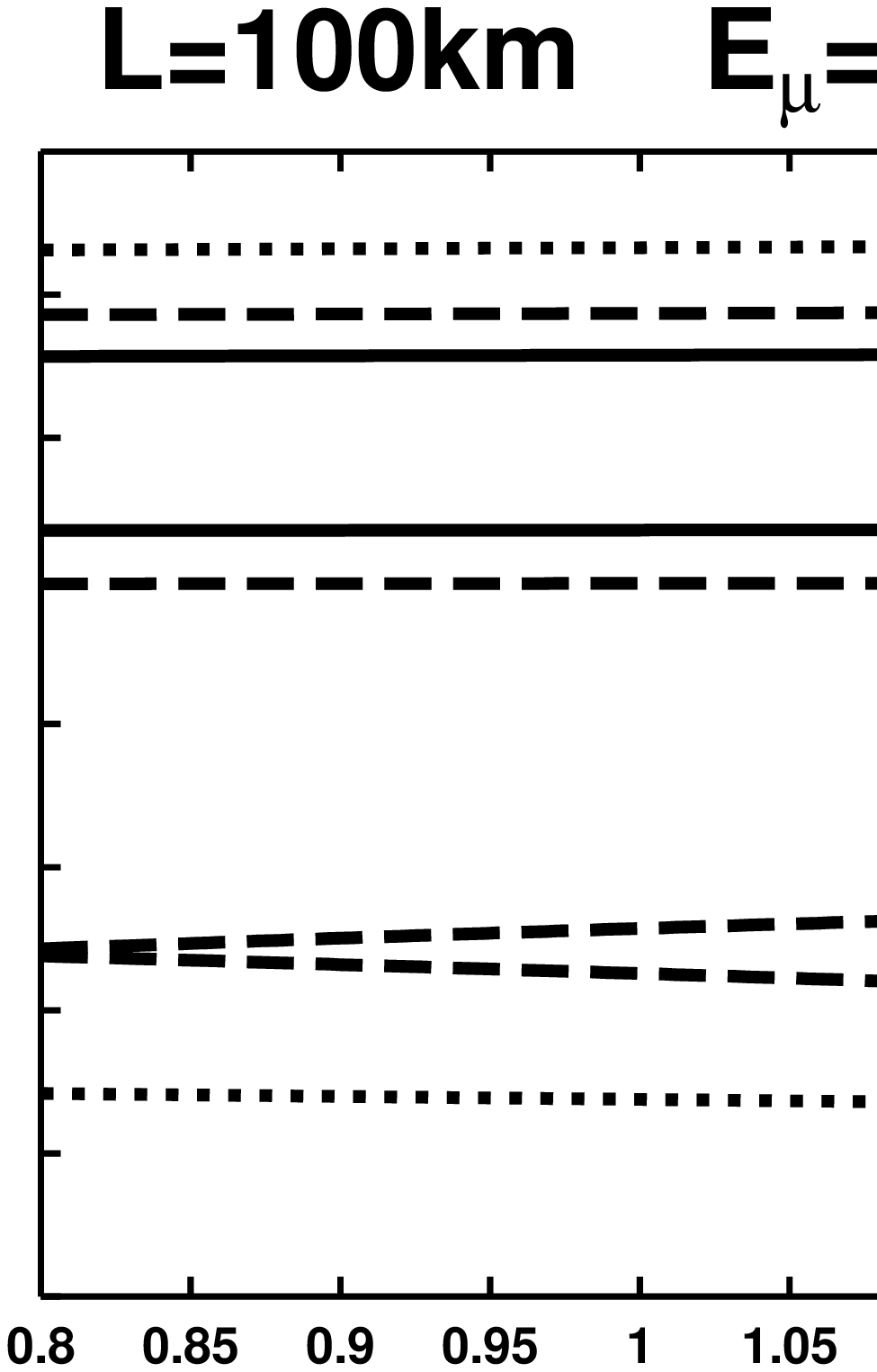,width=8cm}
\vglue -8.1cm \hglue -0.7cm \epsfig{file=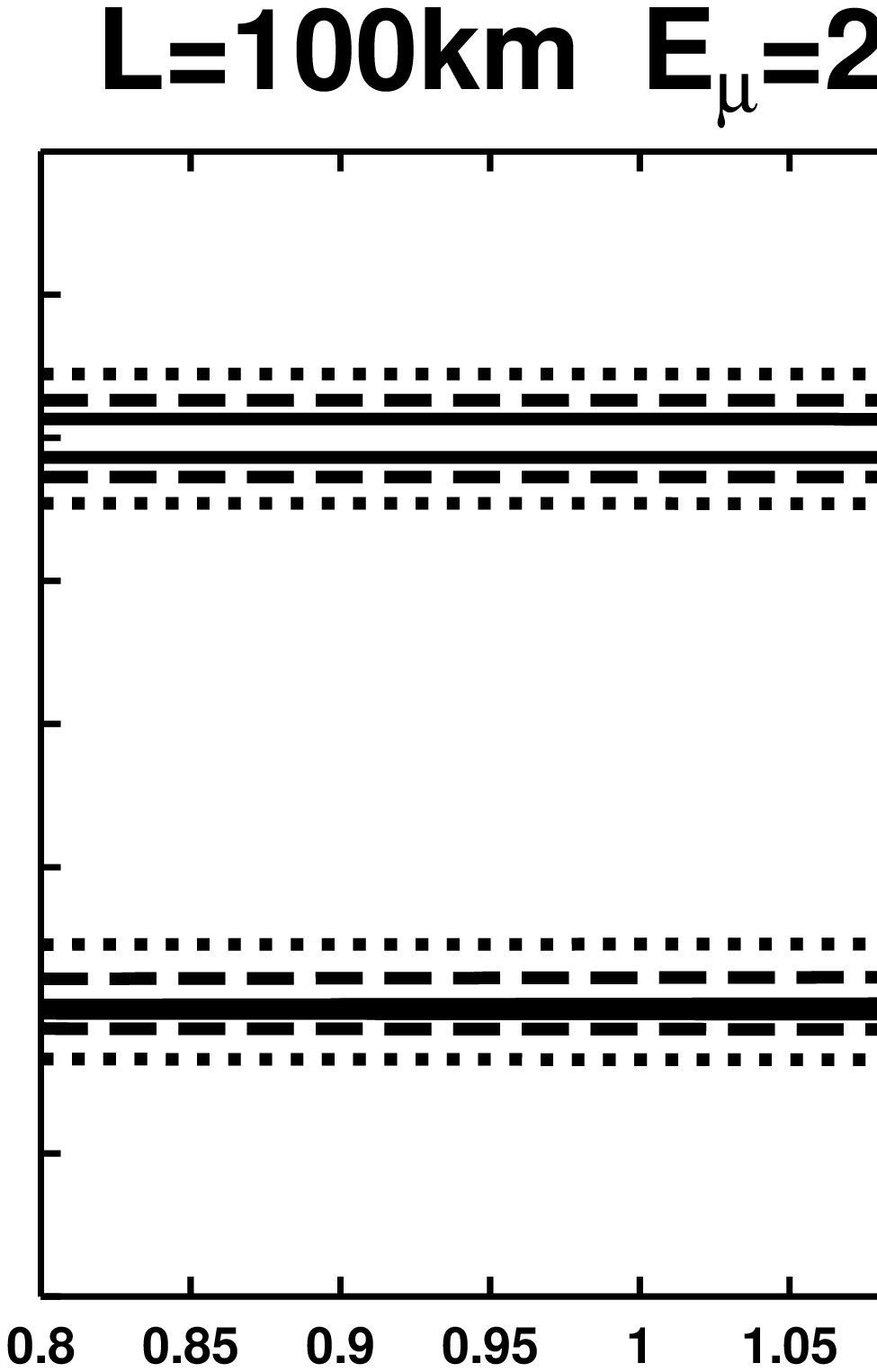,width=8cm}
\vglue -8.1cm \hglue 4.5cm \epsfig{file=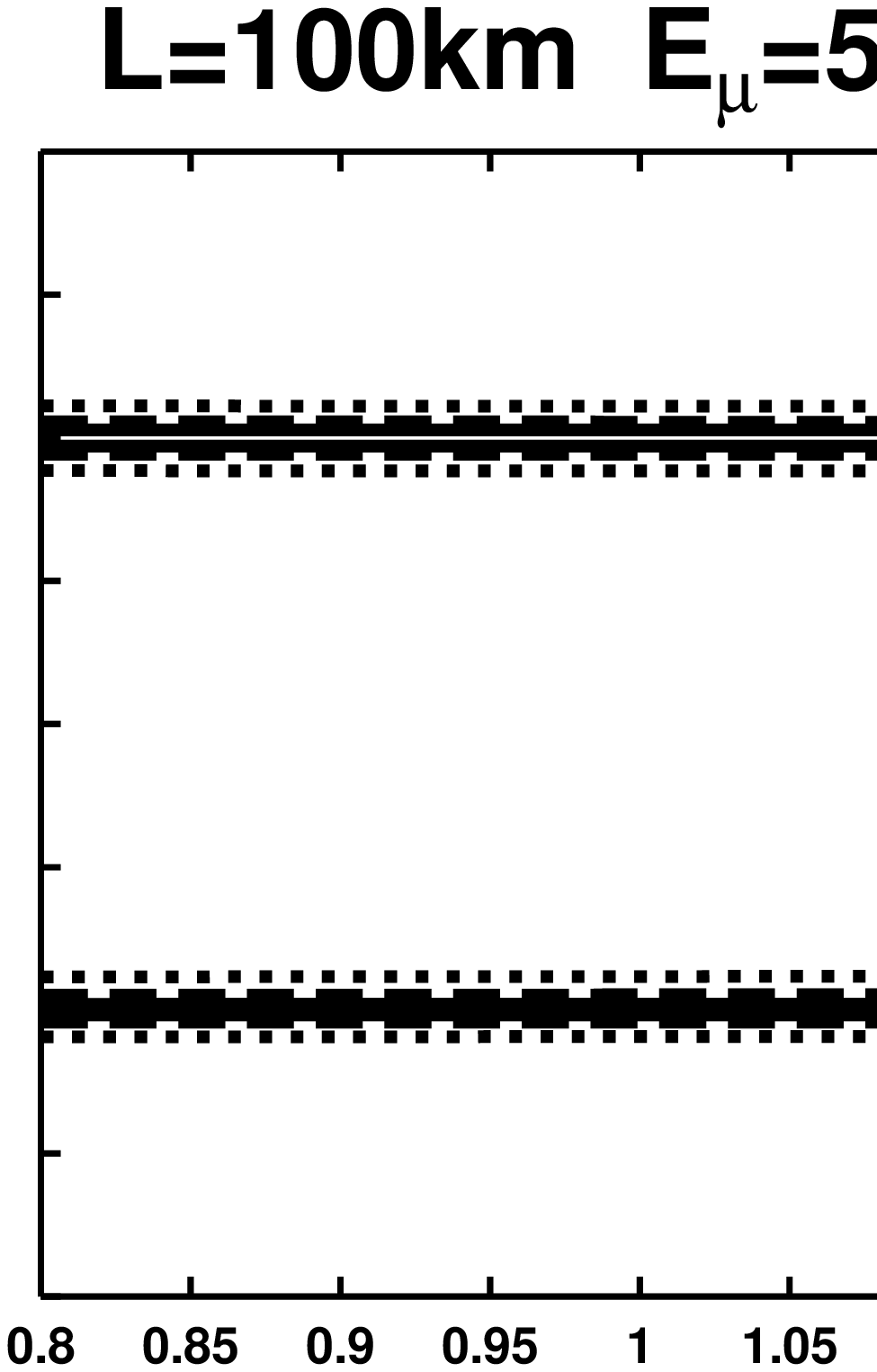,width=8cm}

\vglue -2.4cm
\hglue -6.0cm 
\epsfig{file=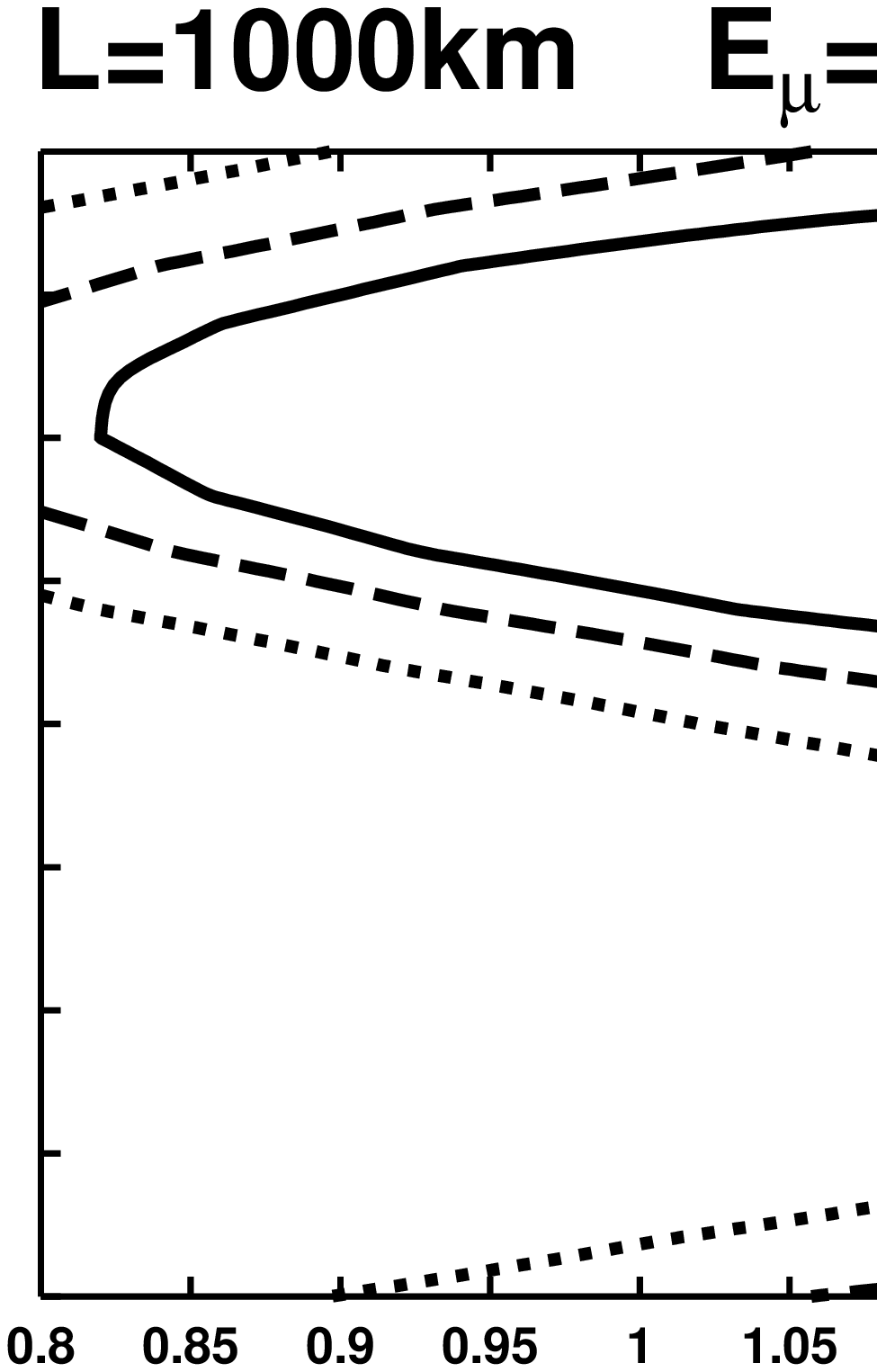,width=8cm}
\vglue -8.1cm \hglue -0.7cm \epsfig{file=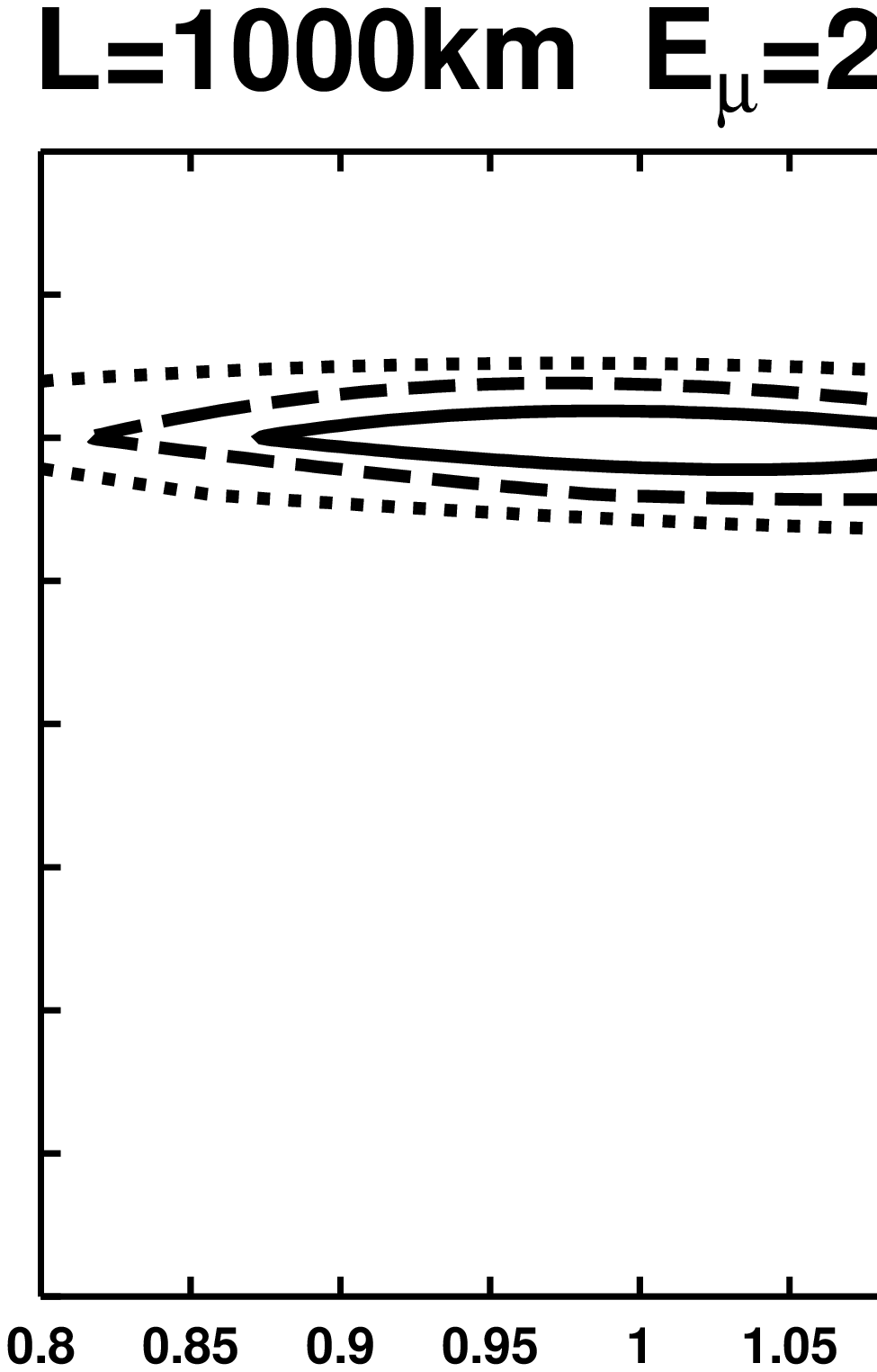,width=8cm}
\vglue -8.1cm \hglue 4.5cm \epsfig{file=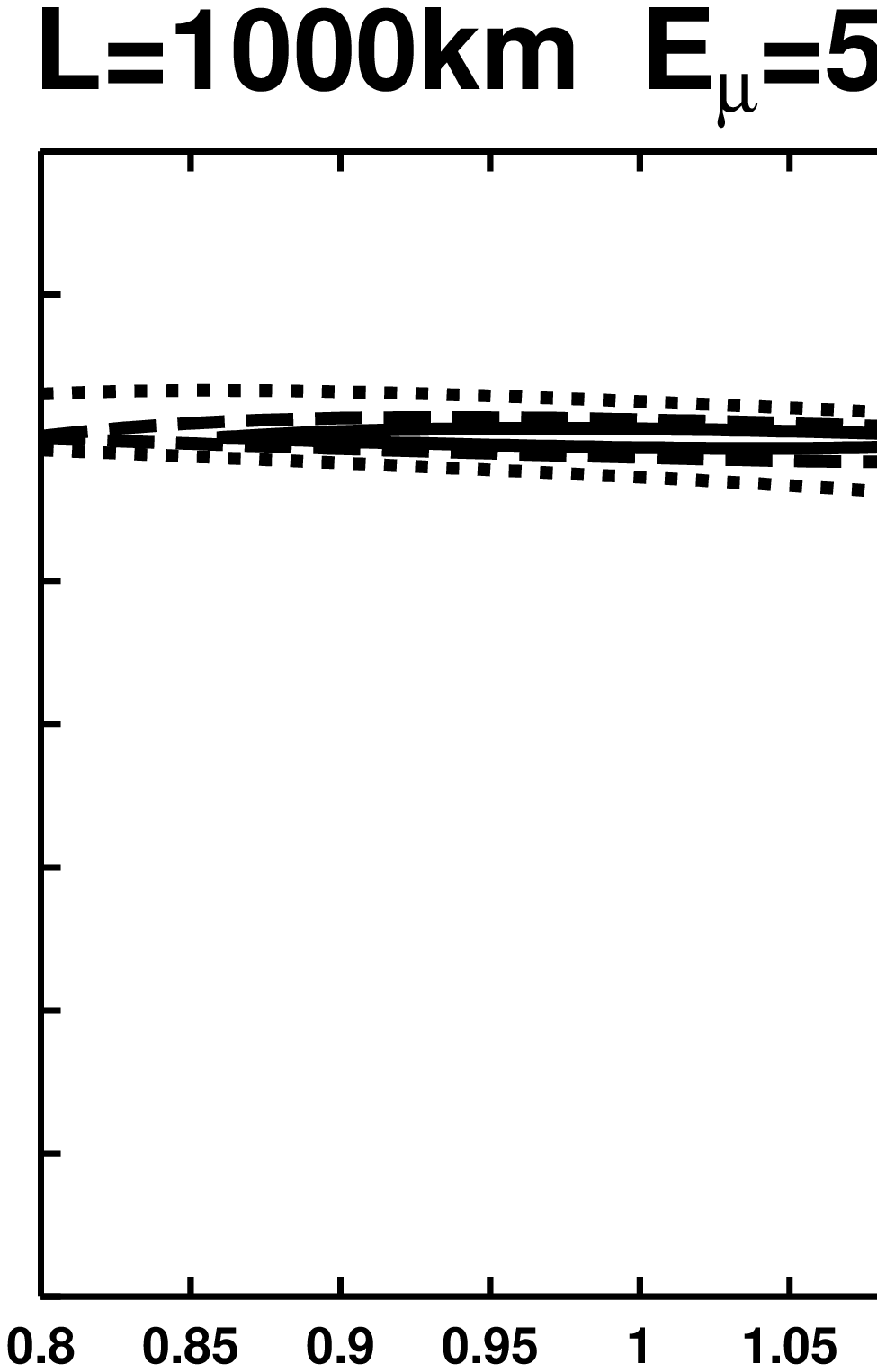,width=8cm}

\vglue -2.4cm
\hglue -6.0cm 
\epsfig{file=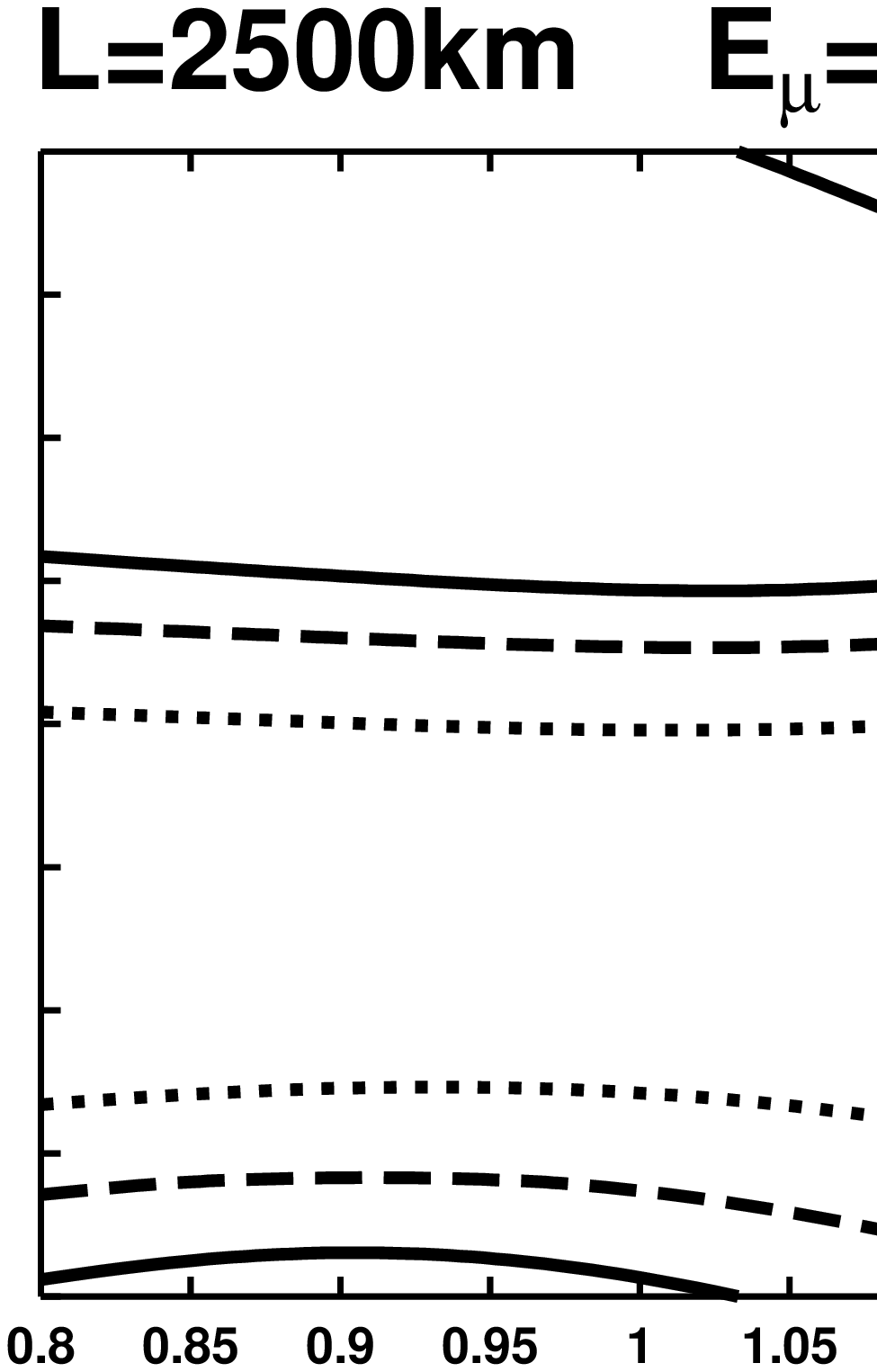,width=8cm}
\vglue -8.1cm \hglue -0.7cm \epsfig{file=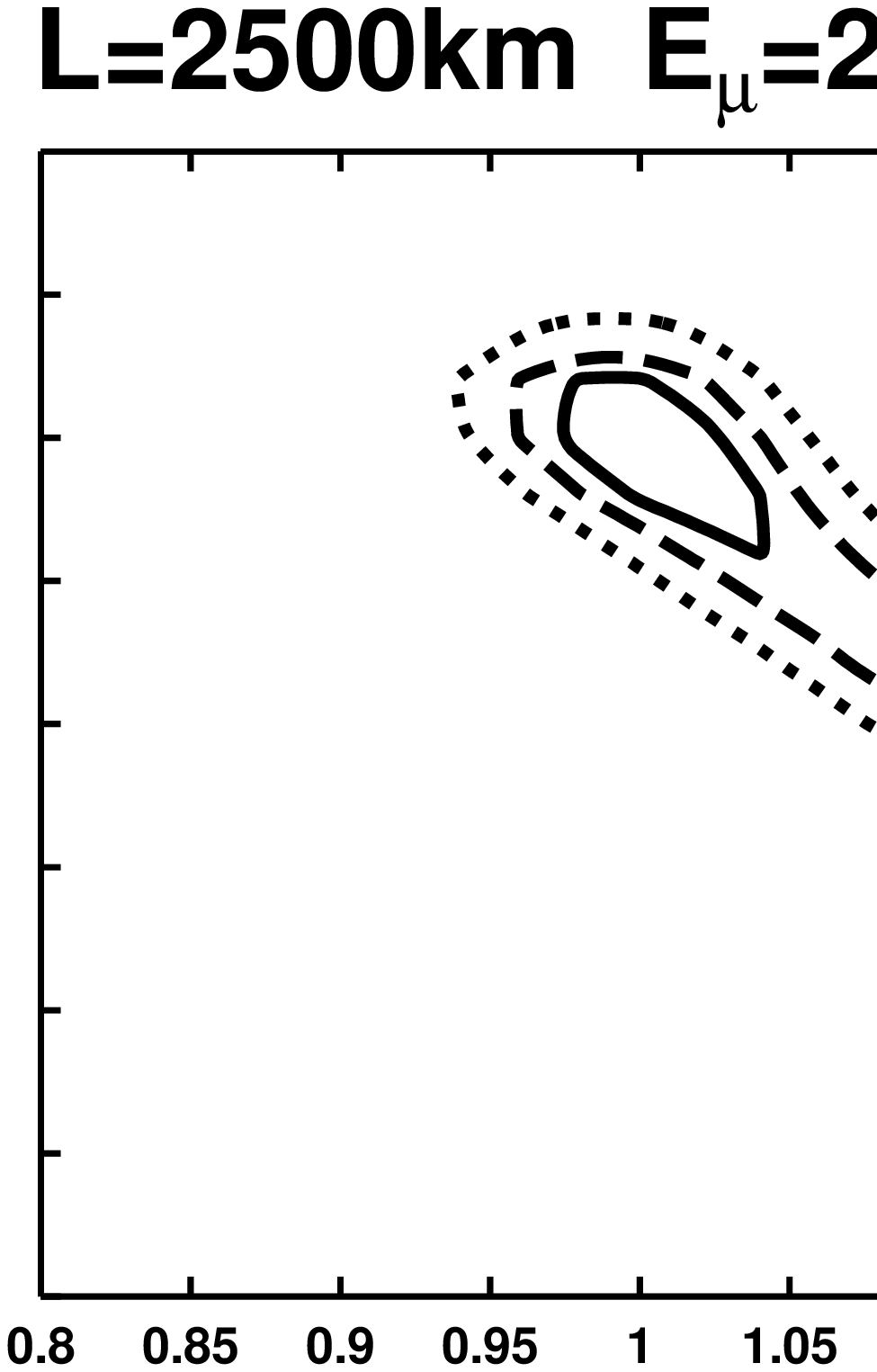,width=8cm}
\vglue -8.1cm \hglue 4.5cm \epsfig{file=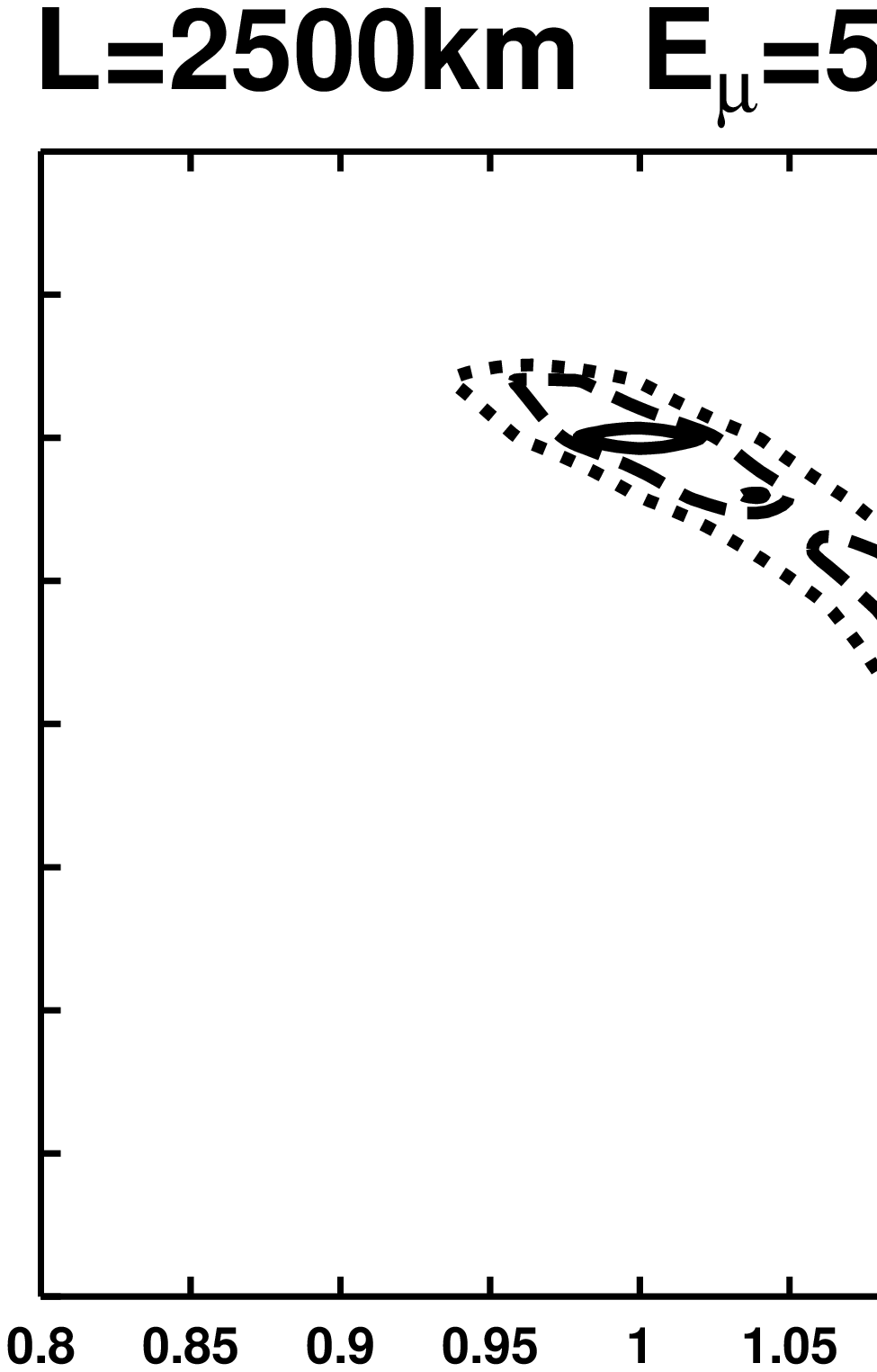,width=8cm}

\vglue -2.4cm
\hglue -6.0cm 
\epsfig{file=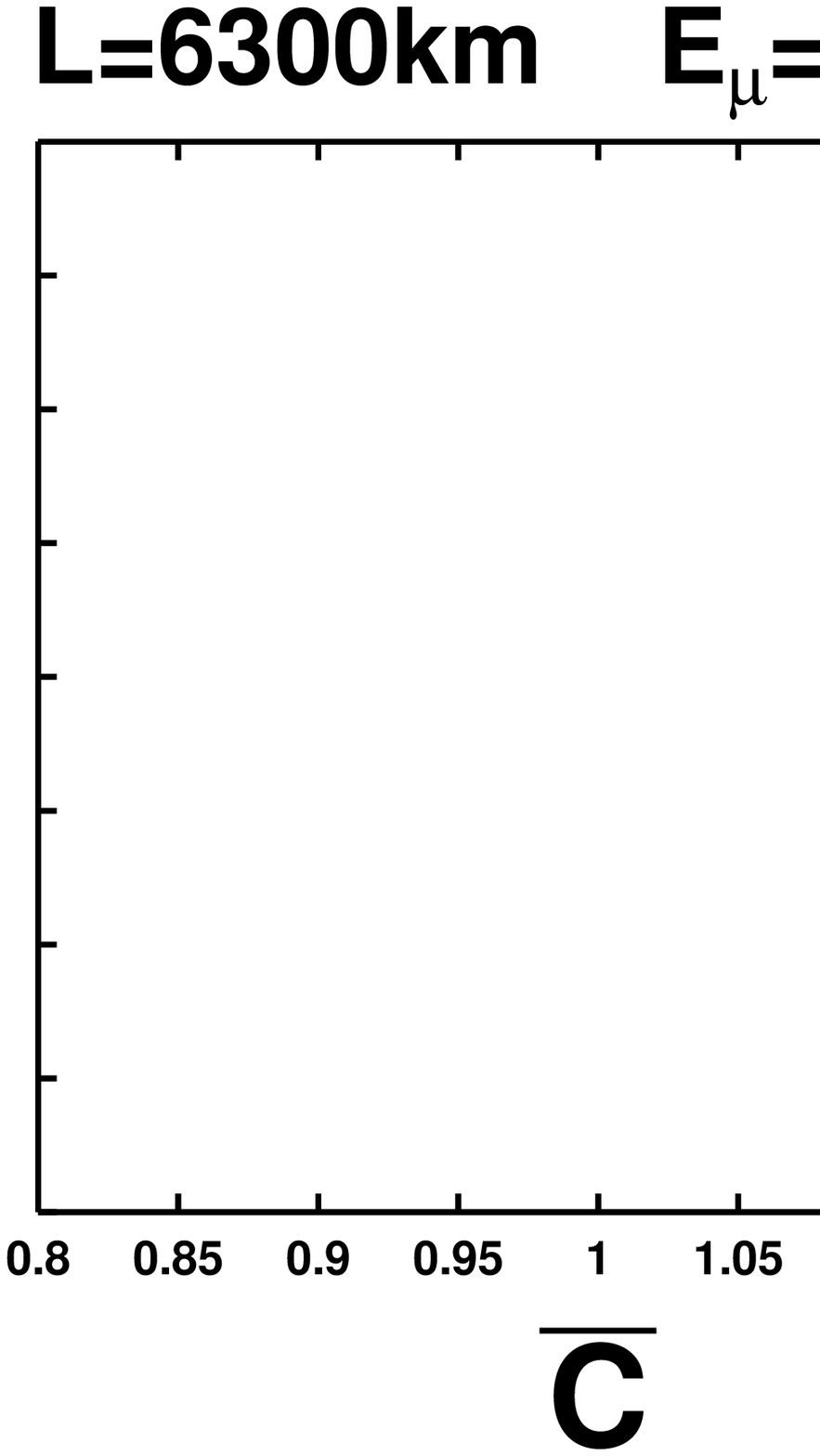,width=8cm}
\vglue -8.1cm \hglue -0.7cm \epsfig{file=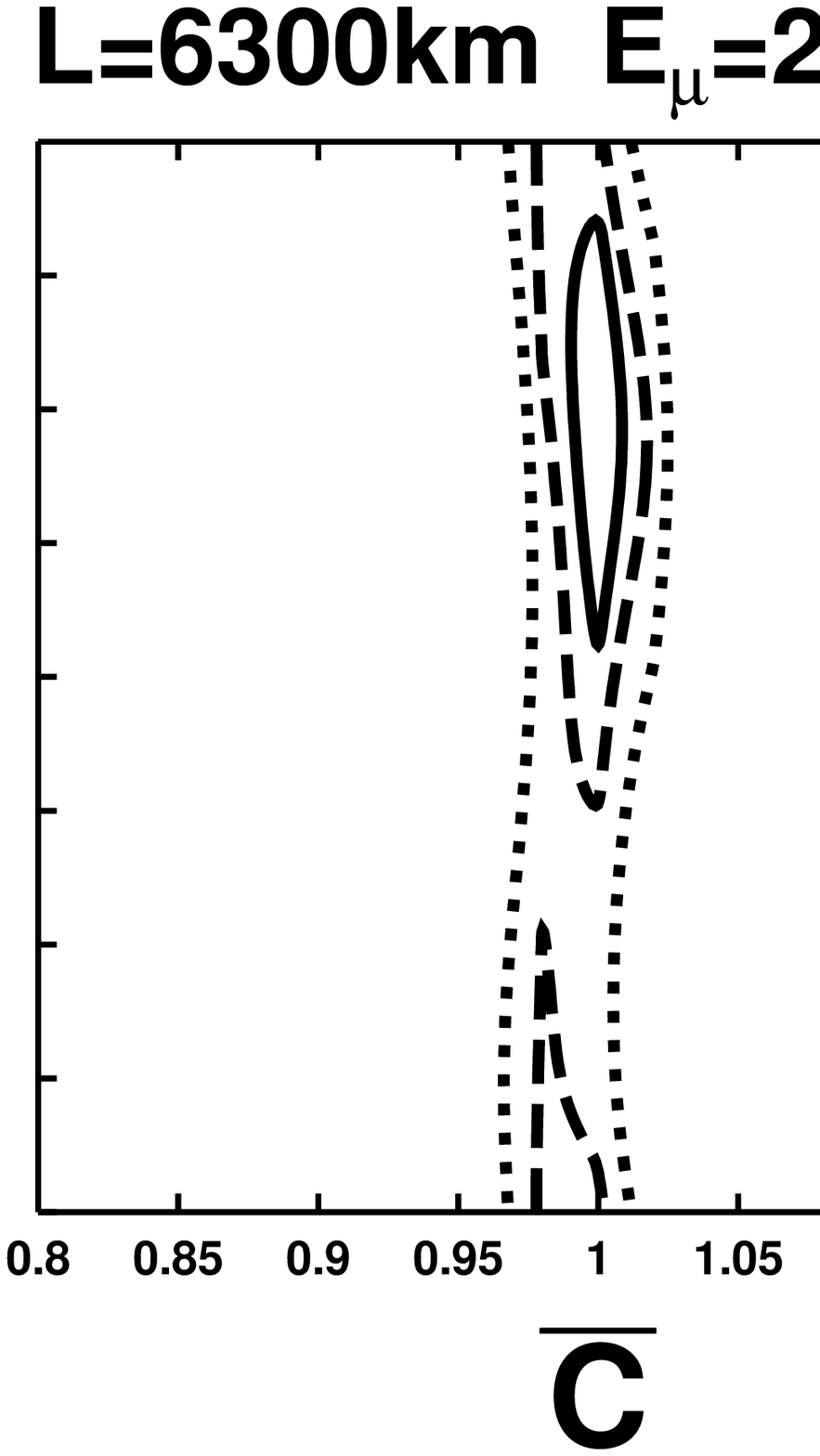,width=8cm}
\vglue -8.1cm \hglue 4.5cm \epsfig{file=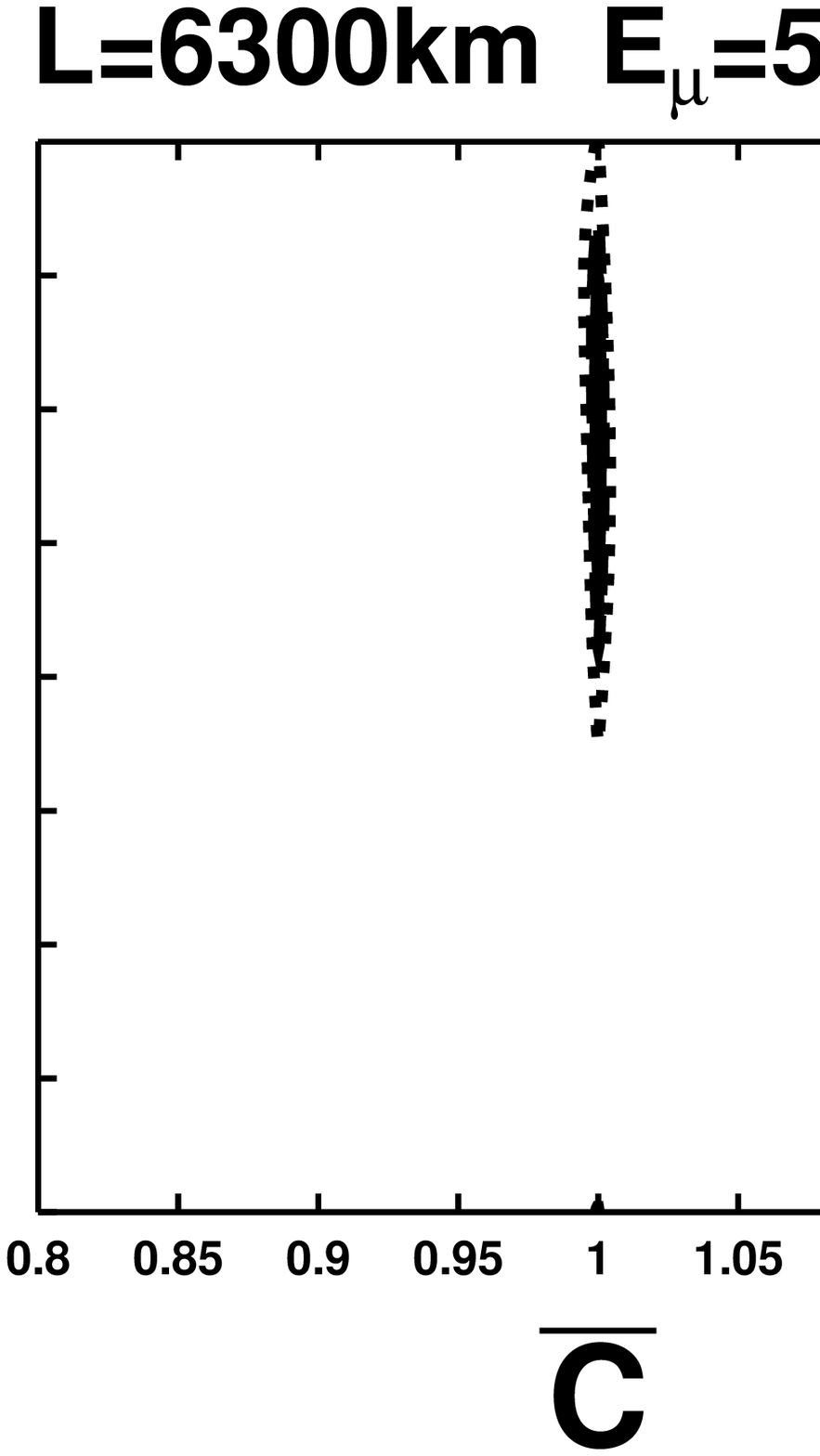,width=8cm}
\vglue -2.0cm\hglue -23.3cm
\epsfig{file=fig0113.eps,width=22cm}
\vglue -21.5cm\hglue 6.3cm
\epsfig{file=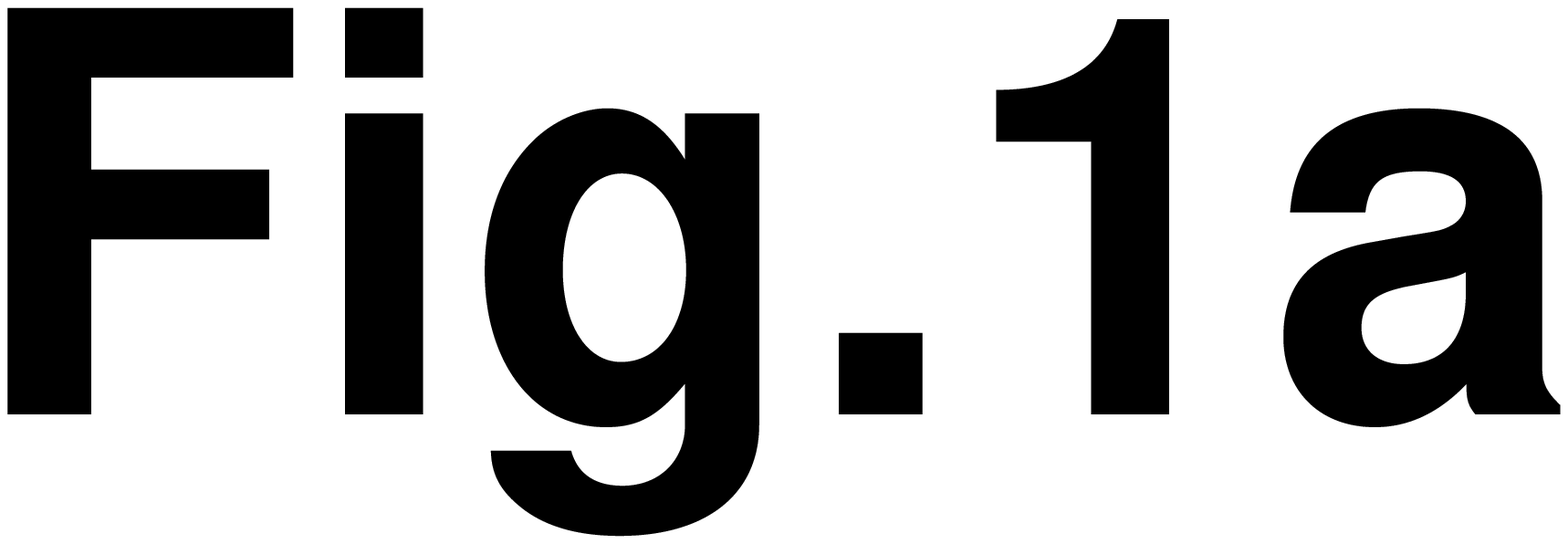,width=4cm}
\newpage
\vglue -2.5cm
\hglue -4.0cm 
\epsfig{file=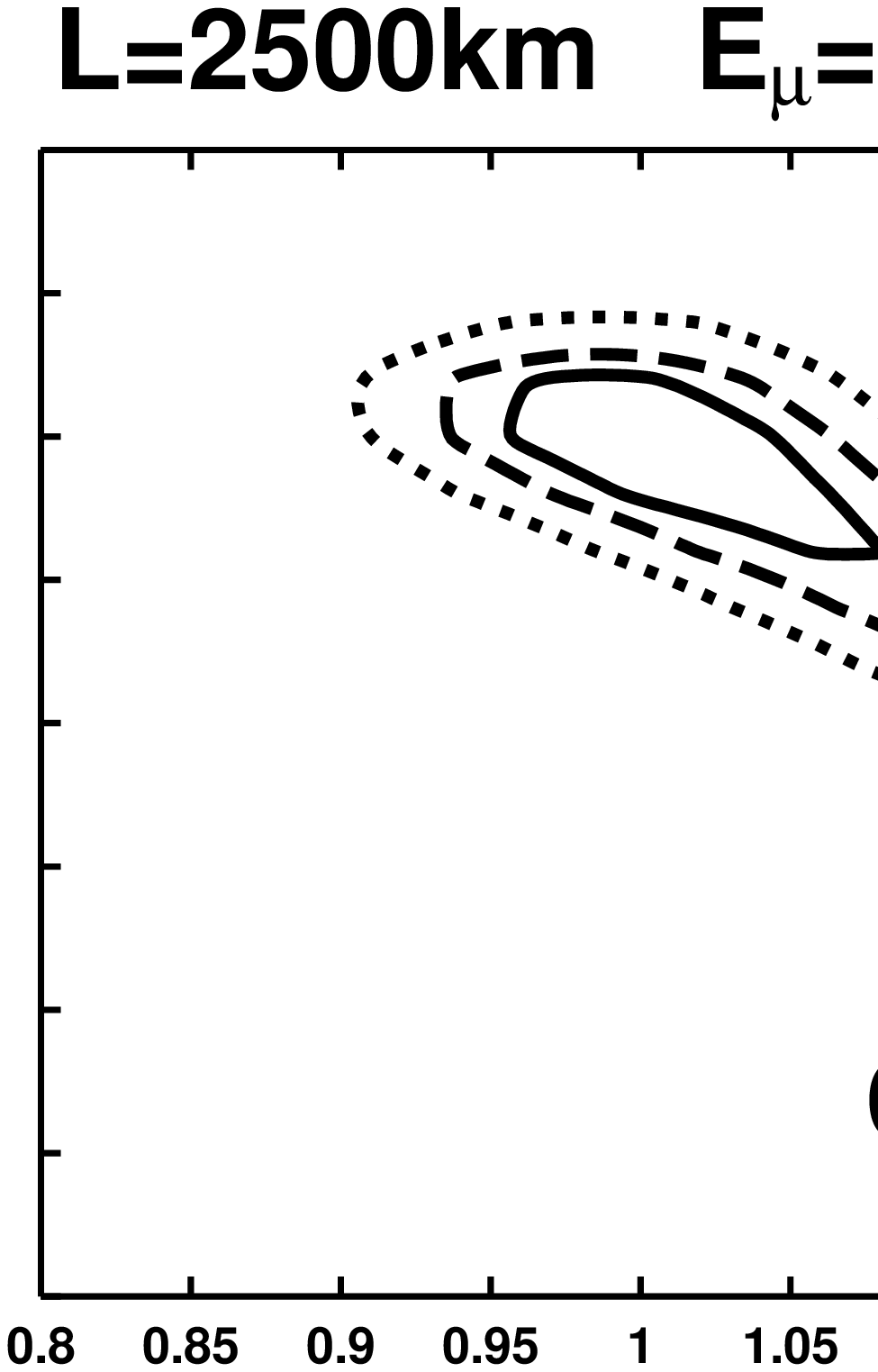,width=8cm}
\vglue -8.1cm \hglue 1.3cm \epsfig{file=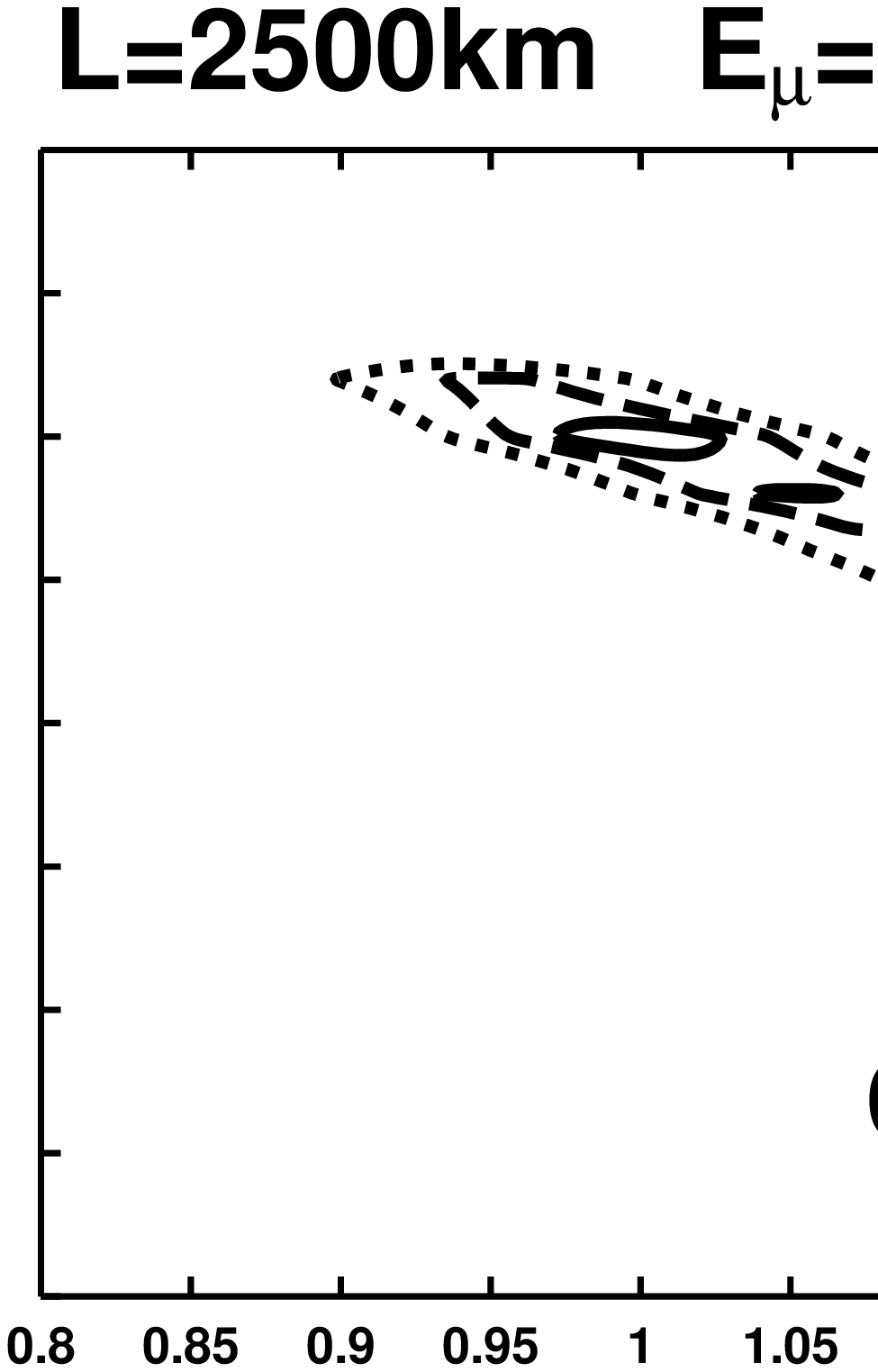,width=8cm}

\vglue -2.4cm
\hglue -4.0cm 
\epsfig{file=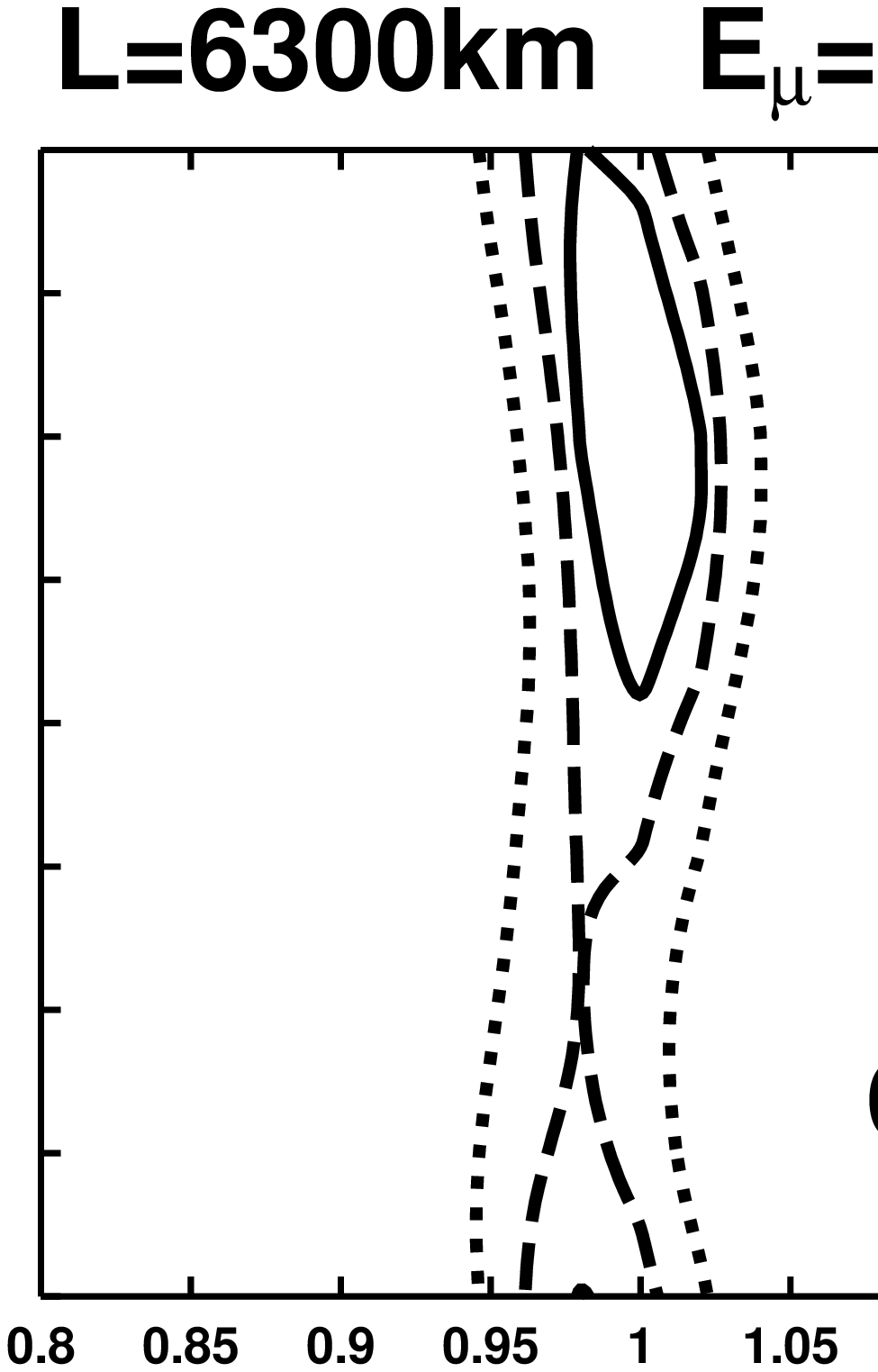,width=8cm}
\vglue -8.1cm \hglue 1.3cm \epsfig{file=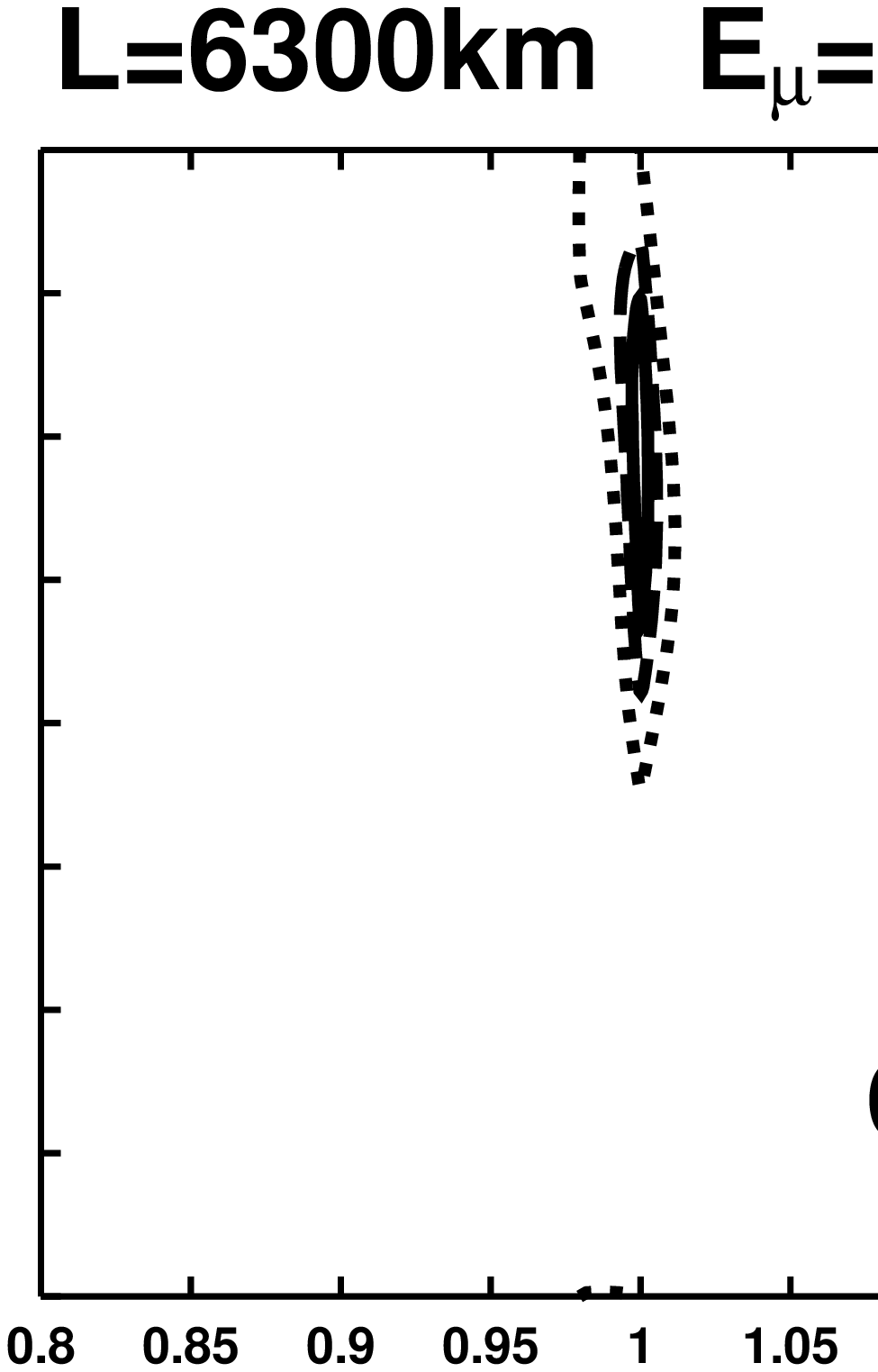,width=8cm}

\vglue -2.4cm
\hglue -4.0cm 
\epsfig{file=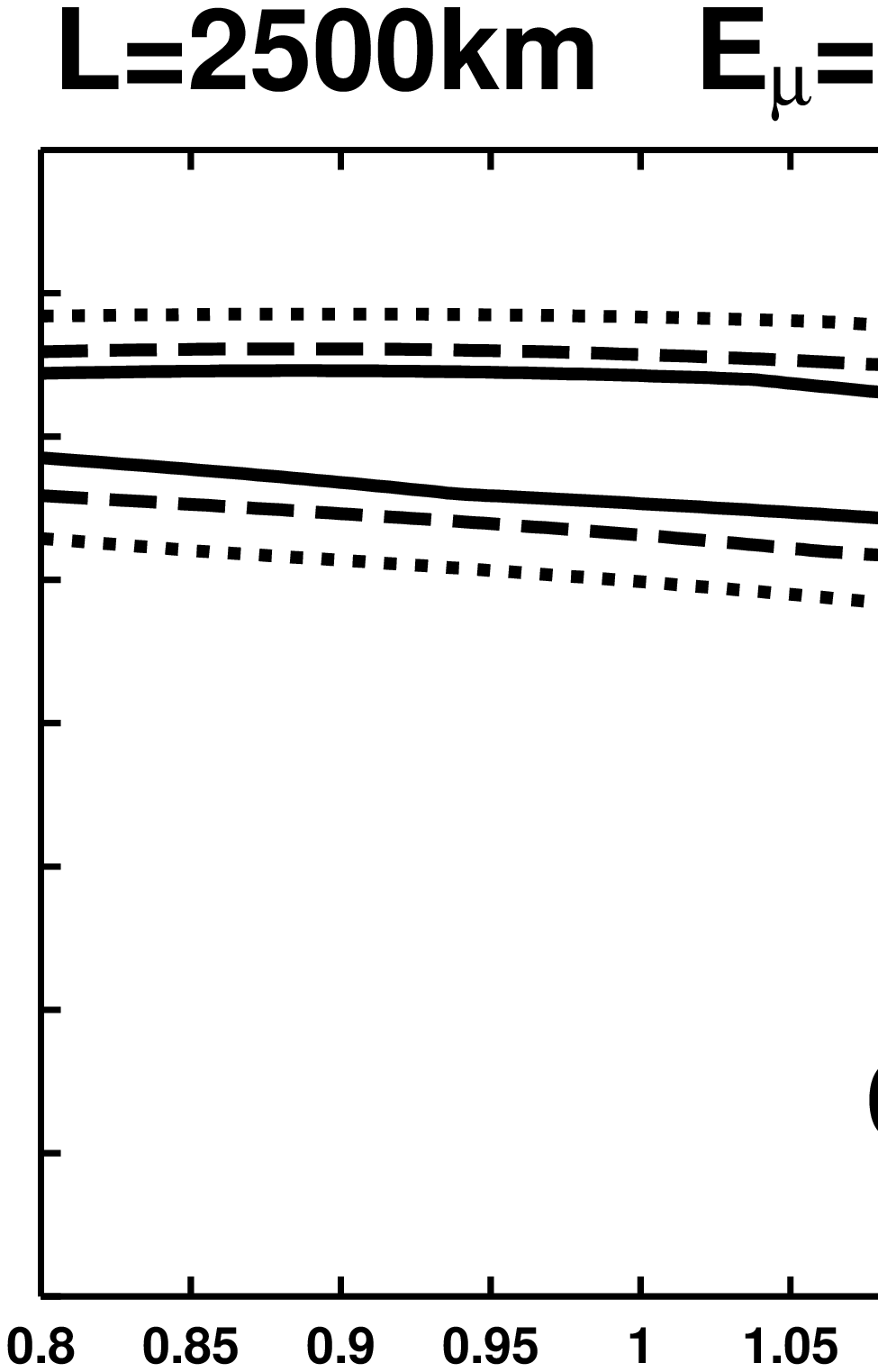,width=8cm}
\vglue -8.1cm \hglue 1.3cm \epsfig{file=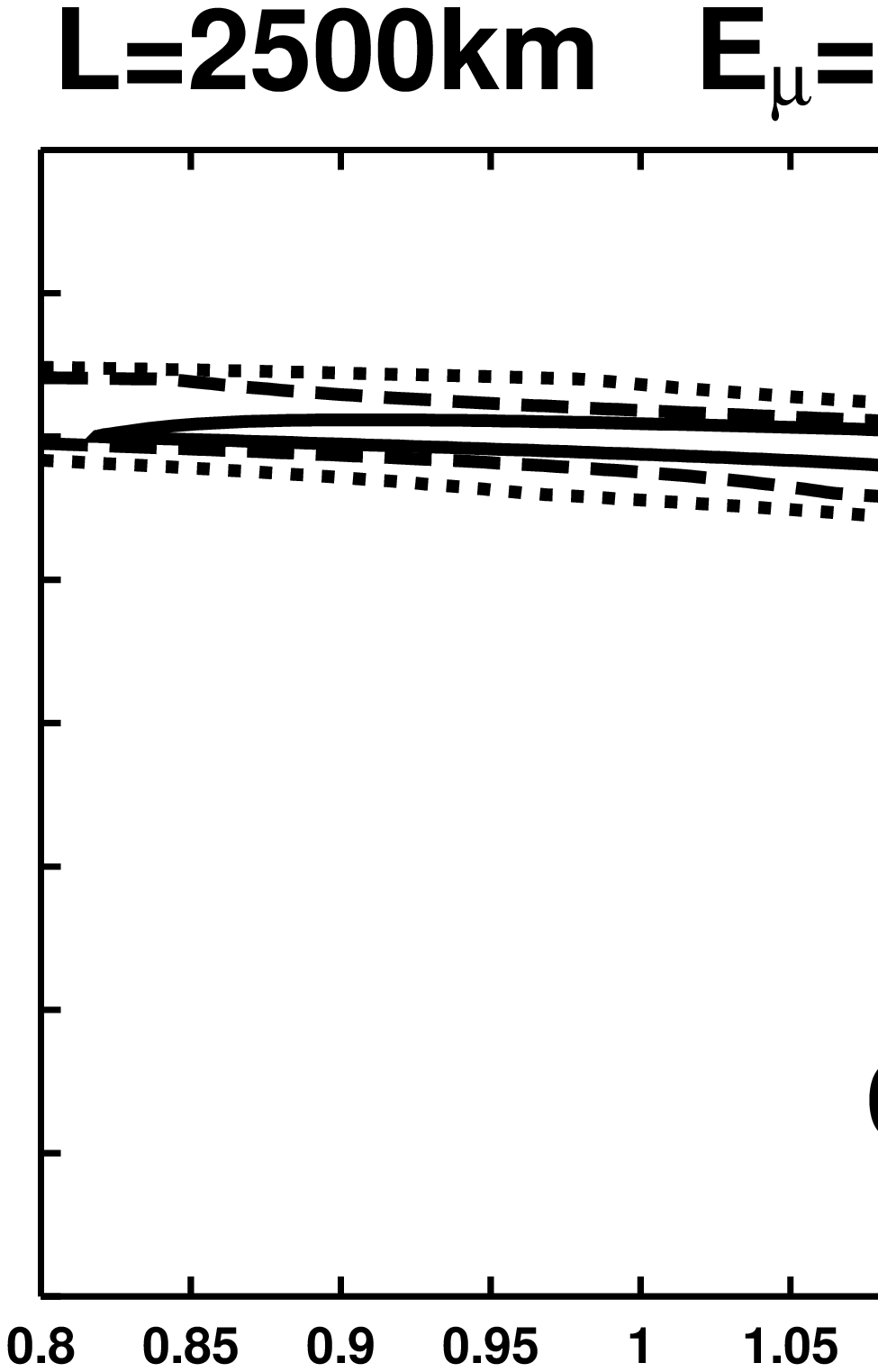,width=8cm}

\vglue -2.4cm
\hglue -4.0cm 
\epsfig{file=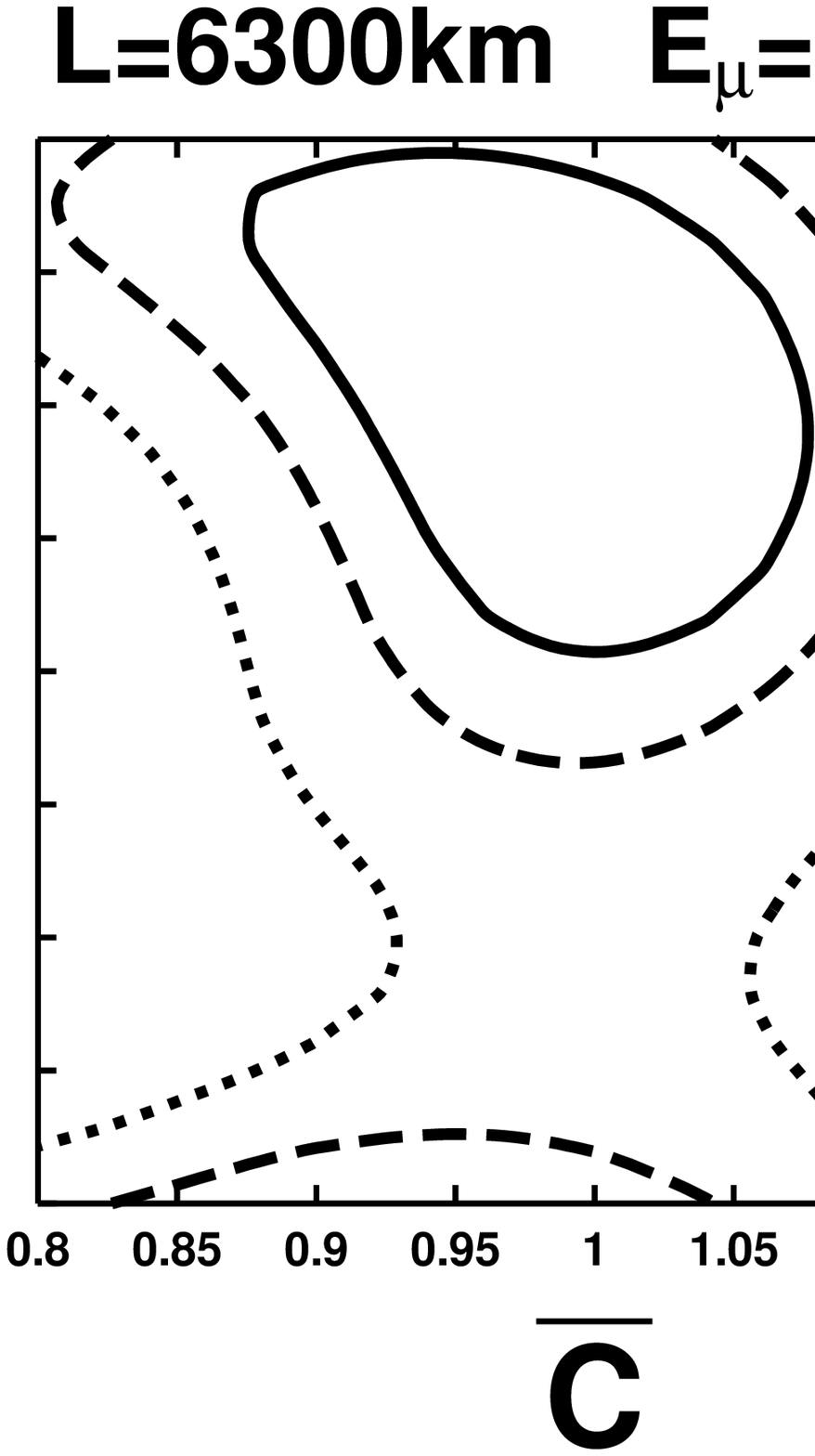,width=8cm}
\vglue -8.1cm \hglue 1.3cm \epsfig{file=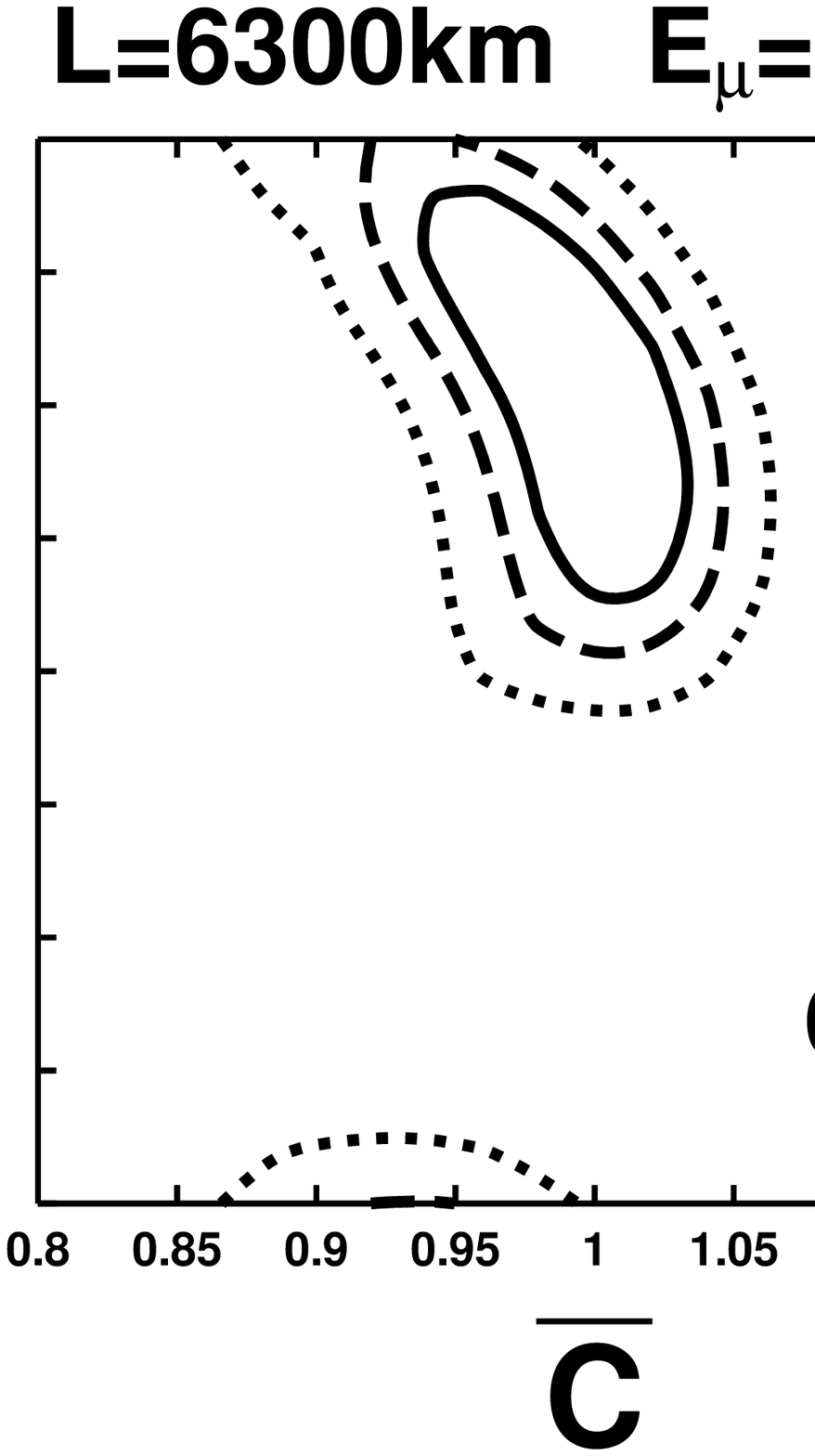,width=8cm}
\vglue -2.0cm\hglue -22.3cm
\epsfig{file=fig0123.eps,width=22cm}
\vglue -21.5cm\hglue 7.3cm
\epsfig{file=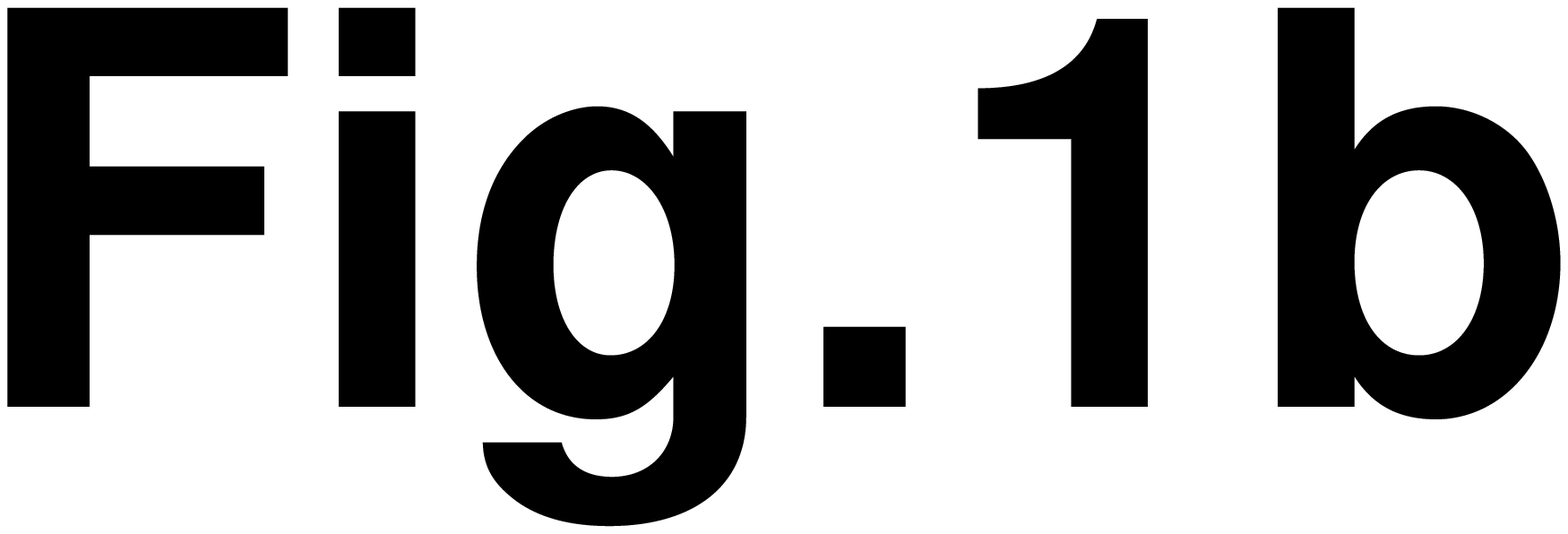,width=4cm}
\newpage
\vglue -2.5cm
\hglue -6.0cm 
\epsfig{file=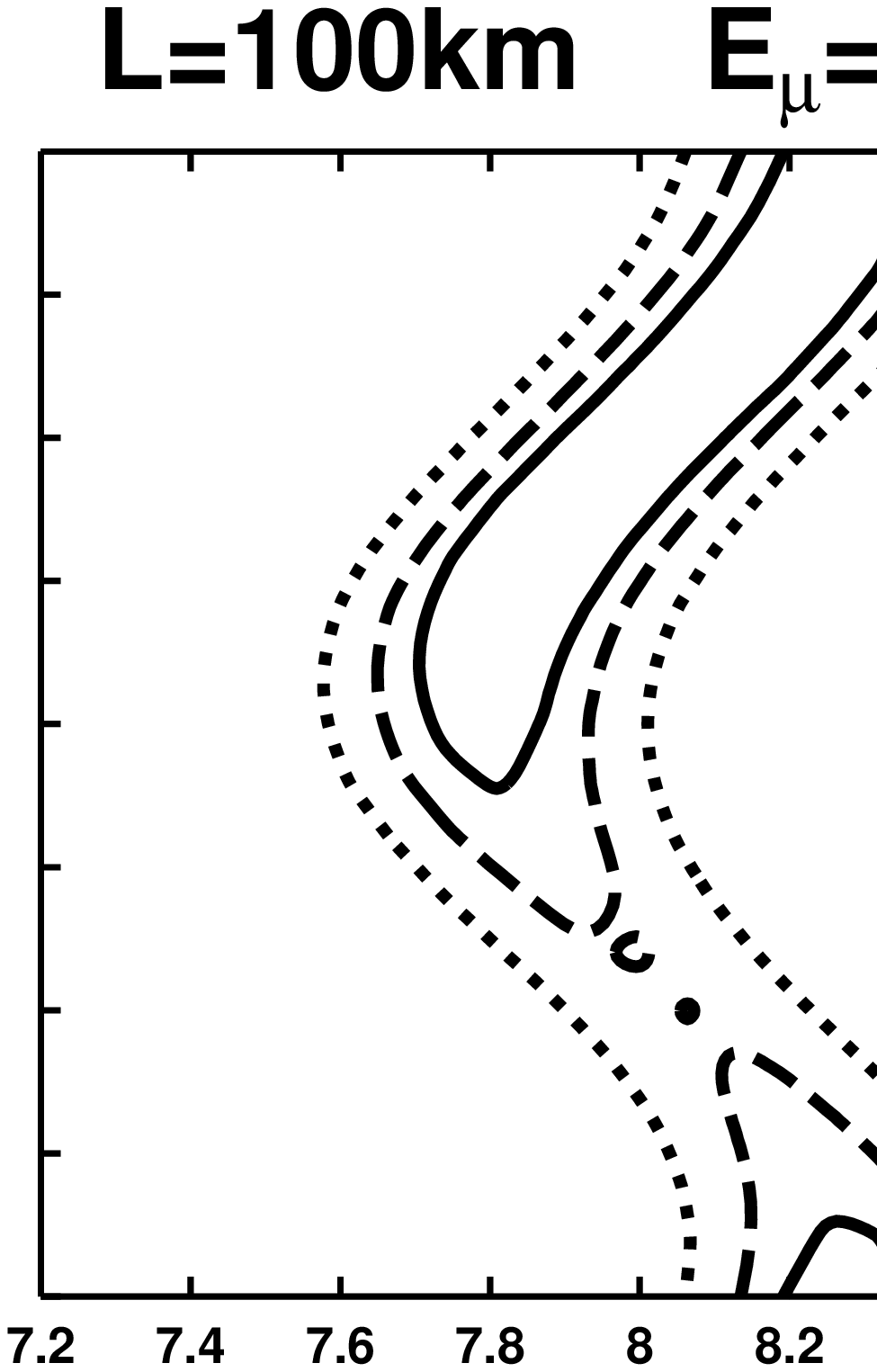,width=8cm}
\vglue -8.1cm \hglue -0.7cm \epsfig{file=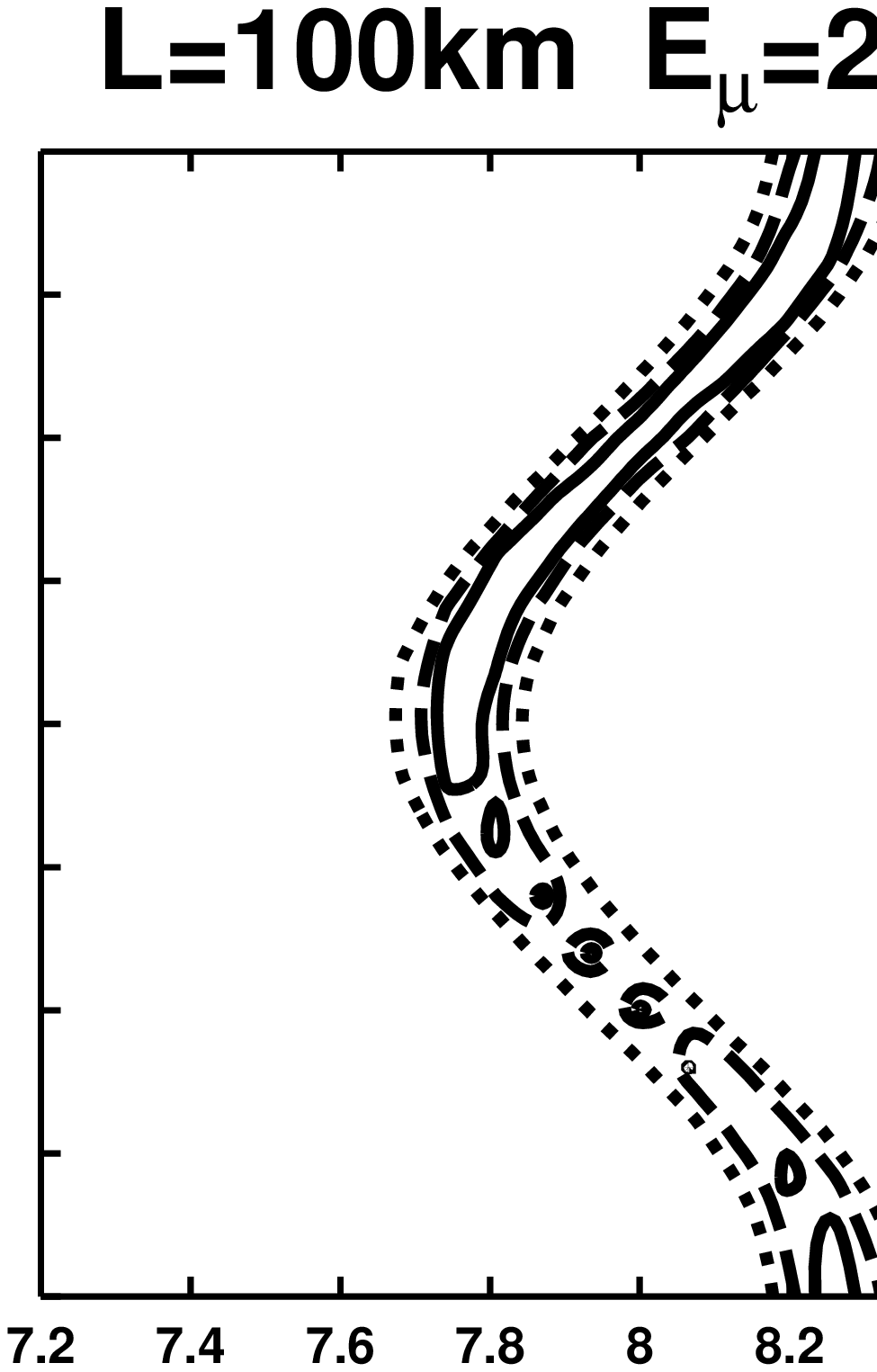,width=8cm}
\vglue -8.1cm \hglue 4.5cm \epsfig{file=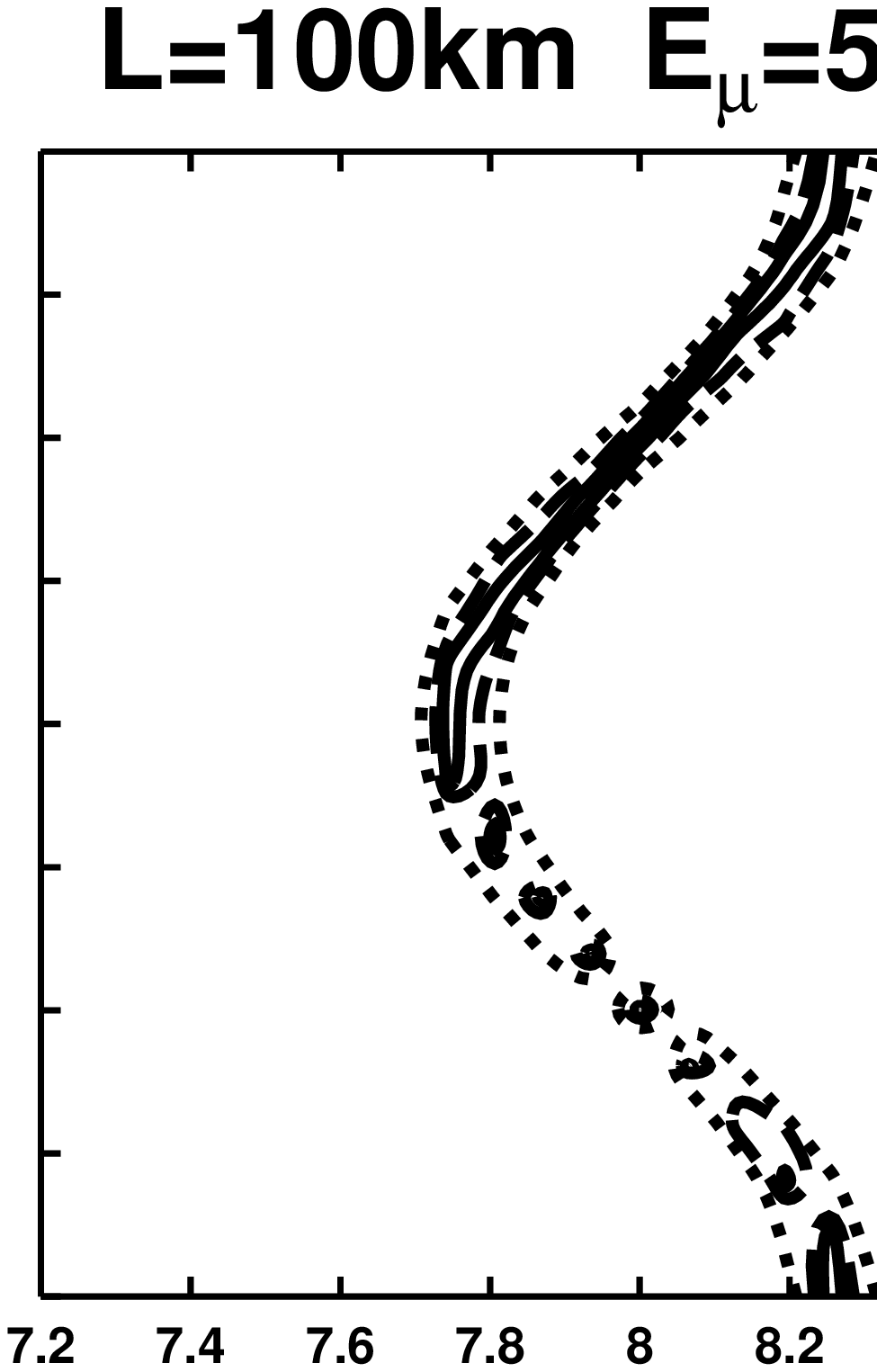,width=8cm}

\vglue -2.4cm
\hglue -6.0cm 
\epsfig{file=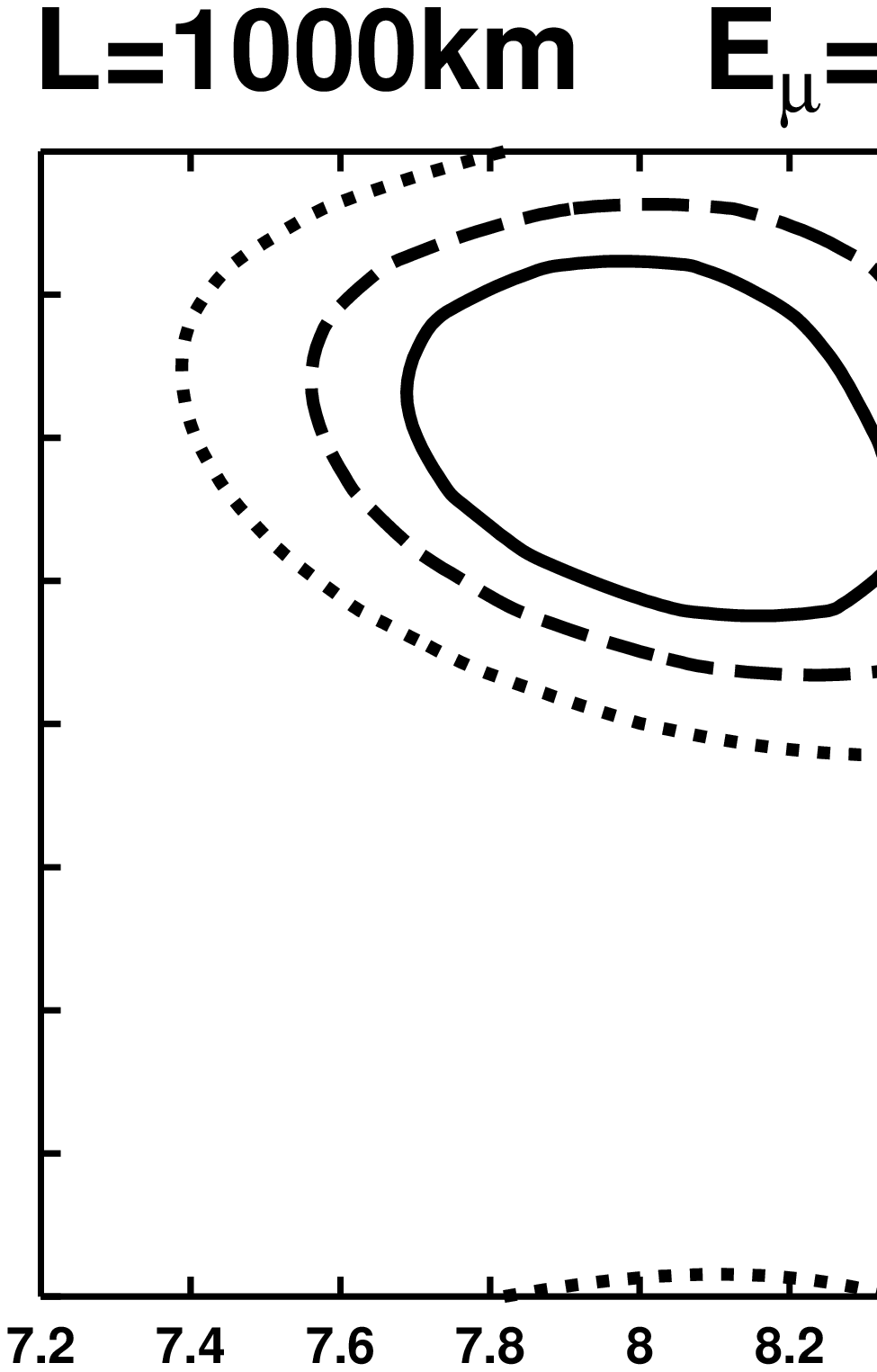,width=8cm}
\vglue -8.1cm \hglue -0.7cm \epsfig{file=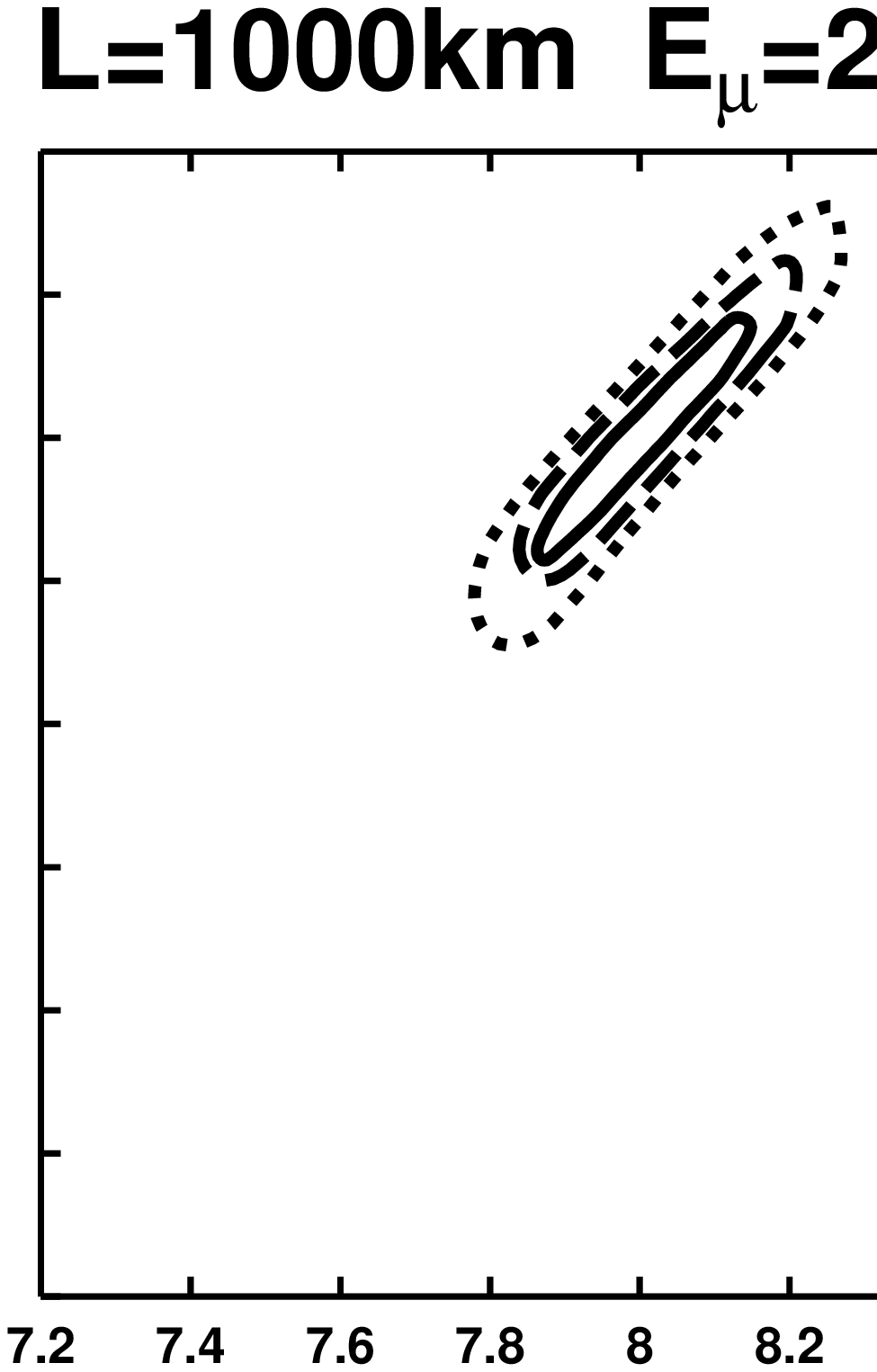,width=8cm}
\vglue -8.1cm \hglue 4.5cm \epsfig{file=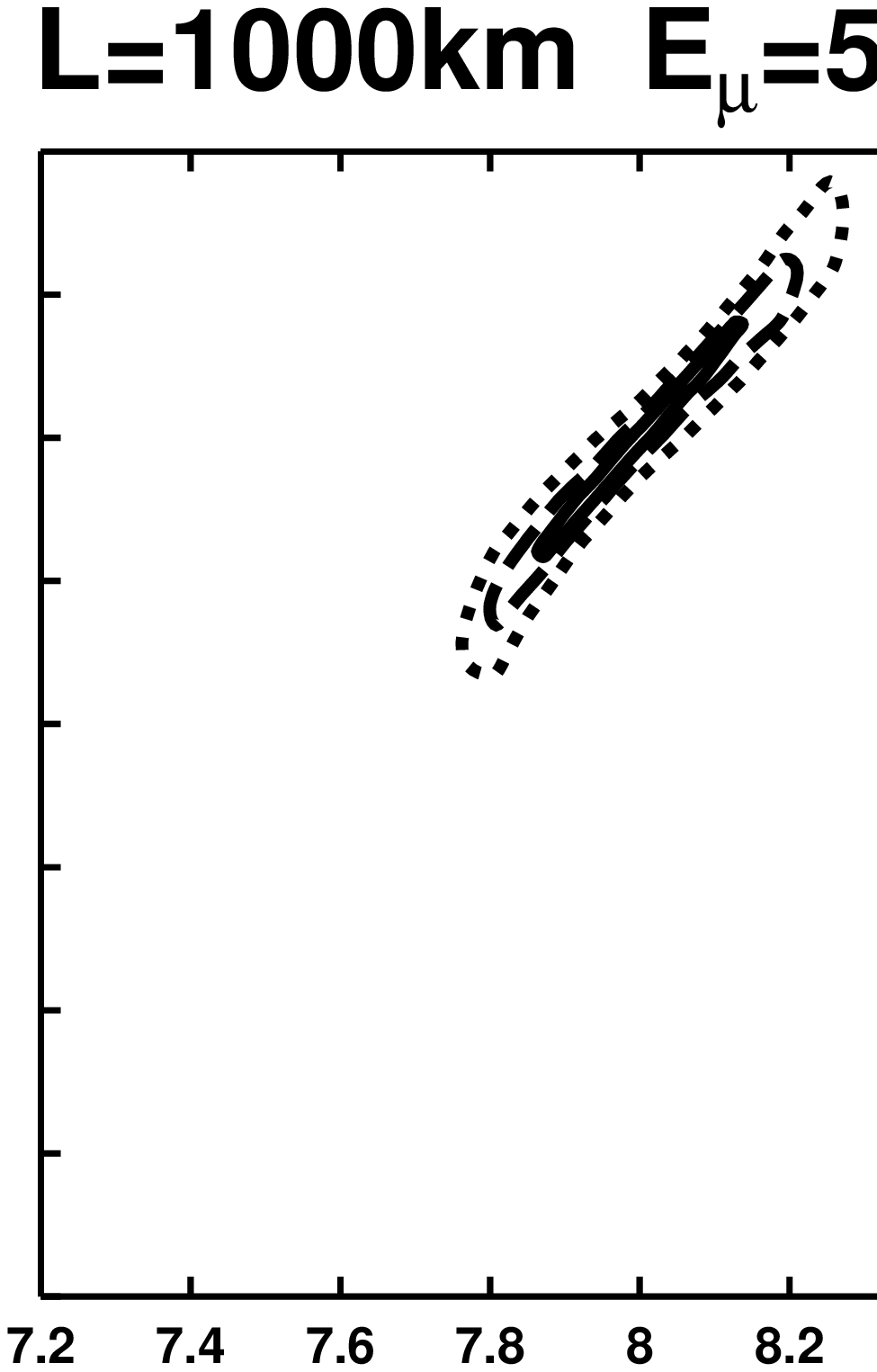,width=8cm}

\vglue -2.4cm
\hglue -6.0cm 
\epsfig{file=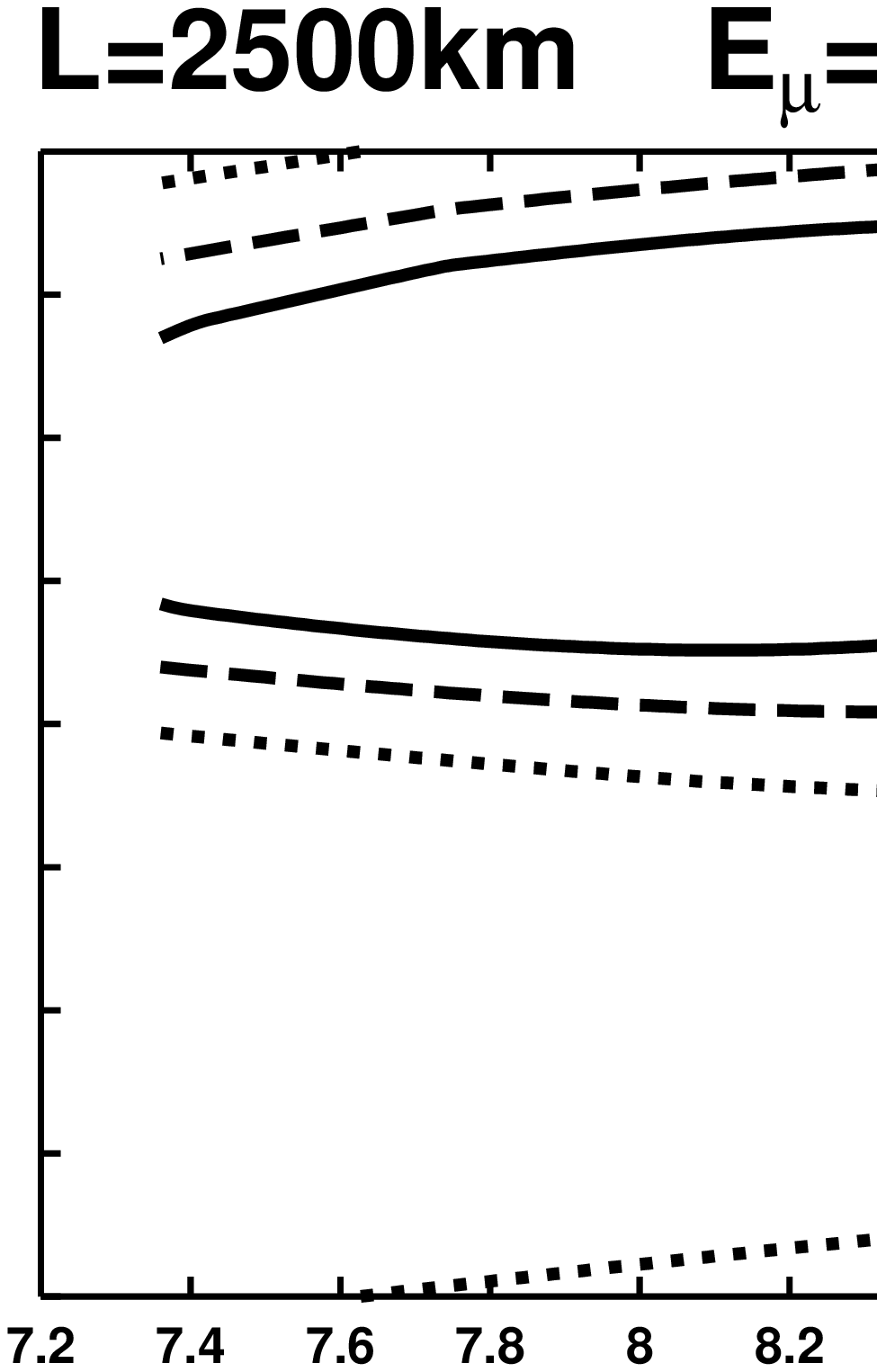,width=8cm}
\vglue -8.1cm \hglue -0.7cm \epsfig{file=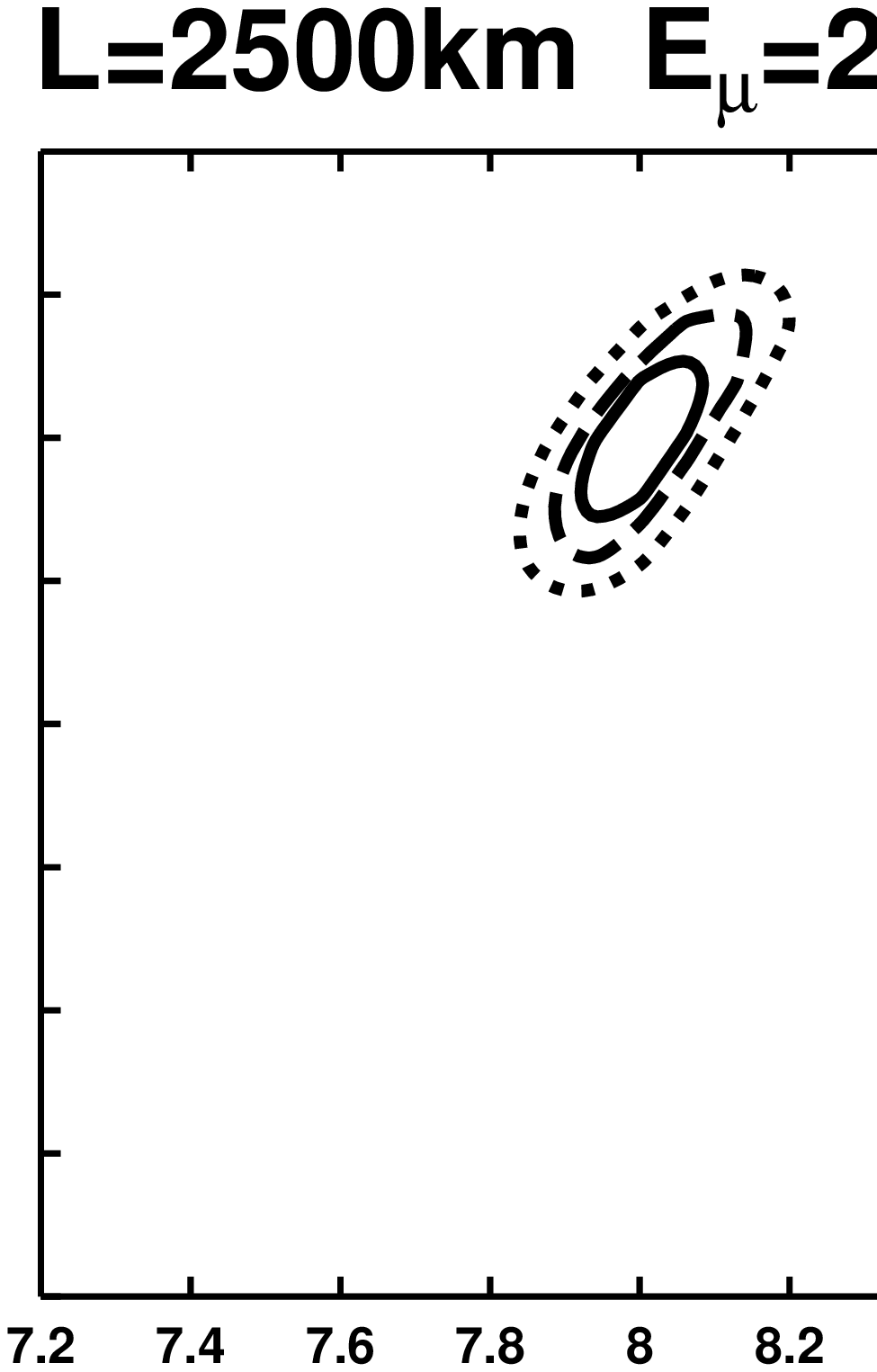,width=8cm}
\vglue -8.1cm \hglue 4.5cm \epsfig{file=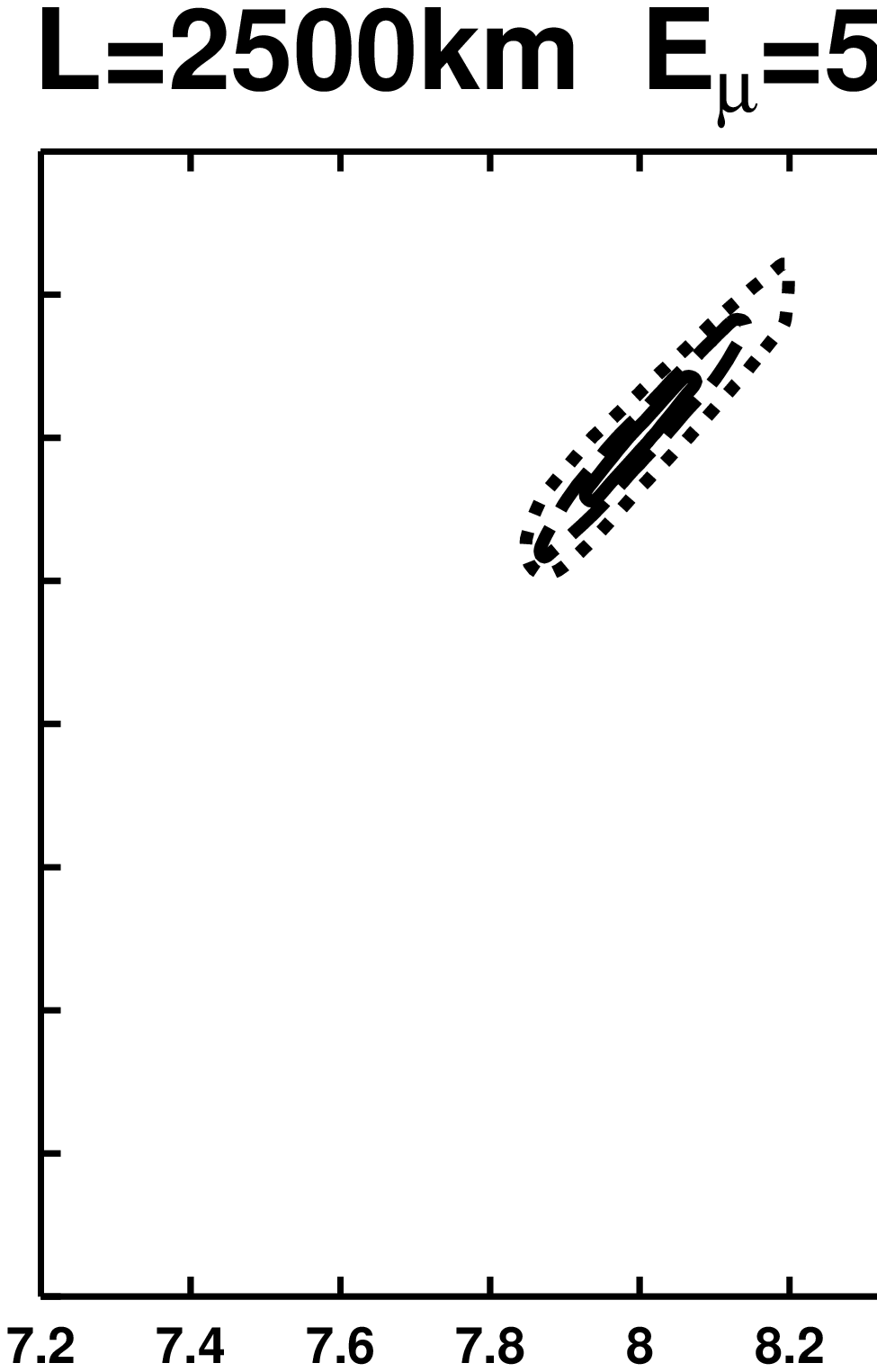,width=8cm}

\vglue -2.4cm
\hglue -6.0cm 
\epsfig{file=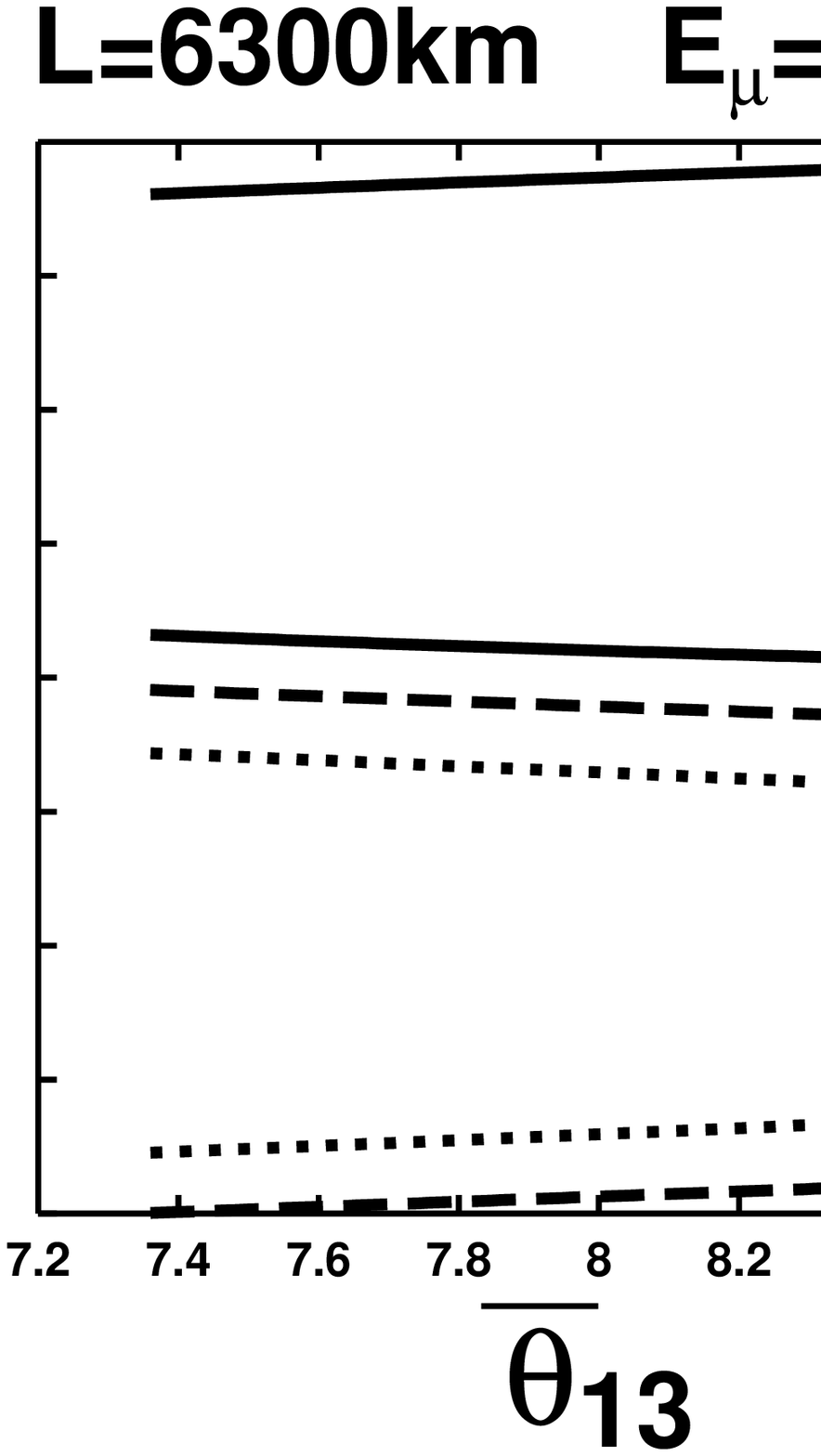,width=8cm}
\vglue -8.1cm \hglue -0.7cm \epsfig{file=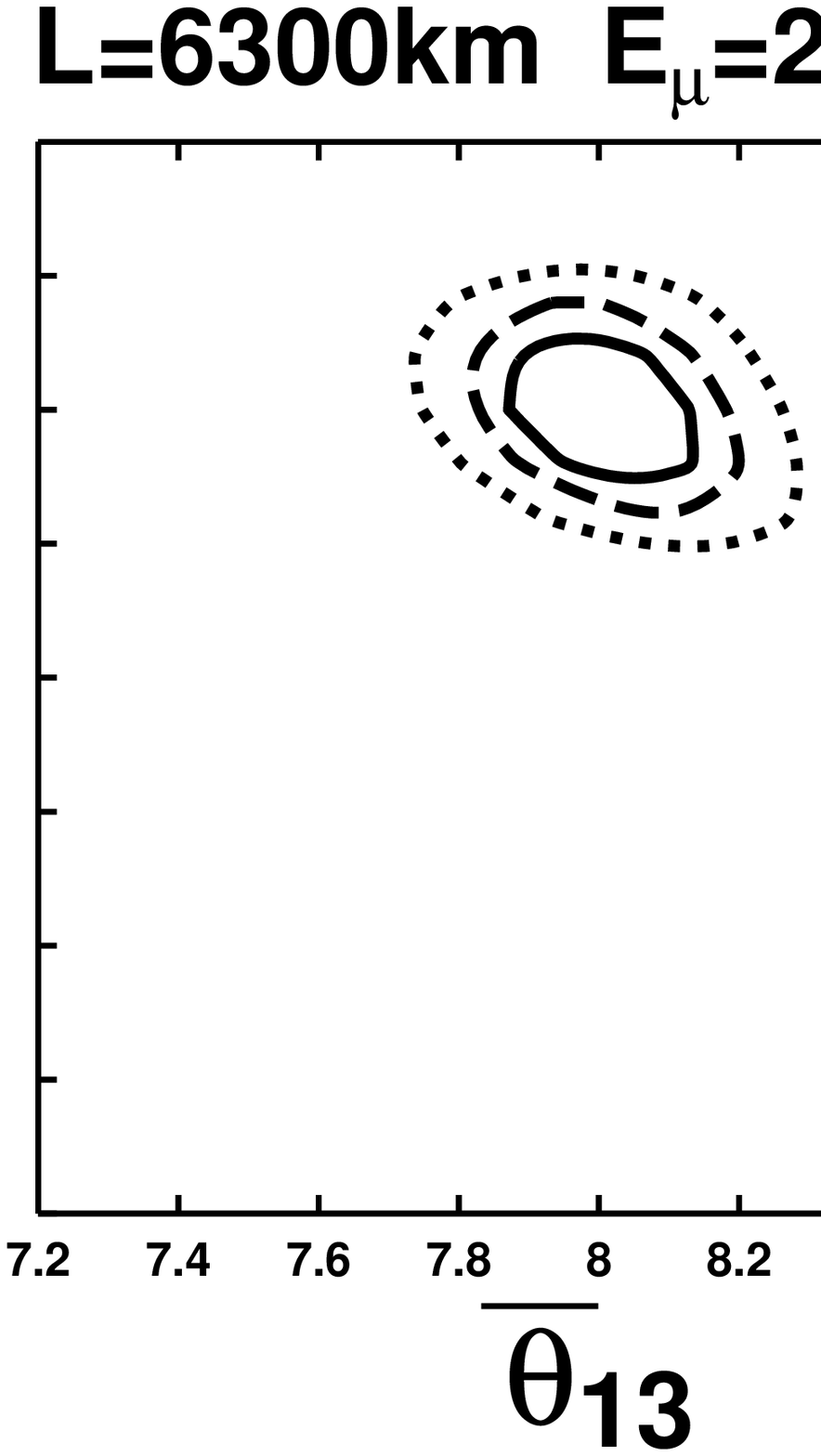,width=8cm}
\vglue -8.1cm \hglue 4.5cm \epsfig{file=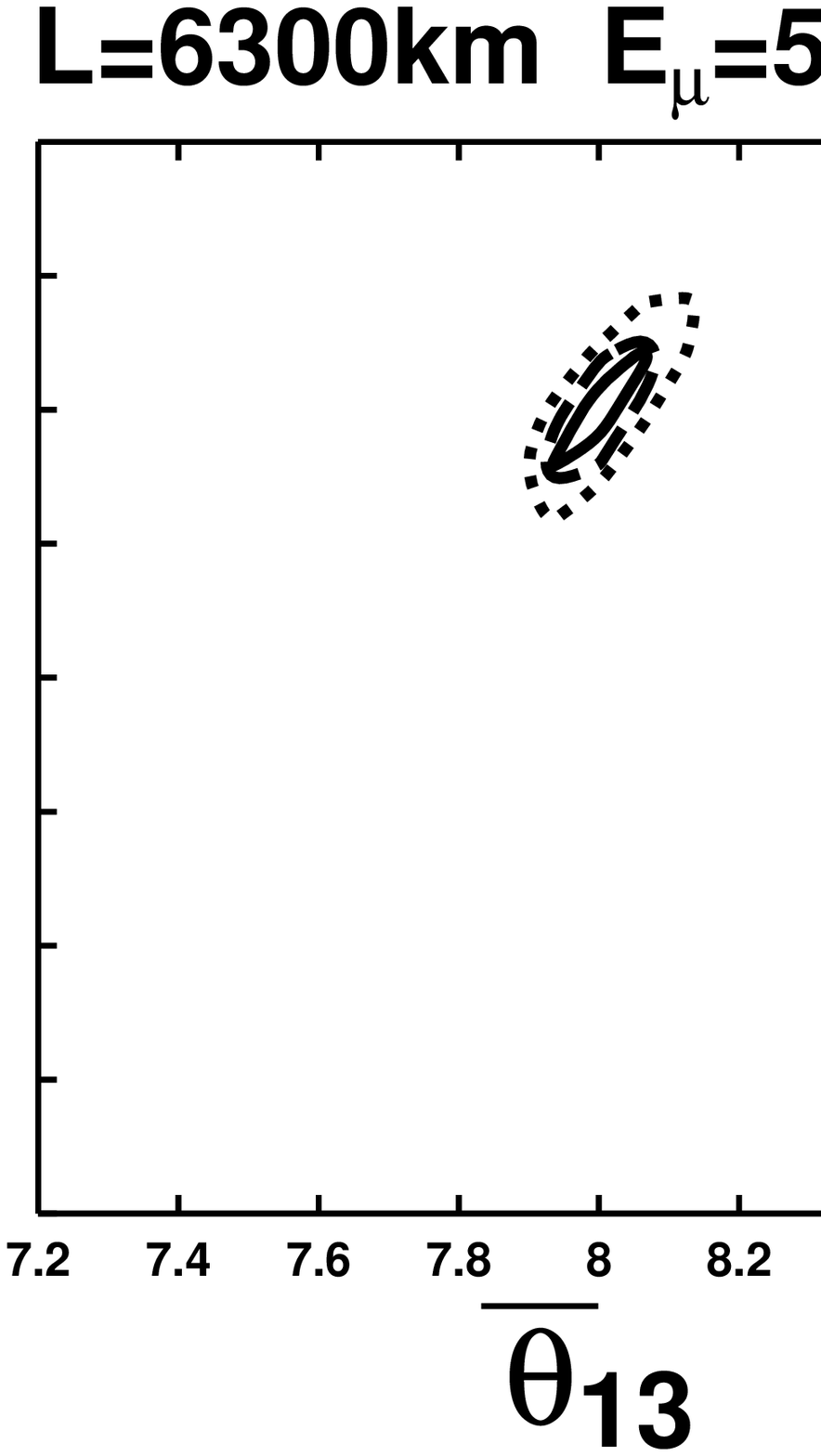,width=8cm}
\vglue -2.0cm\hglue -23.3cm
\epsfig{file=fig0213.eps,width=22cm}
\vglue -21.5cm\hglue 6.3cm
\epsfig{file=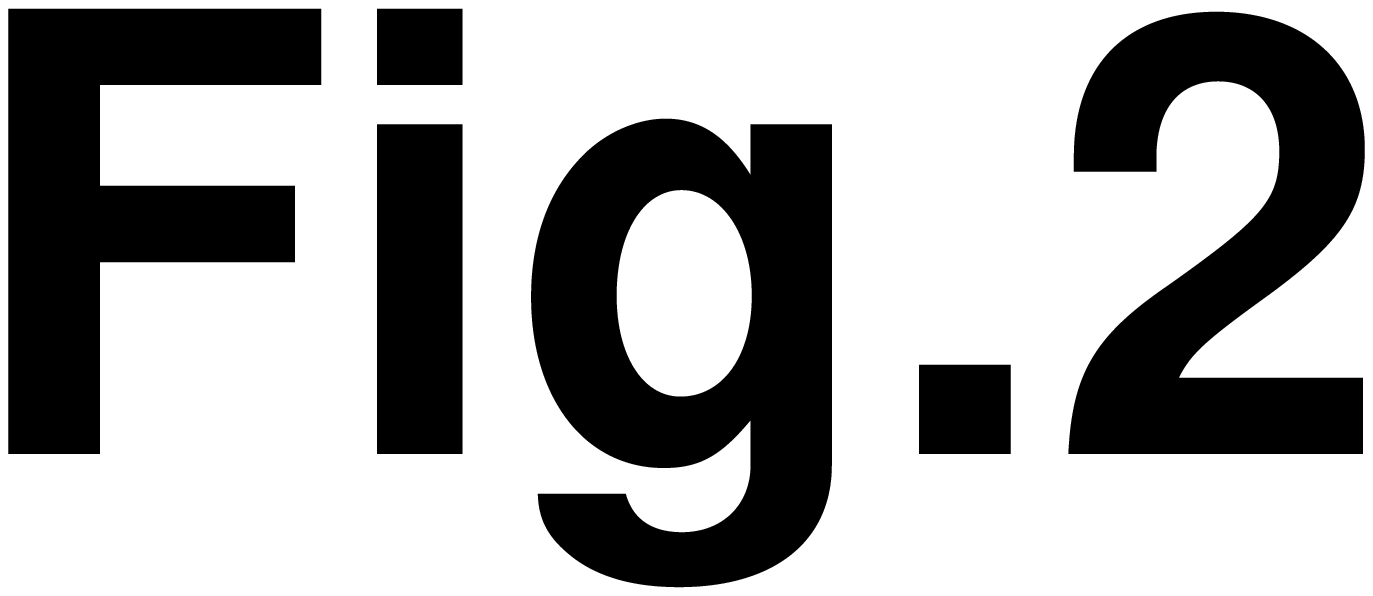,width=4cm}
\newpage
\vglue -2.5cm
\hglue -6.0cm 
\epsfig{file=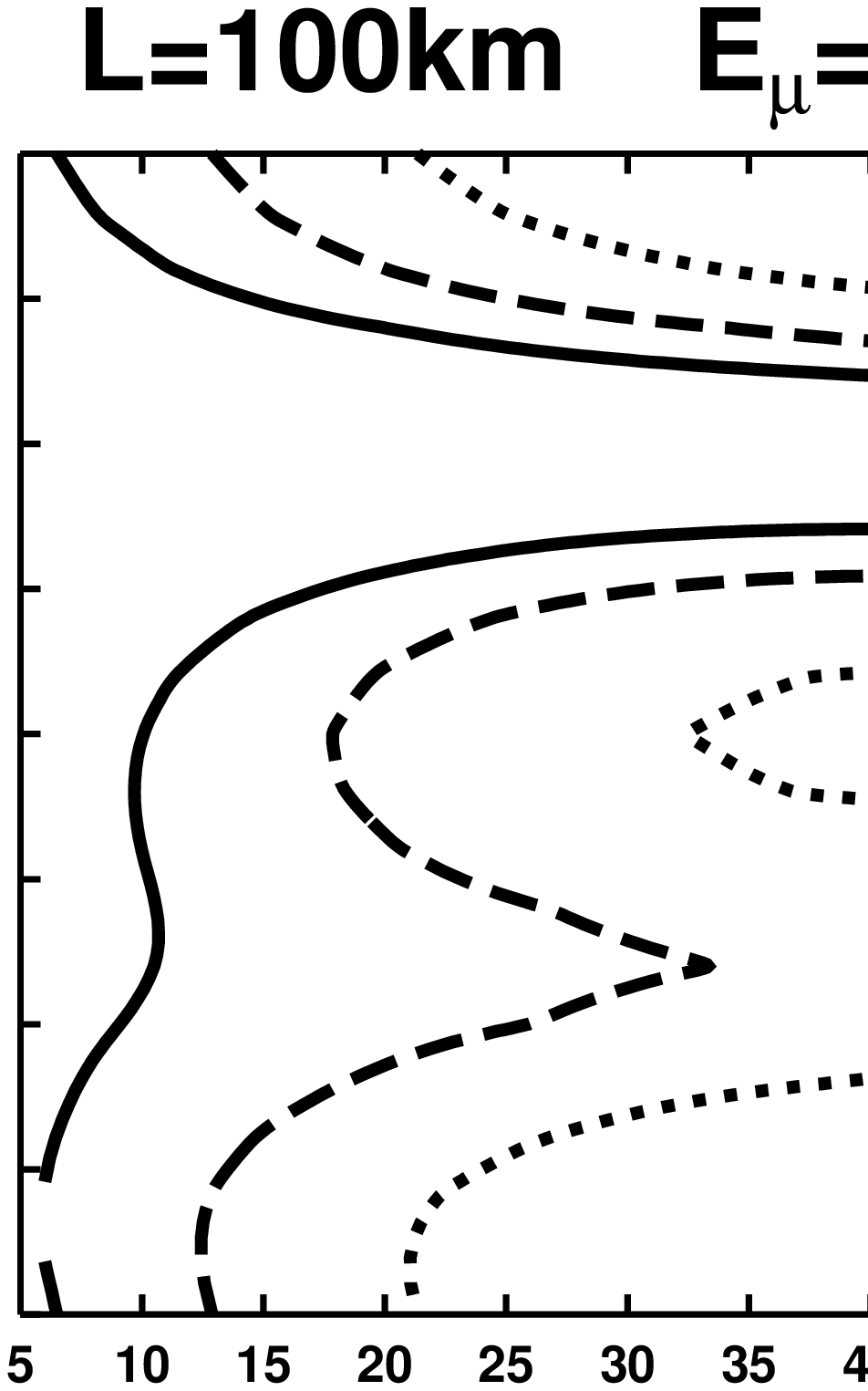,width=8cm}
\vglue -8.1cm \hglue -0.7cm \epsfig{file=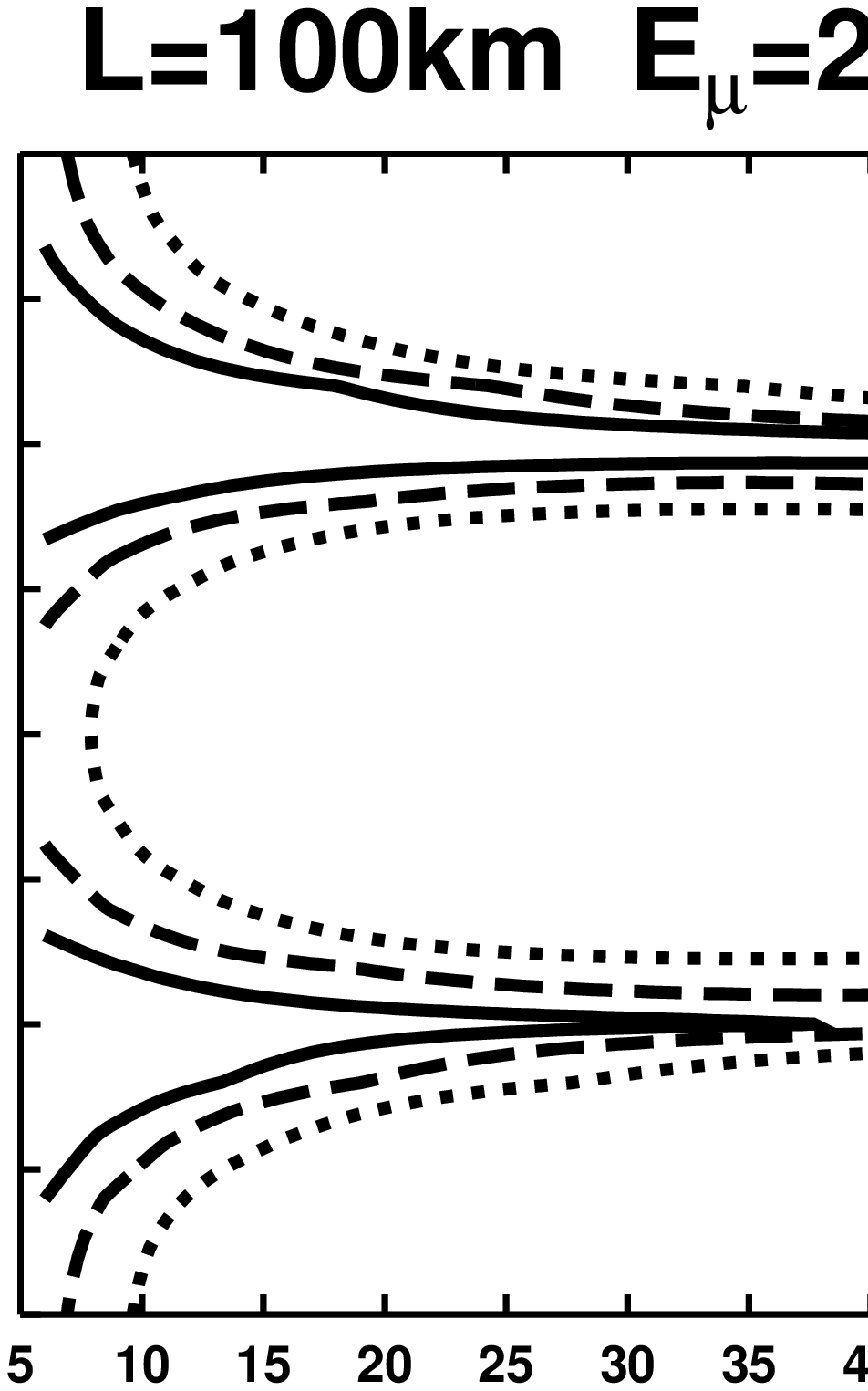,width=8cm}
\vglue -8.1cm \hglue 4.5cm \epsfig{file=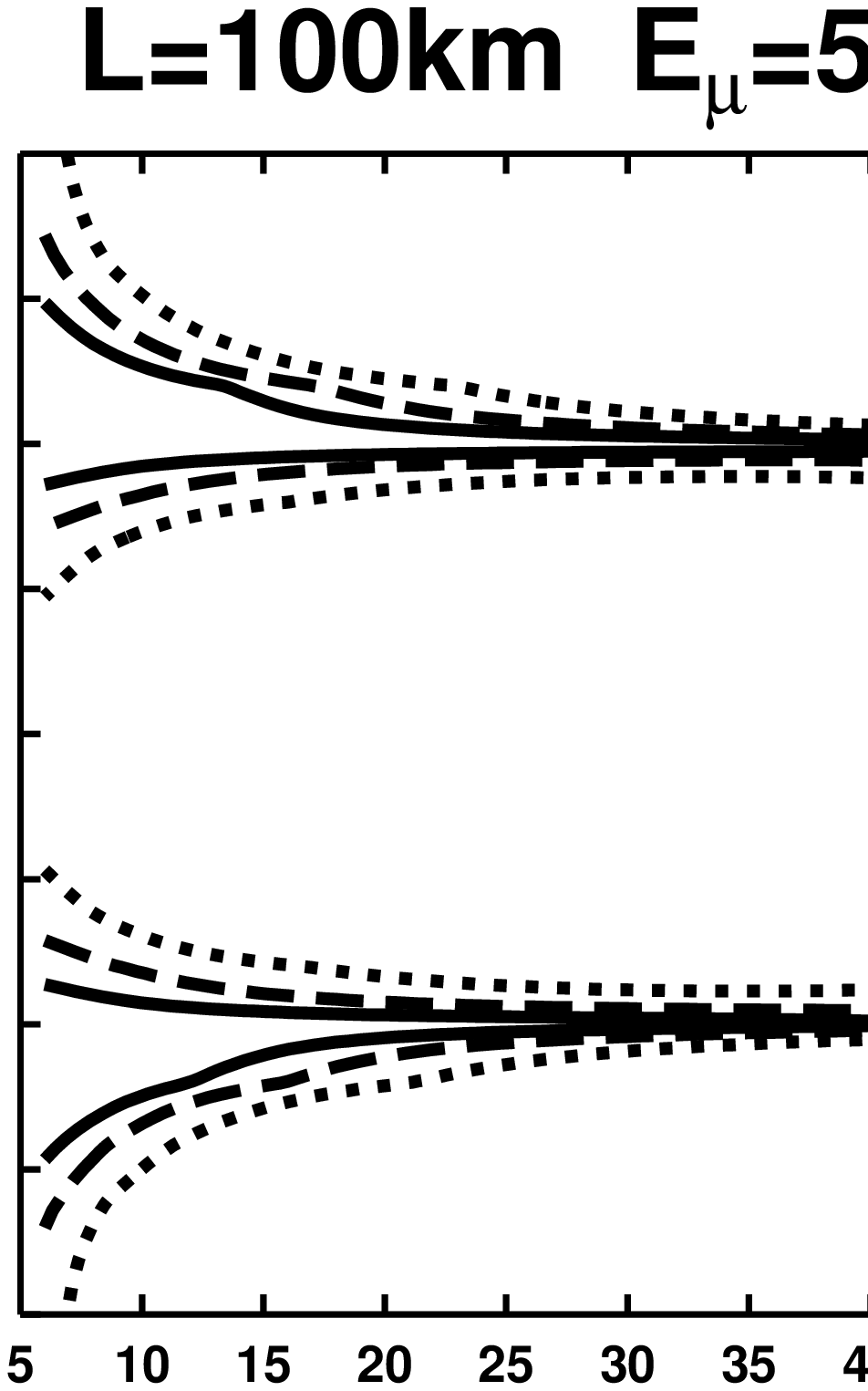,width=8cm}

\vglue -2.4cm
\hglue -6.0cm 
\epsfig{file=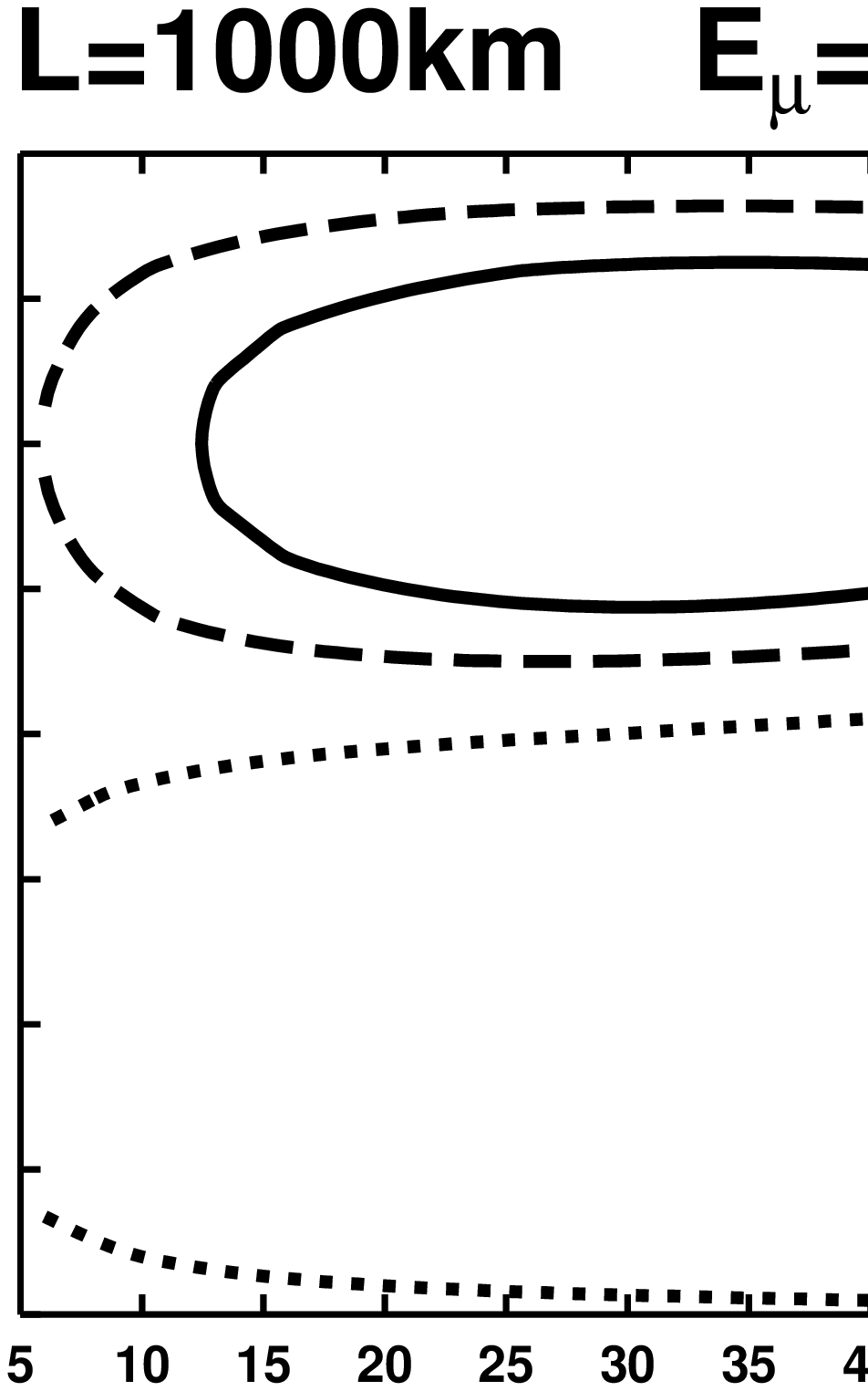,width=8cm}
\vglue -8.1cm \hglue -0.7cm \epsfig{file=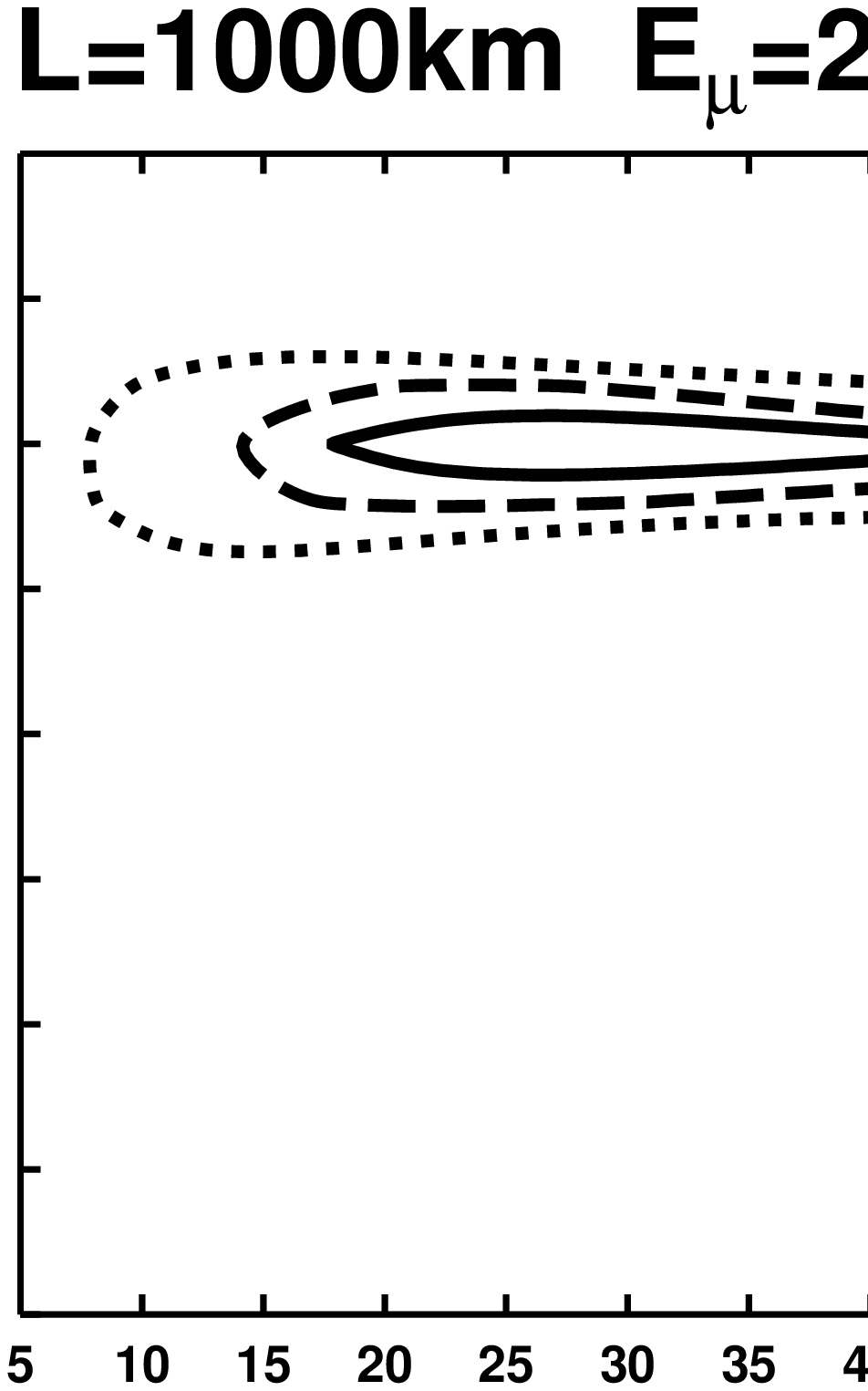,width=8cm}
\vglue -8.1cm \hglue 4.5cm \epsfig{file=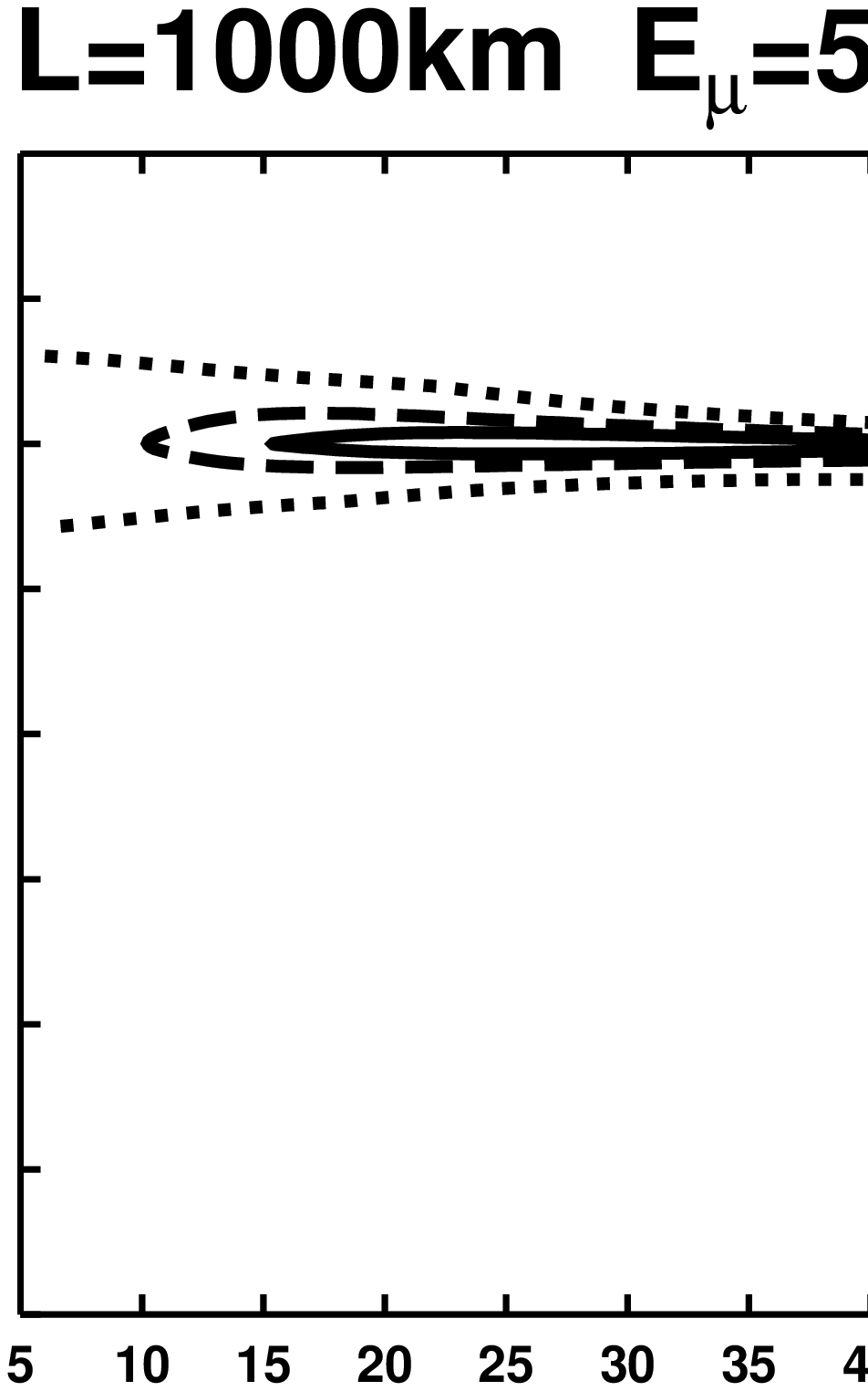,width=8cm}

\vglue -2.4cm
\hglue -6.0cm 
\epsfig{file=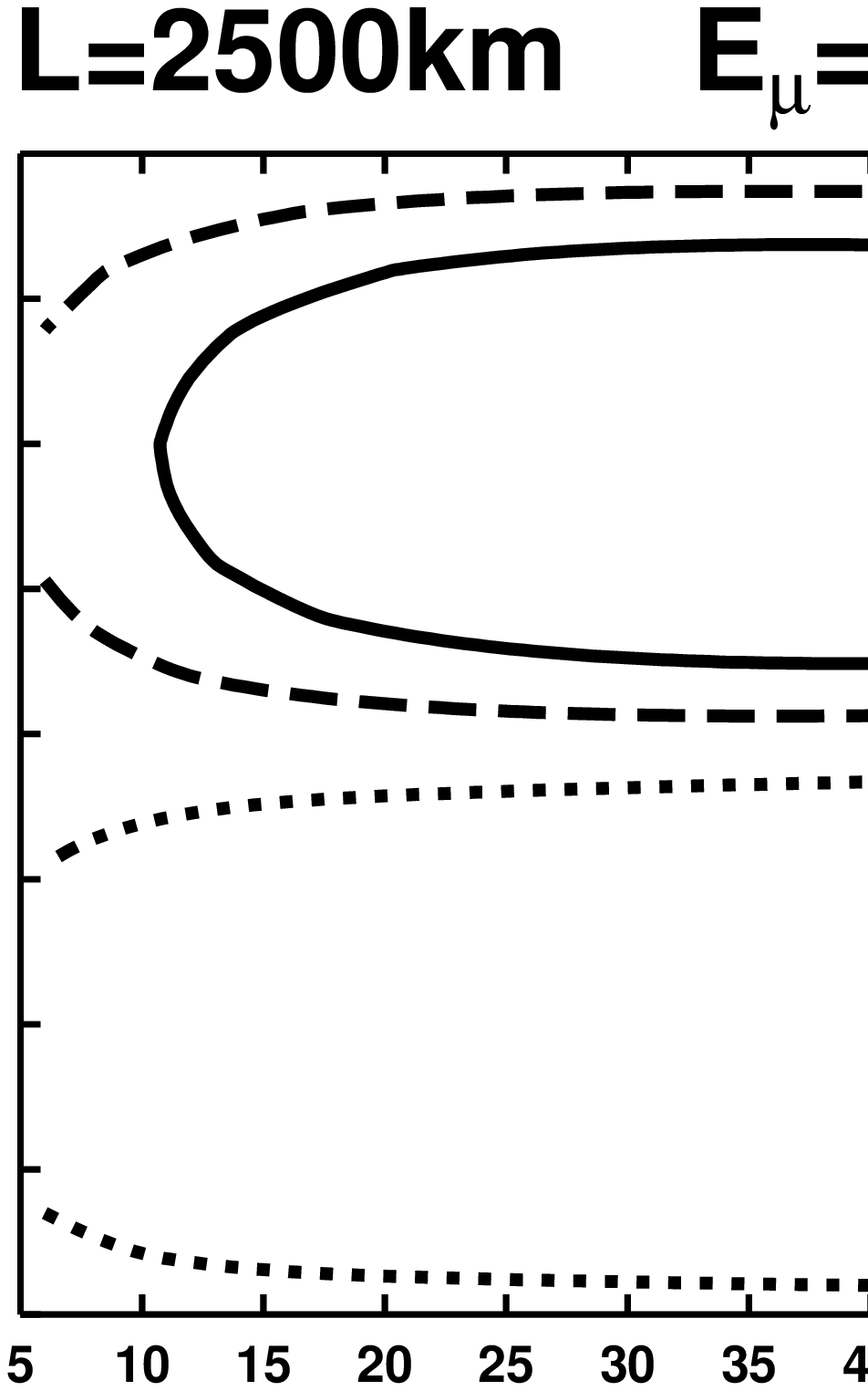,width=8cm}
\vglue -8.1cm \hglue -0.7cm \epsfig{file=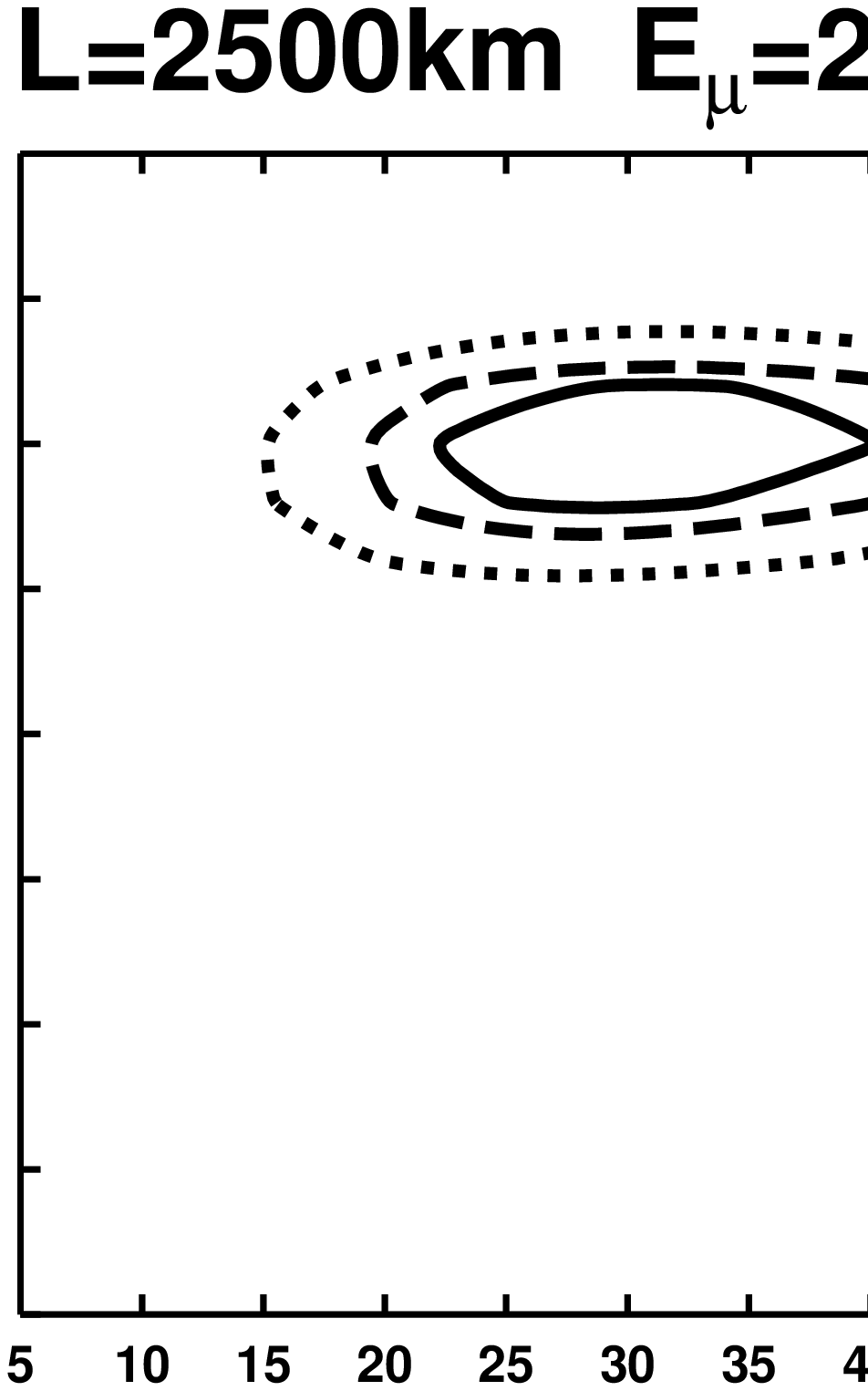,width=8cm}
\vglue -8.1cm \hglue 4.5cm \epsfig{file=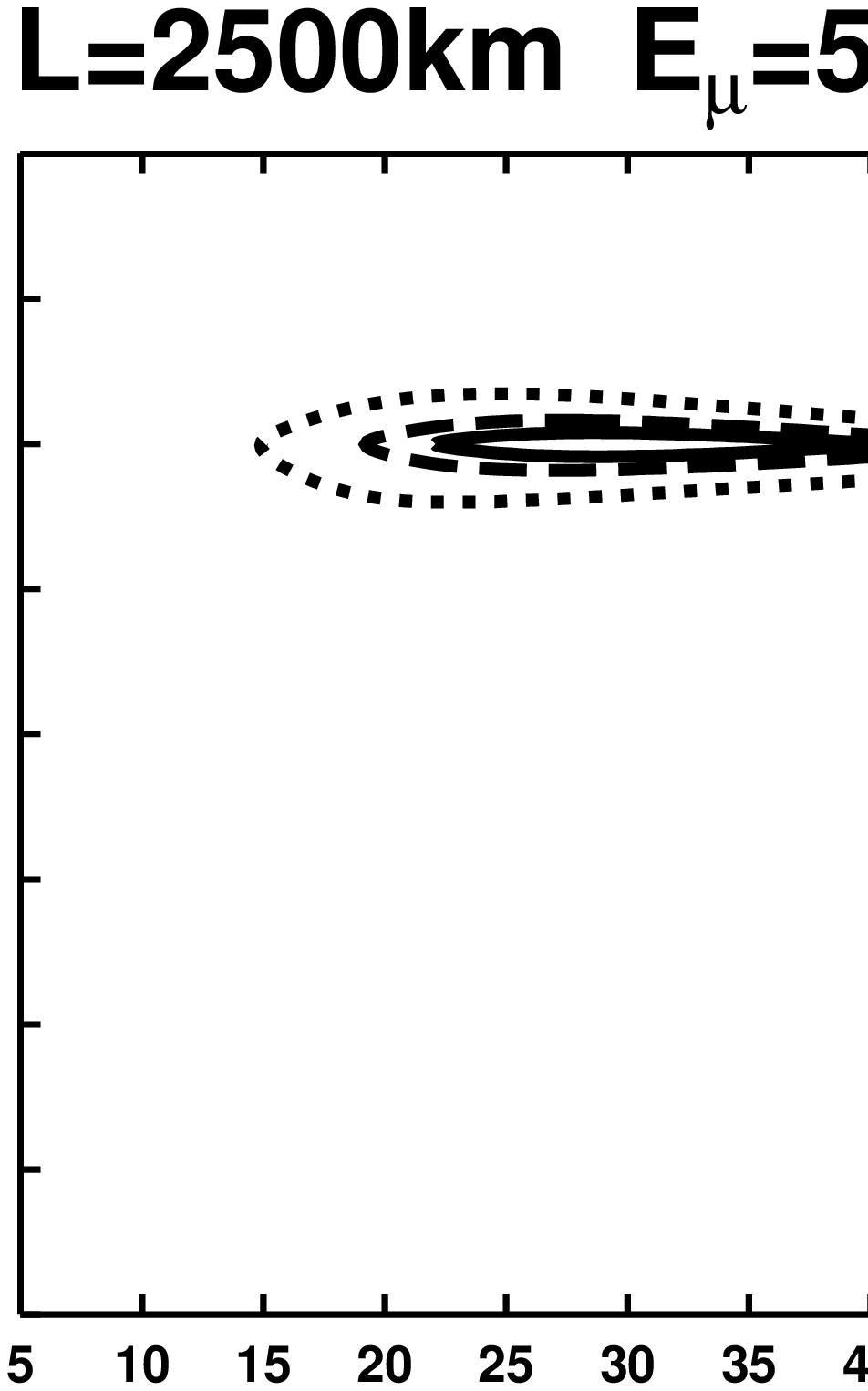,width=8cm}

\vglue -2.4cm
\hglue -6.0cm 
\epsfig{file=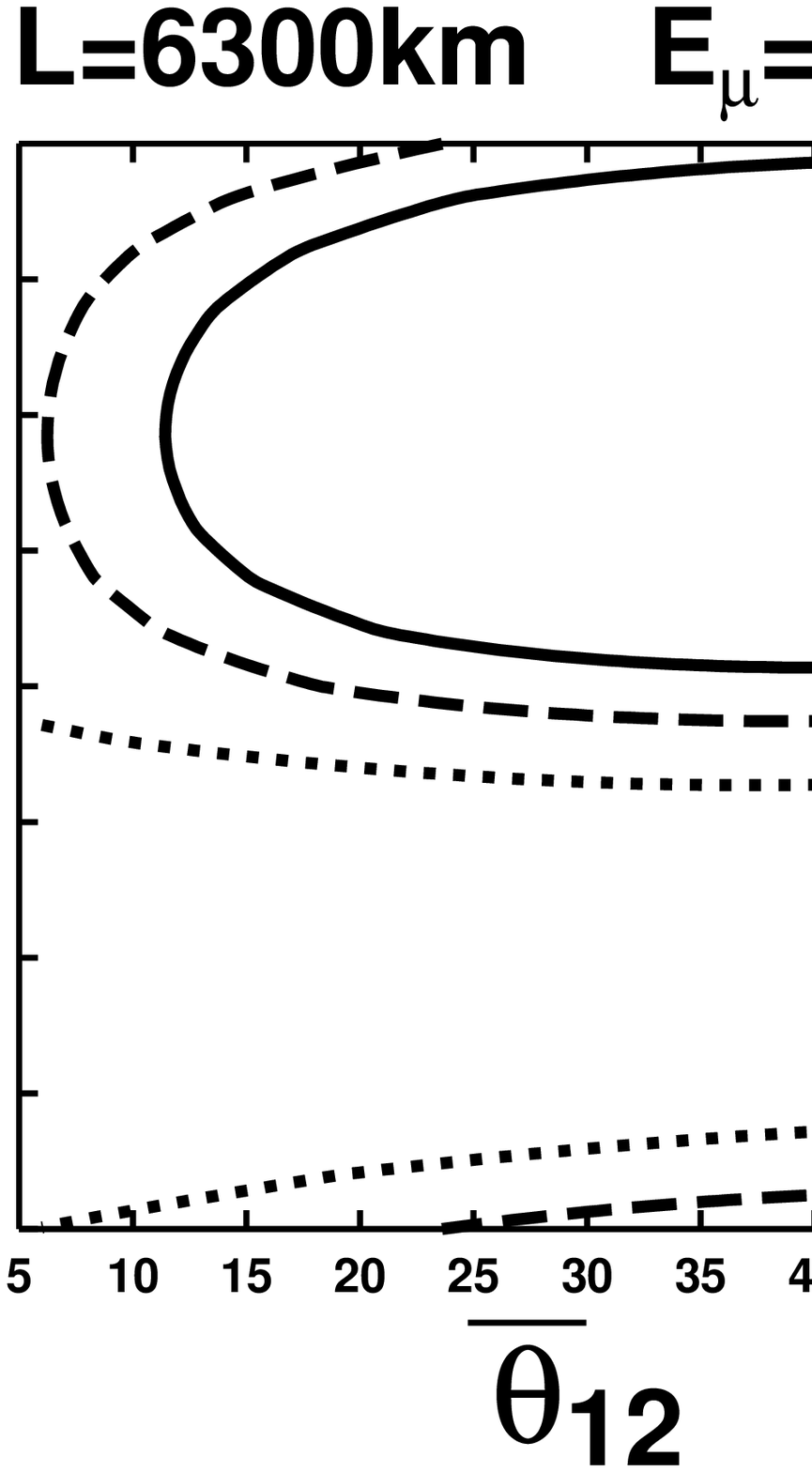,width=8cm}
\vglue -8.1cm \hglue -0.7cm \epsfig{file=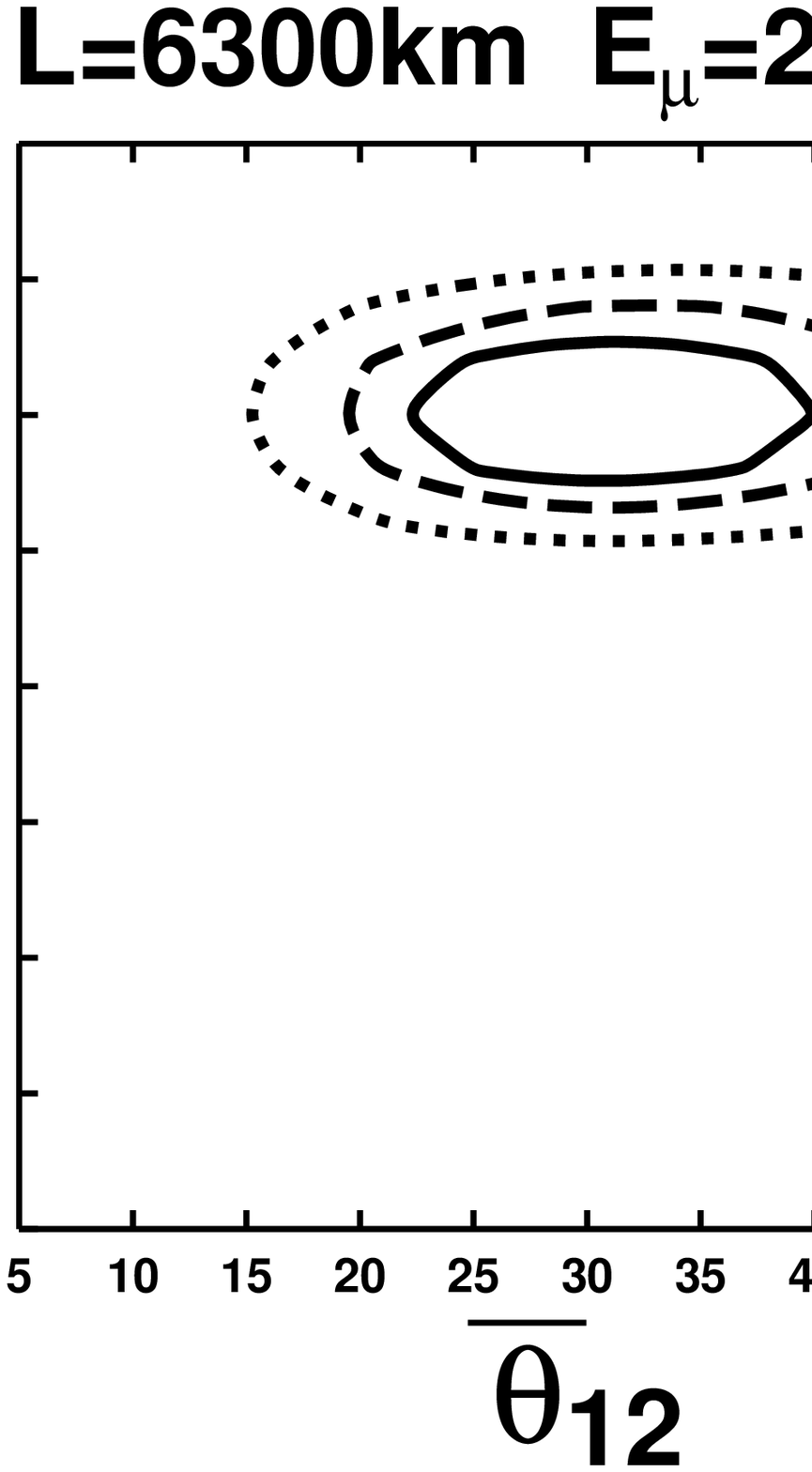,width=8cm}
\vglue -8.1cm \hglue 4.5cm \epsfig{file=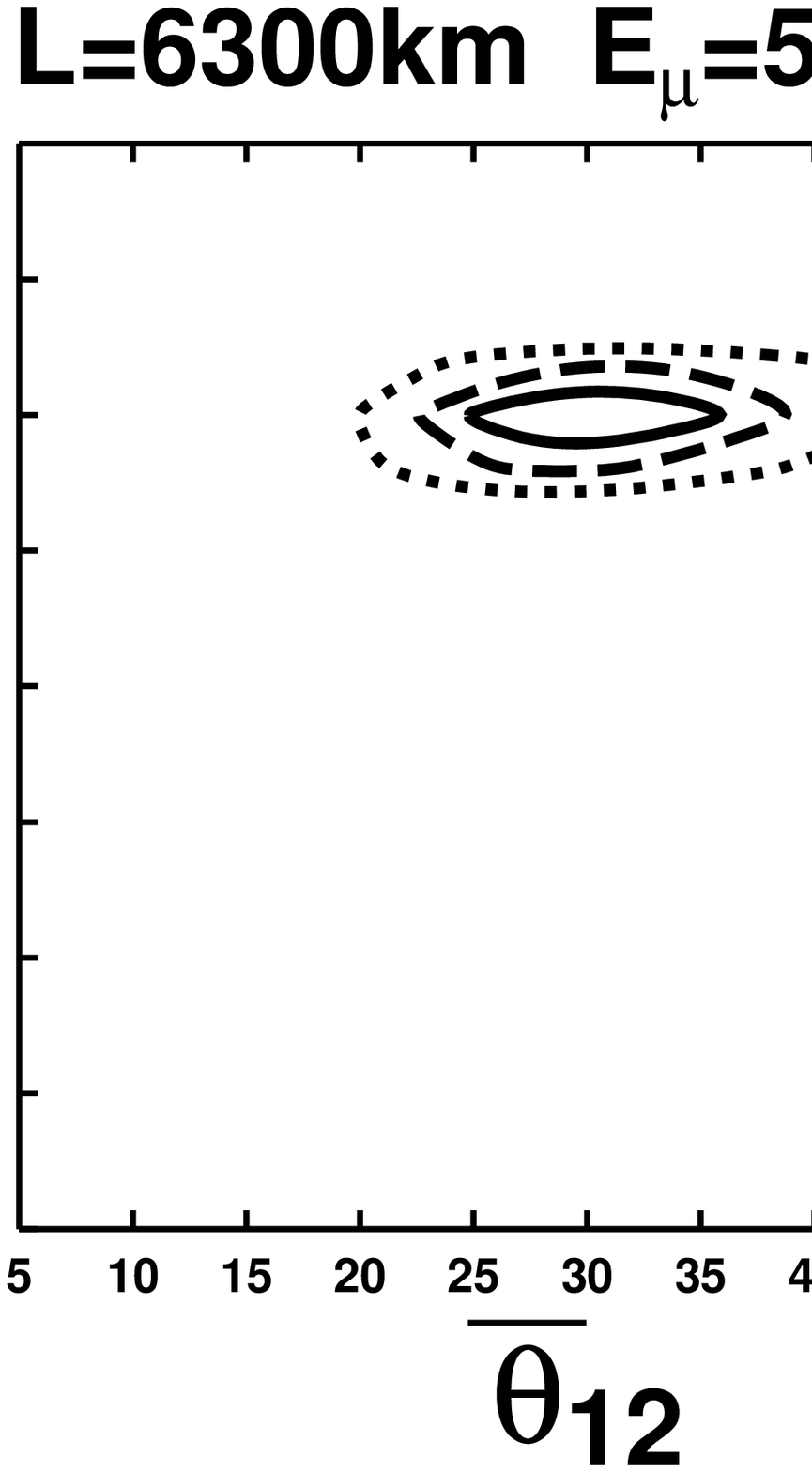,width=8cm}
\vglue -2.0cm\hglue -23.3cm
\epsfig{file=fig0313.eps,width=22cm}
\vglue -21.5cm\hglue 6.3cm
\epsfig{file=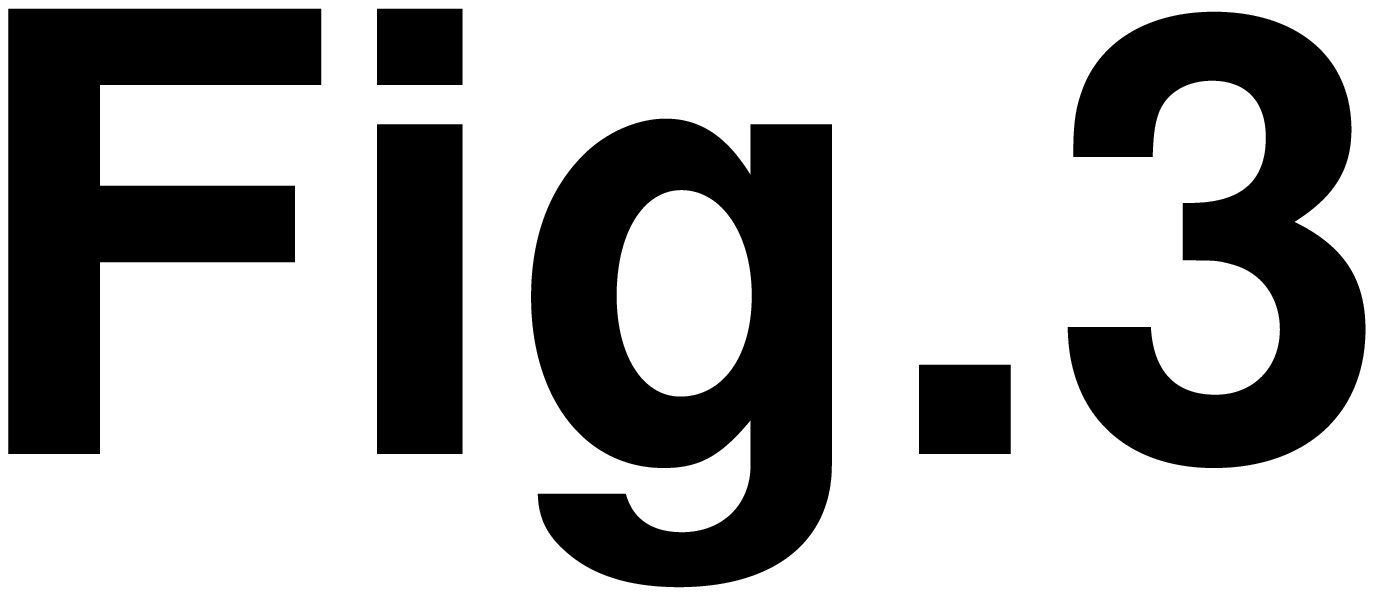,width=4cm}
\newpage
\vglue -2.5cm
\hglue -6.0cm 
\epsfig{file=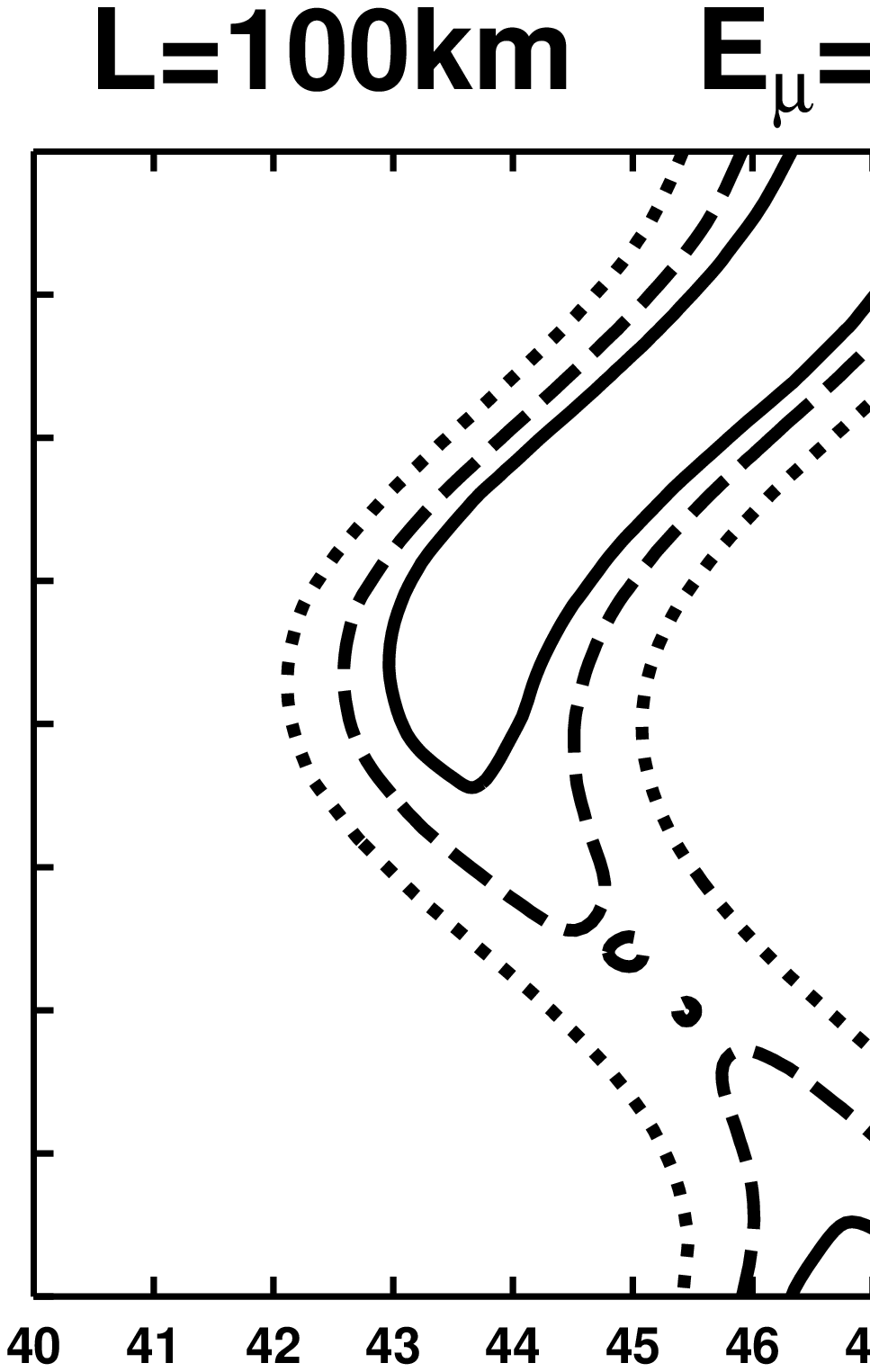,width=8cm}
\vglue -8.1cm \hglue -0.7cm \epsfig{file=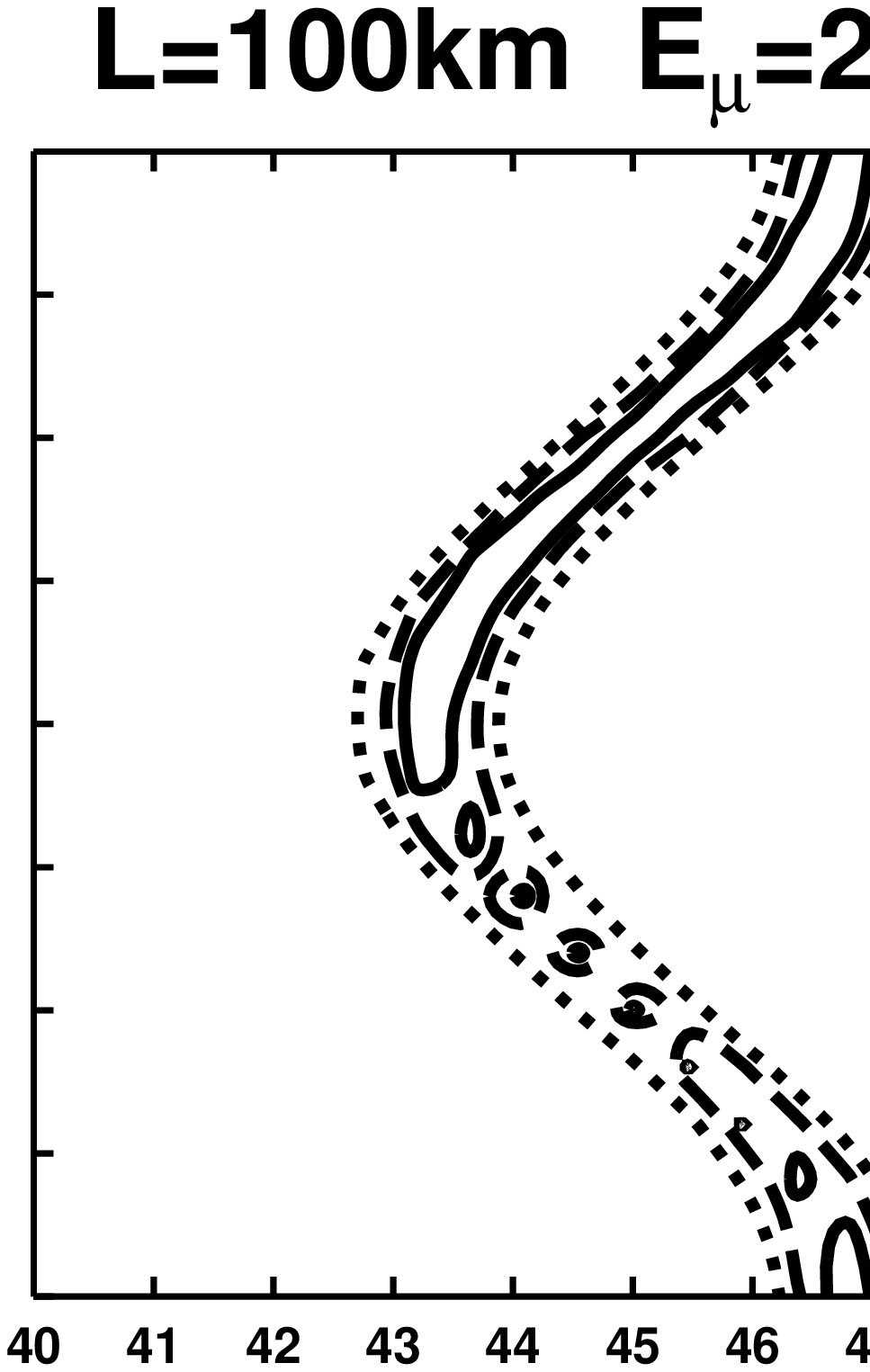,width=8cm}
\vglue -8.1cm \hglue 4.5cm \epsfig{file=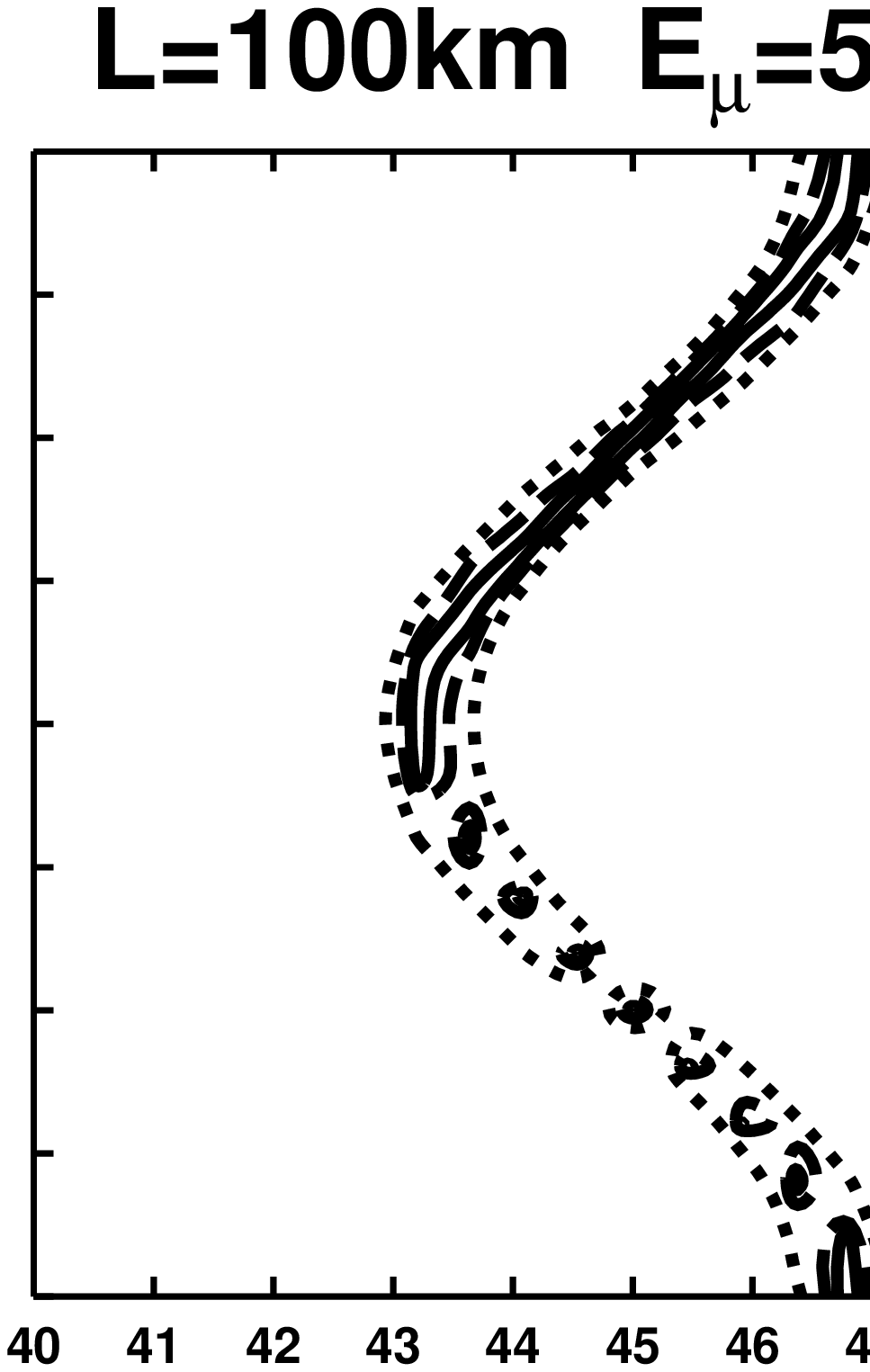,width=8cm}

\vglue -2.4cm
\hglue -6.0cm 
\epsfig{file=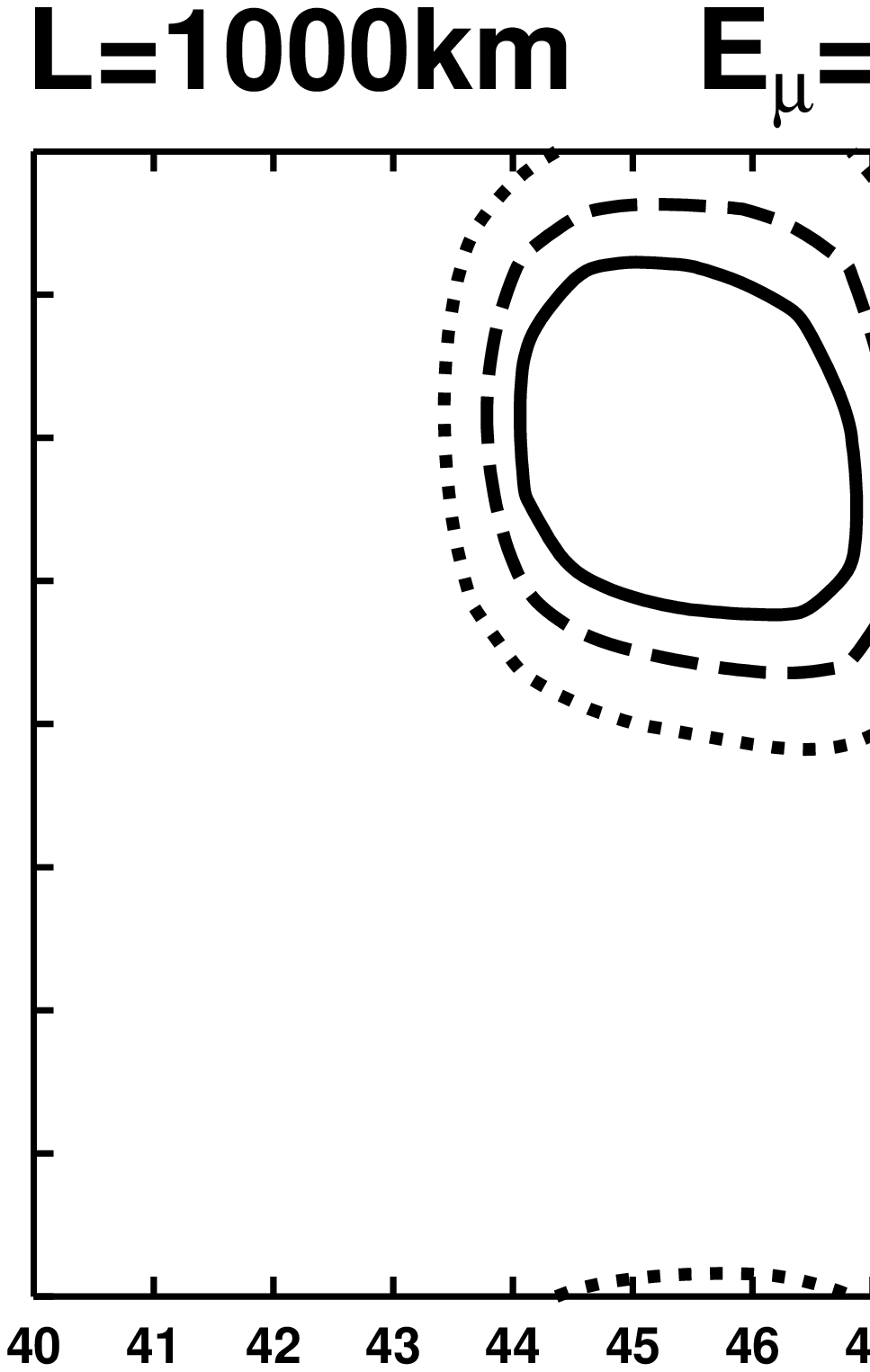,width=8cm}
\vglue -8.1cm \hglue -0.7cm \epsfig{file=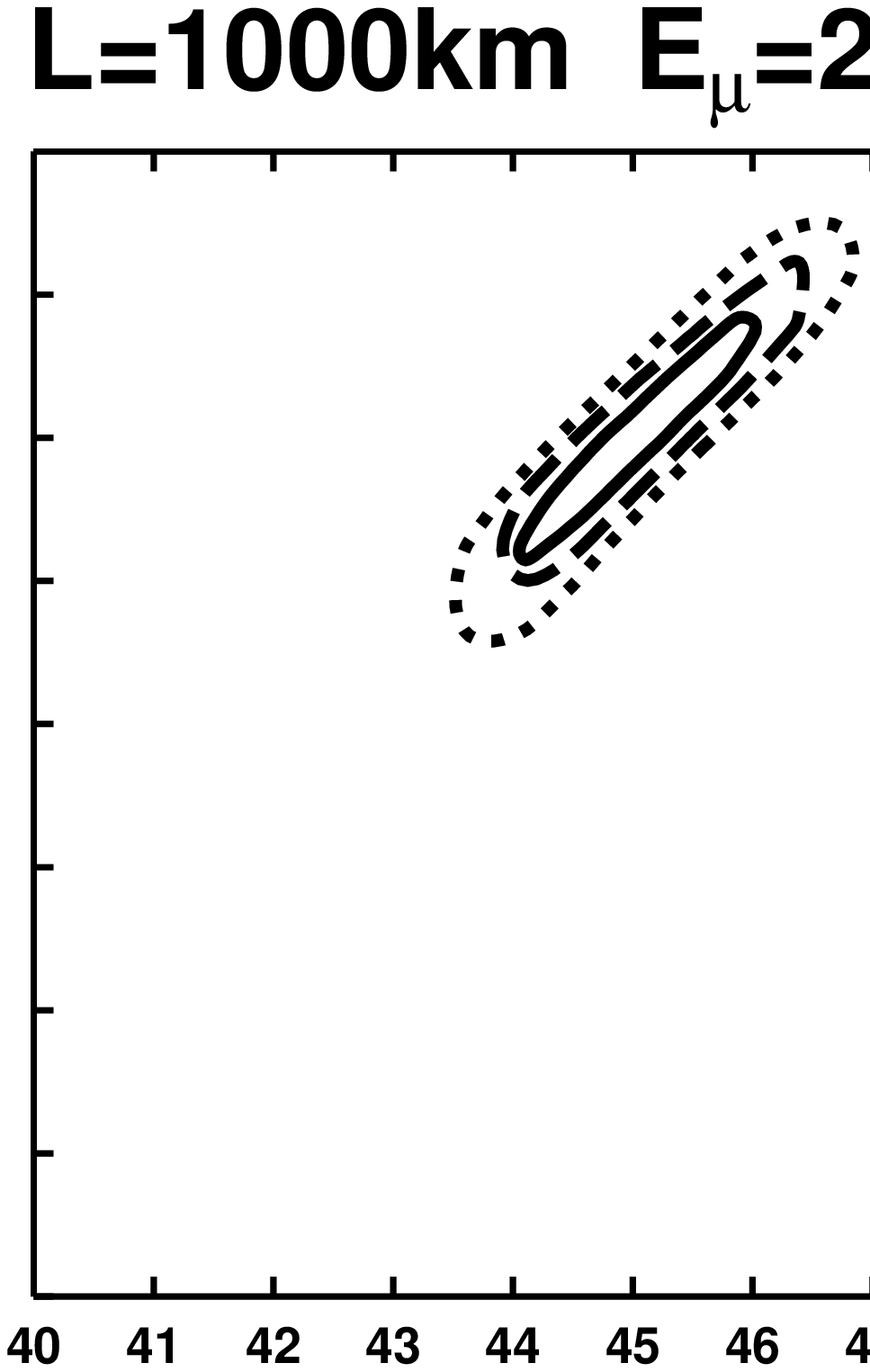,width=8cm}
\vglue -8.1cm \hglue 4.5cm \epsfig{file=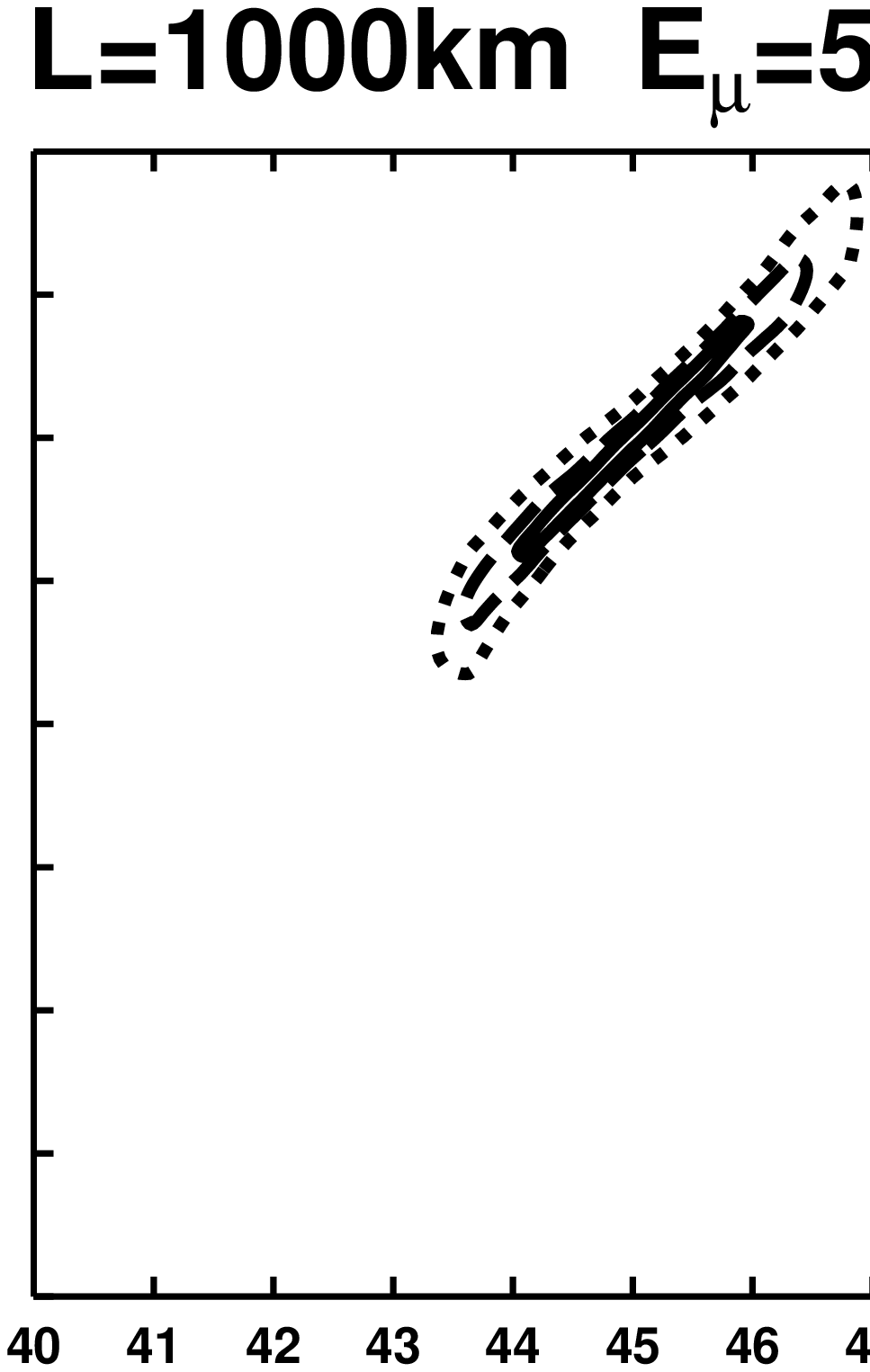,width=8cm}

\vglue -2.4cm
\hglue -6.0cm 
\epsfig{file=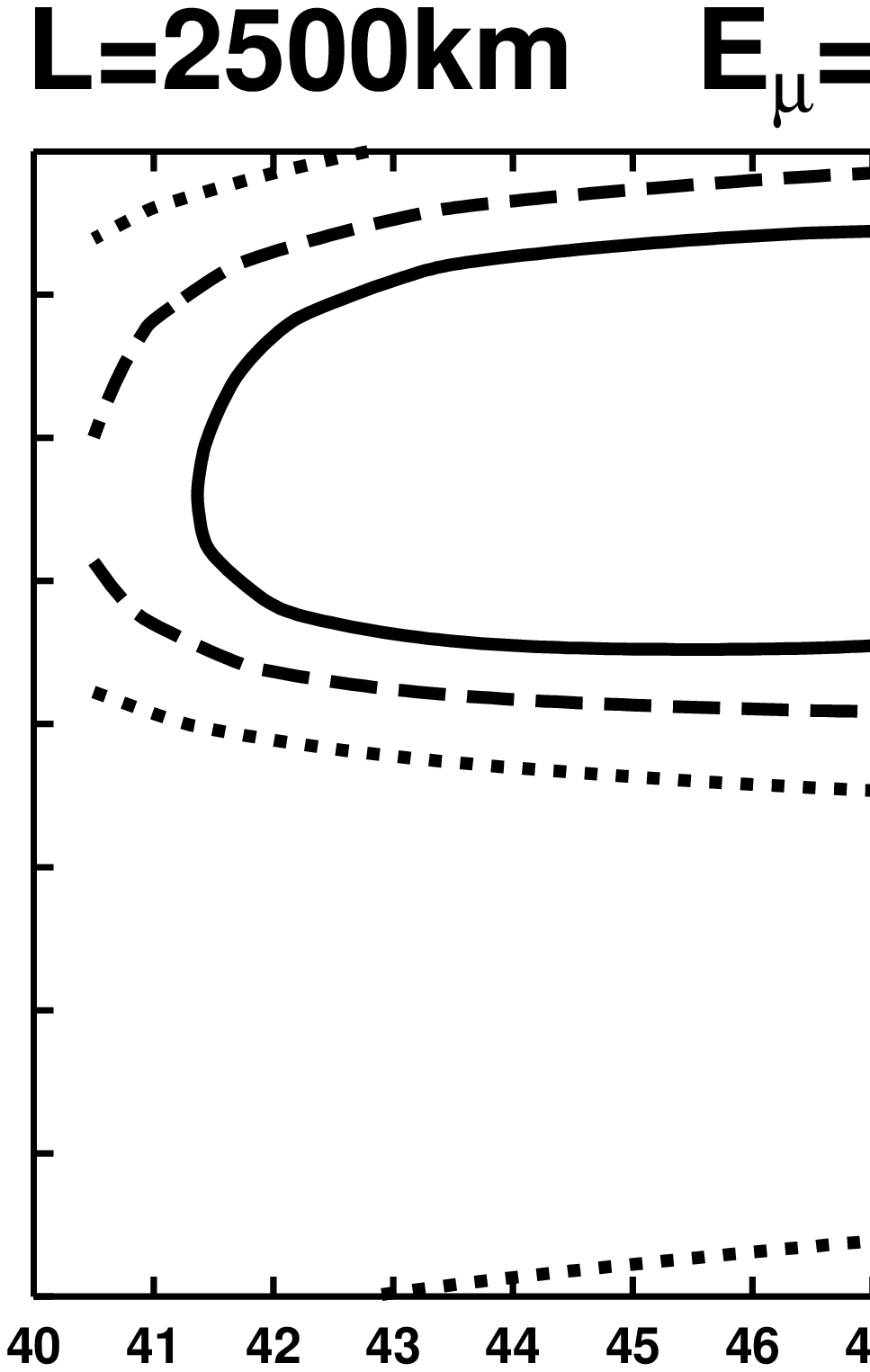,width=8cm}
\vglue -8.1cm \hglue -0.7cm \epsfig{file=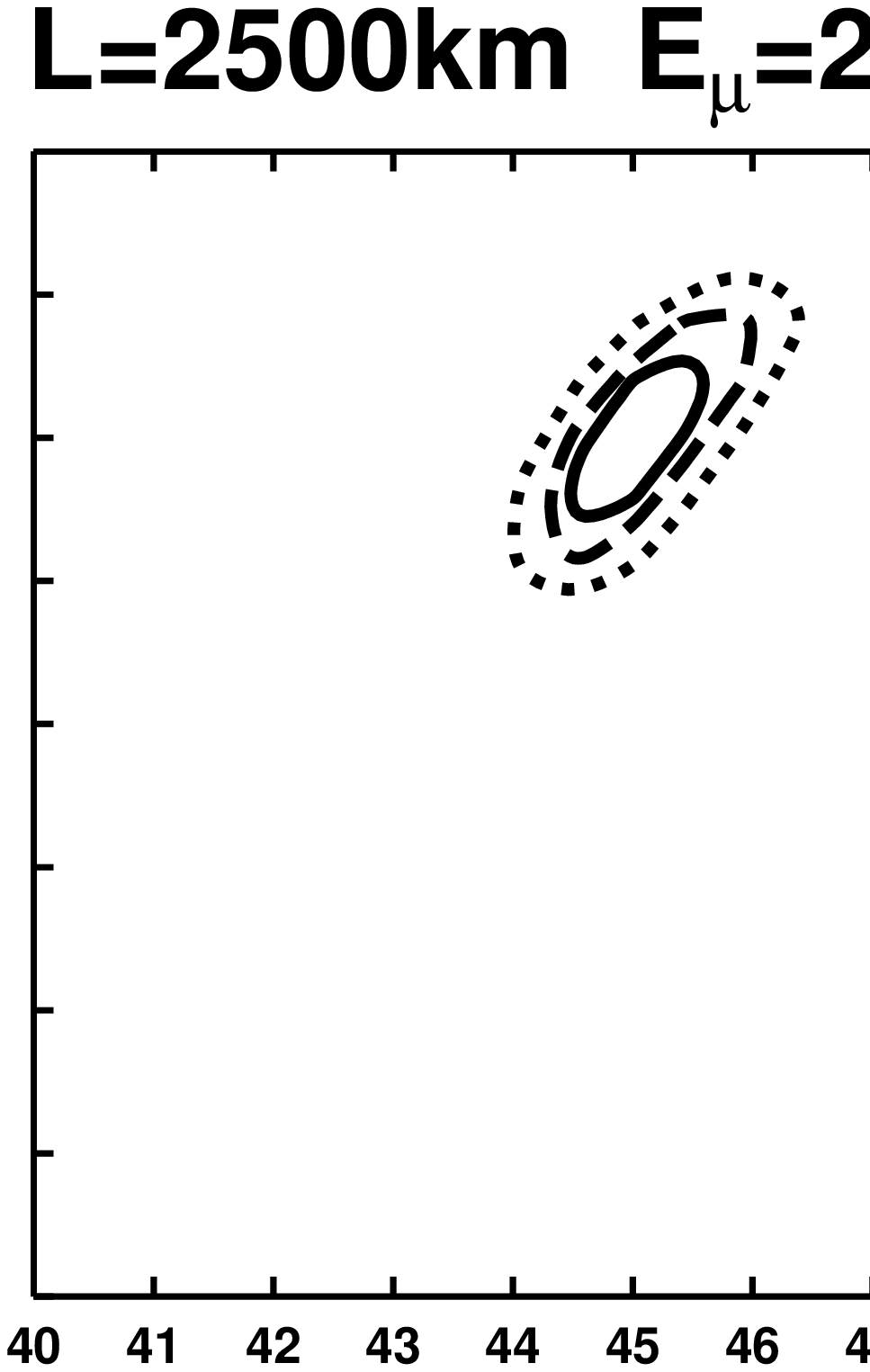,width=8cm}
\vglue -8.1cm \hglue 4.5cm \epsfig{file=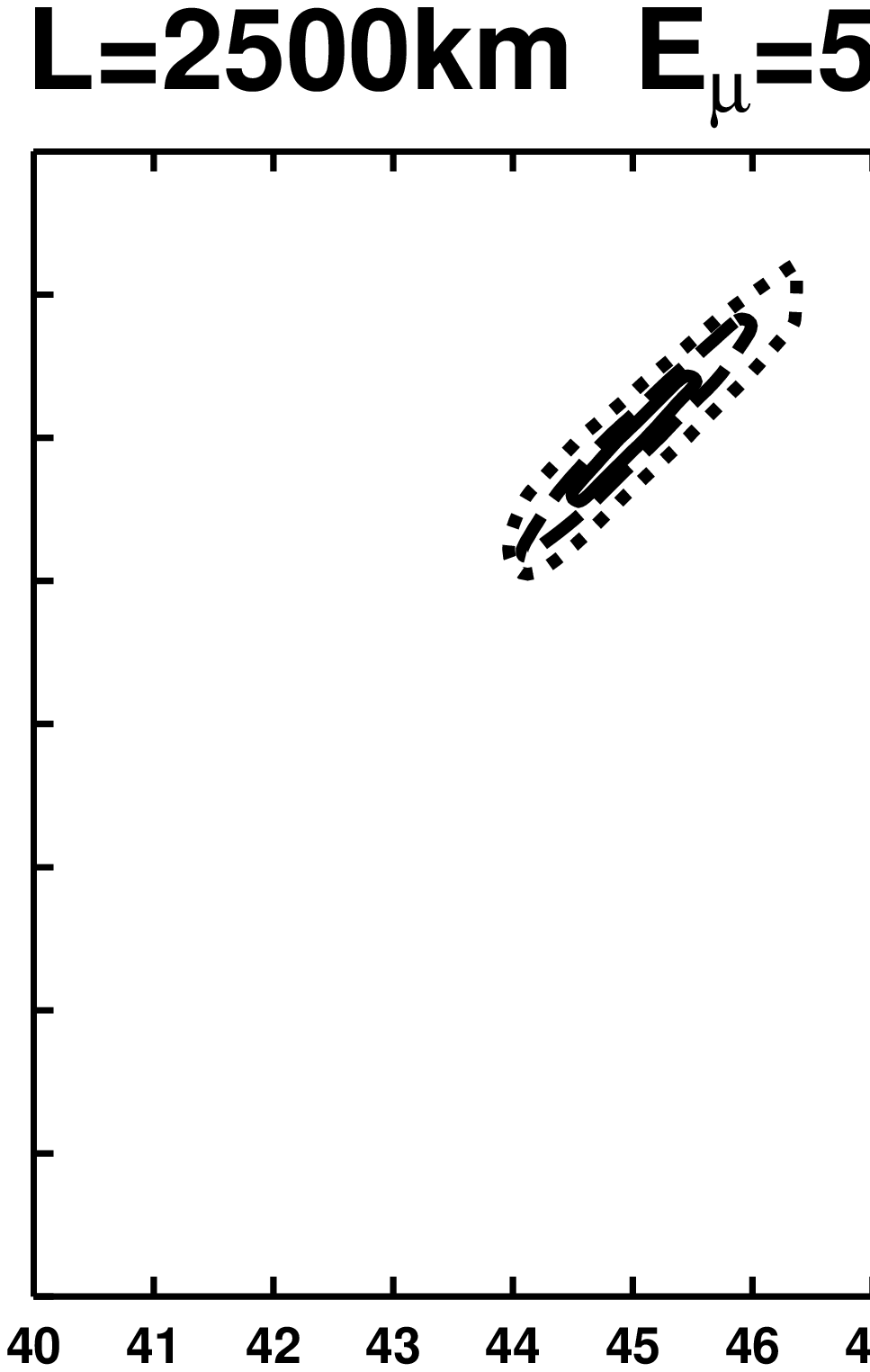,width=8cm}

\vglue -2.4cm
\hglue -6.0cm 
\epsfig{file=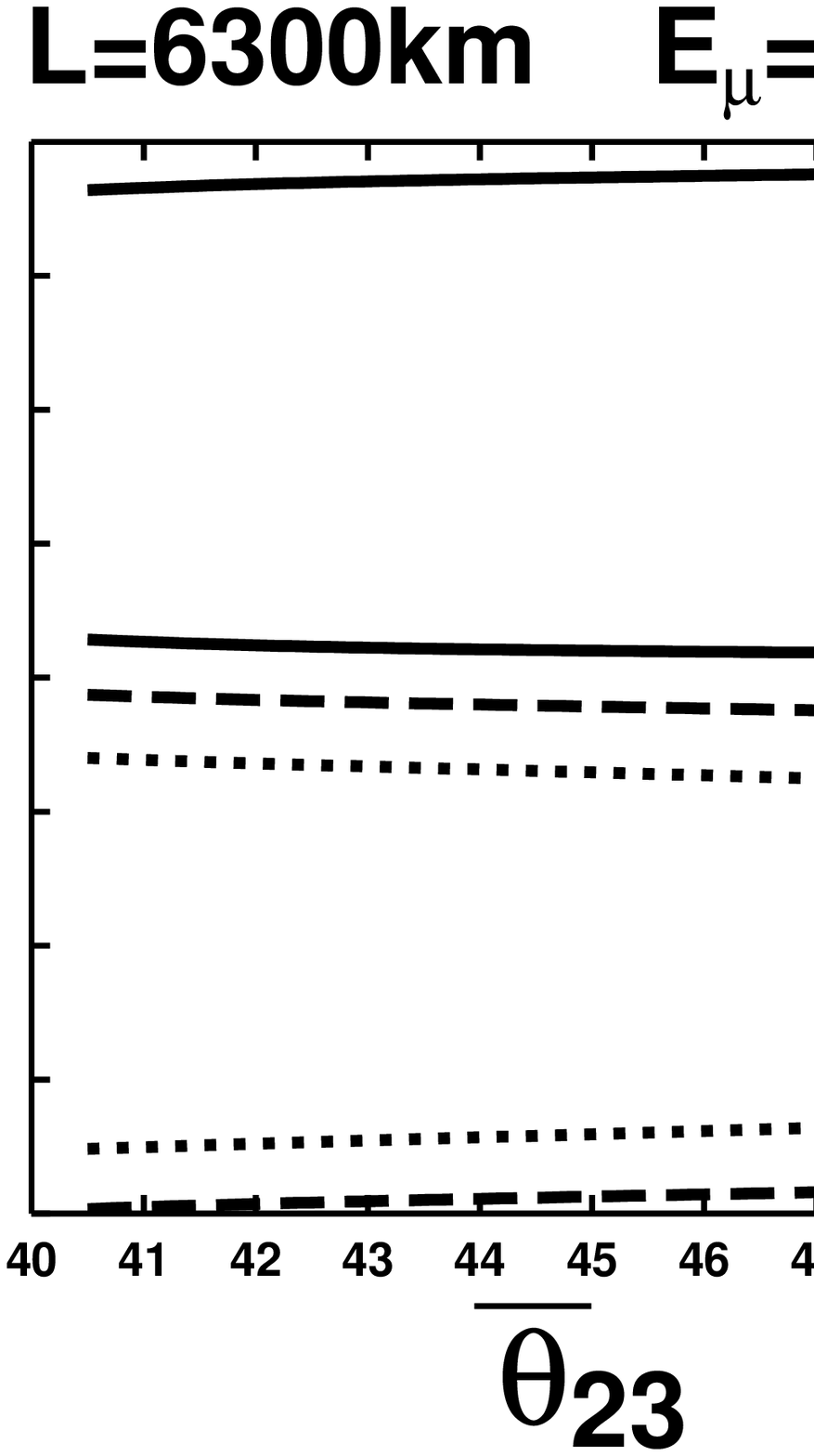,width=8cm}
\vglue -8.1cm \hglue -0.7cm \epsfig{file=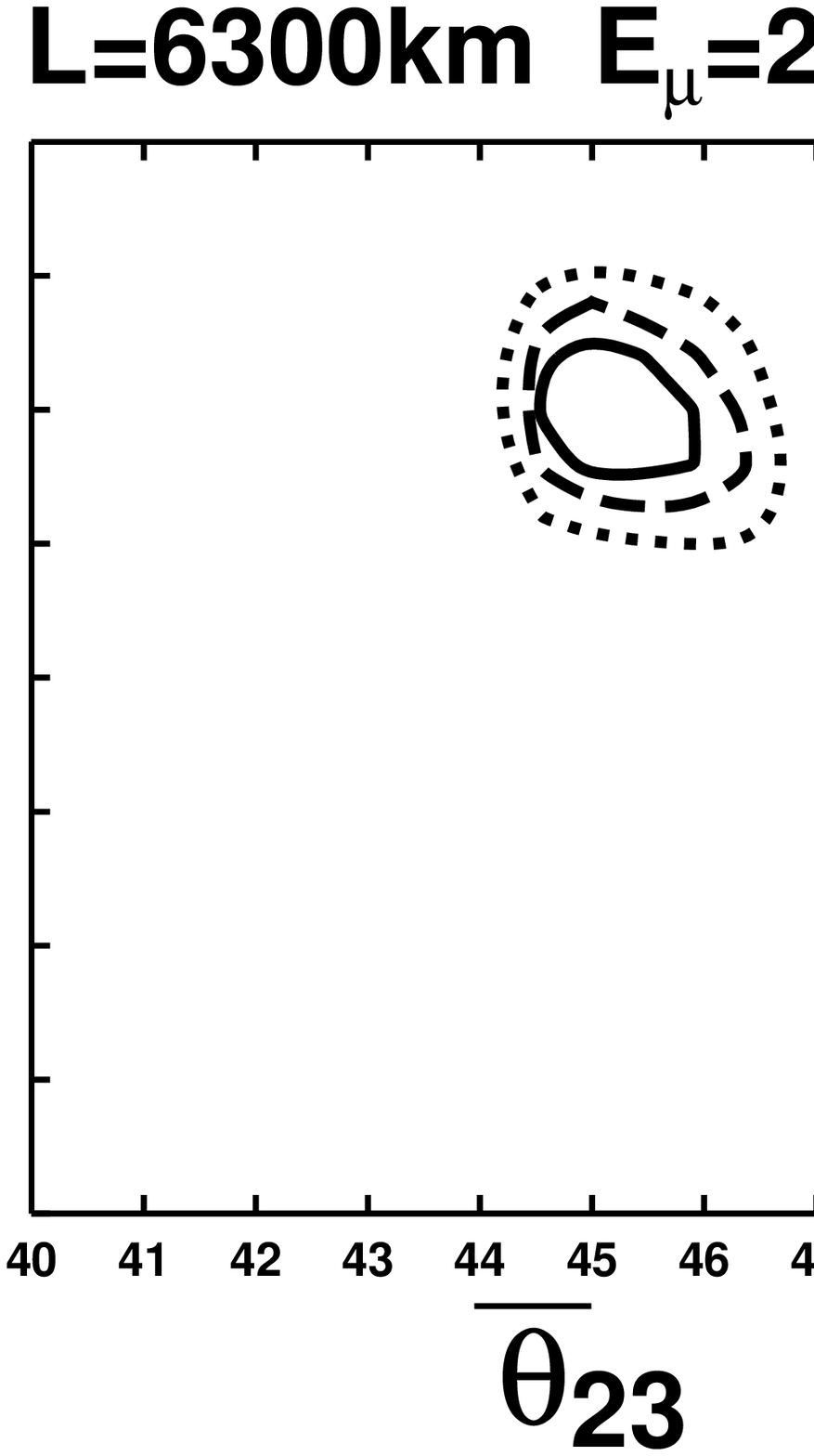,width=8cm}
\vglue -8.1cm \hglue 4.5cm \epsfig{file=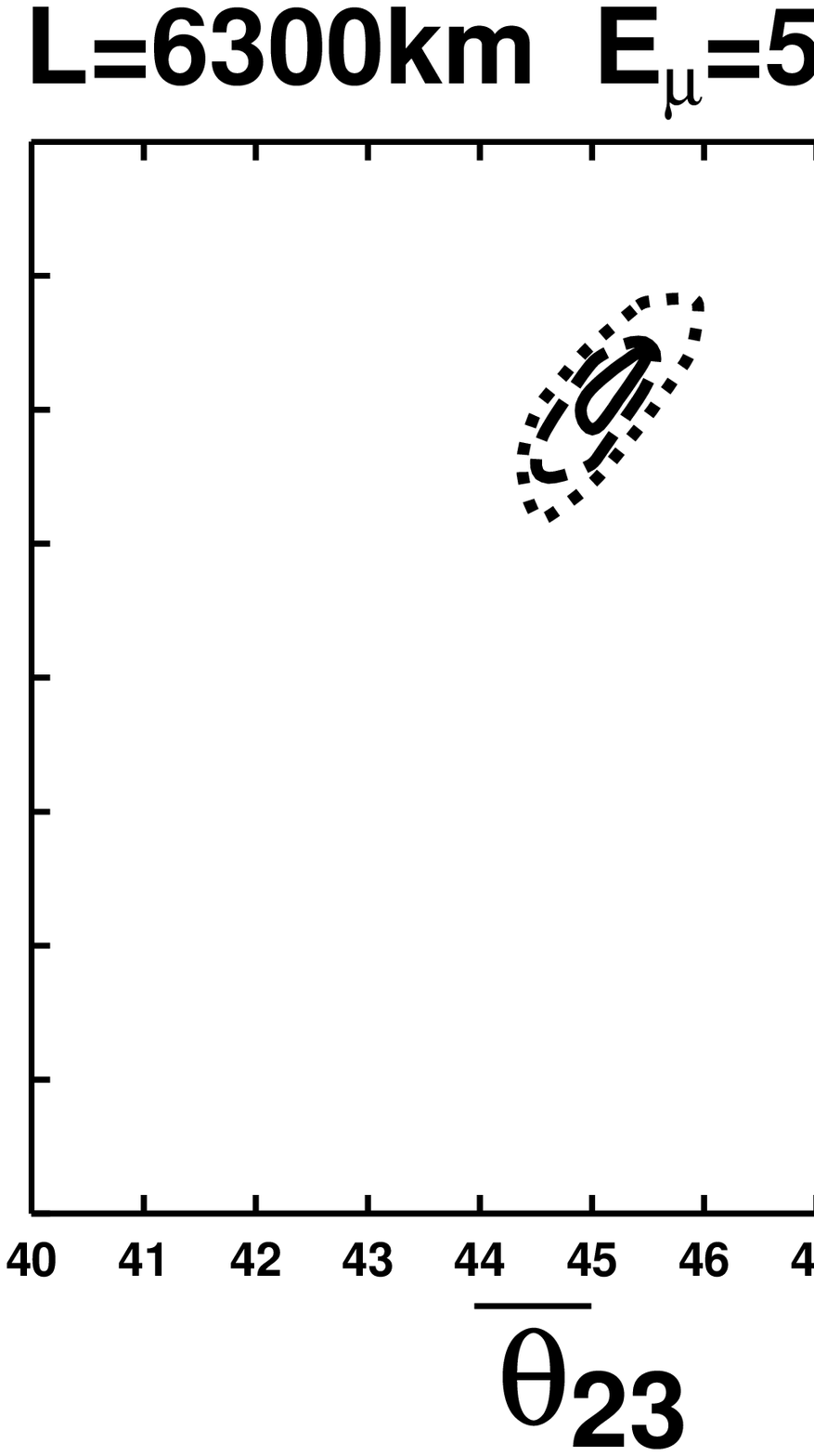,width=8cm}
\vglue -2.0cm\hglue -23.3cm
\epsfig{file=fig0413.eps,width=22cm}
\vglue -21.5cm\hglue 6.3cm
\epsfig{file=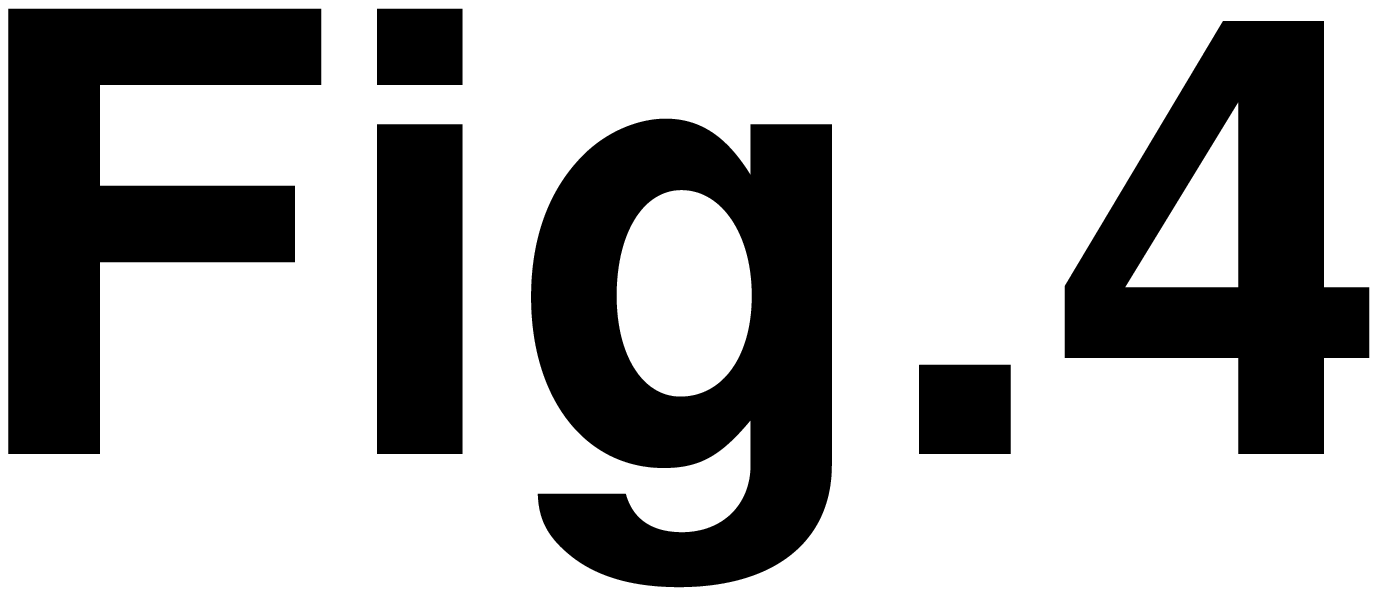,width=4cm}
\newpage
\vglue -2.5cm
\hglue -6.0cm 
\epsfig{file=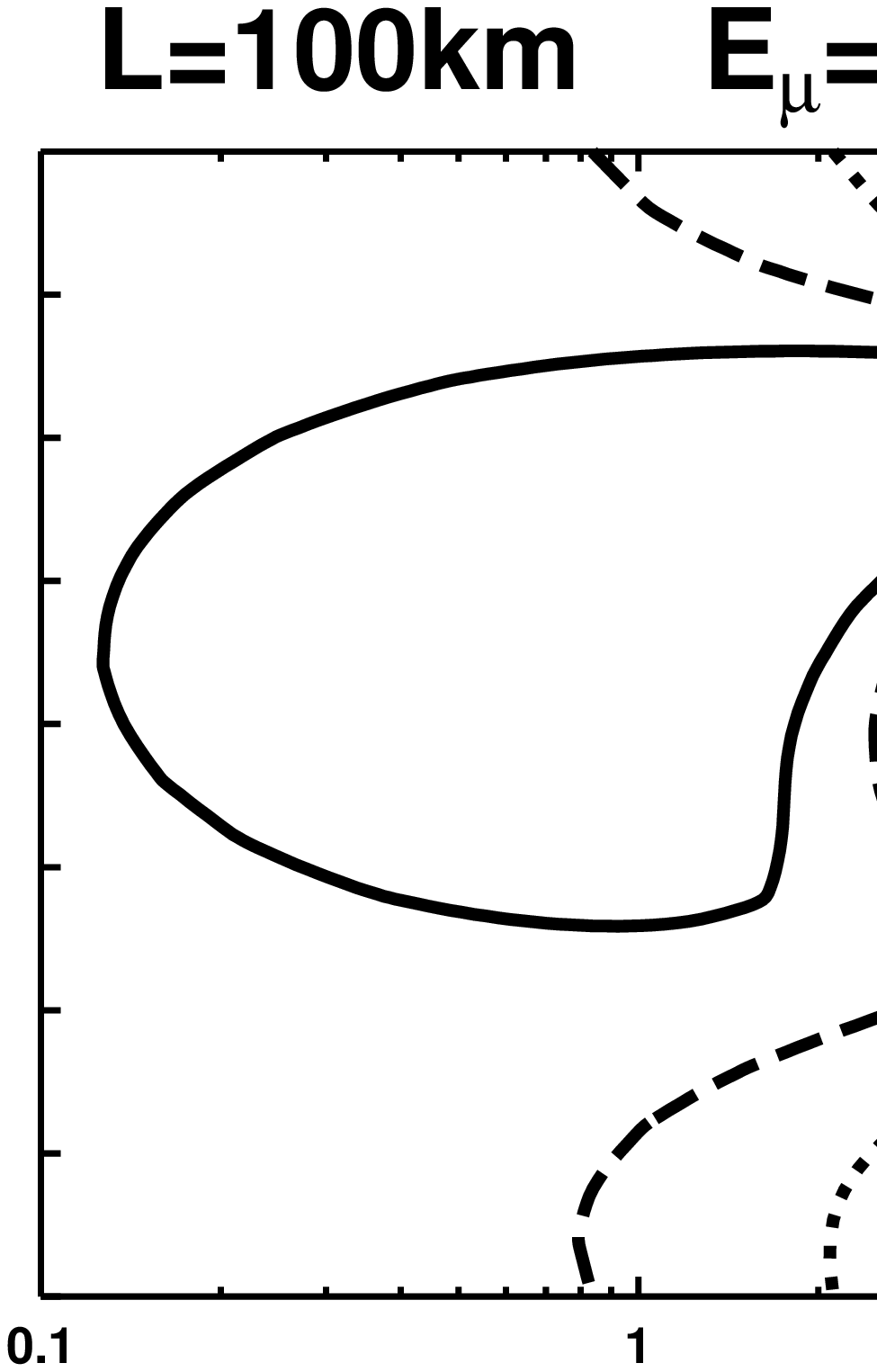,width=8cm}
\vglue -8.1cm \hglue -0.7cm \epsfig{file=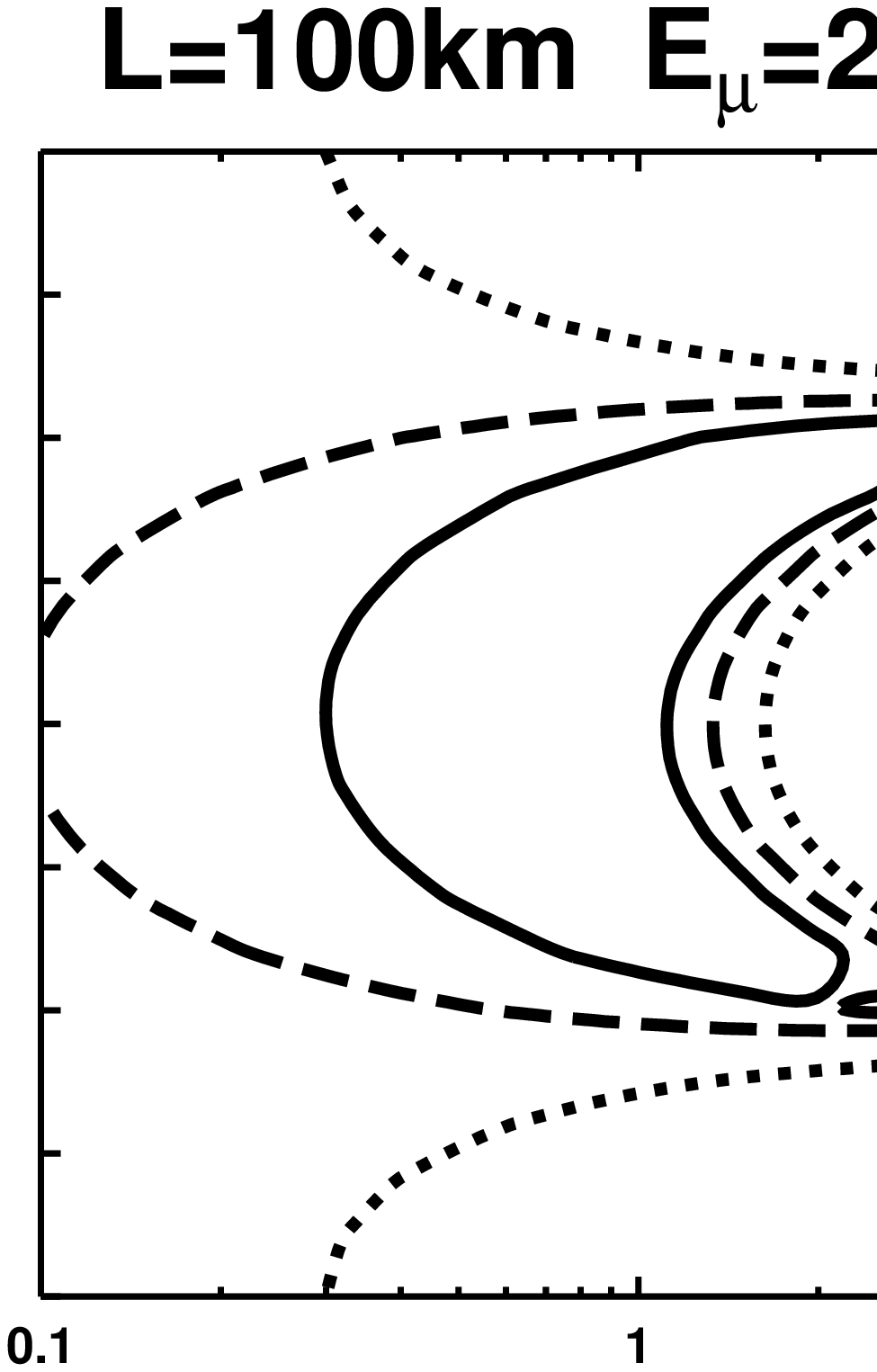,width=8cm}
\vglue -8.1cm \hglue 4.5cm \epsfig{file=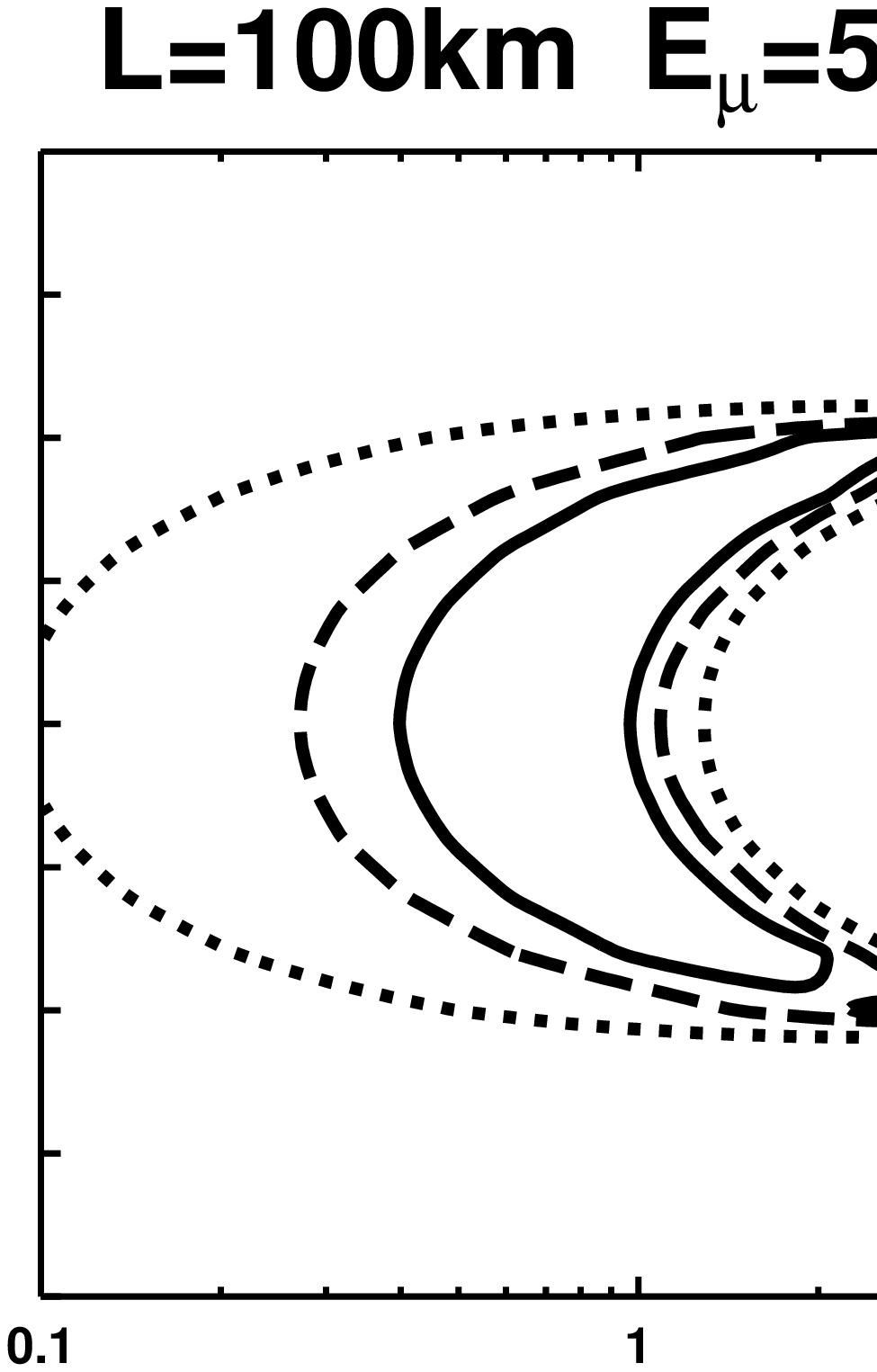,width=8cm}

\vglue -2.4cm
\hglue -6.0cm 
\epsfig{file=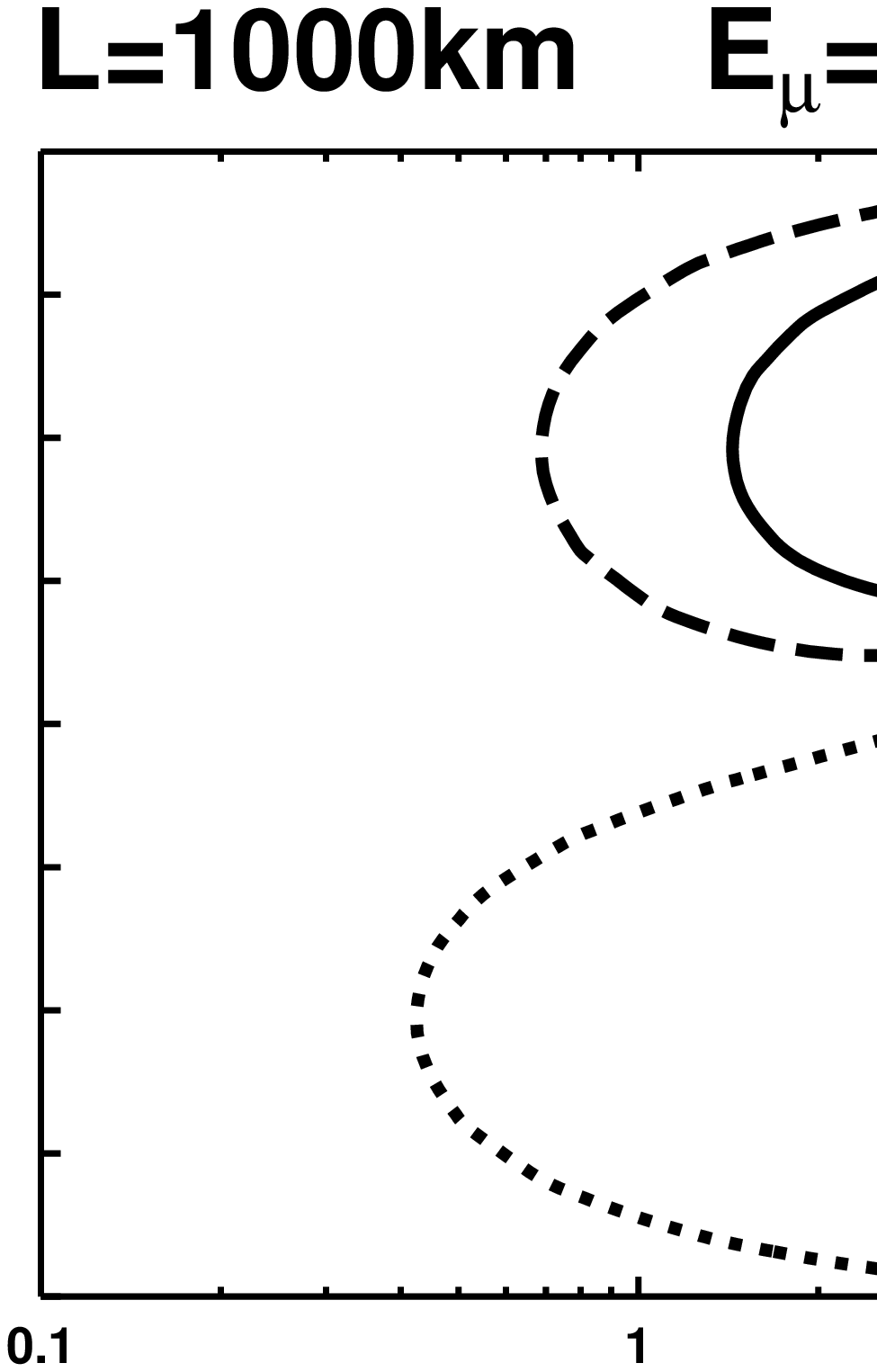,width=8cm}
\vglue -8.1cm \hglue -0.7cm \epsfig{file=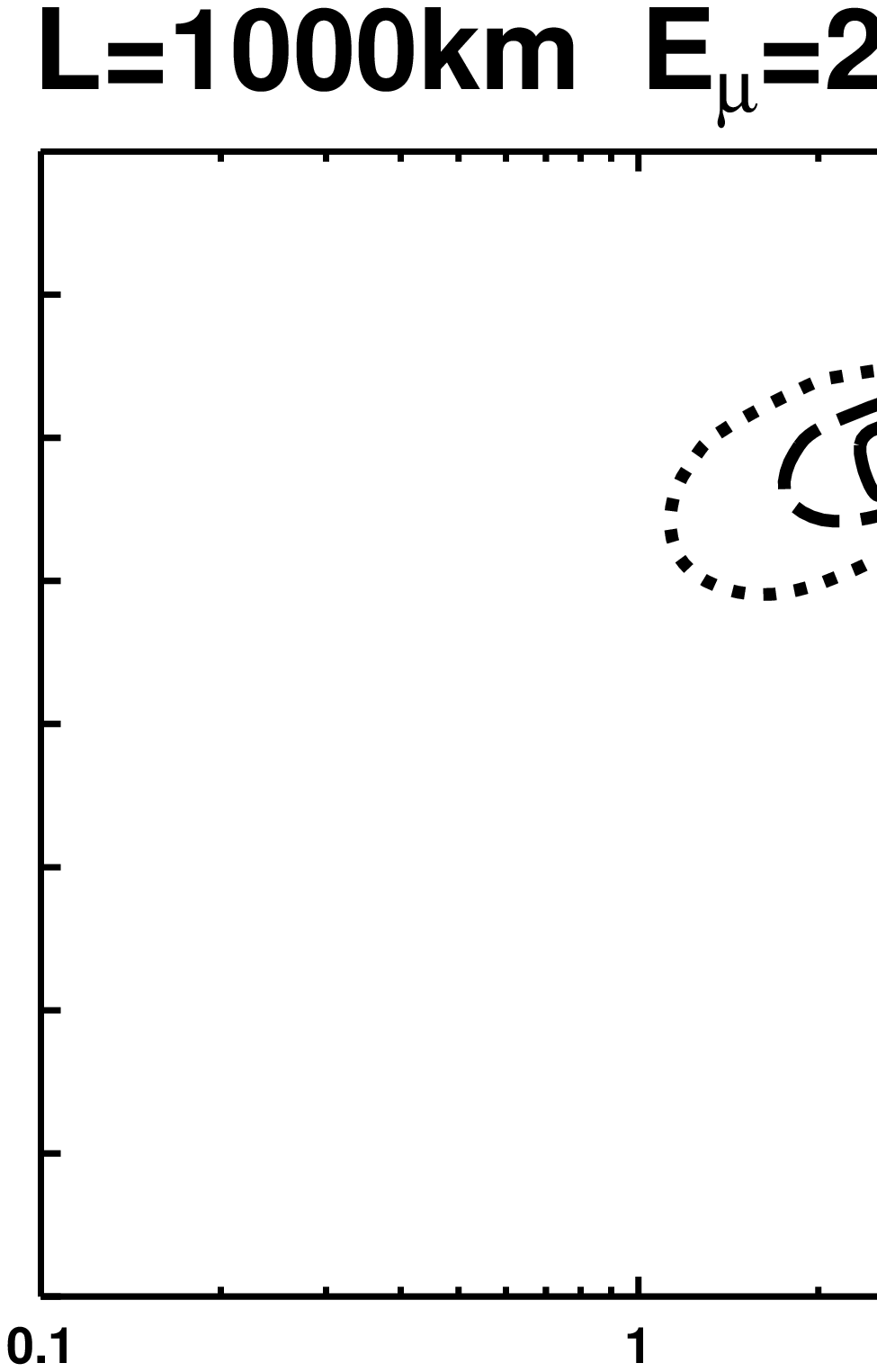,width=8cm}
\vglue -8.1cm \hglue 4.5cm \epsfig{file=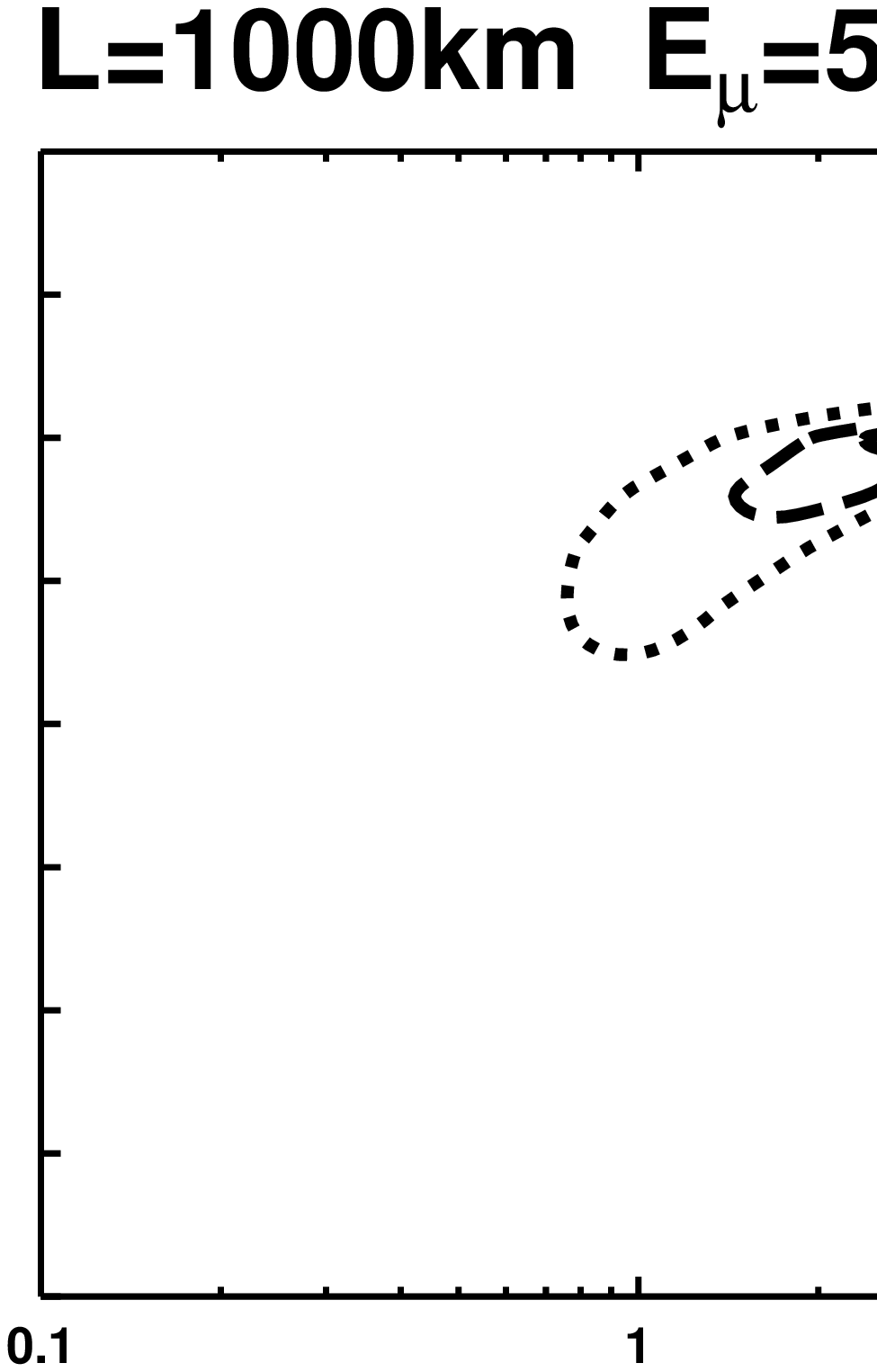,width=8cm}

\vglue -2.4cm
\hglue -6.0cm 
\epsfig{file=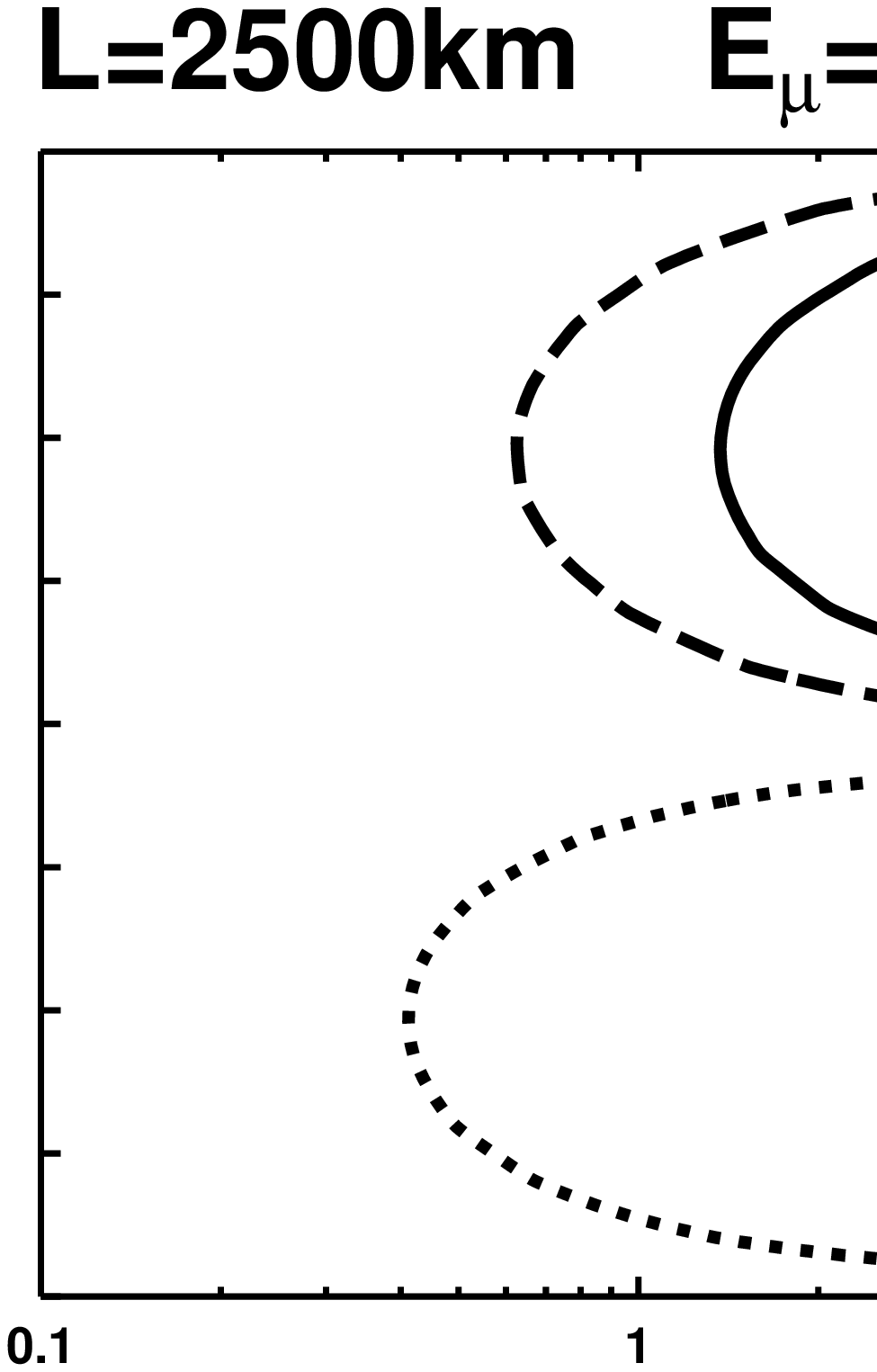,width=8cm}
\vglue -8.1cm \hglue -0.7cm \epsfig{file=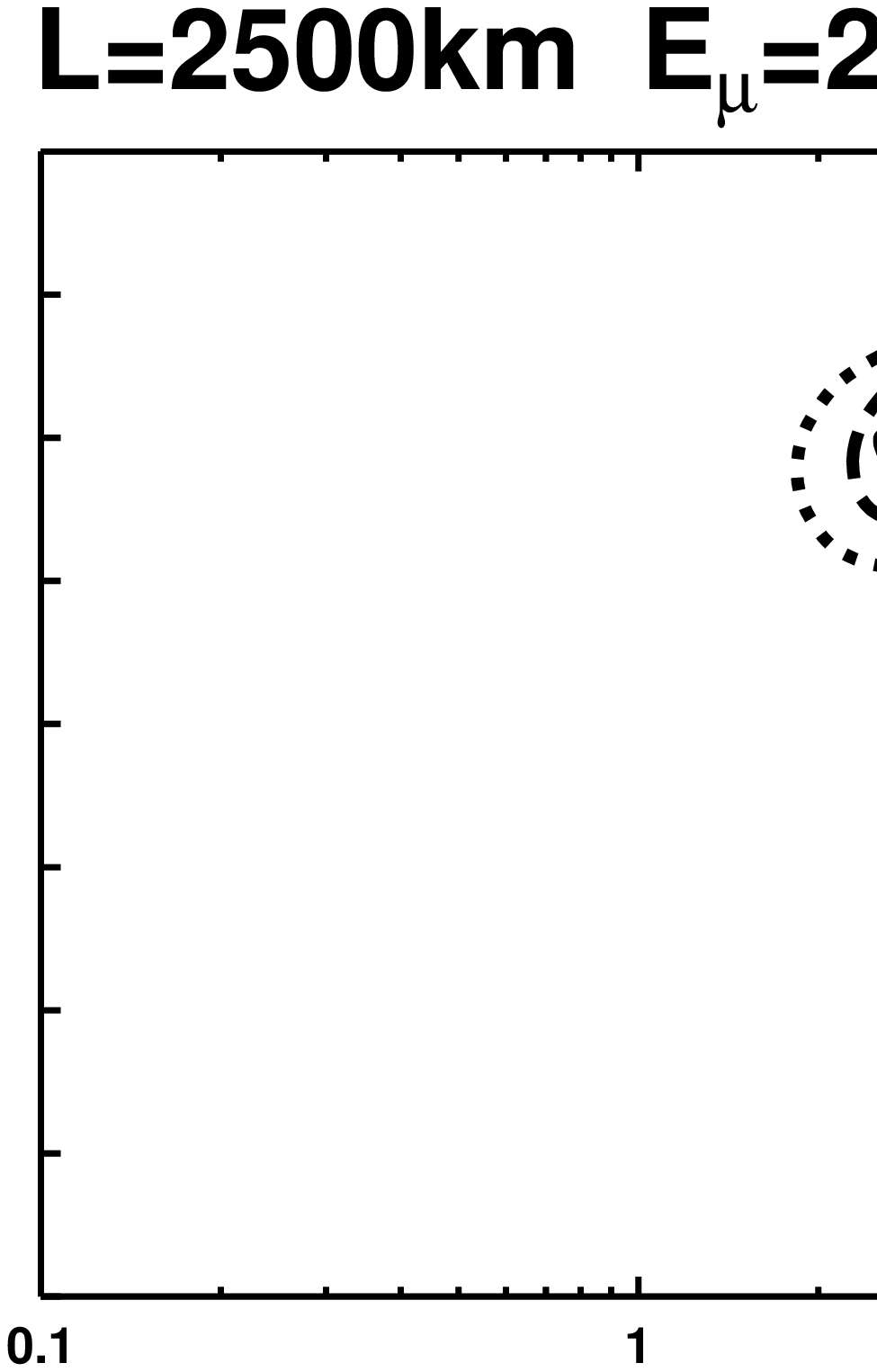,width=8cm}
\vglue -8.1cm \hglue 4.5cm \epsfig{file=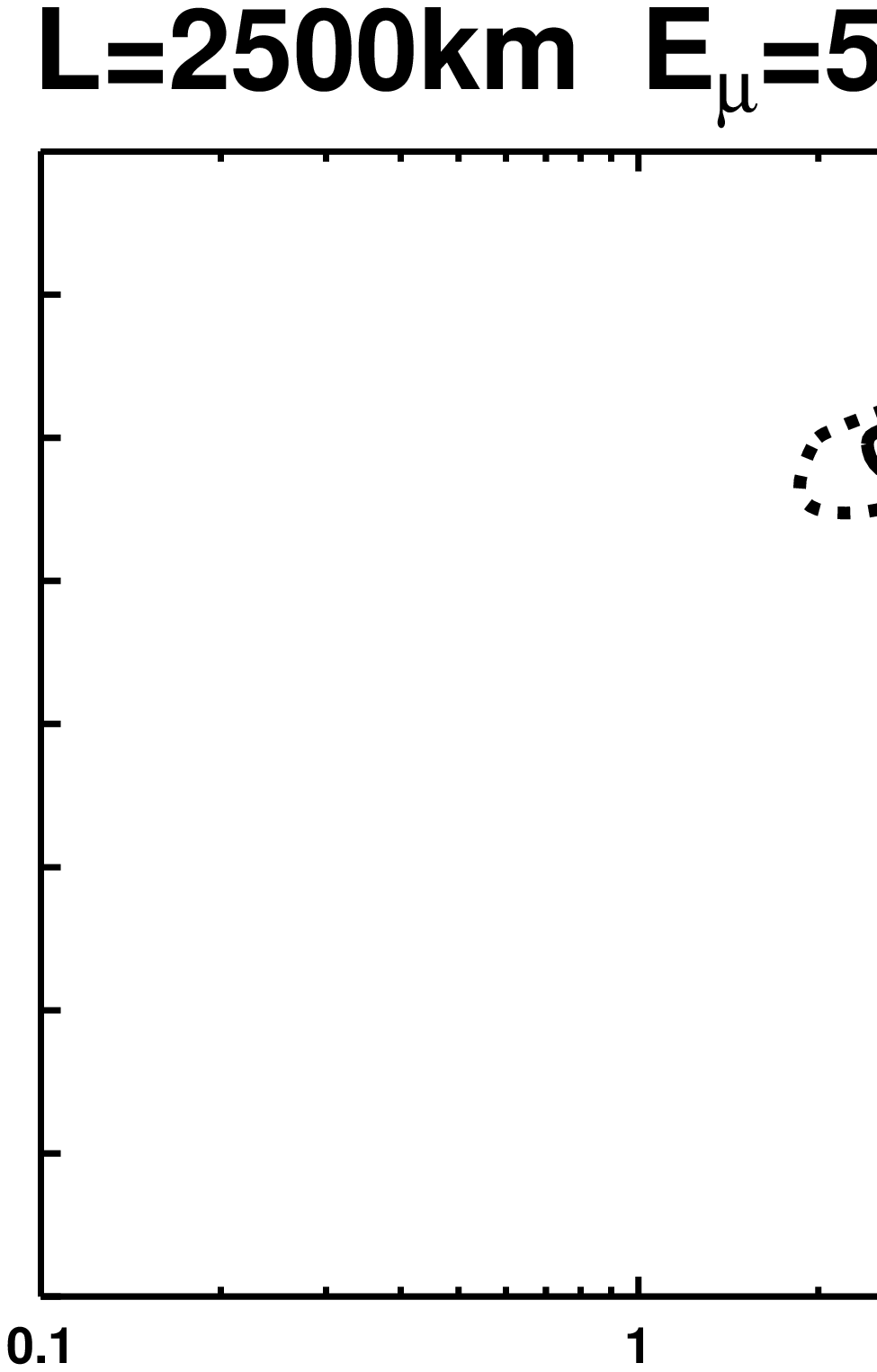,width=8cm}

\vglue -2.4cm
\hglue -6.0cm 
\epsfig{file=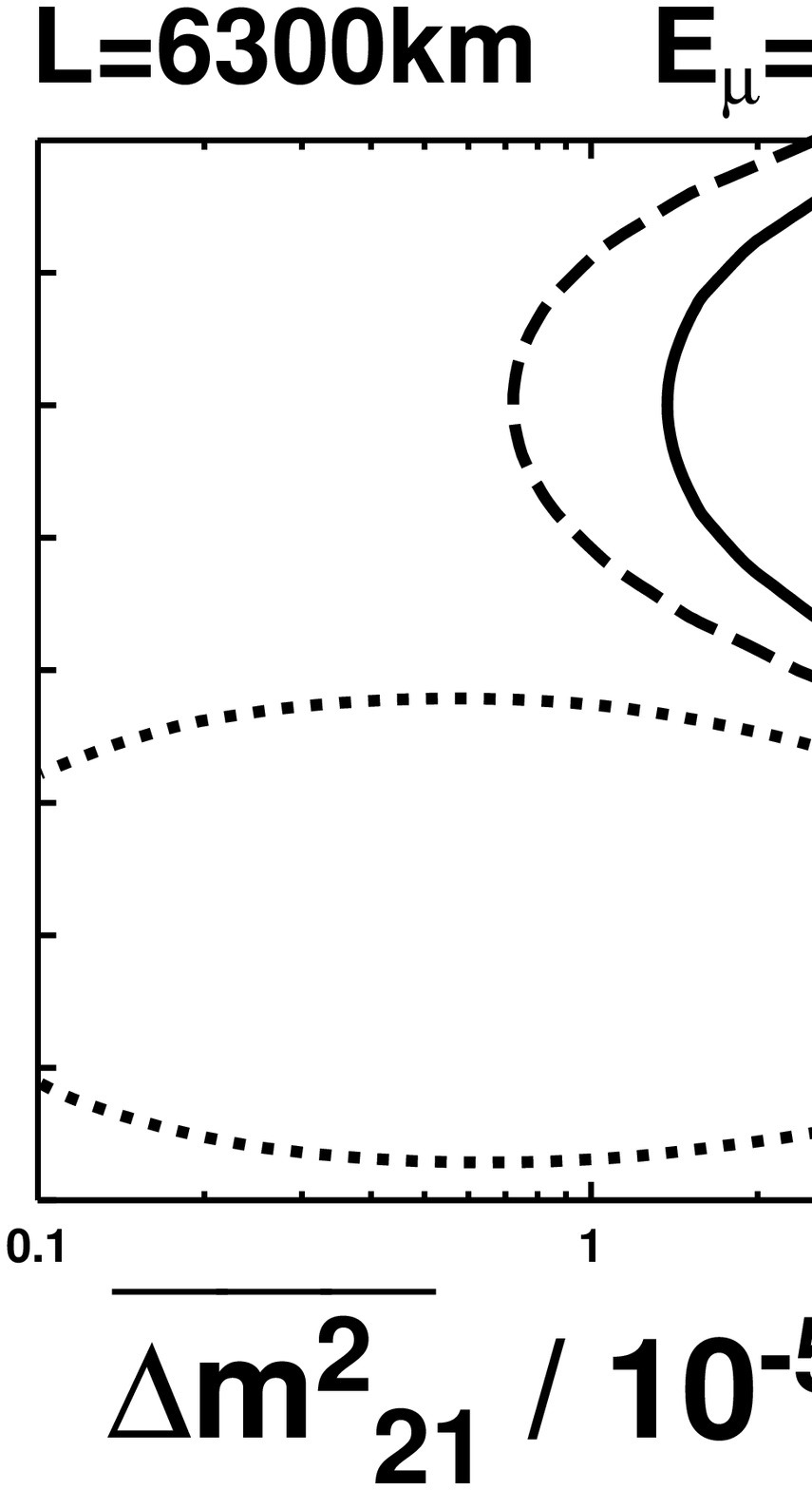,width=8cm}
\vglue -8.1cm \hglue -0.7cm \epsfig{file=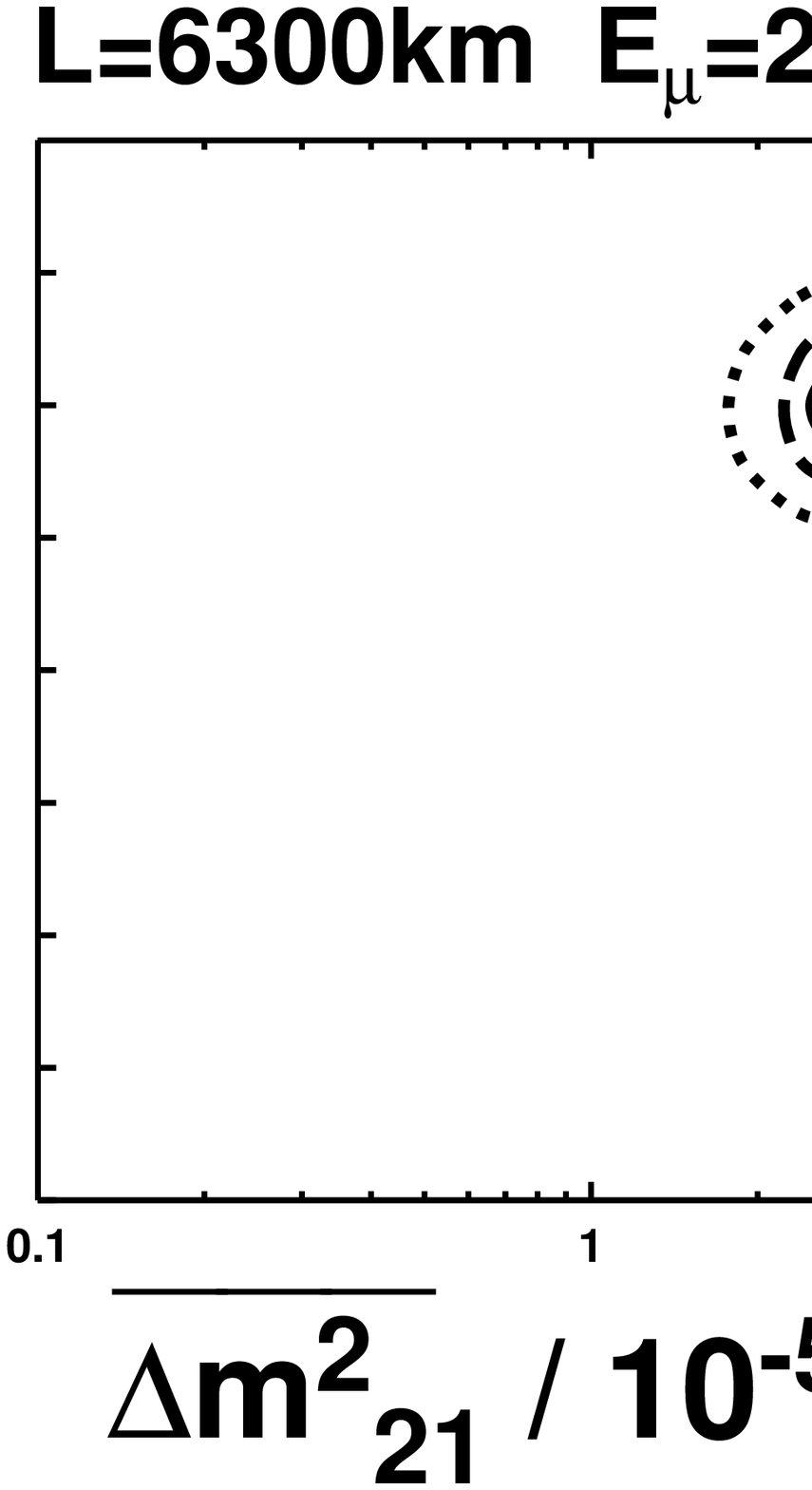,width=8cm}
\vglue -8.1cm \hglue 4.5cm \epsfig{file=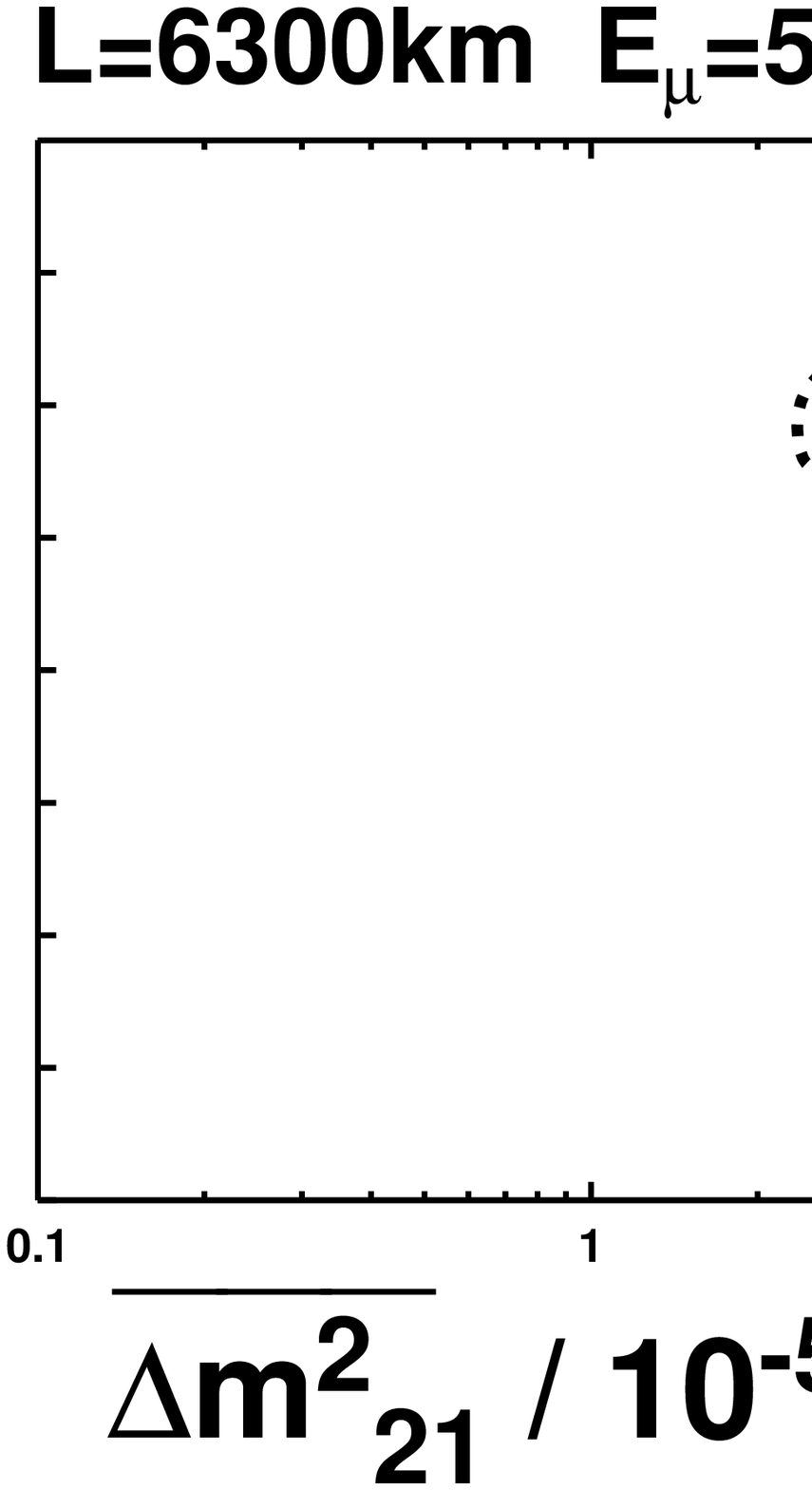,width=8cm}
\vglue -2.0cm\hglue -23.3cm
\epsfig{file=fig0513.eps,width=22cm}
\vglue -21.5cm\hglue 6.3cm
\epsfig{file=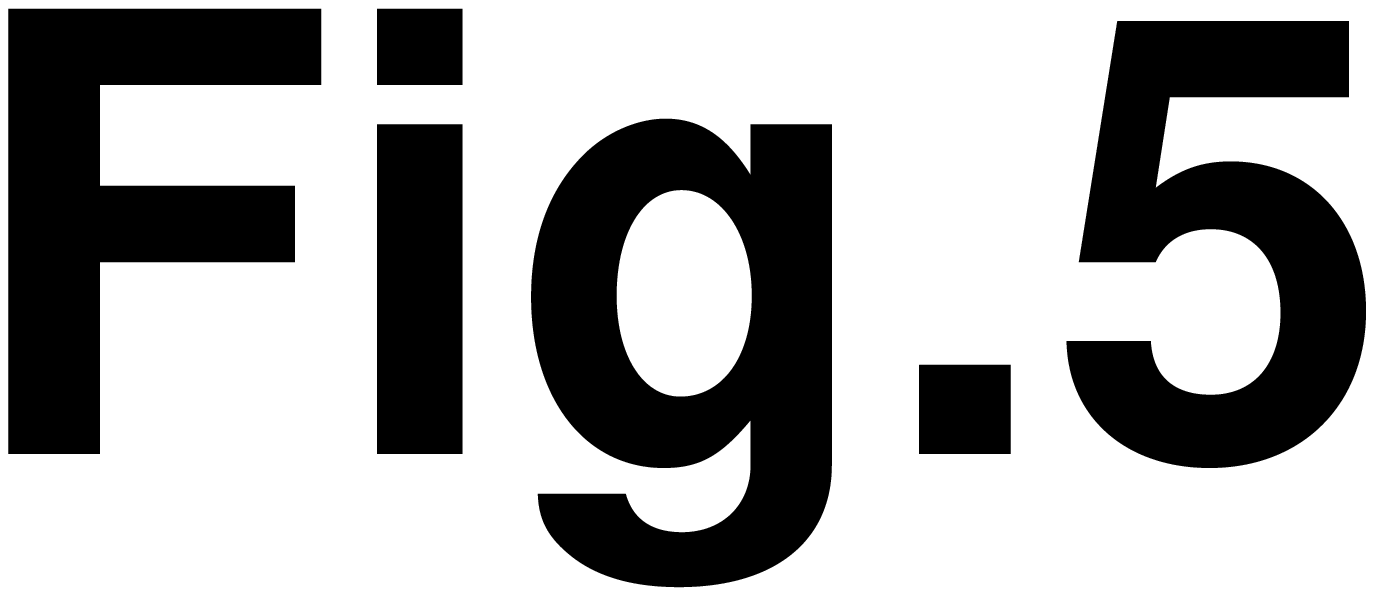,width=4cm}
\newpage
\vglue -2.5cm
\hglue -6.0cm 
\epsfig{file=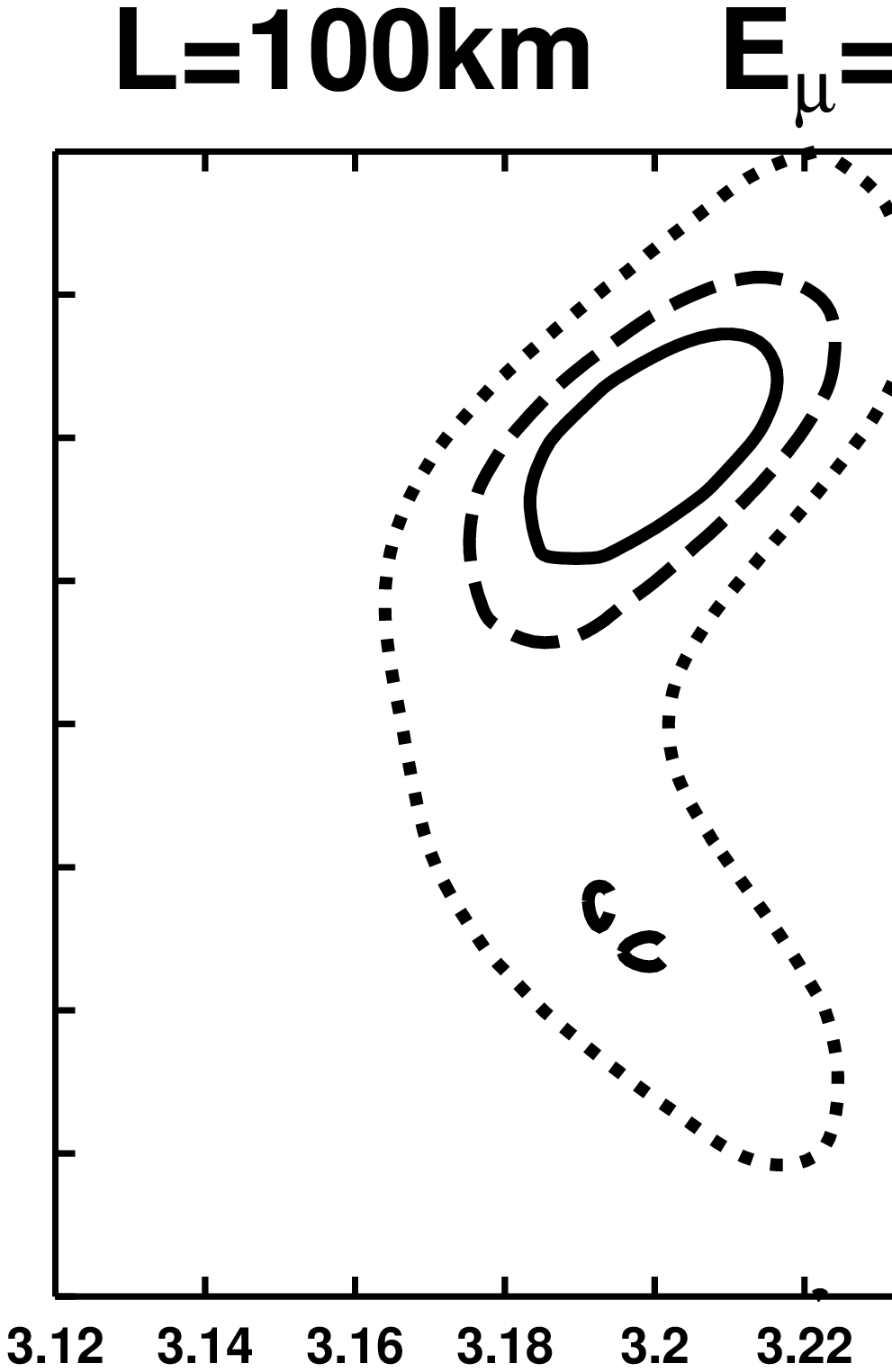,width=8cm}
\vglue -8.1cm \hglue -0.7cm \epsfig{file=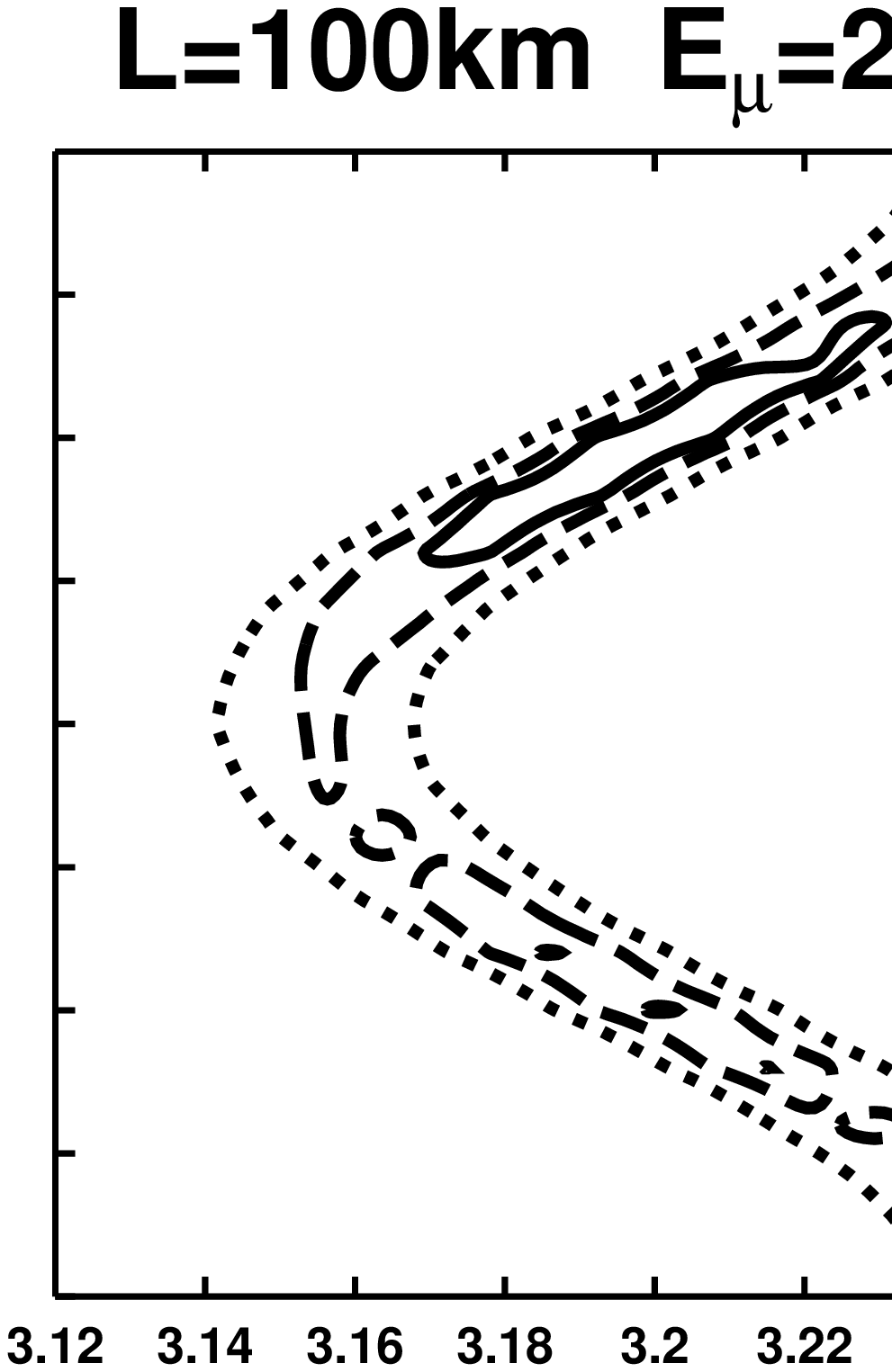,width=8cm}
\vglue -8.1cm \hglue 4.5cm \epsfig{file=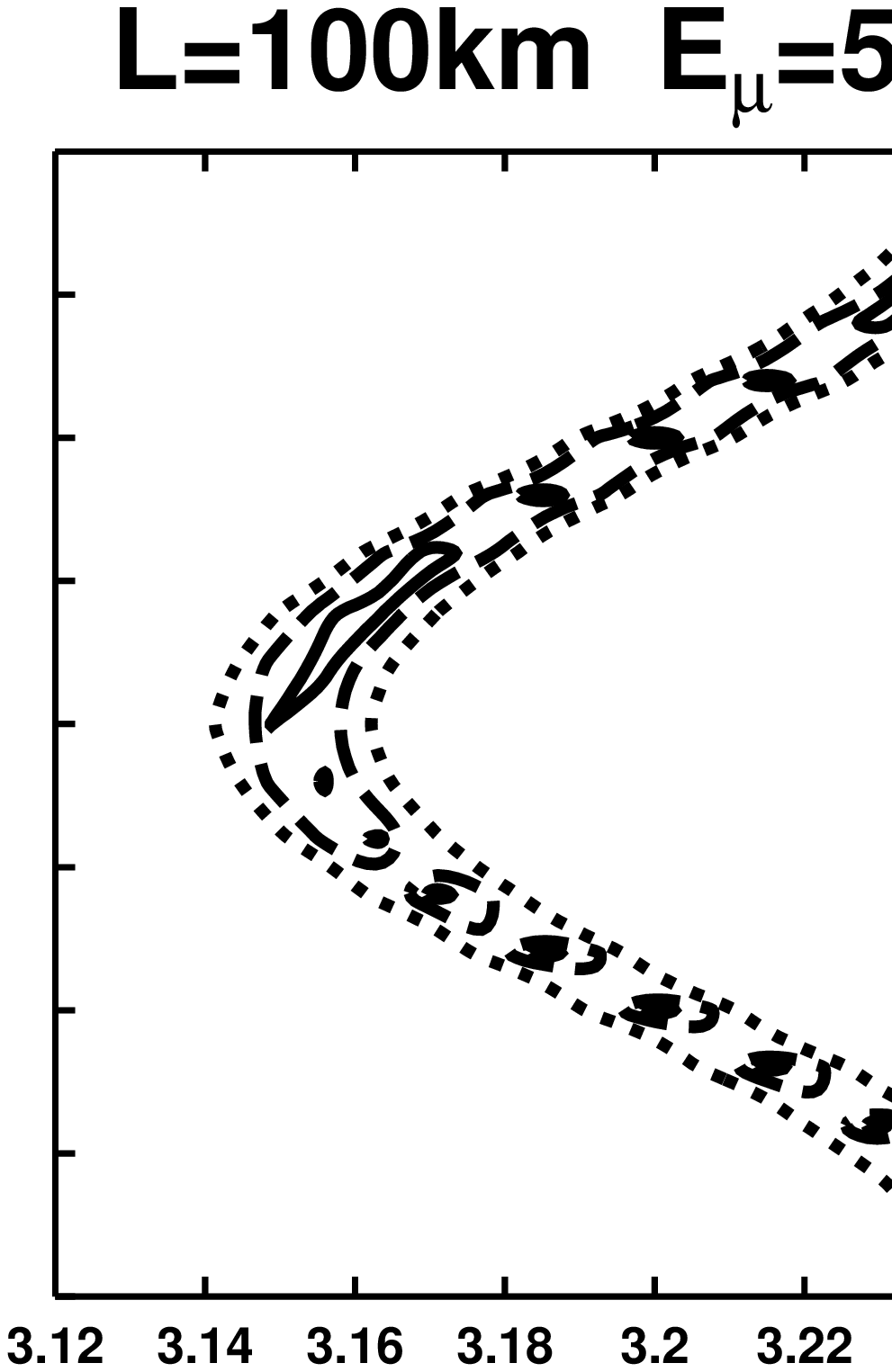,width=8cm}

\vglue -2.4cm
\hglue -6.0cm 
\epsfig{file=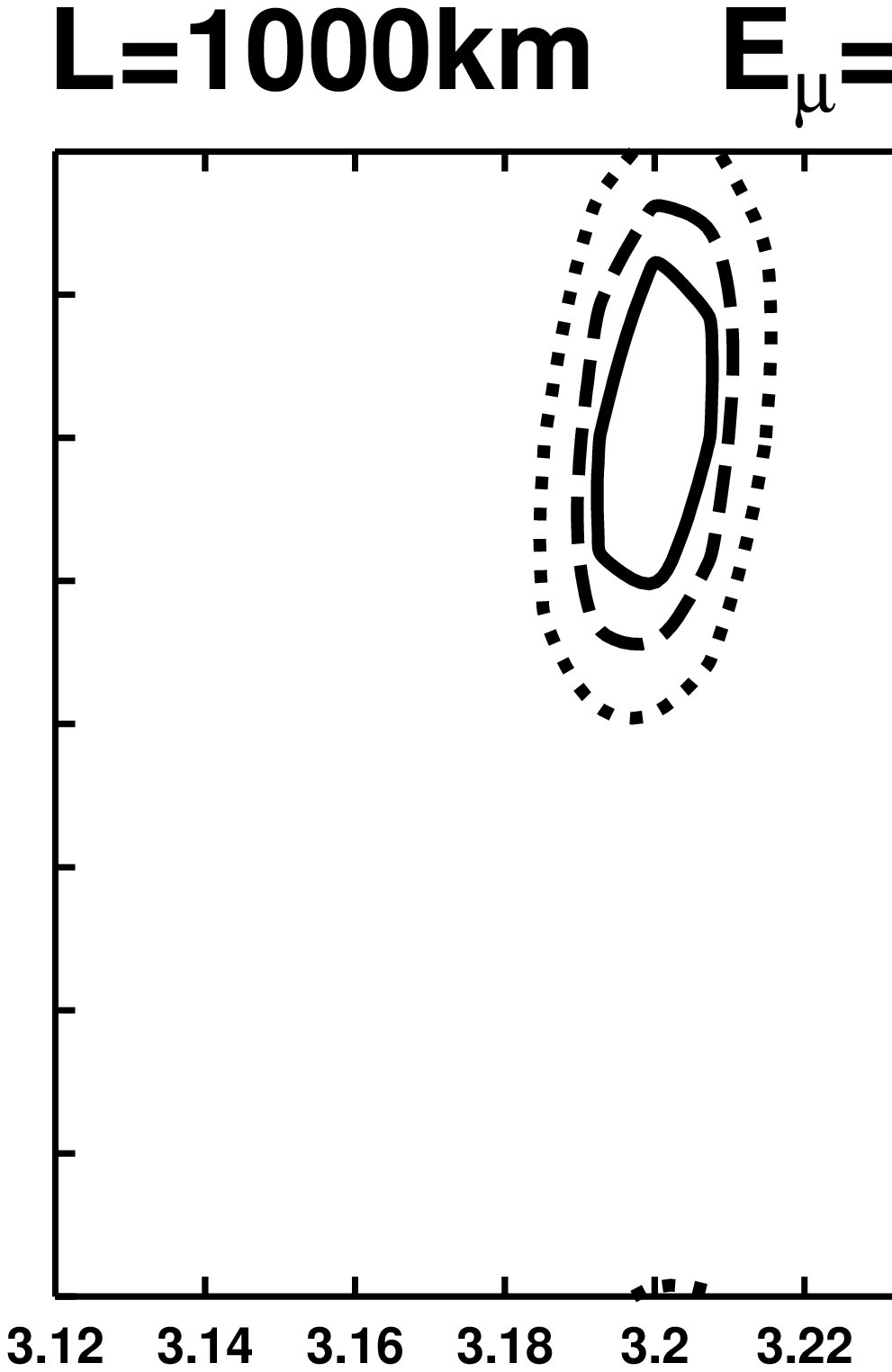,width=8cm}
\vglue -8.1cm \hglue -0.7cm \epsfig{file=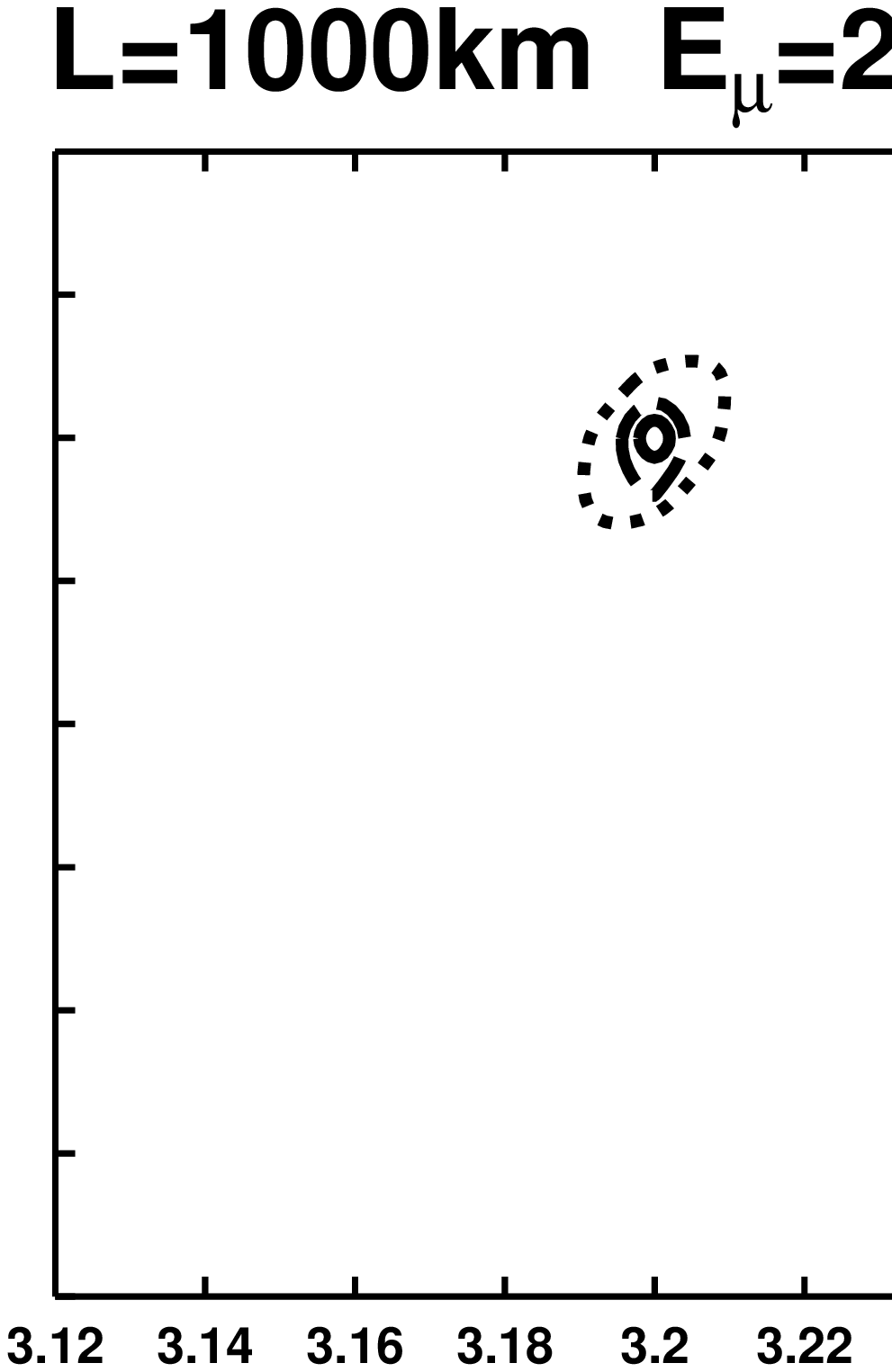,width=8cm}
\vglue -8.1cm \hglue 4.5cm \epsfig{file=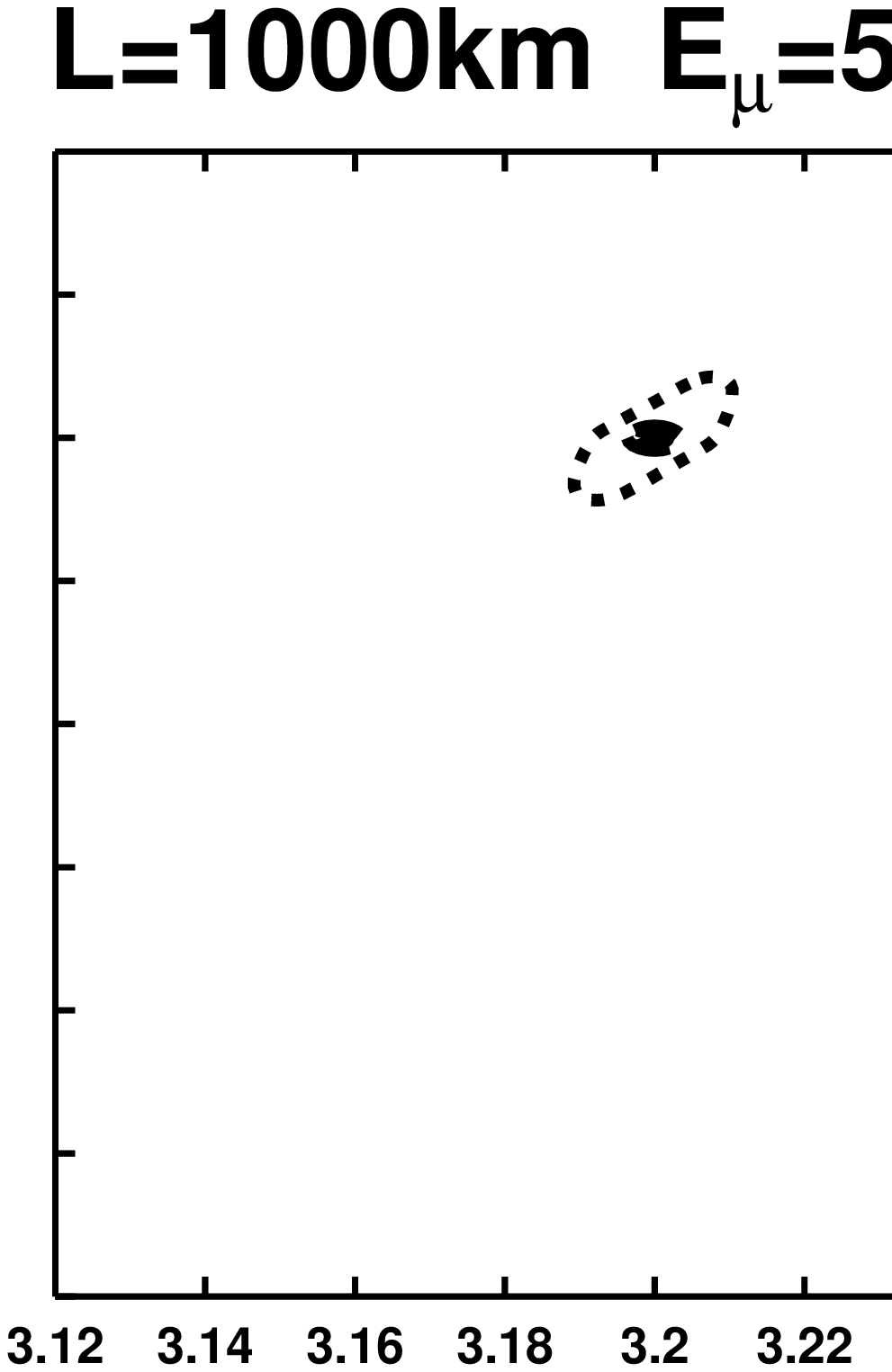,width=8cm}

\vglue -2.4cm
\hglue -6.0cm 
\epsfig{file=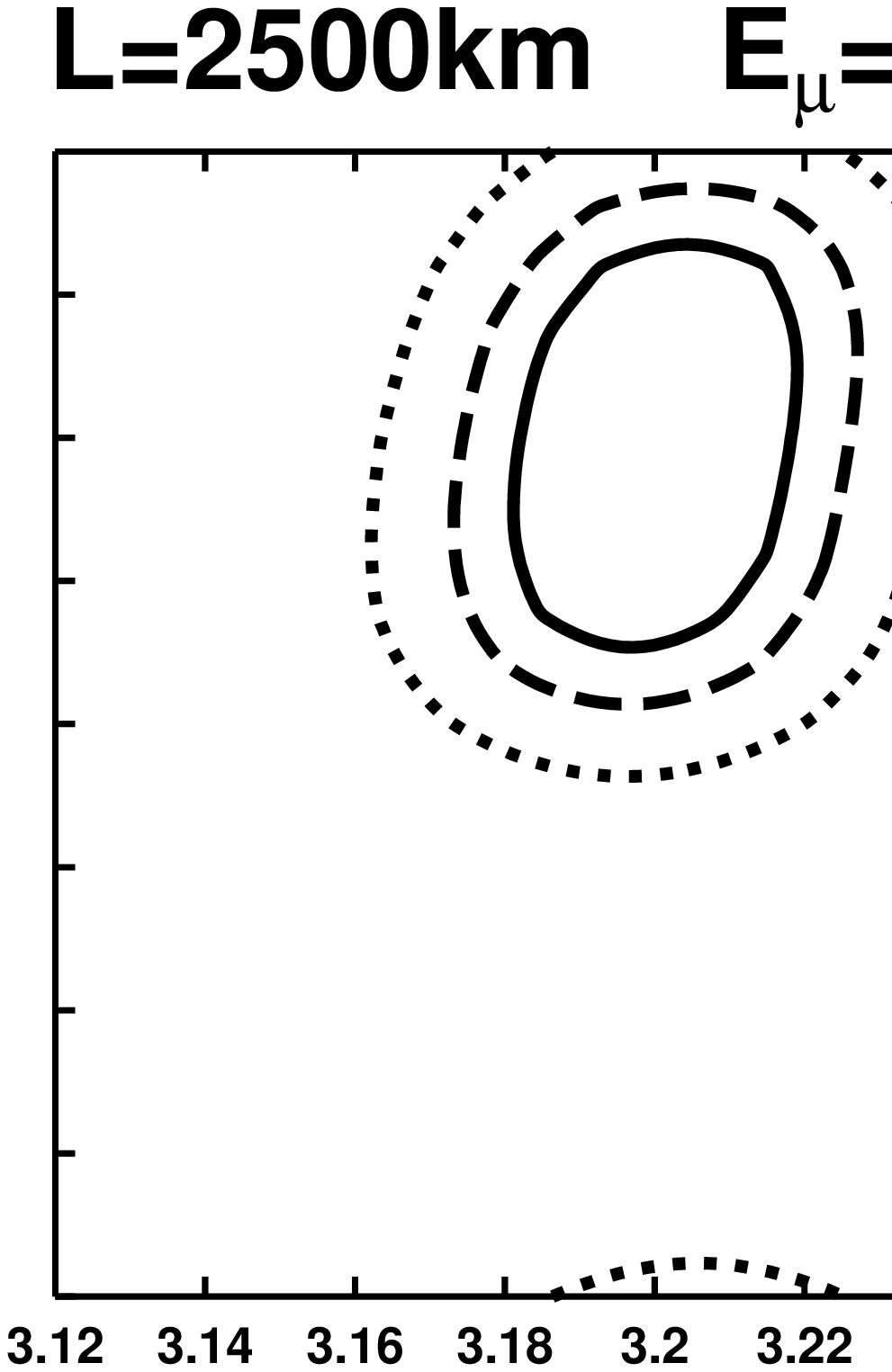,width=8cm}
\vglue -8.1cm \hglue -0.7cm \epsfig{file=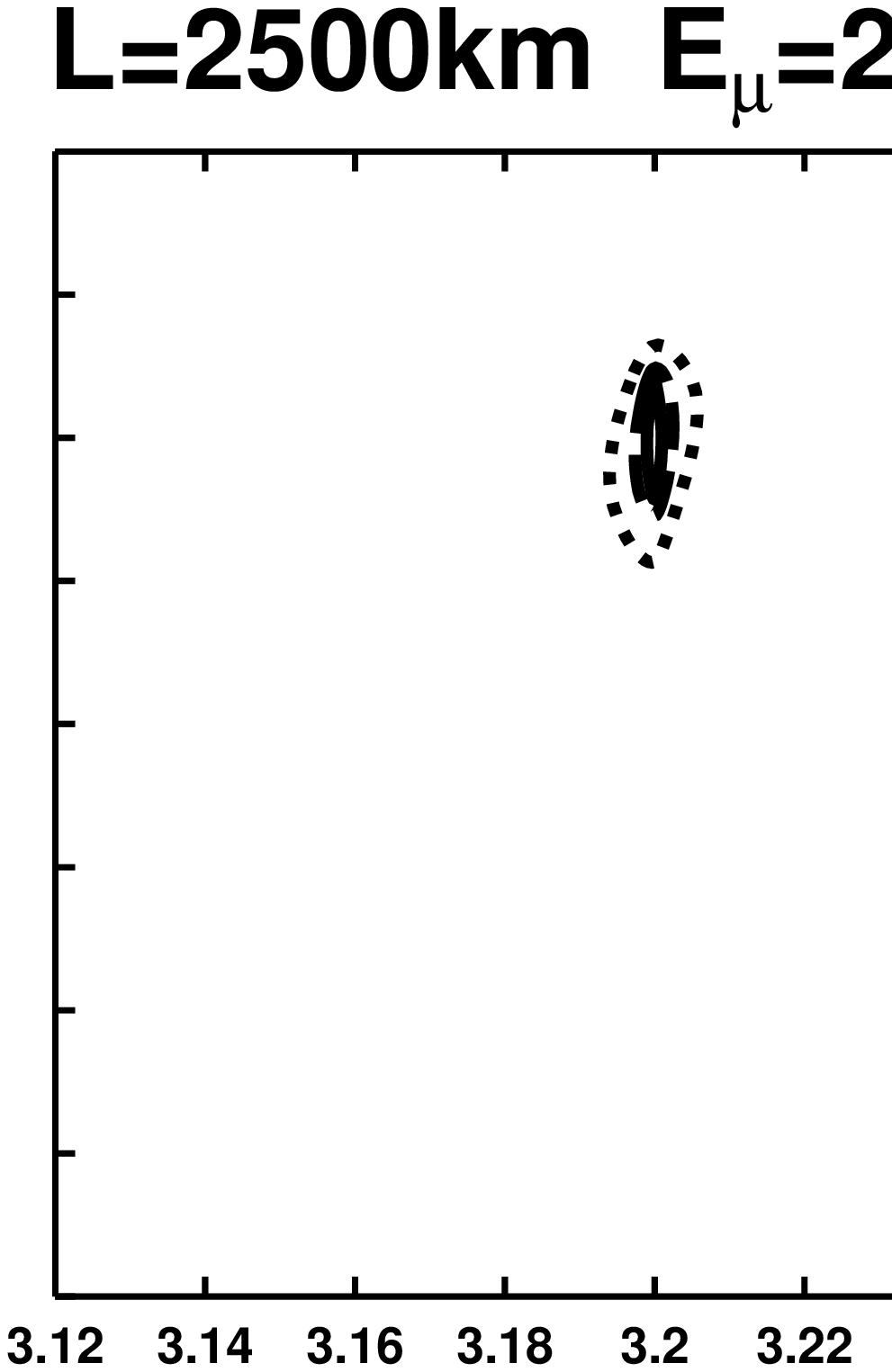,width=8cm}
\vglue -8.1cm \hglue 4.5cm \epsfig{file=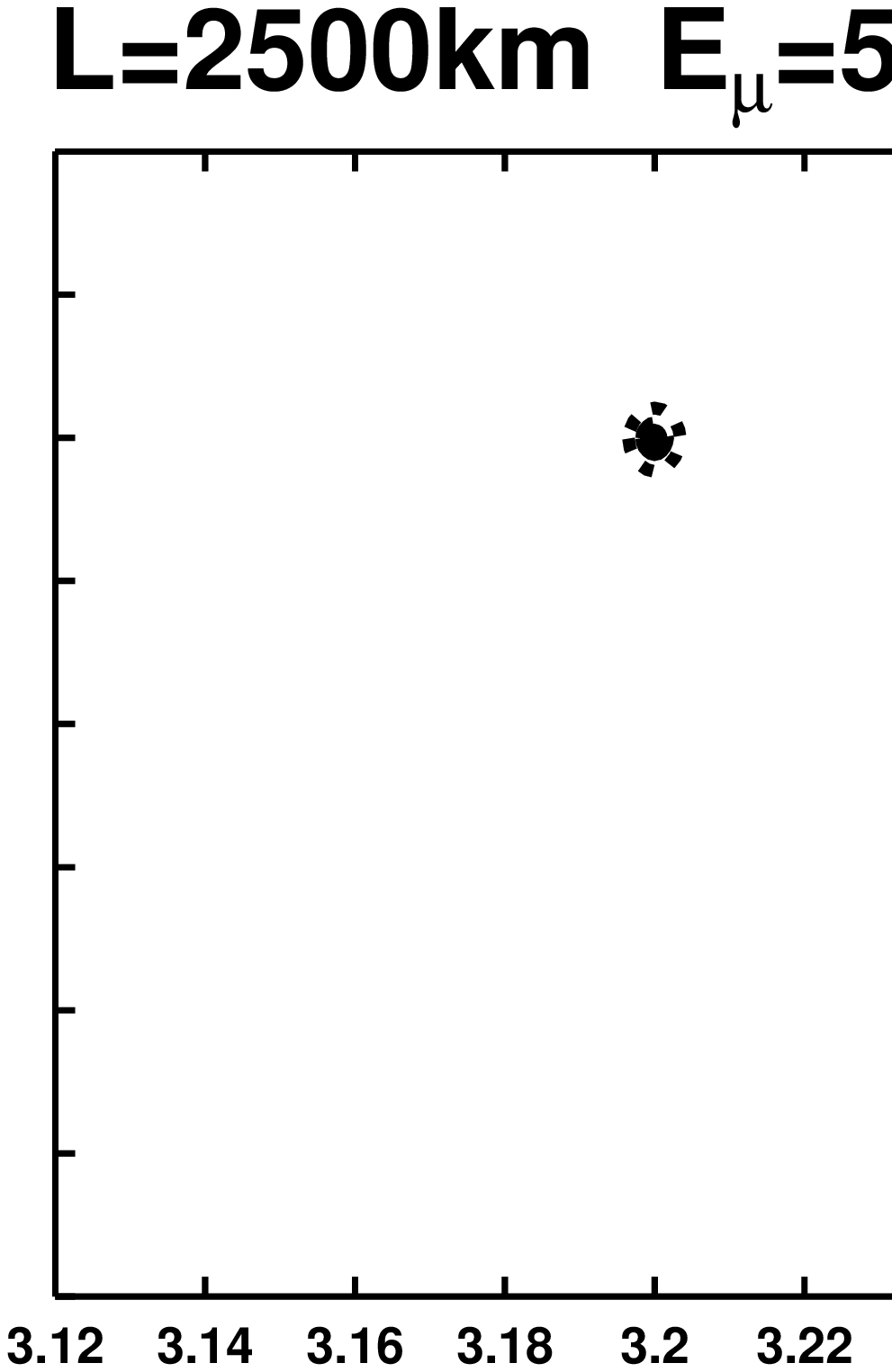,width=8cm}

\vglue -2.4cm
\hglue -6.0cm 
\epsfig{file=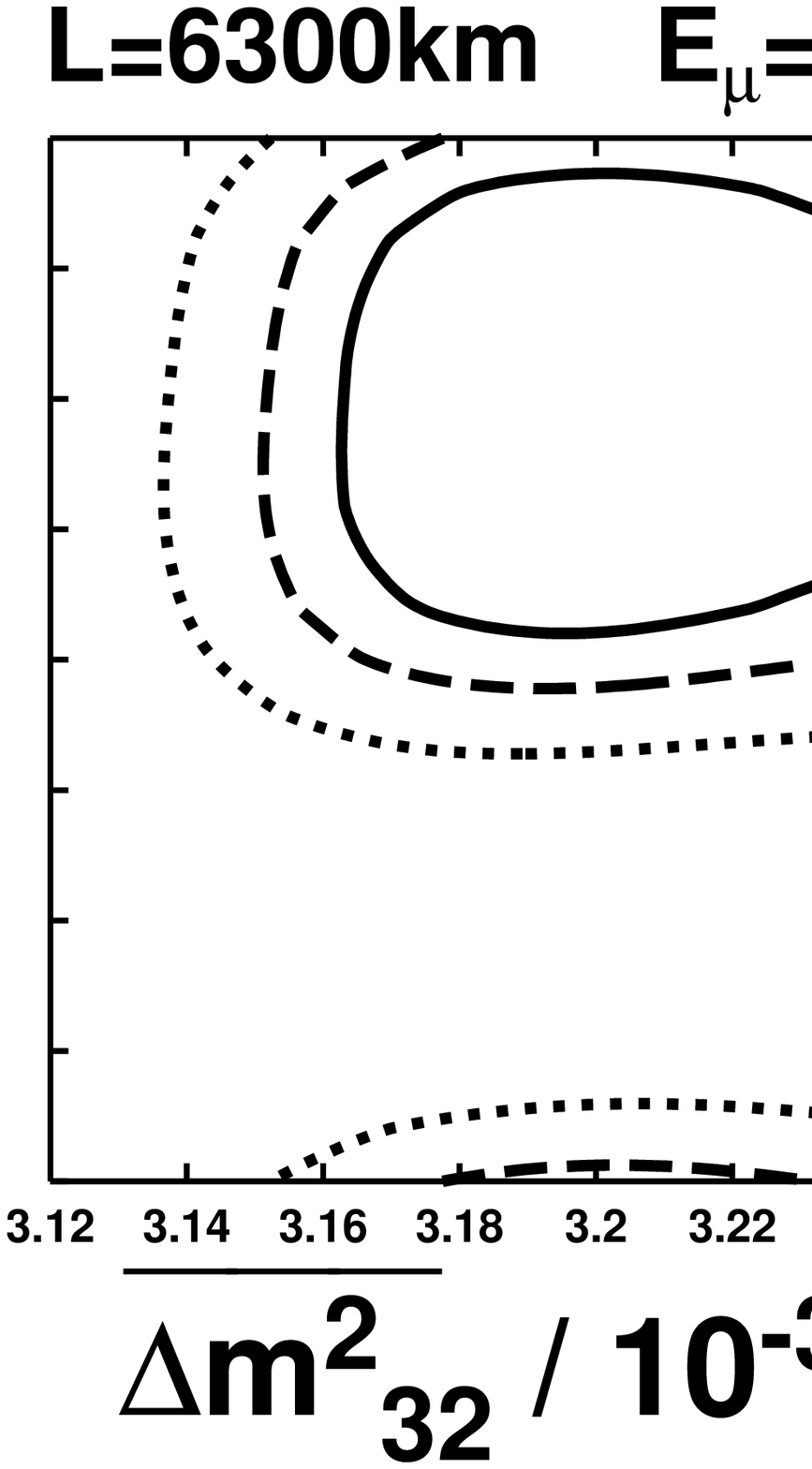,width=8cm}
\vglue -8.1cm \hglue -0.7cm \epsfig{file=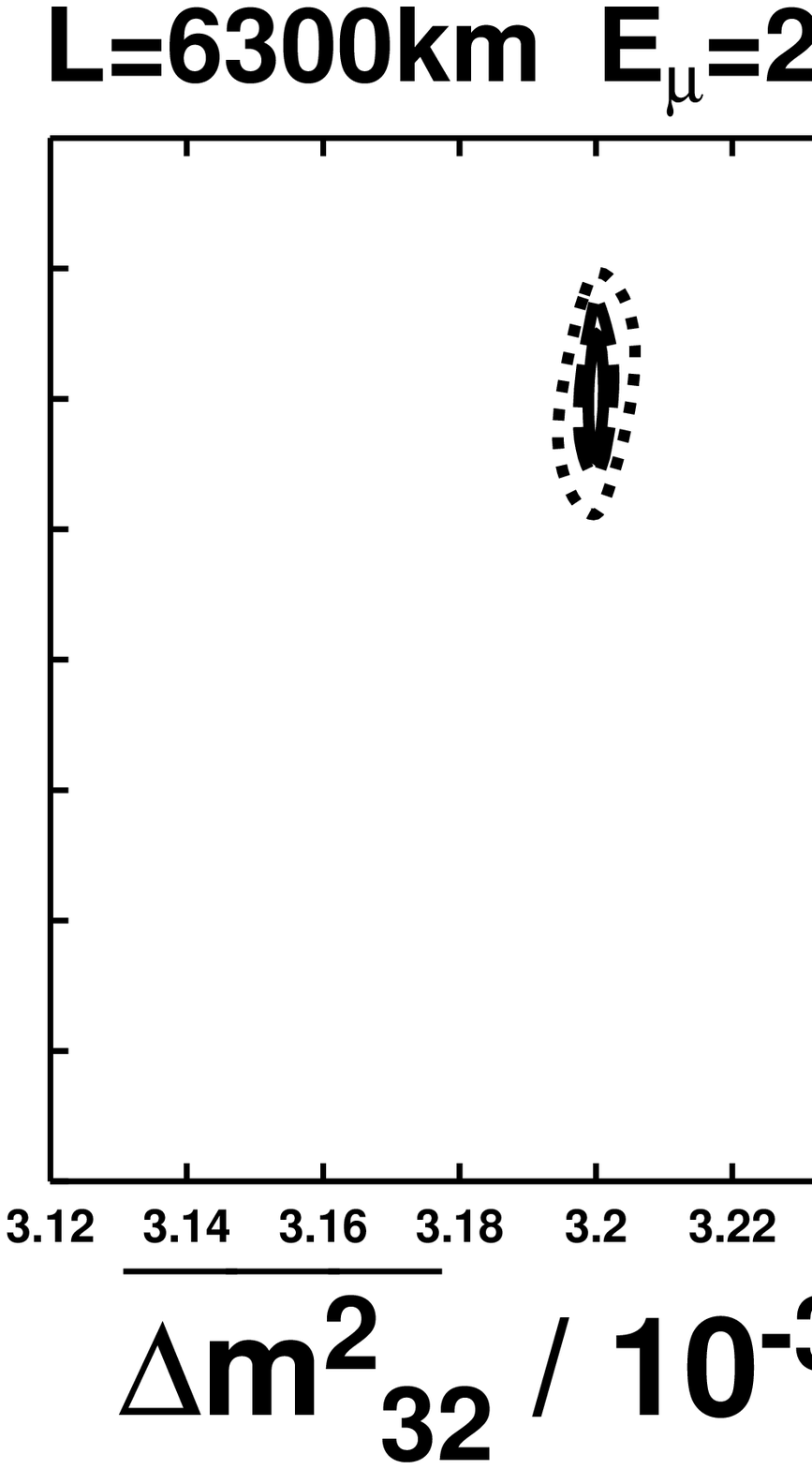,width=8cm}
\vglue -8.1cm \hglue 4.5cm \epsfig{file=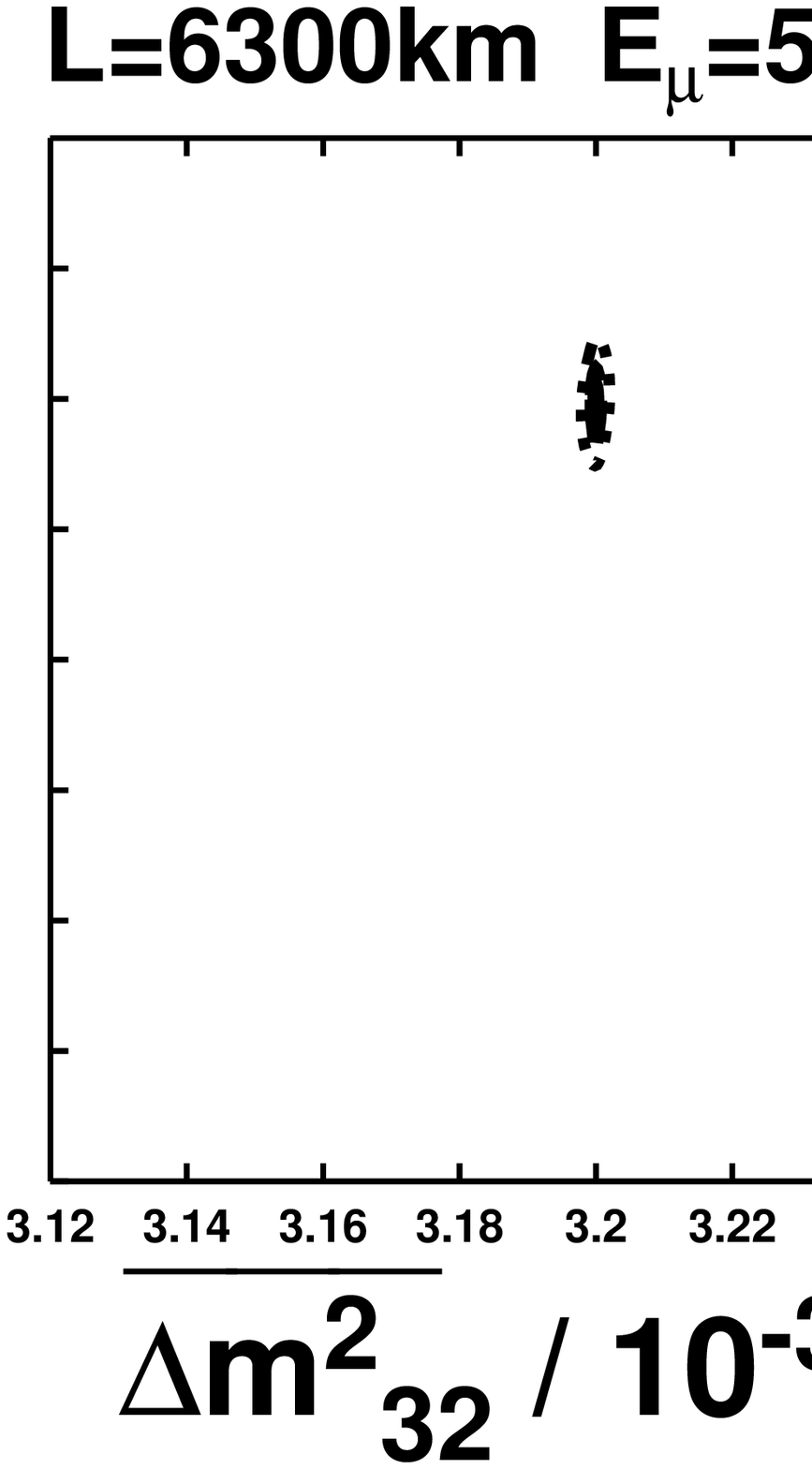,width=8cm}
\vglue -2.0cm\hglue -23.3cm
\epsfig{file=fig0613.eps,width=22cm}
\vglue -21.5cm\hglue 6.3cm
\epsfig{file=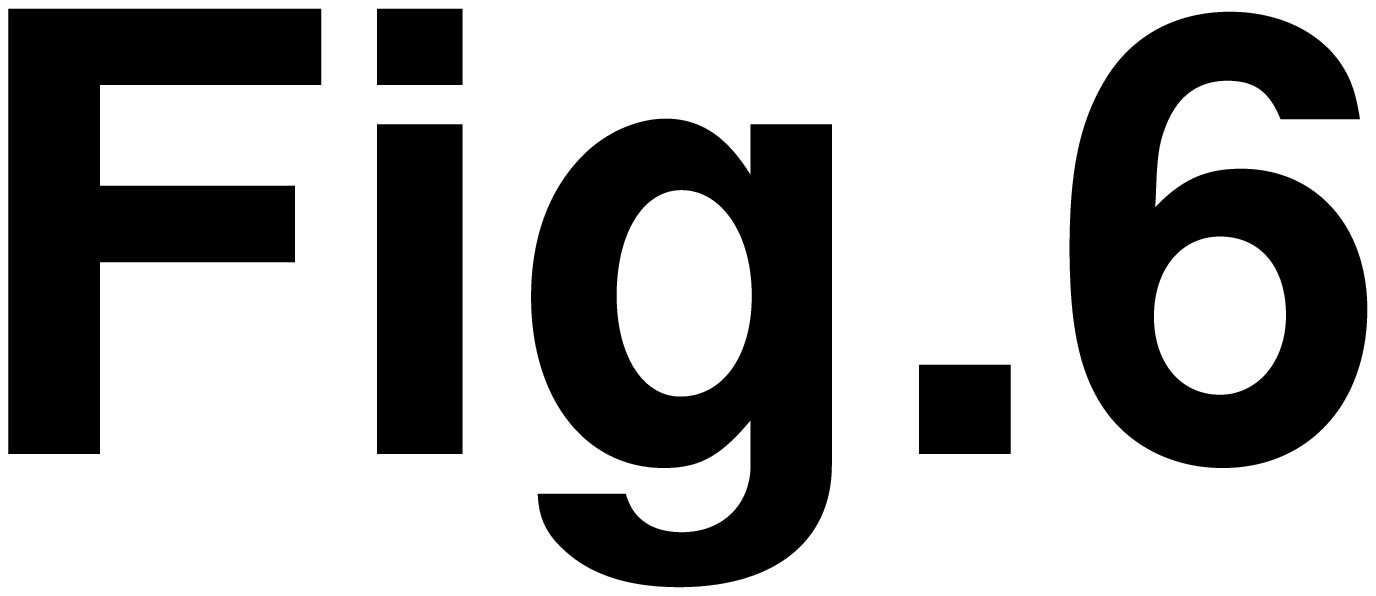,width=4cm}
\newpage
\vglue -5.10cm \hglue -3.0cm
\epsfig{file=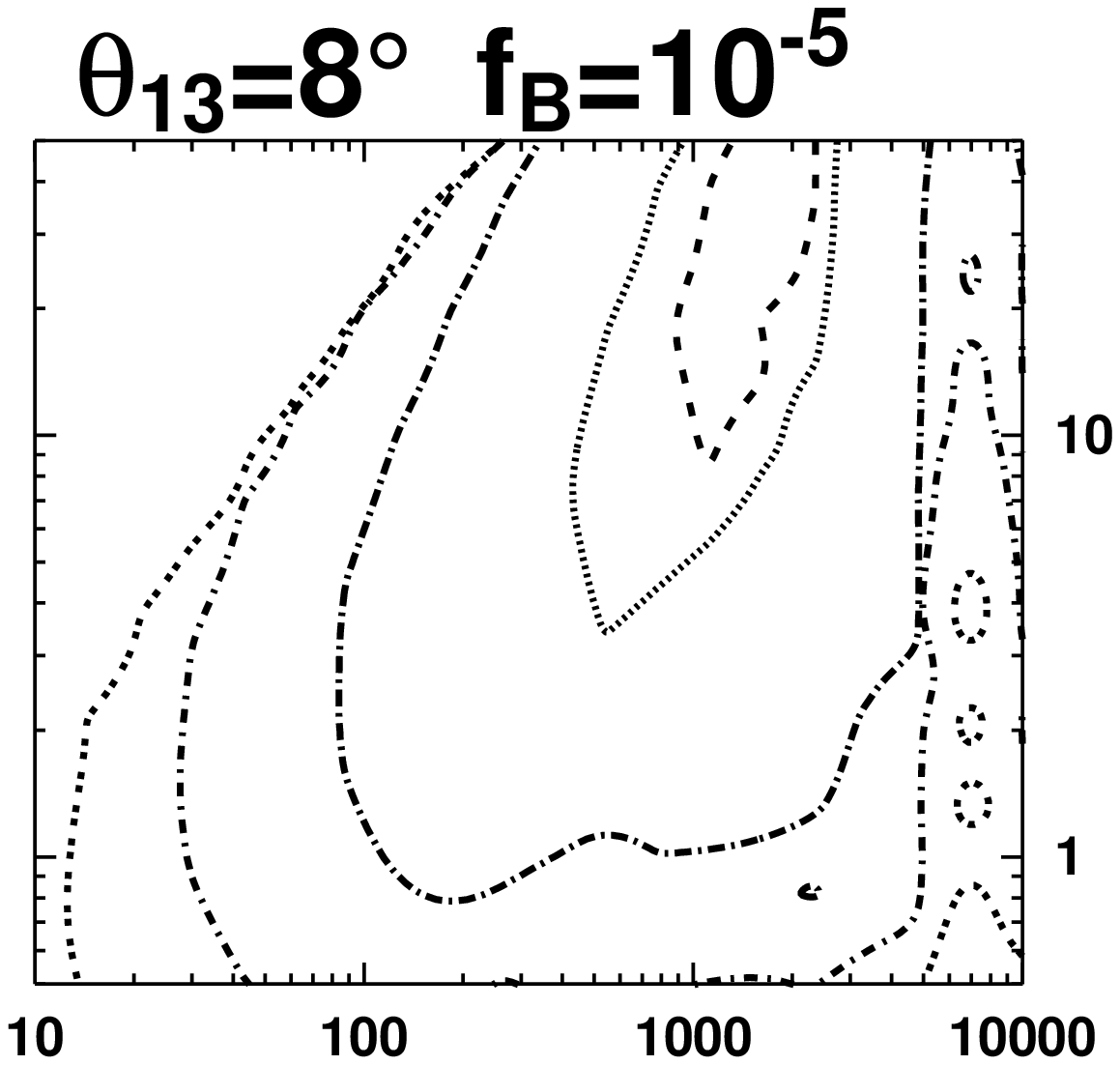,width=11cm}
\vglue -12.4cm \hglue 4.8cm
\epsfig{file=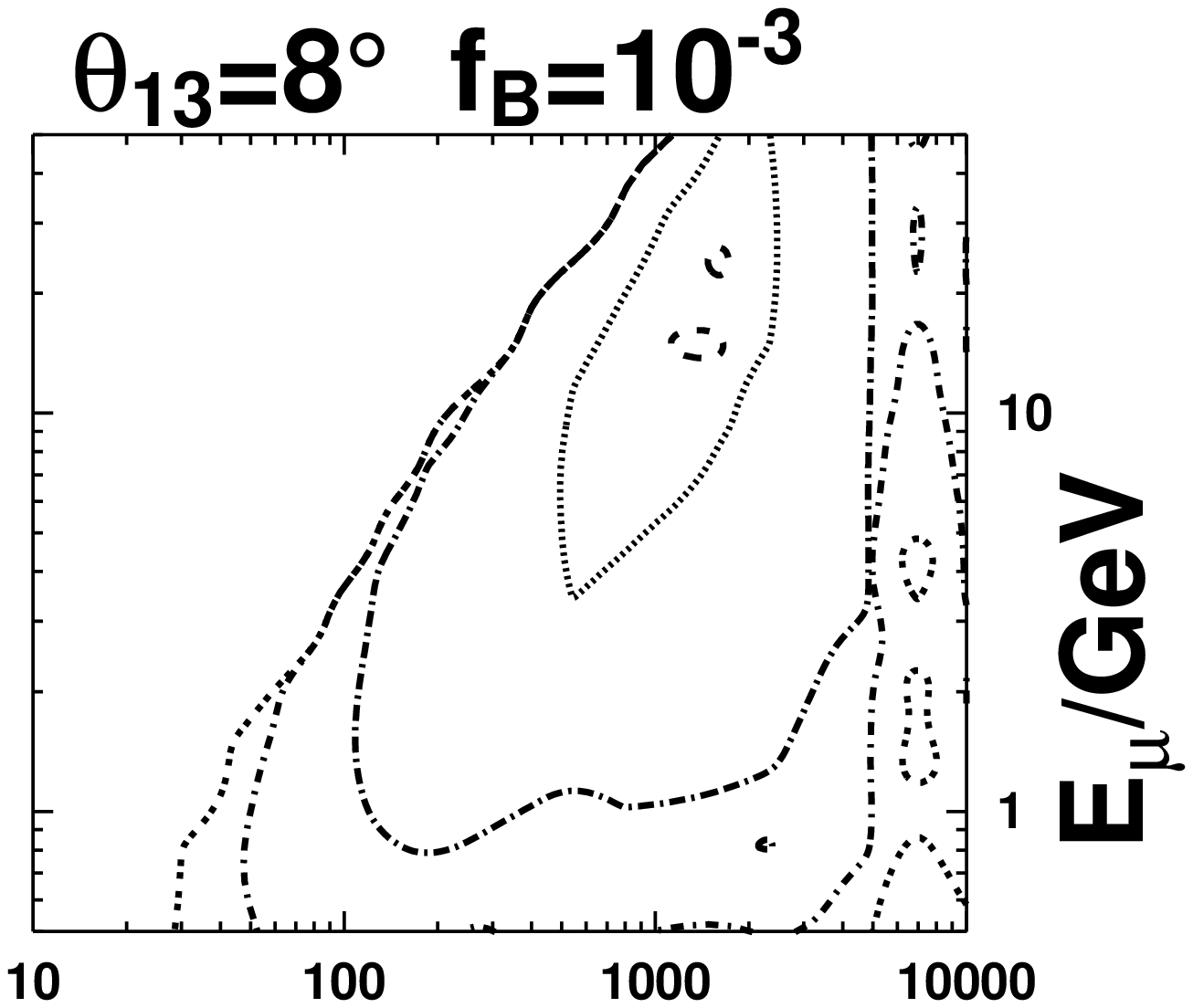,width=11cm}
\vglue -5.0cm \hglue -3.0cm
\epsfig{file=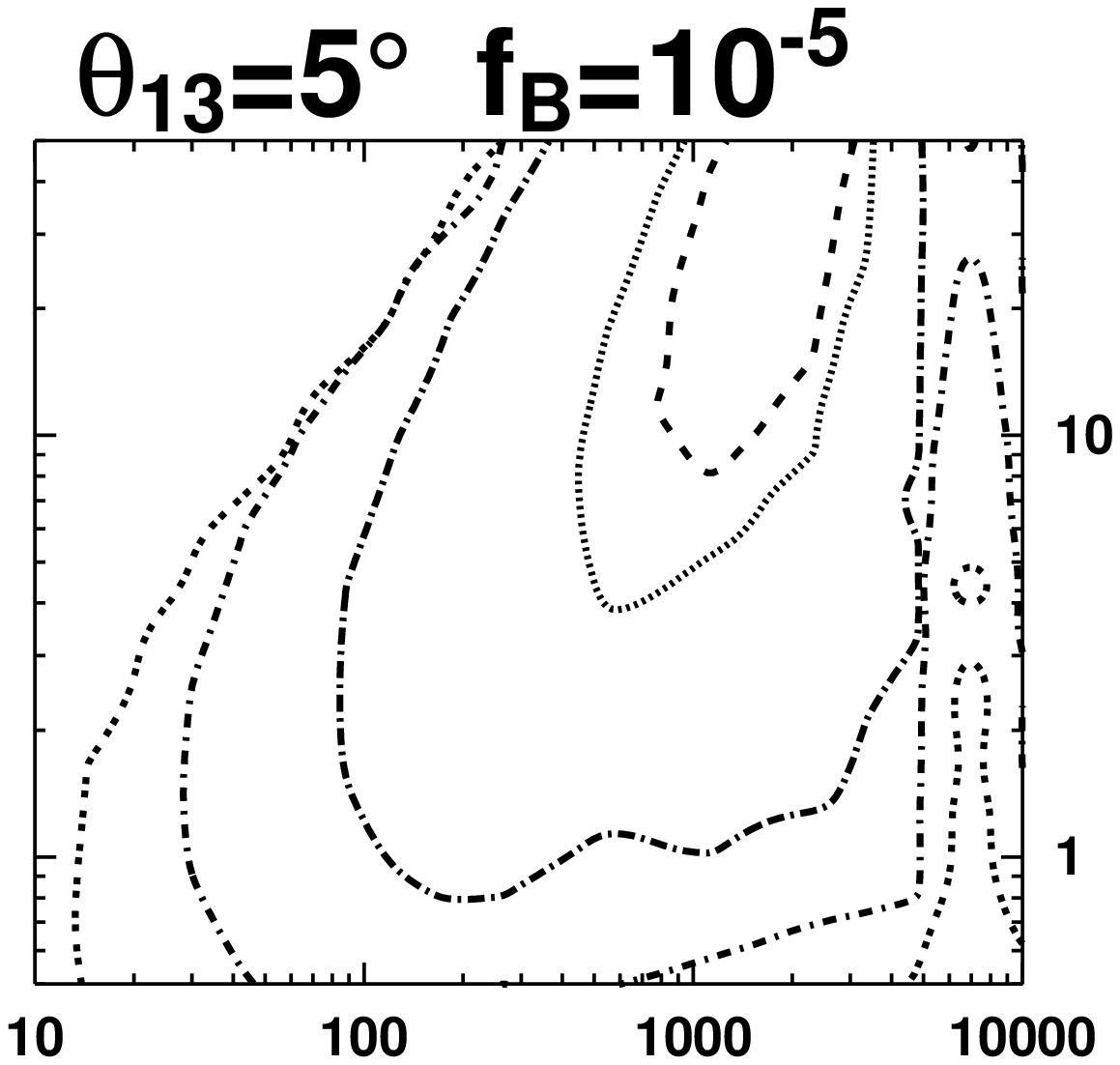,width=11cm}
\vglue -12.5cm \hglue 4.8cm
\epsfig{file=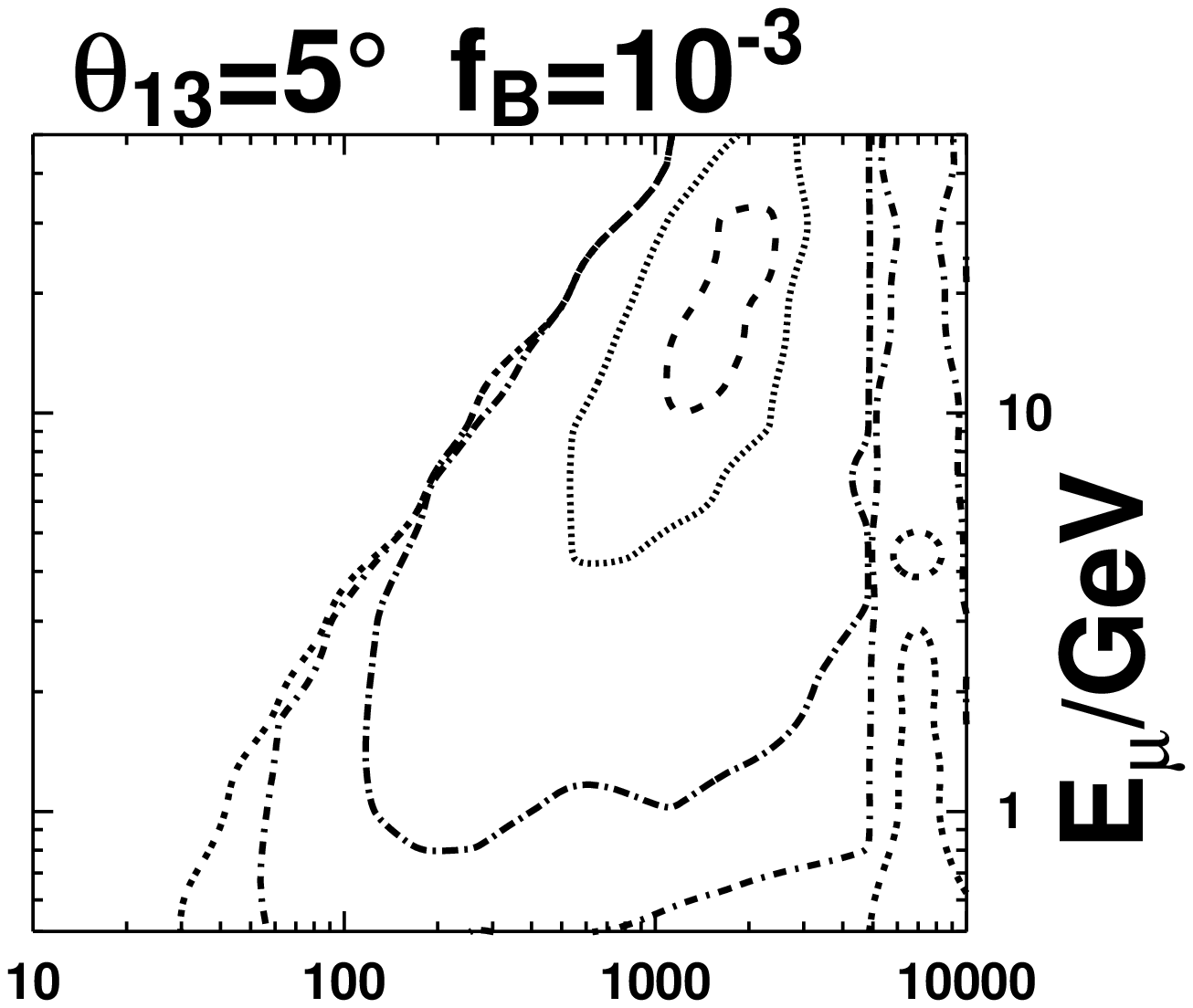,width=11cm}
\vglue -5.0cm \hglue -3.0cm
\epsfig{file=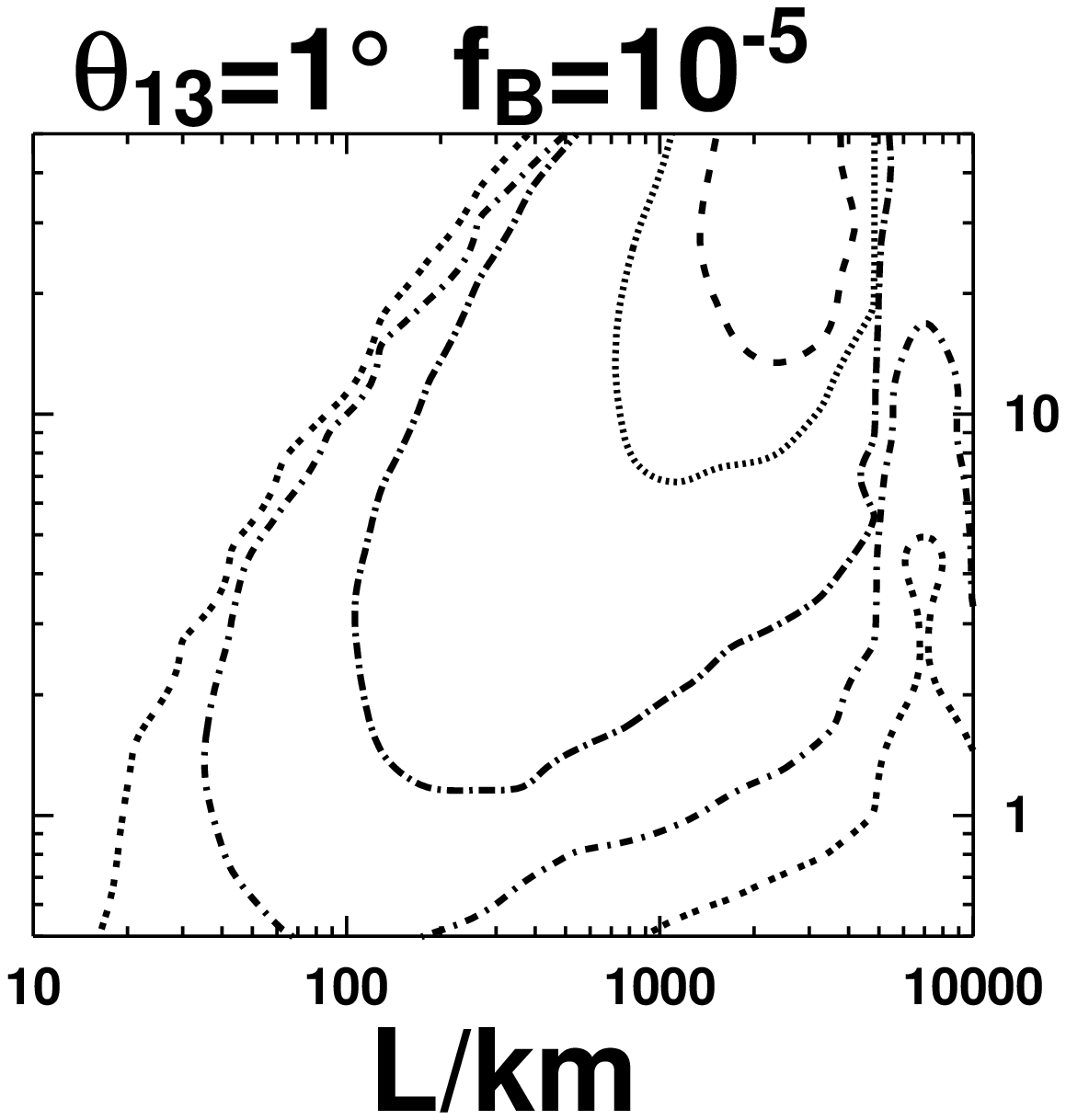,width=11cm}
\vglue -12.5cm \hglue 4.8cm
\epsfig{file=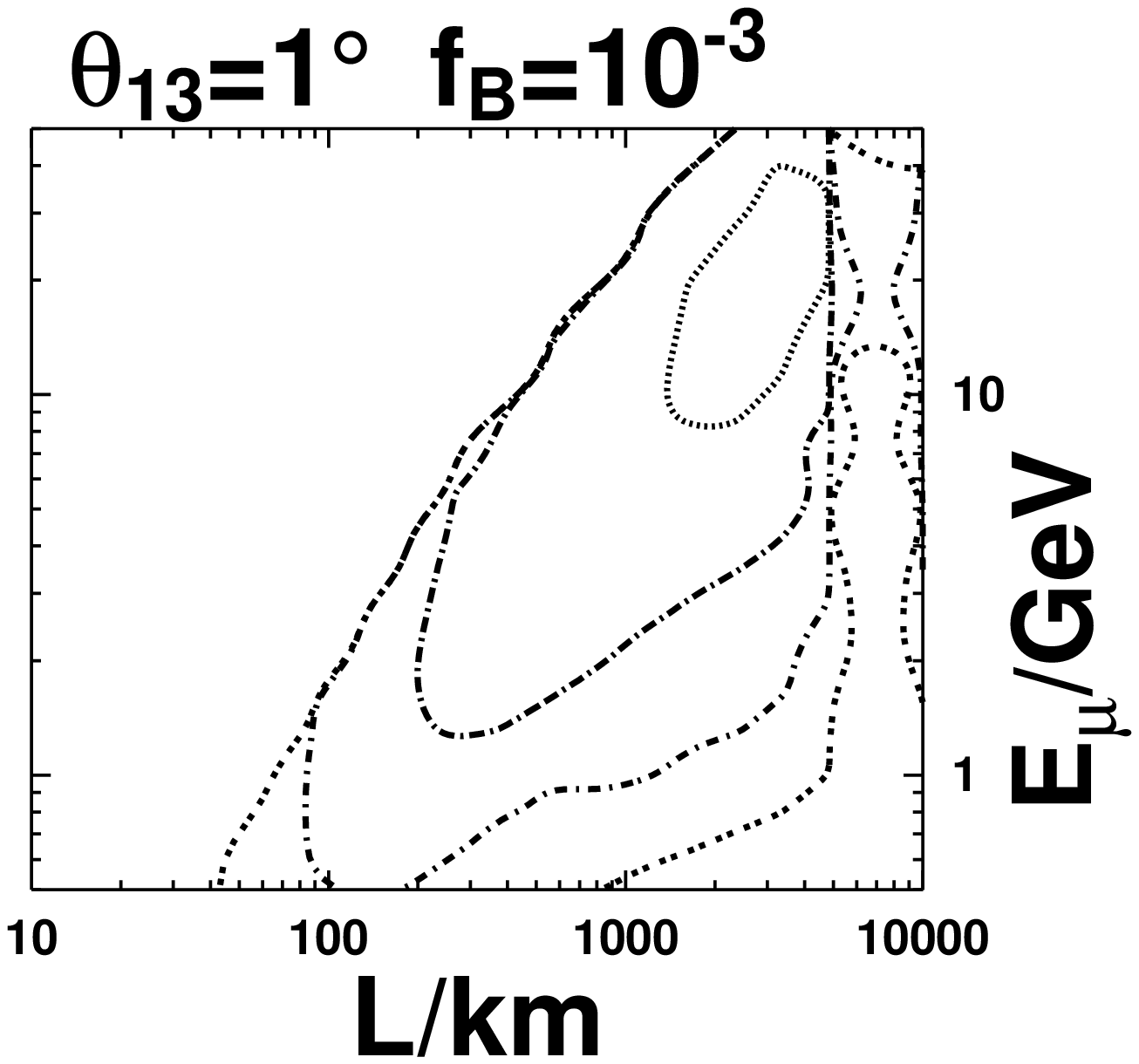,width=11cm}
\vglue -4.5cm\hglue -14.3cm
\epsfig{file=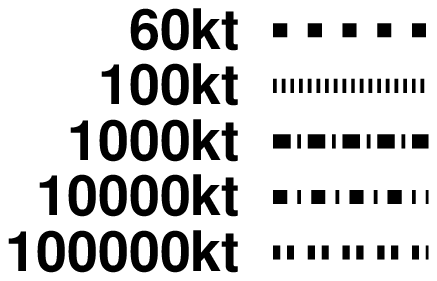,width=18cm}
\vglue -16.5cm\hglue 6.3cm
\epsfig{file=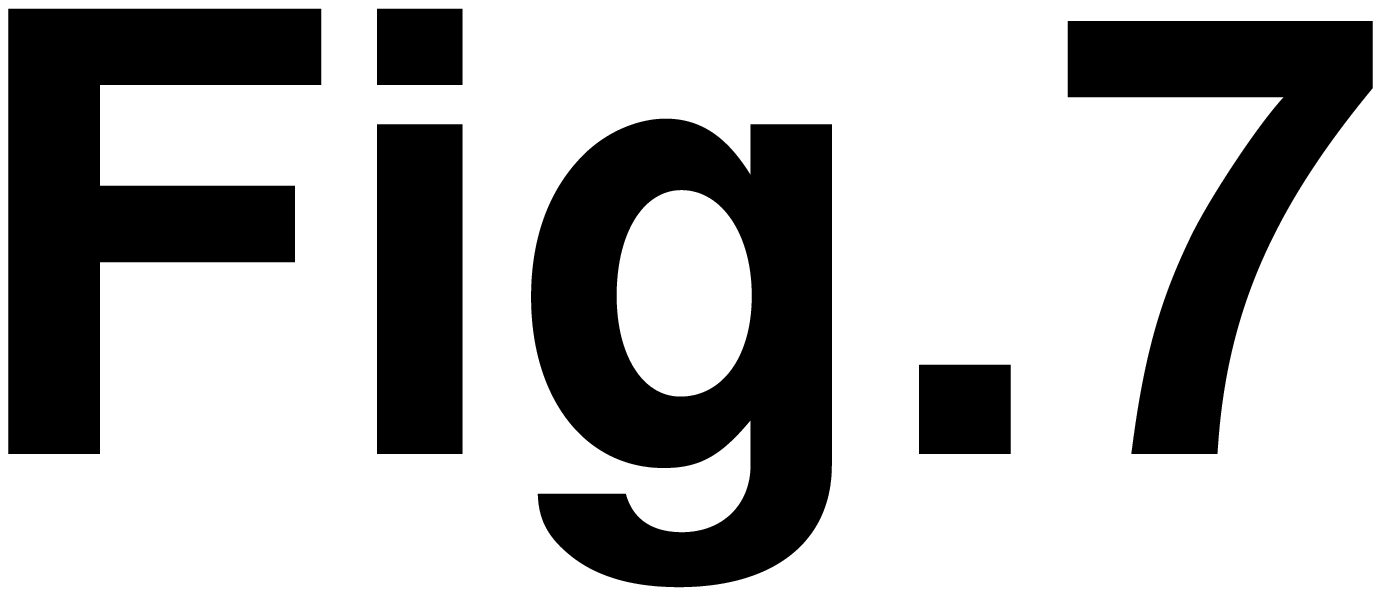,width=4cm}
\newpage
\vglue -1.60cm \hglue -3.0cm
\epsfig{file=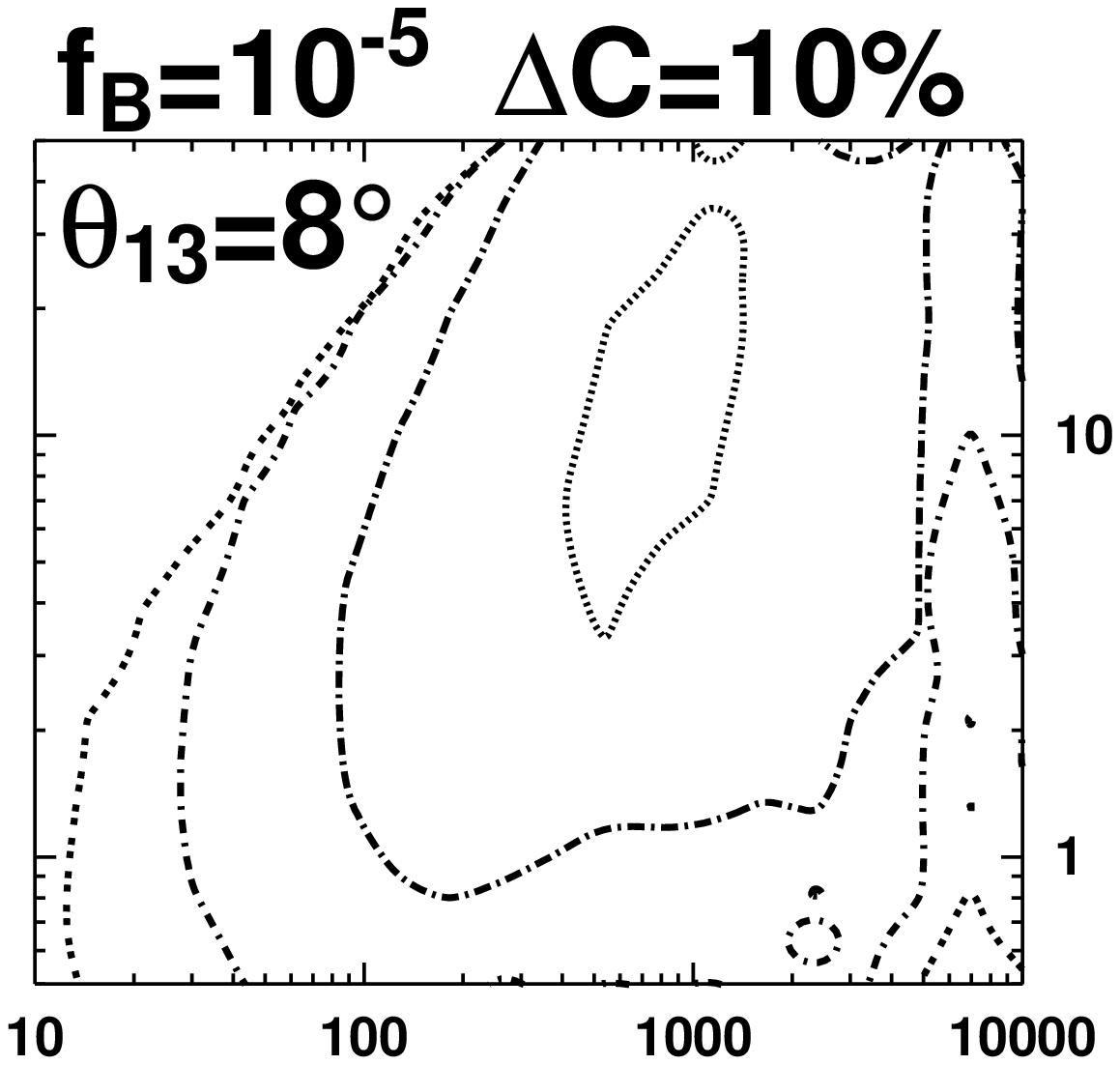,width=11cm}
\vglue -12.4cm \hglue 4.8cm
\epsfig{file=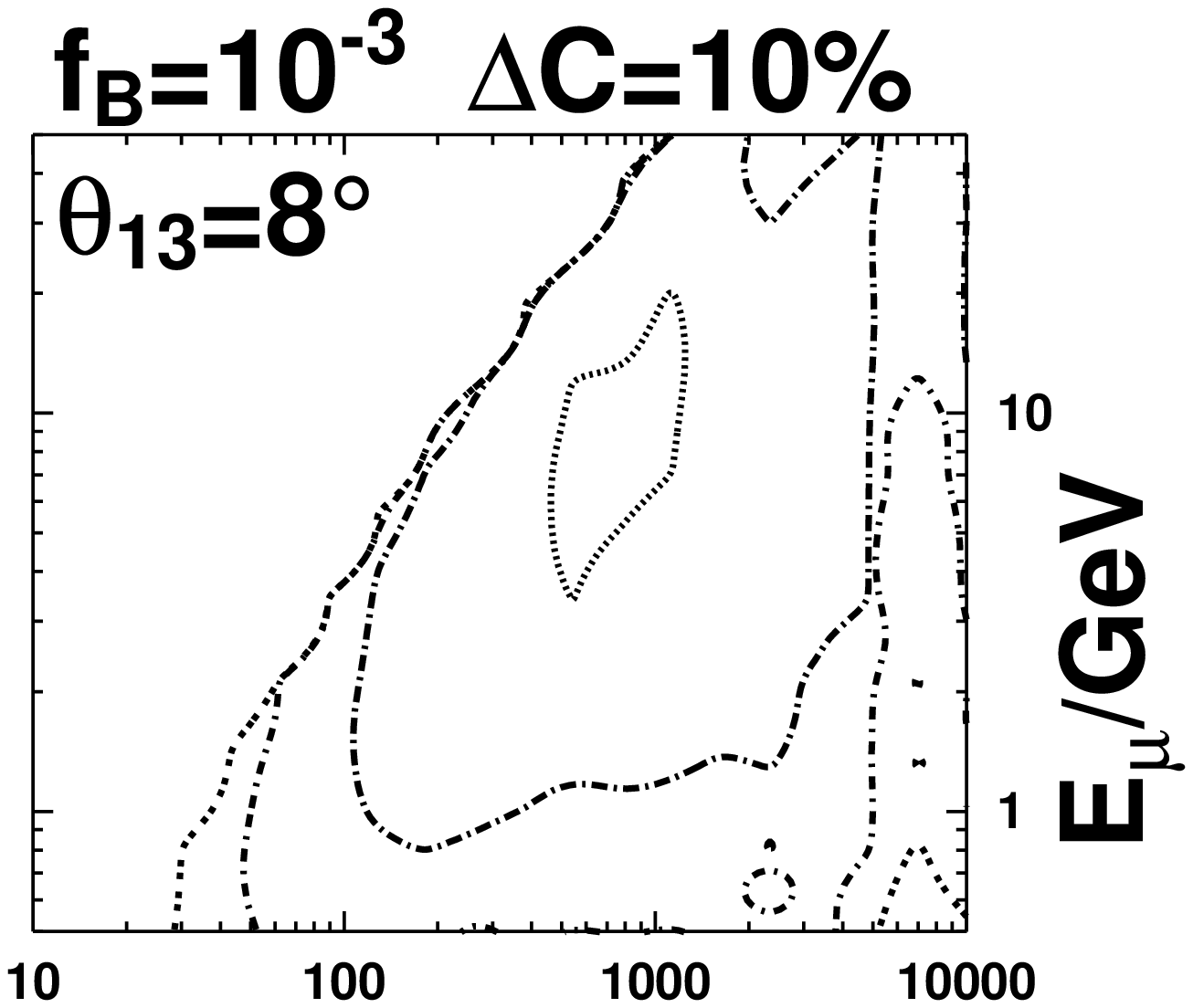,width=11cm}
\vglue -4.5cm \hglue -3.0cm
\epsfig{file=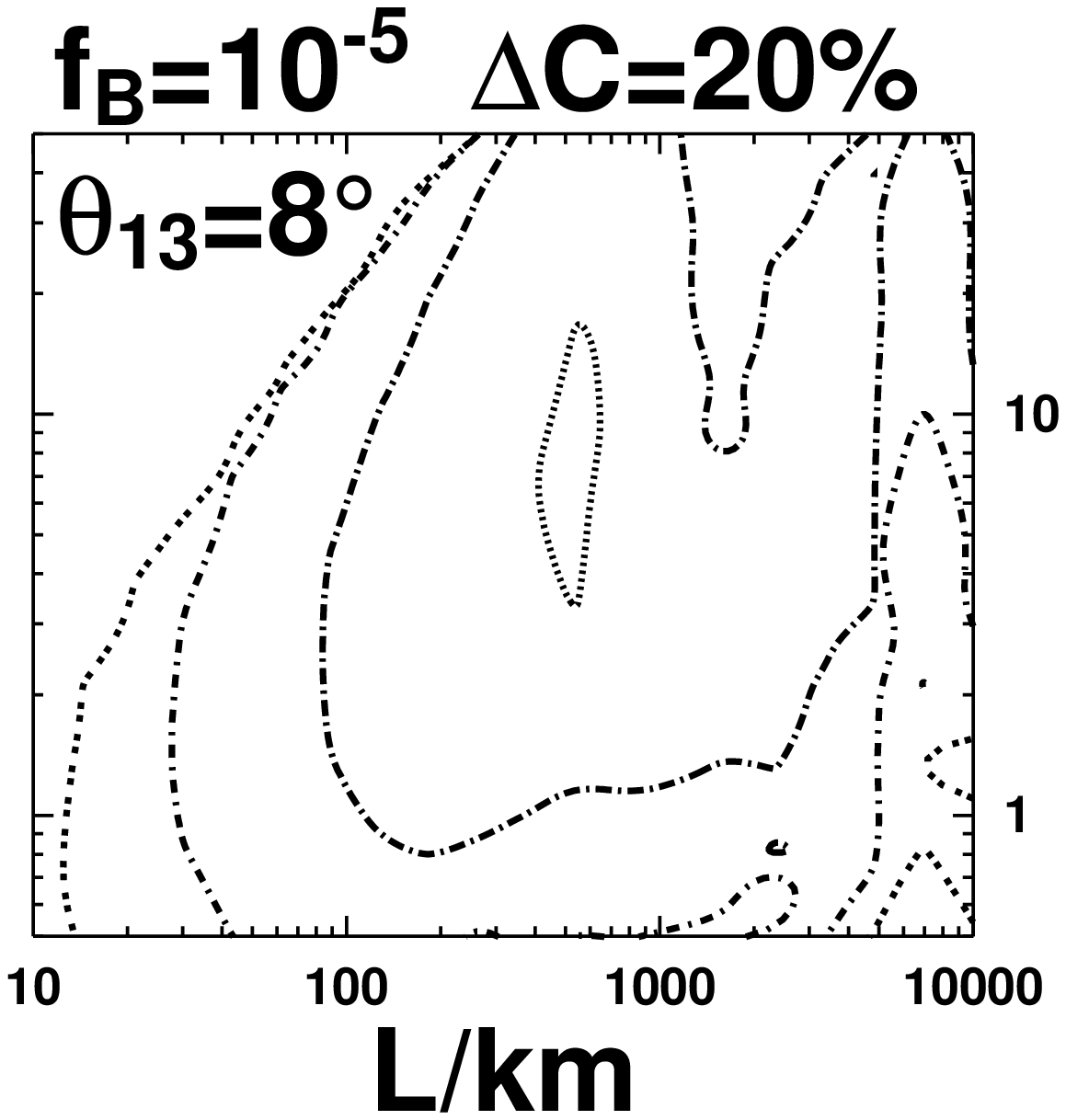,width=11cm}
\vglue -12.5cm \hglue 4.8cm
\epsfig{file=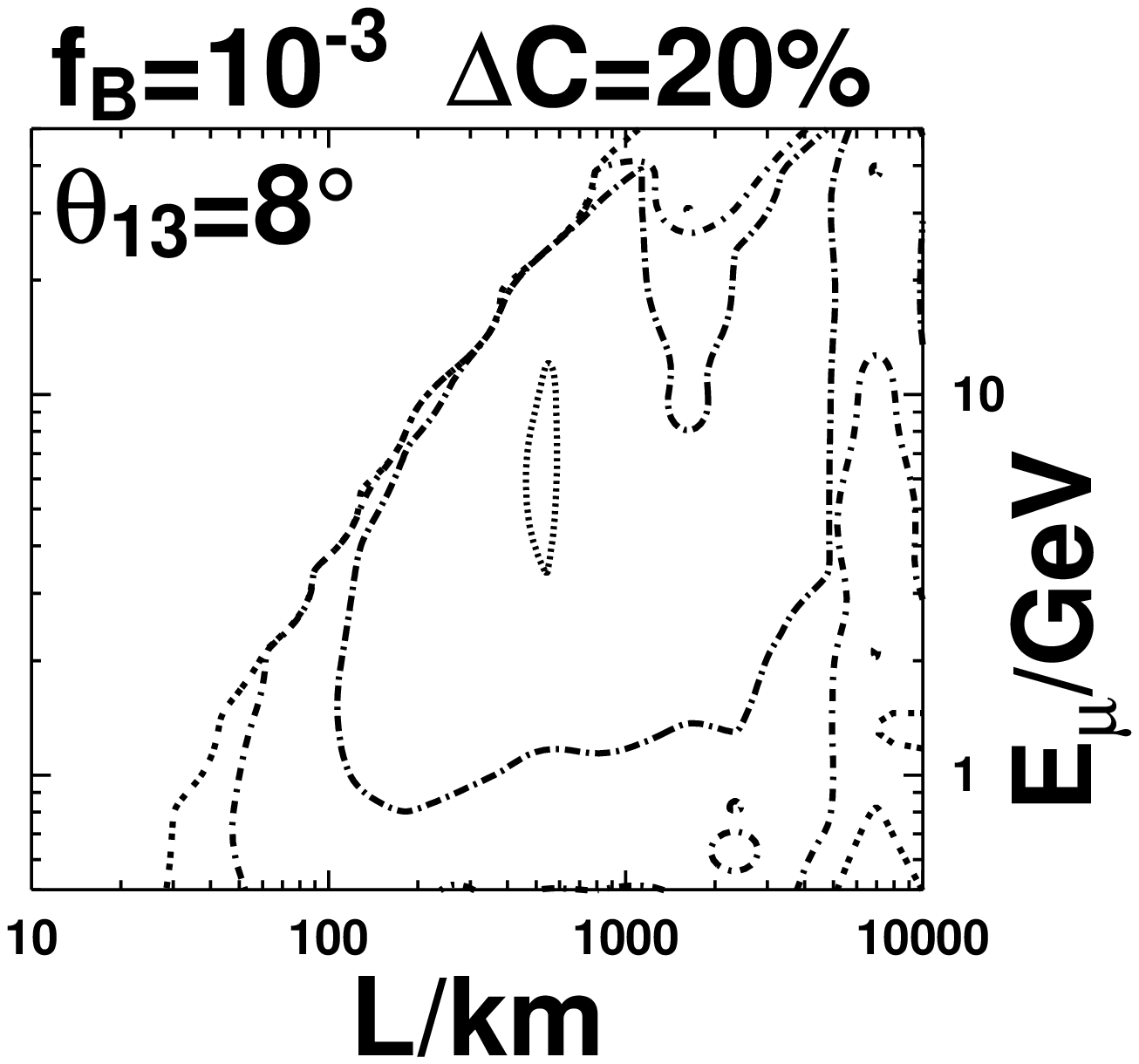,width=11cm}
\vglue -4.5cm\hglue -14.3cm
\epsfig{file=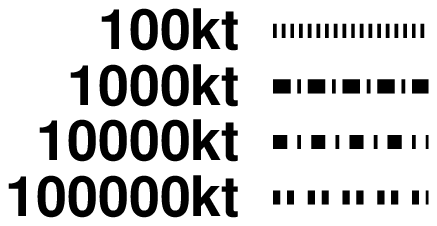,width=18cm}
\vglue -16.5cm\hglue 6.3cm
\epsfig{file=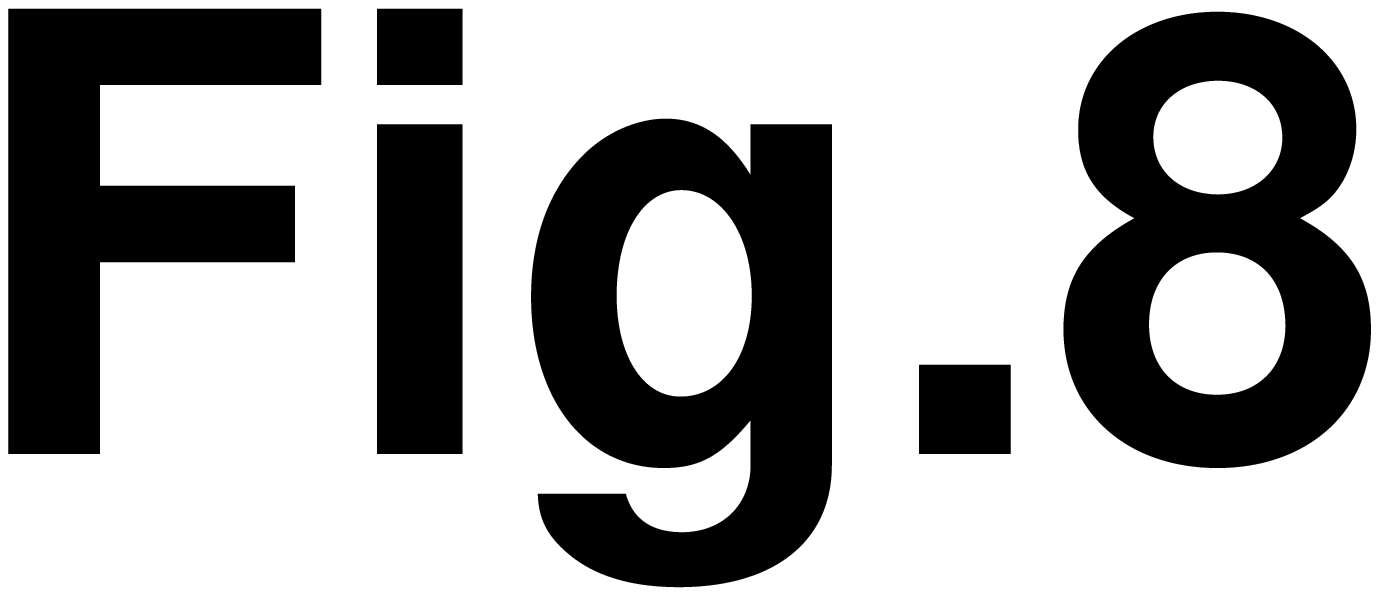,width=4cm}
\newpage
\vglue -2.0cm
\hglue -1.2cm 
\epsfig{file=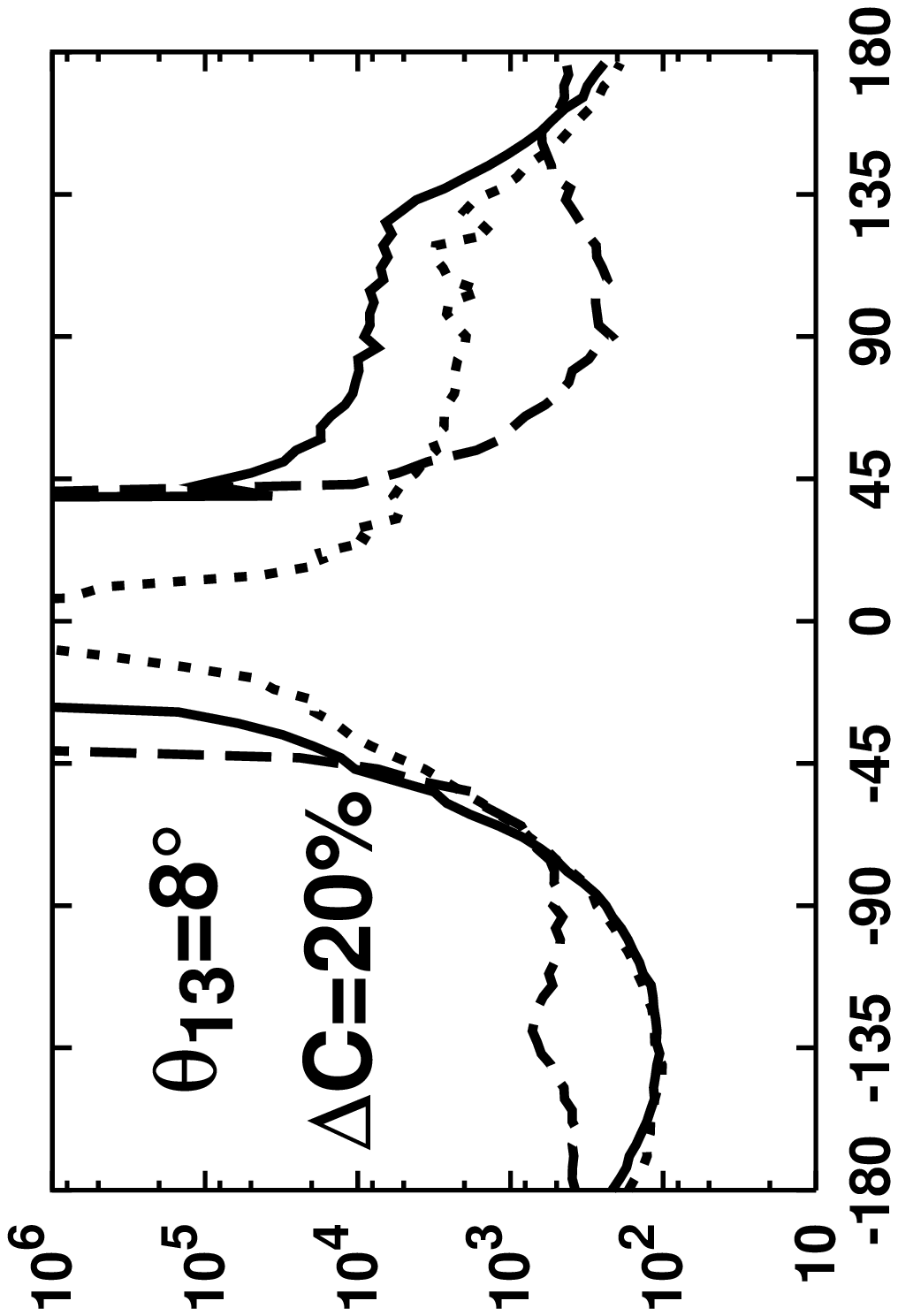,width=5.9cm}
\vglue -8.2cm \hglue 4.2cm \epsfig{file=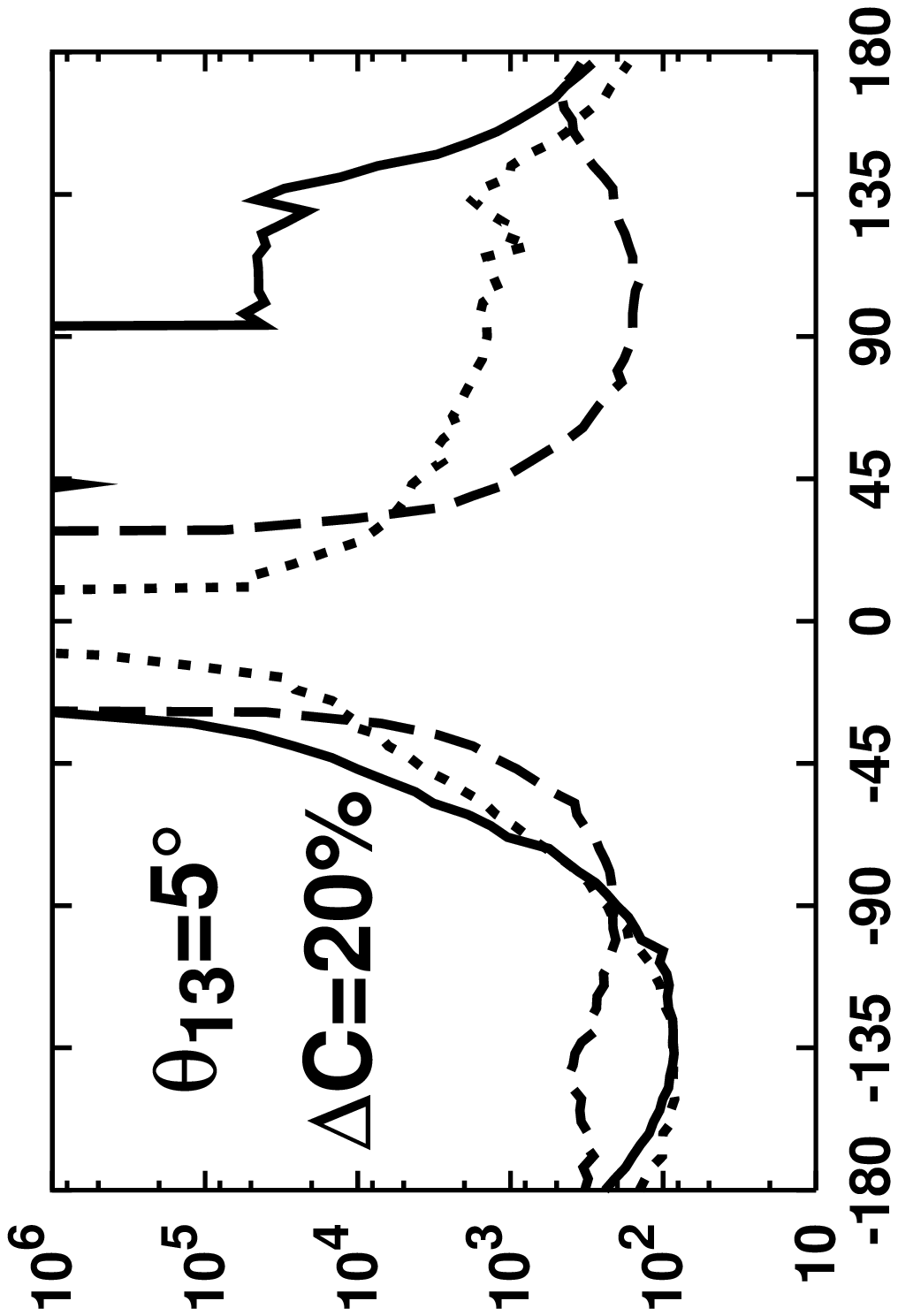,width=5.9cm}
\vglue -8.2cm \hglue 9.5cm \epsfig{file=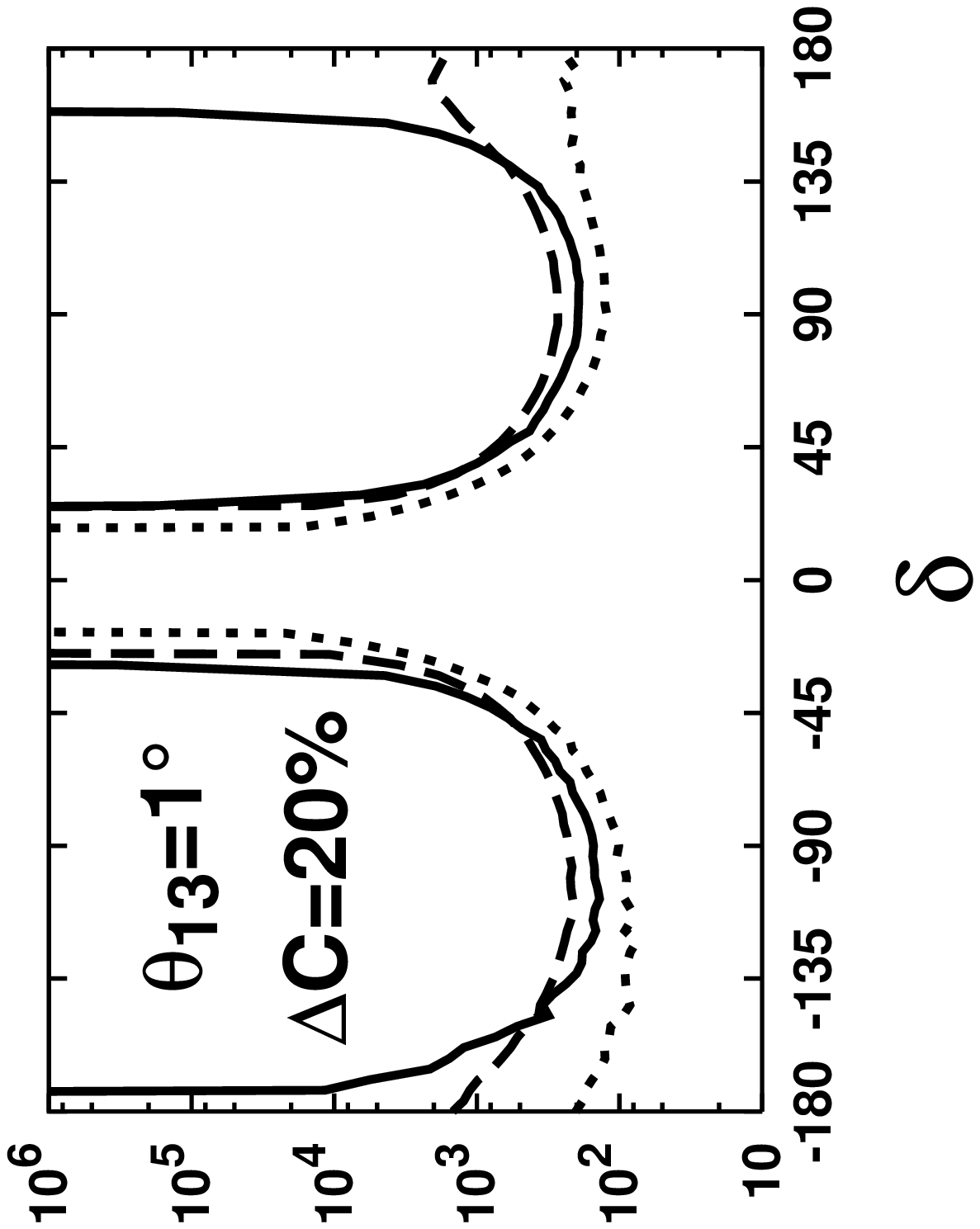,width=5.9cm}

\vglue -1.0cm
\hglue -1.2cm 
\epsfig{file=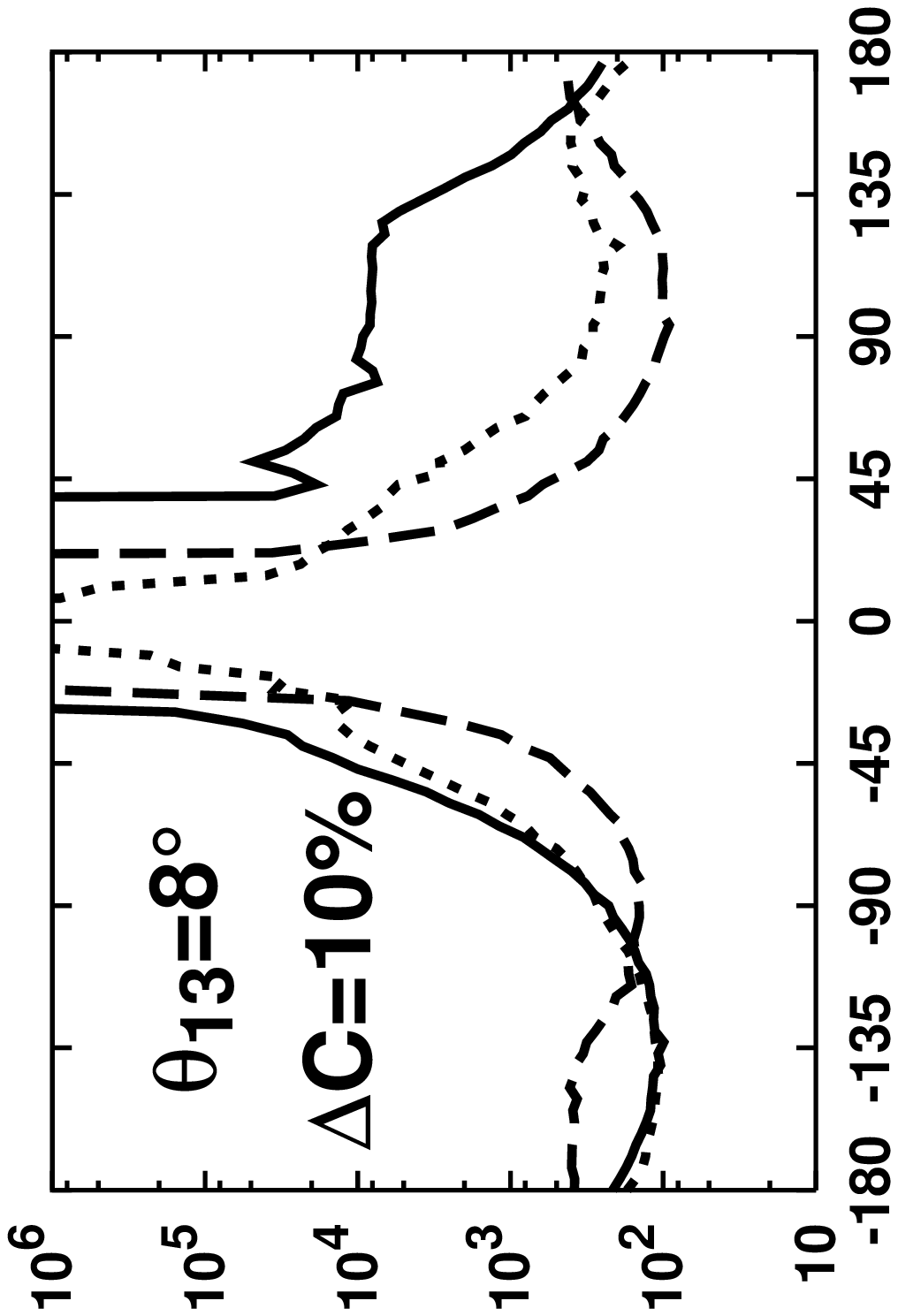,width=5.9cm}
\vglue -8.2cm \hglue 4.2cm \epsfig{file=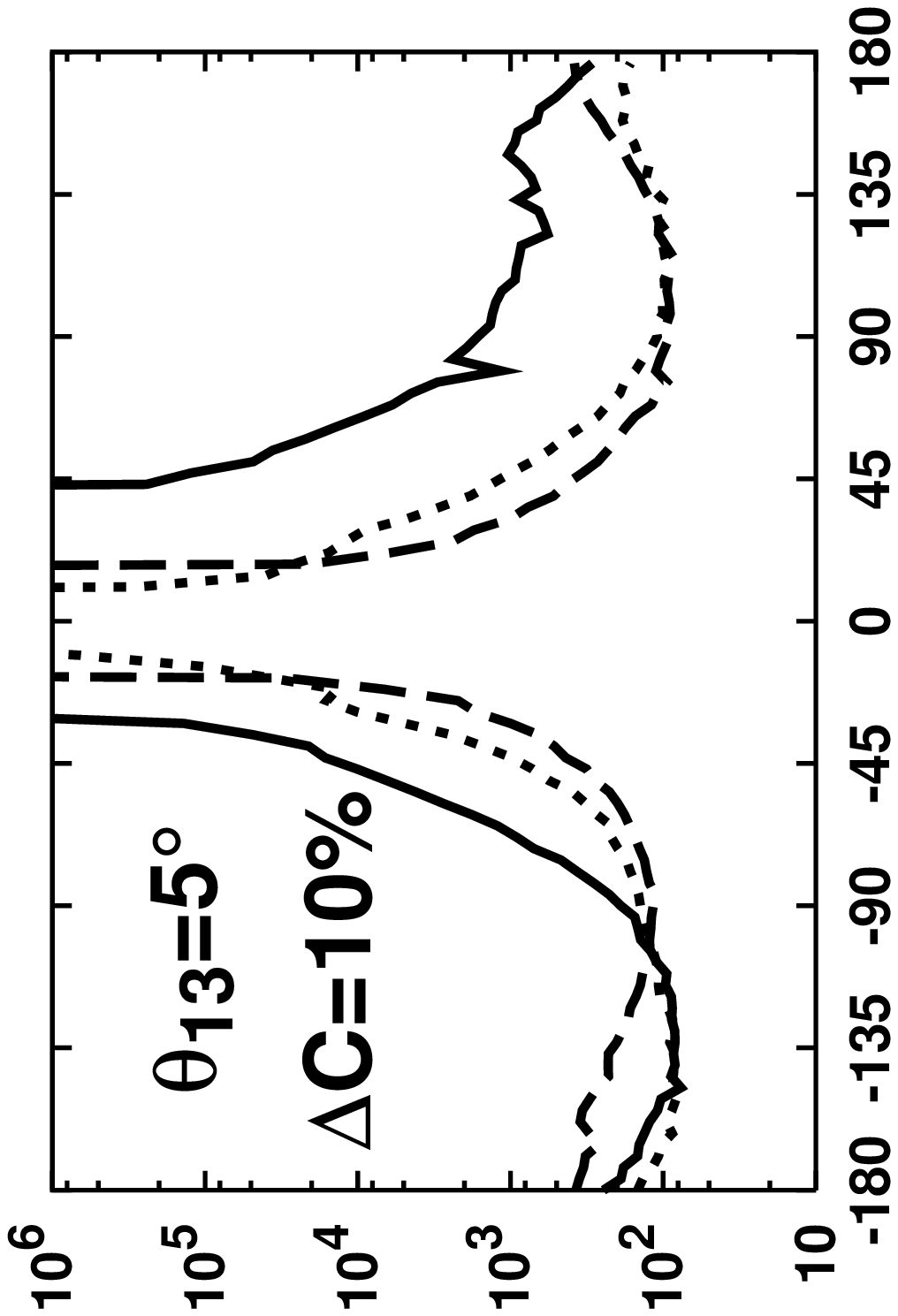,width=5.9cm}
\vglue -8.2cm \hglue 9.5cm \epsfig{file=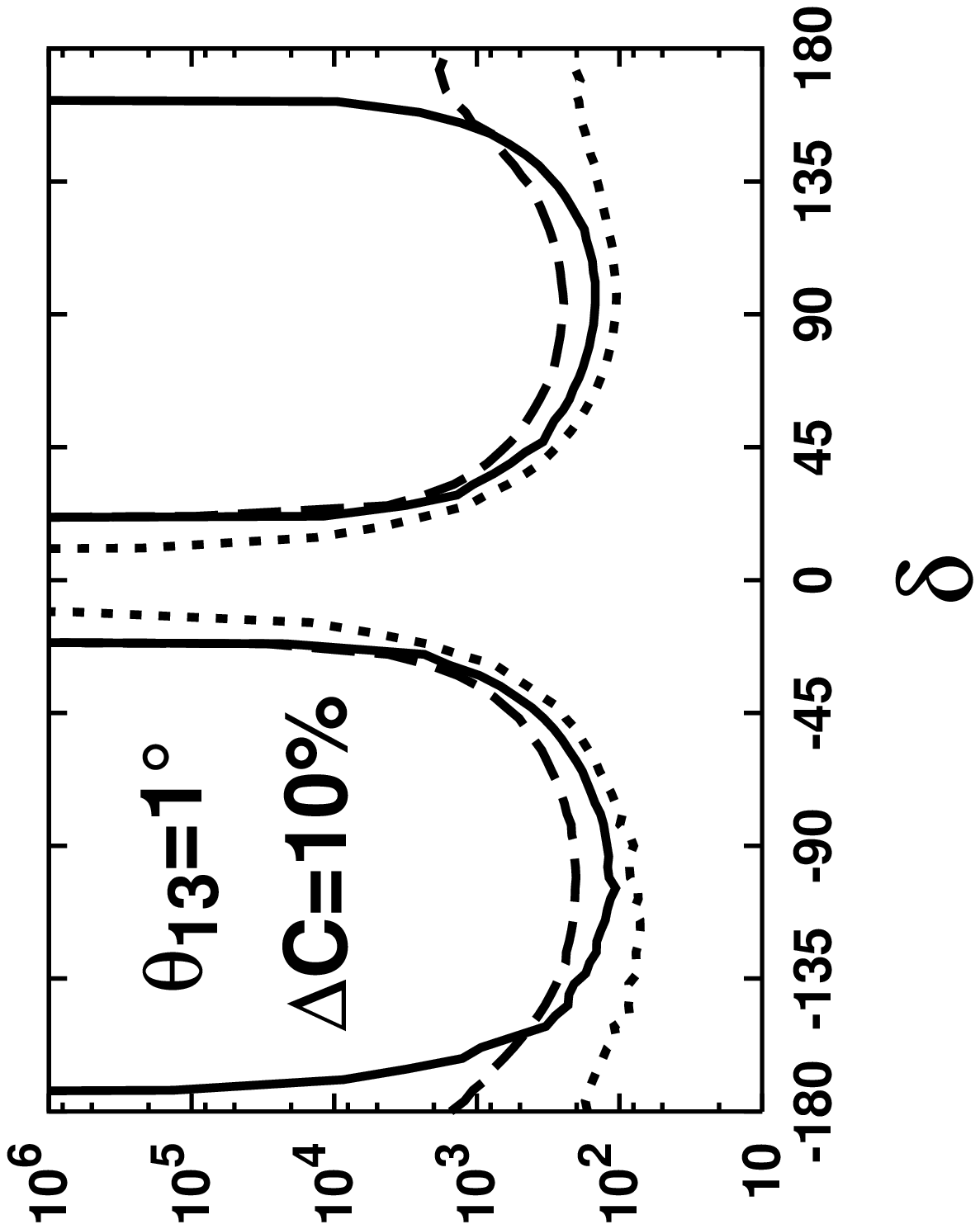,width=5.9cm}

\vglue -1.0cm
\hglue -1.2cm 
\epsfig{file=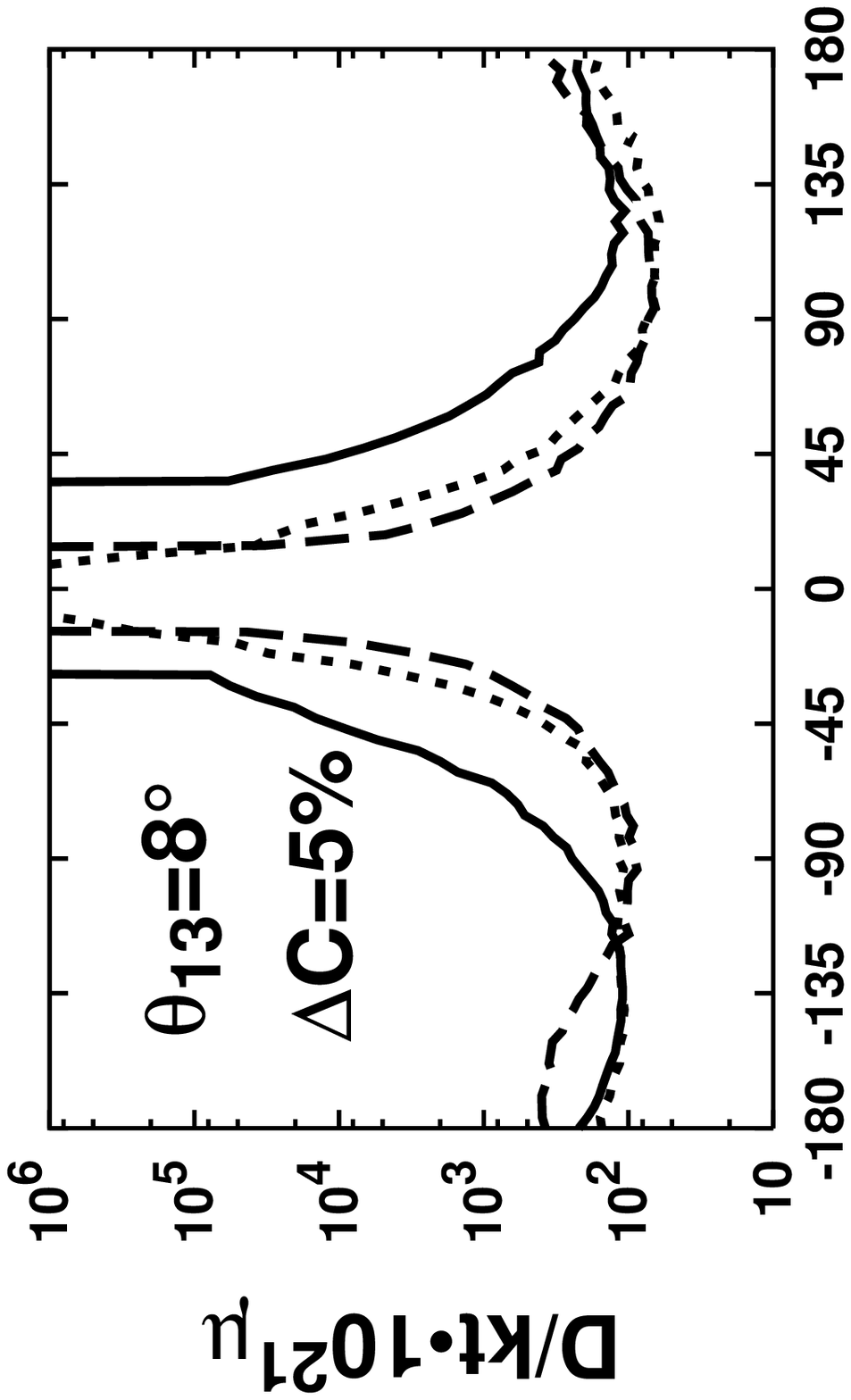,width=5.9cm}
\vglue -8.2cm \hglue 4.2cm \epsfig{file=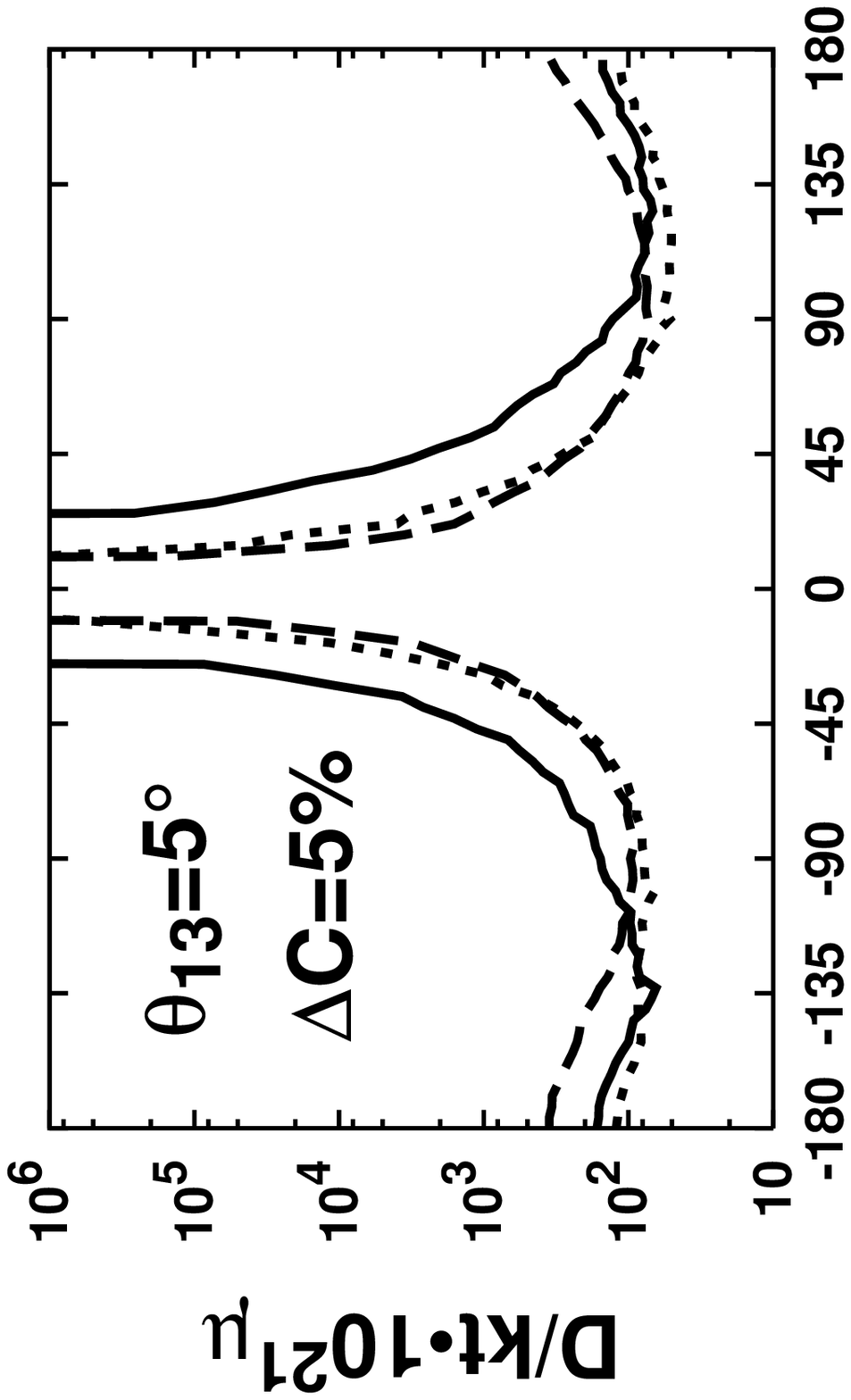,width=5.9cm}
\vglue -8.2cm \hglue 9.5cm \epsfig{file=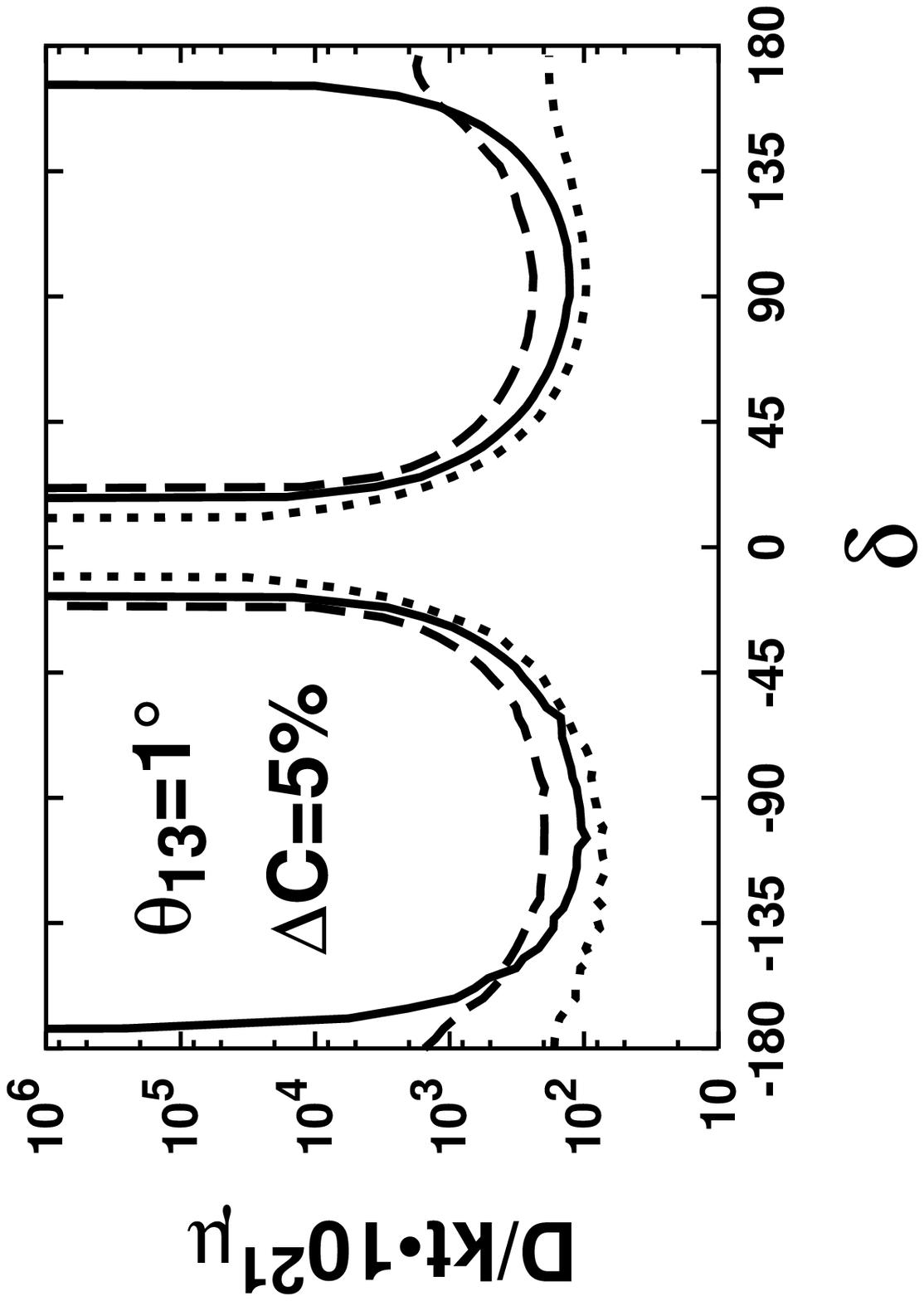,width=5.9cm}
\vglue -0.5cm
\hglue -0.5cm 
\epsfig{file=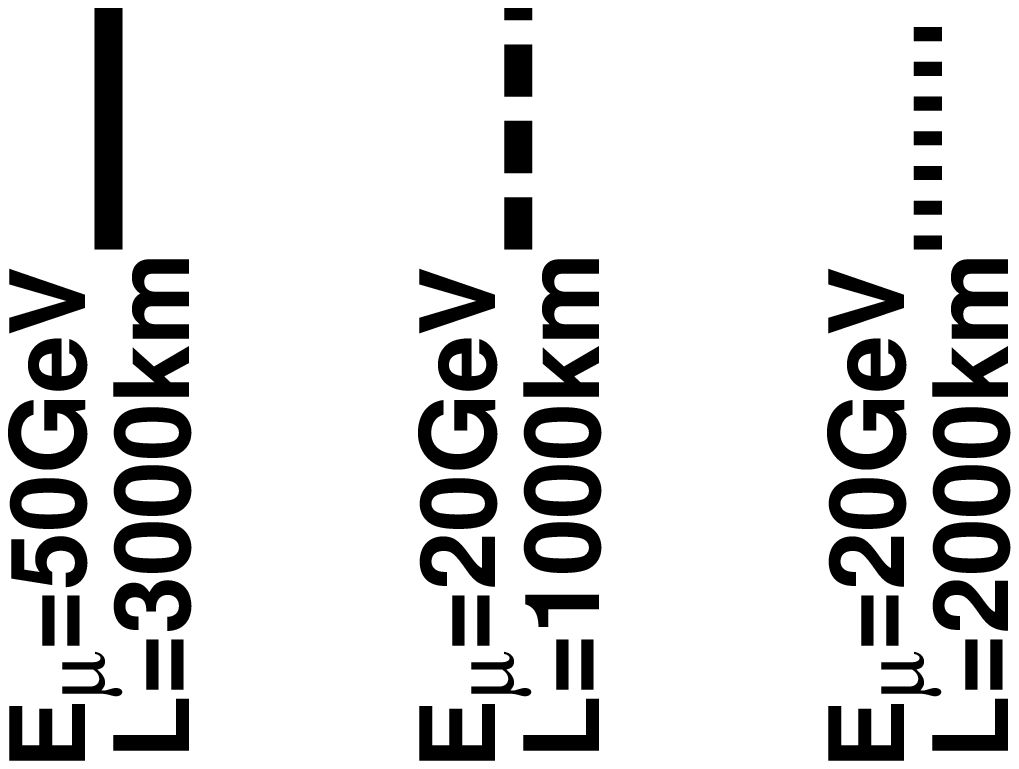,width=5.9cm}
\vglue -7.5cm\hglue 16.3cm
\epsfig{file=fig0911.eps,width=4cm}
\newpage
\vglue -5.60cm \hglue -3.0cm
\epsfig{file=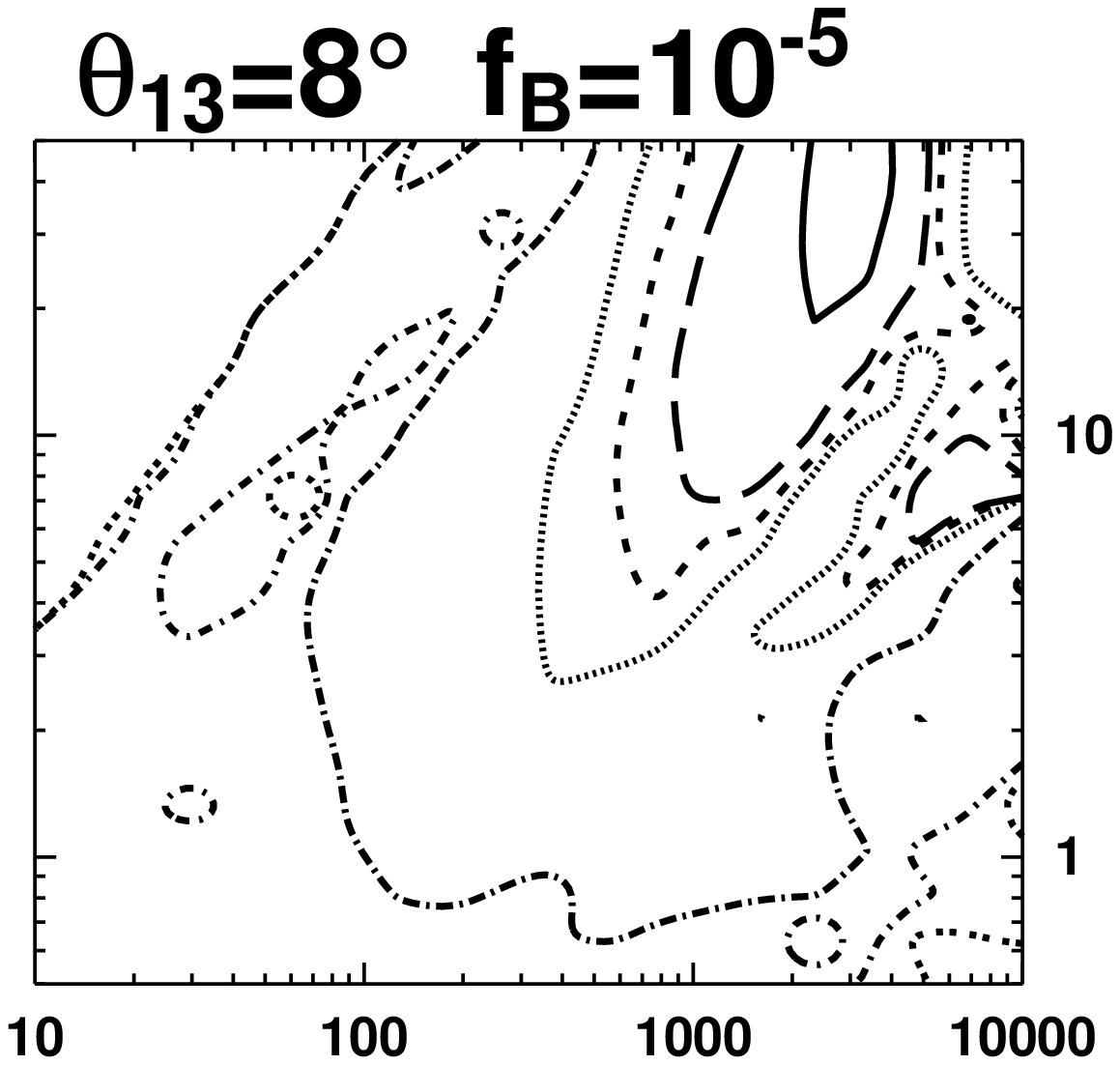,width=11cm}
\vglue -12.4cm \hglue 4.8cm
\epsfig{file=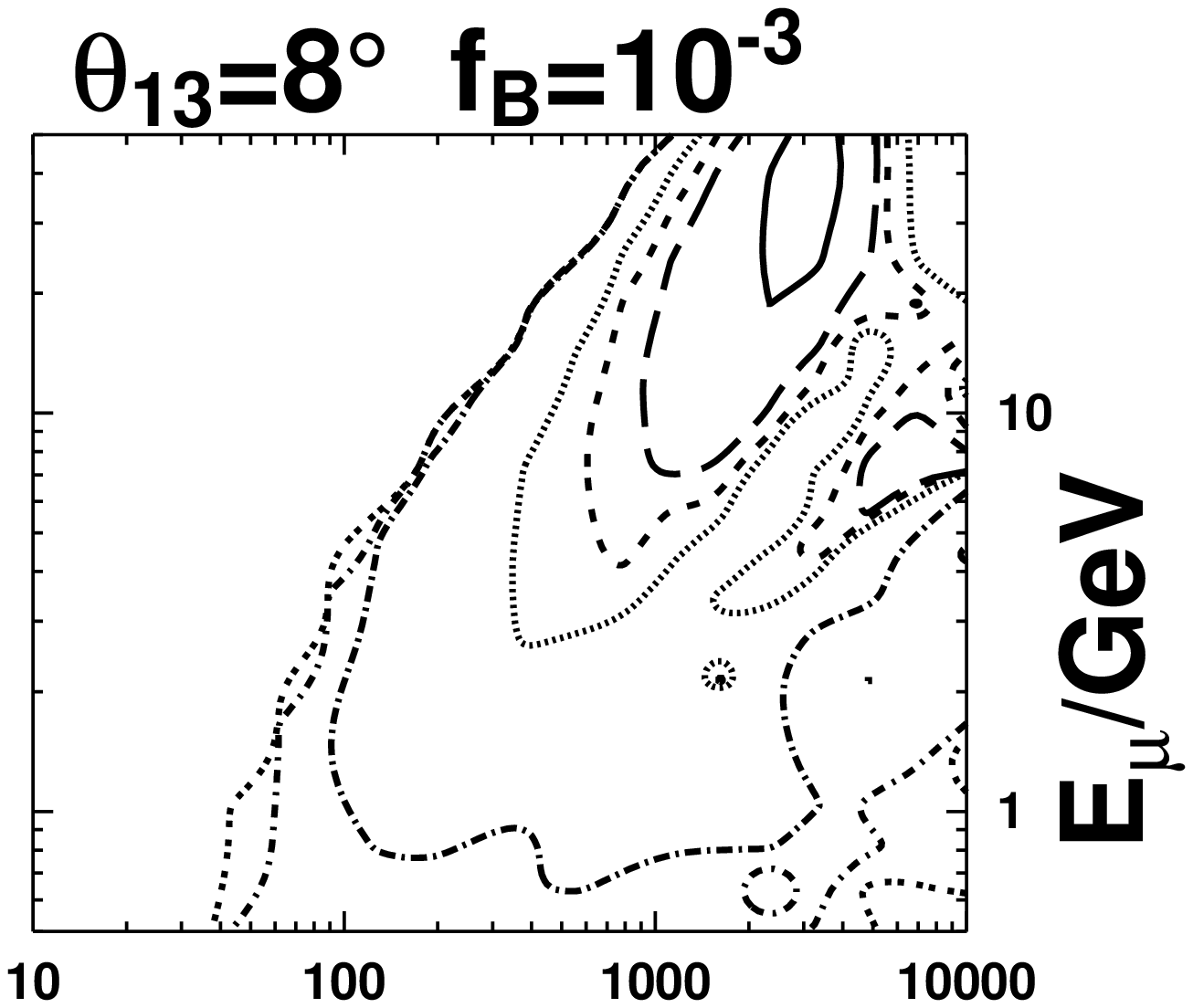,width=11cm}
\vglue -5.0cm \hglue -3.0cm
\epsfig{file=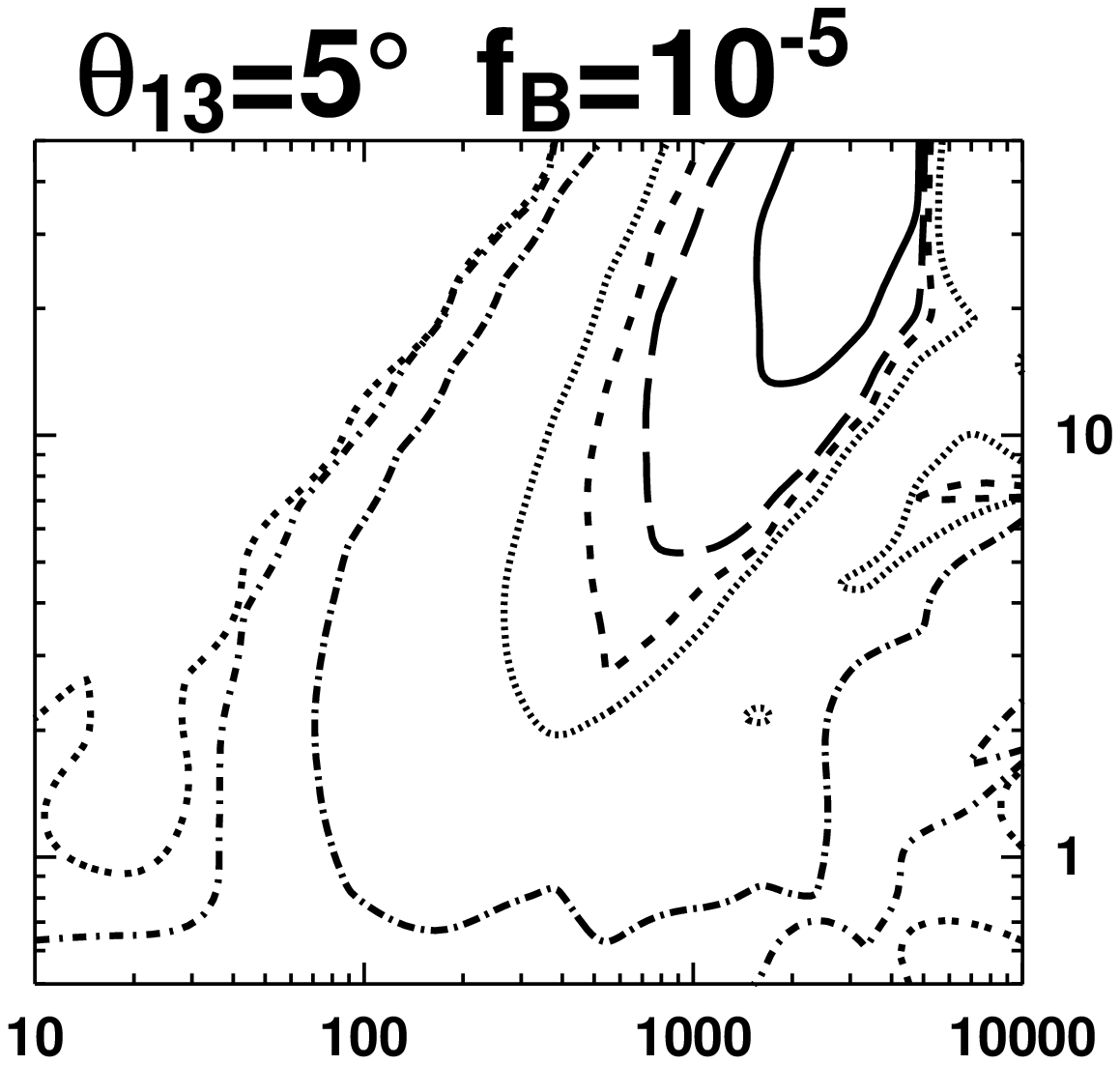,width=11cm}
\vglue -12.5cm \hglue 4.8cm
\epsfig{file=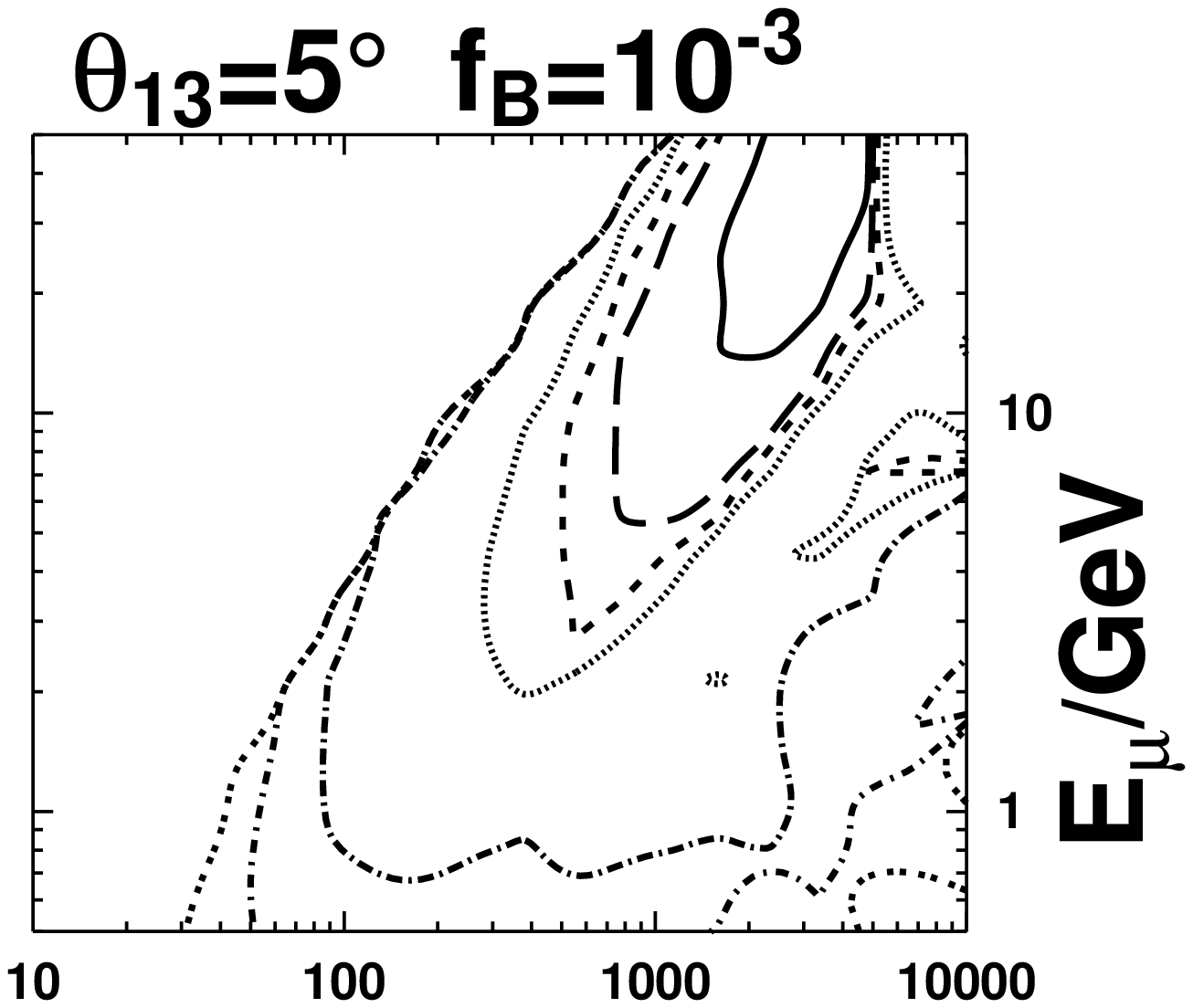,width=11cm}
\vglue -5.0cm \hglue -3.0cm
\epsfig{file=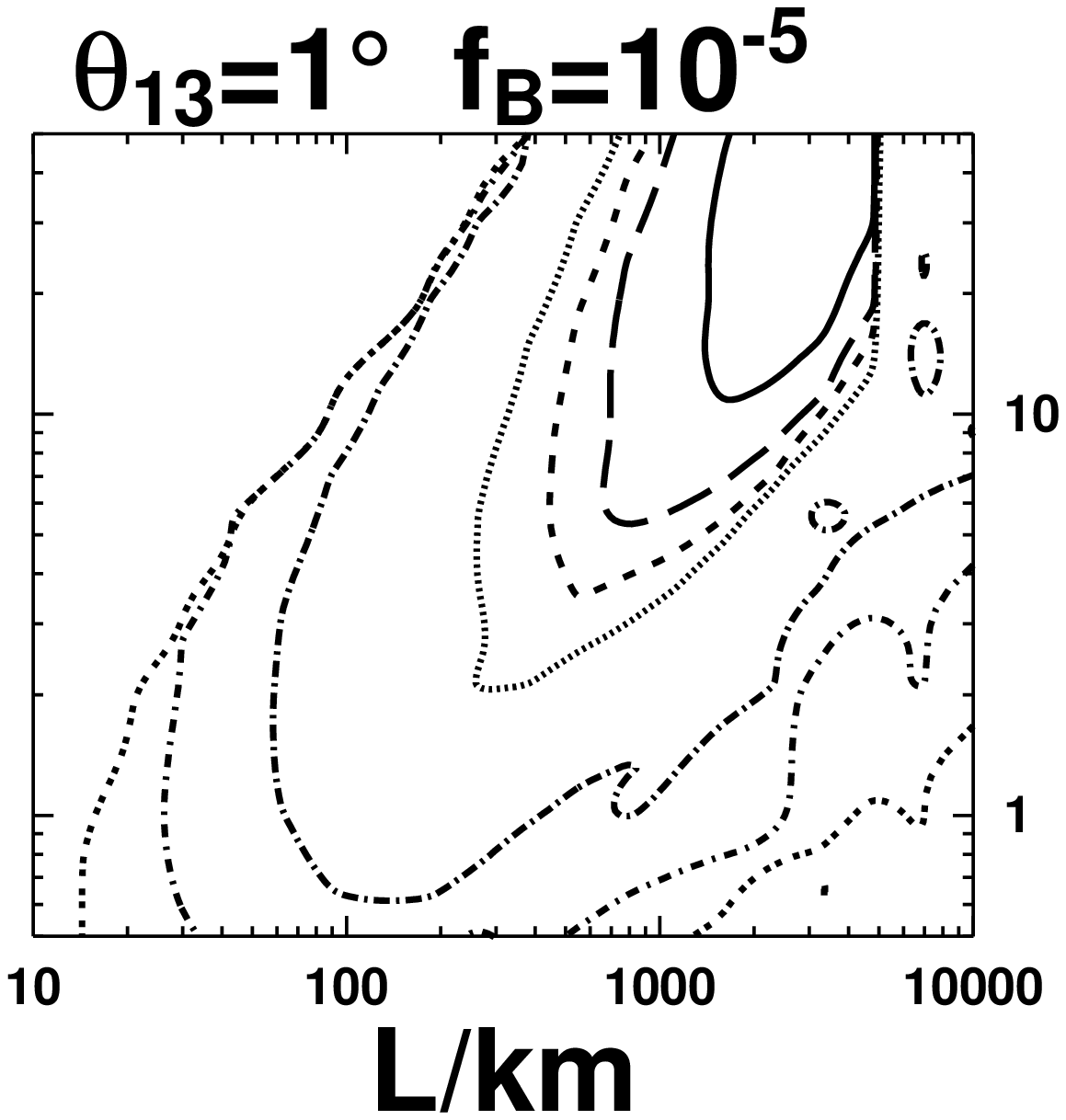,width=11cm}
\vglue -12.5cm \hglue 4.8cm
\epsfig{file=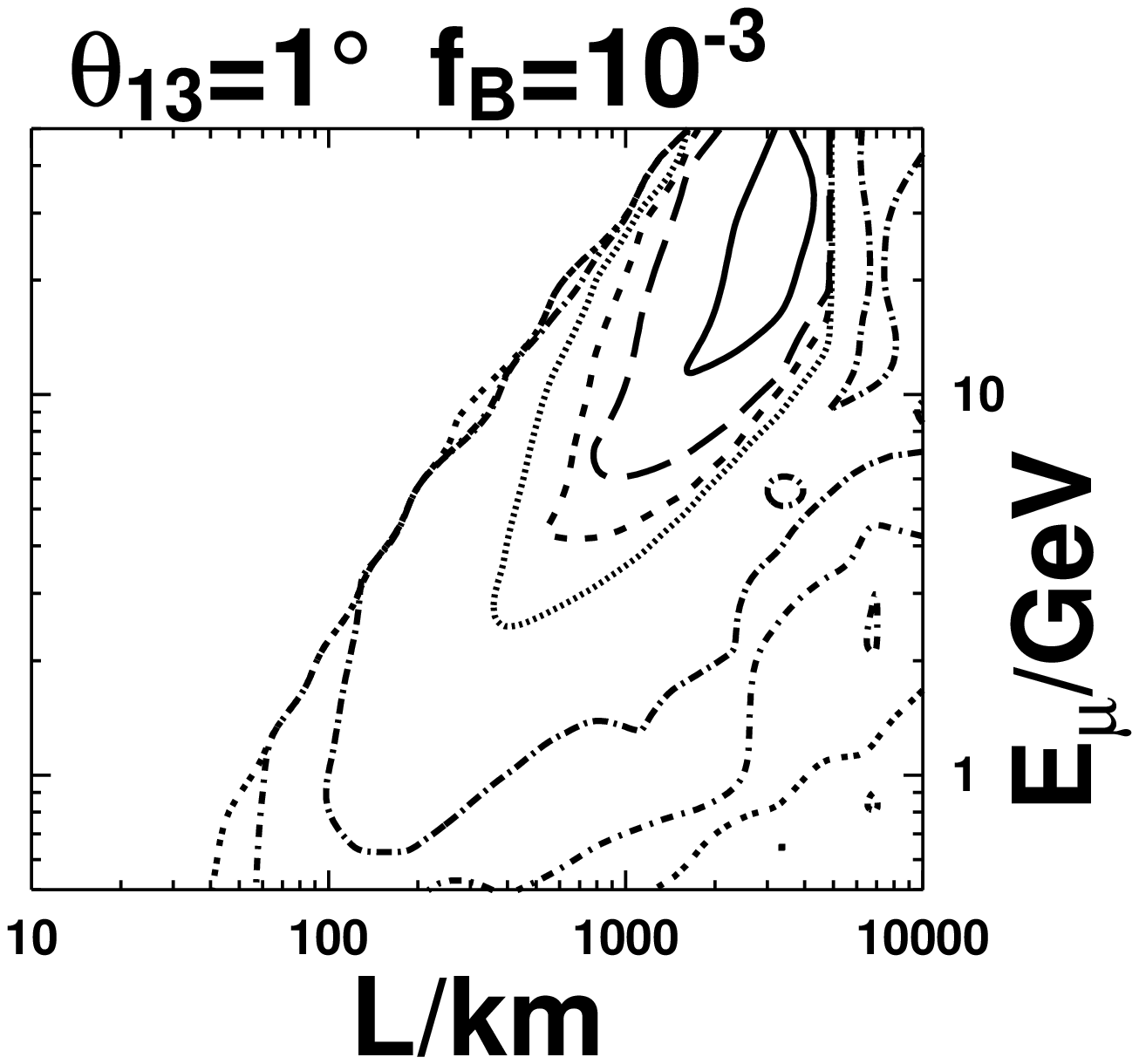,width=11cm}
\vglue -4.0cm\hglue -14.3cm
\epsfig{file=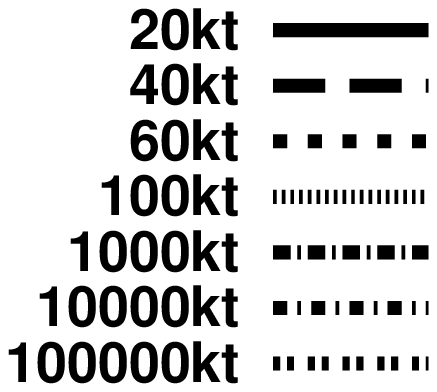,width=18cm}
\vglue -17.5cm\hglue 7.3cm
\epsfig{file=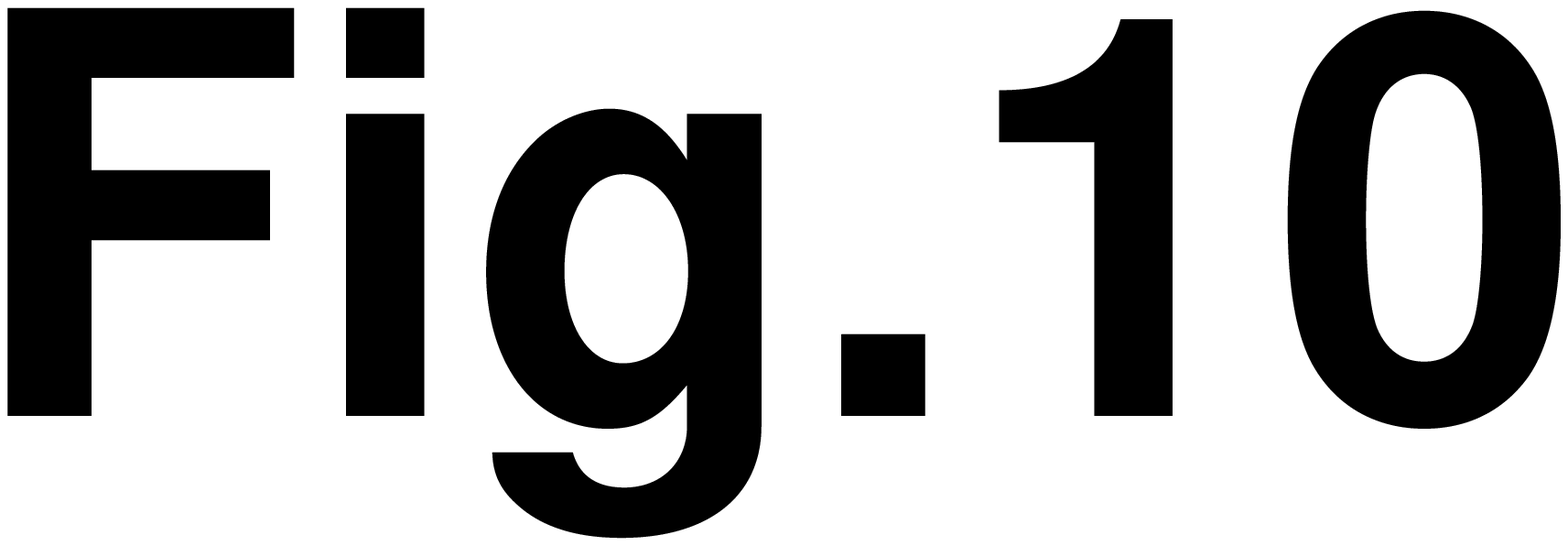,width=4cm}
\newpage
\vglue 1.4cm
\hglue -3.0cm 
\epsfig{file=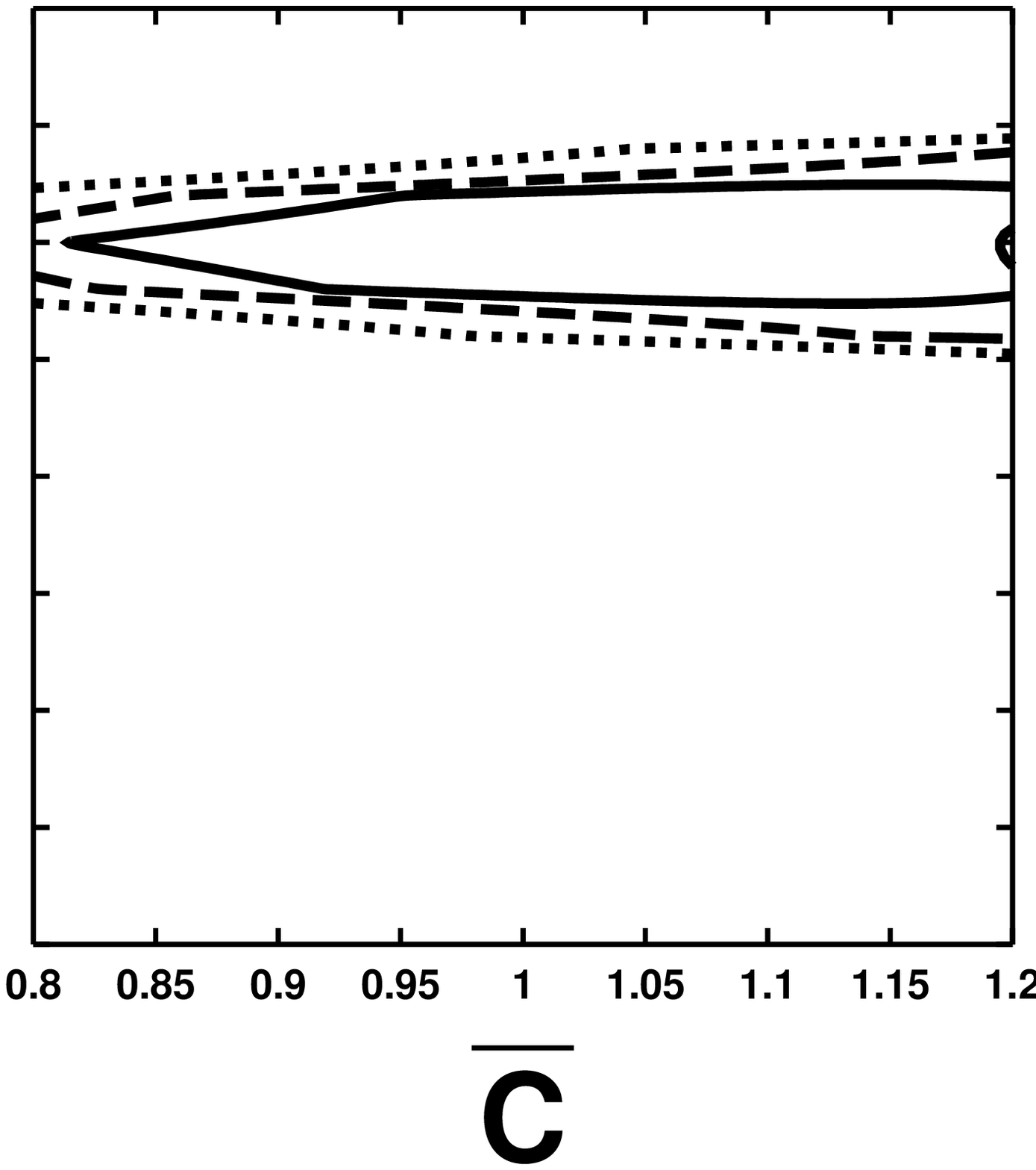,width=8cm}
\vglue -8.1cm \hglue 2.3cm \epsfig{file=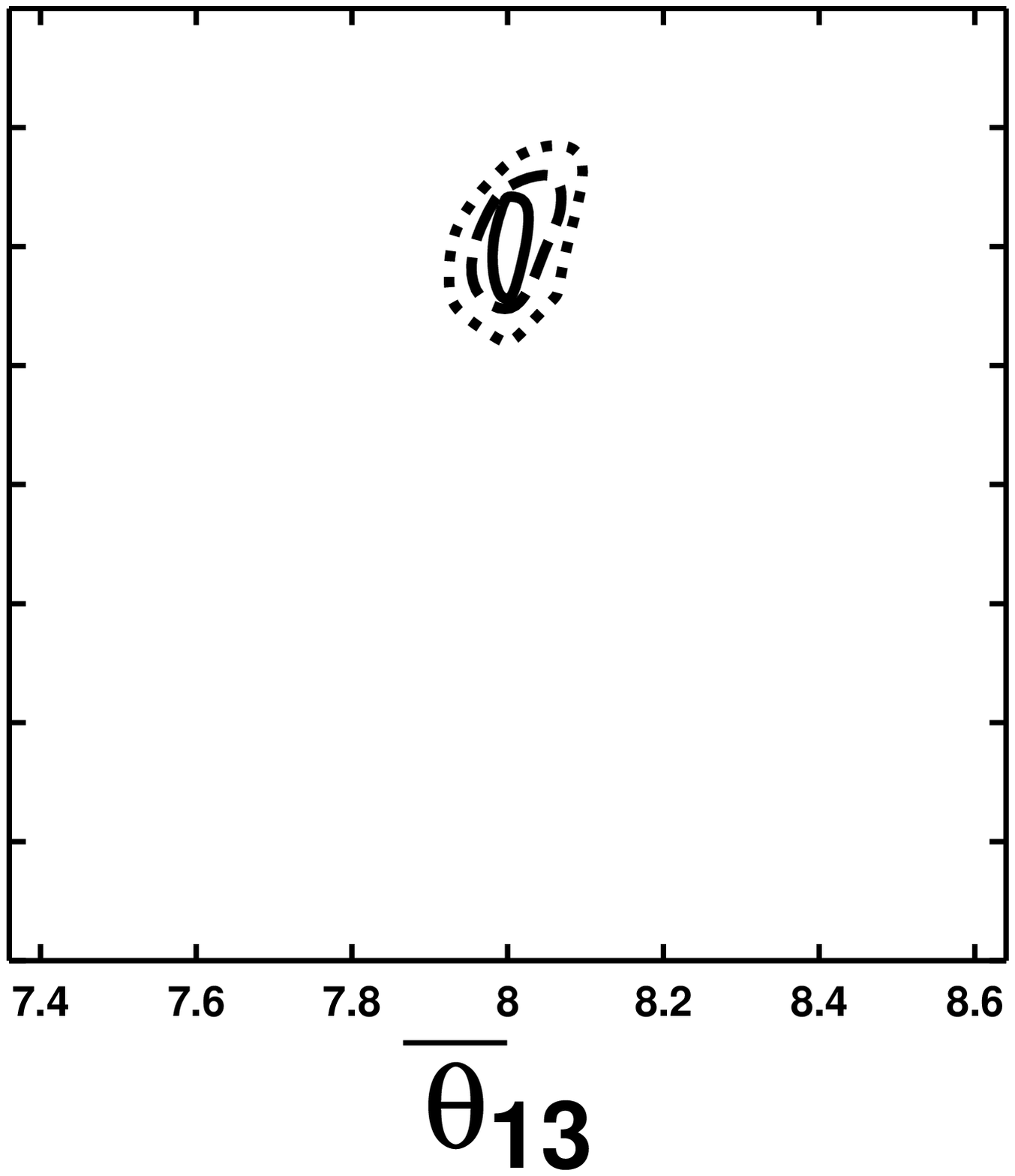,width=8cm}
\vglue -2.6cm
\hglue -3.0cm 
\epsfig{file=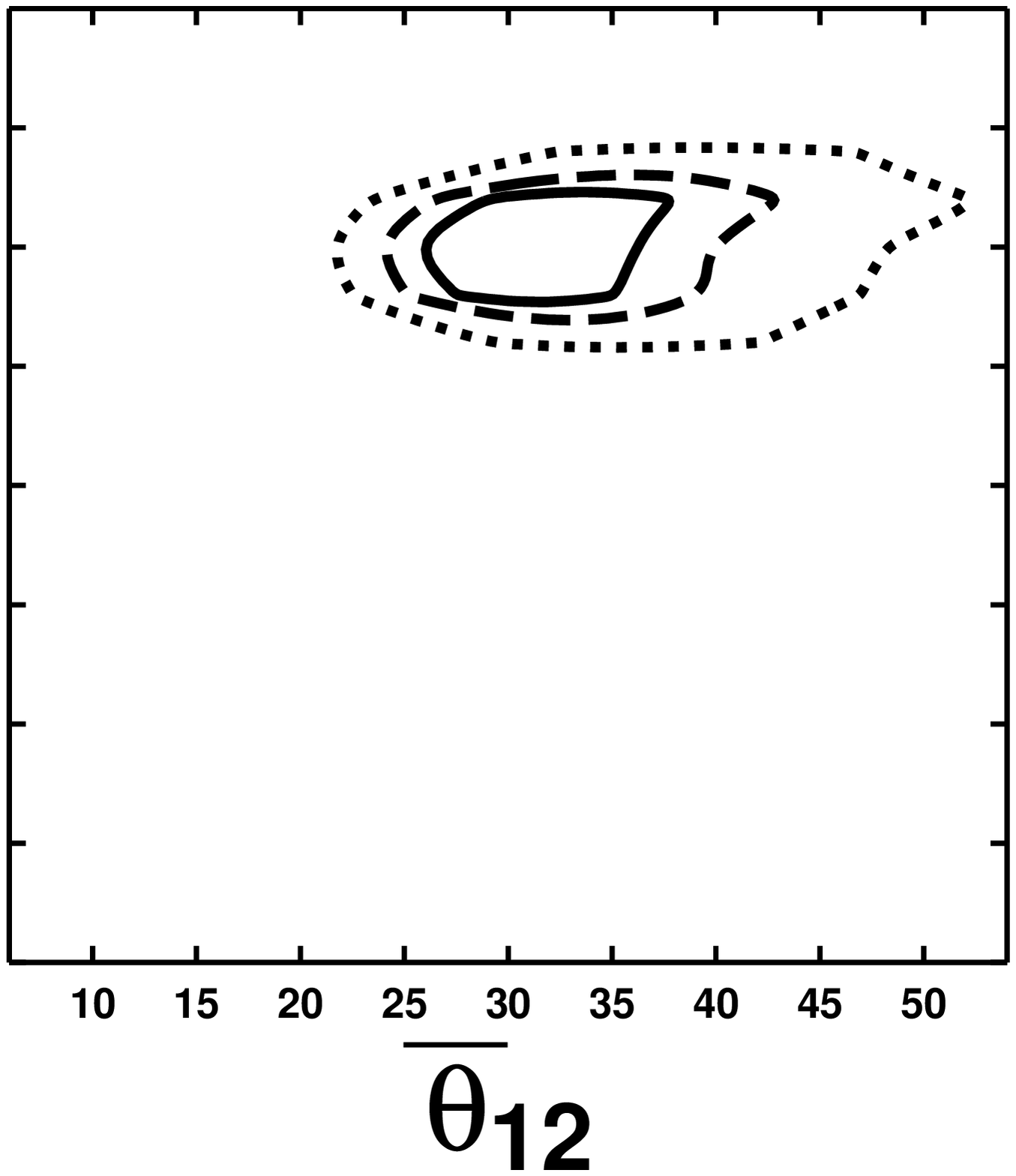,width=8cm}
\vglue -8.1cm \hglue 2.3cm \epsfig{file=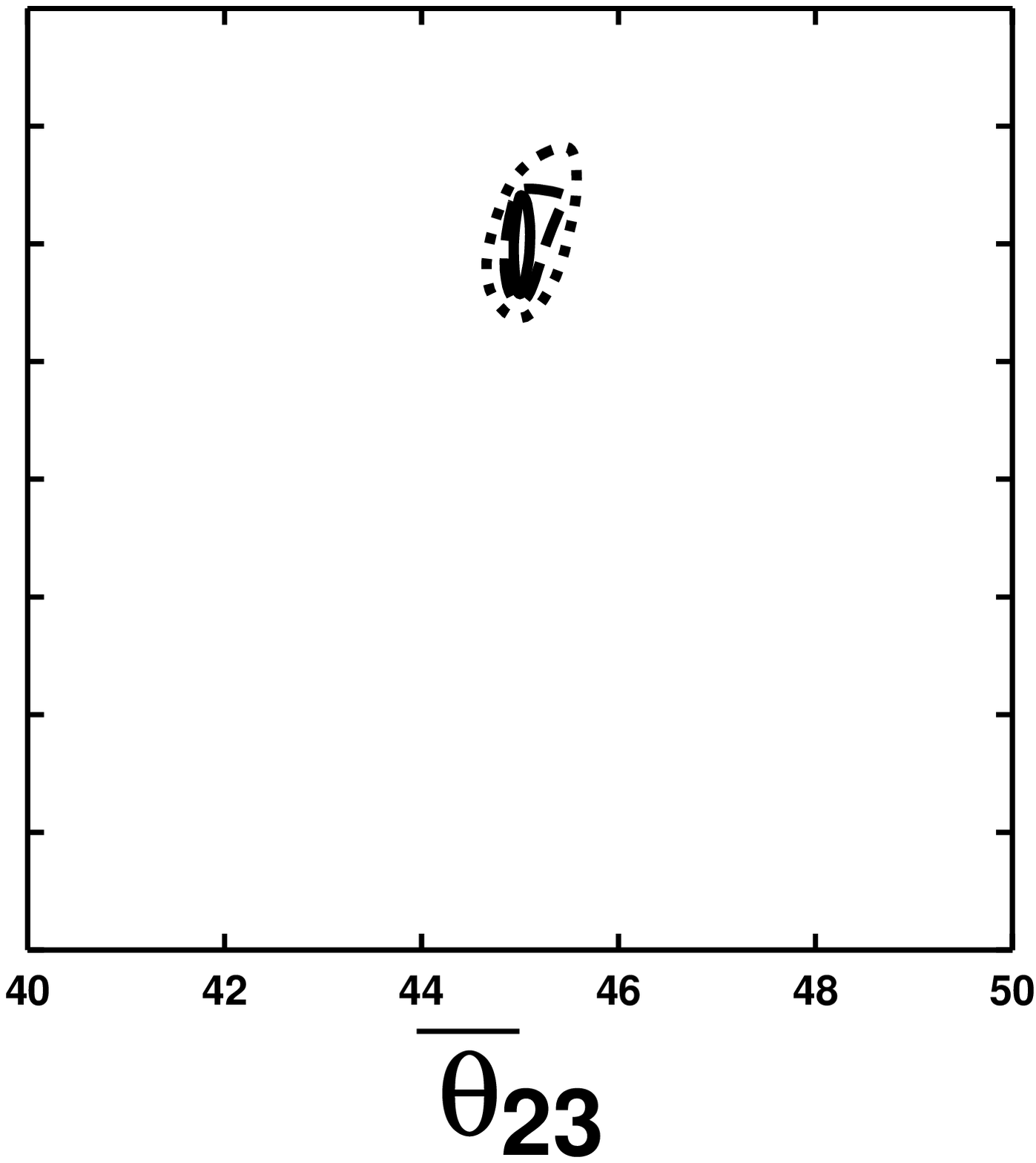,width=8cm}
\vglue -2.6cm
\hglue -3.0cm 
\epsfig{file=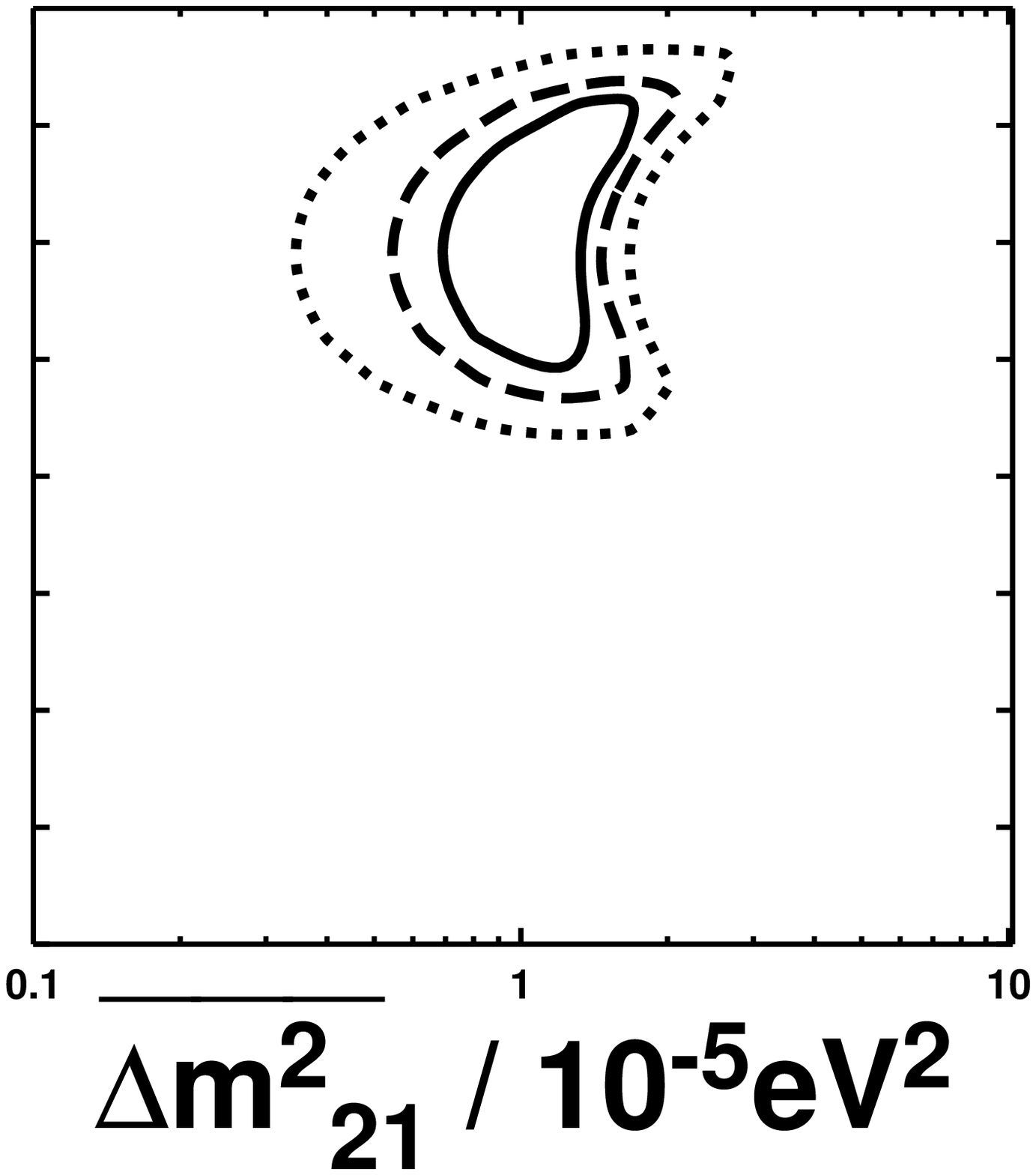,width=8cm}
\vglue -8.1cm \hglue 2.3cm \epsfig{file=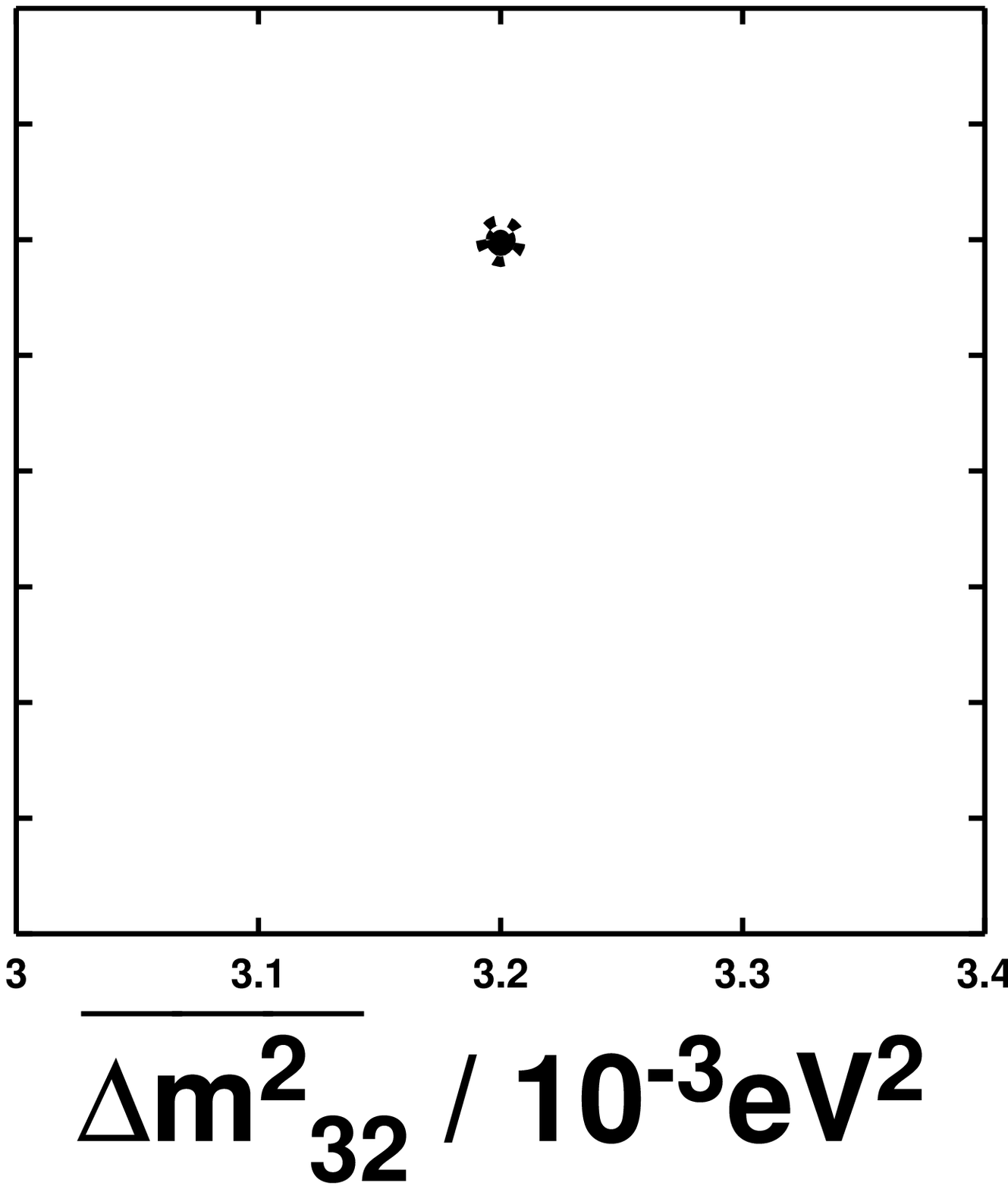,width=8cm}
\vglue -2.0cm\hglue -23.3cm
\epsfig{file=fig1107.eps,width=22cm}
\vglue -21.5cm\hglue 6.3cm
\epsfig{file=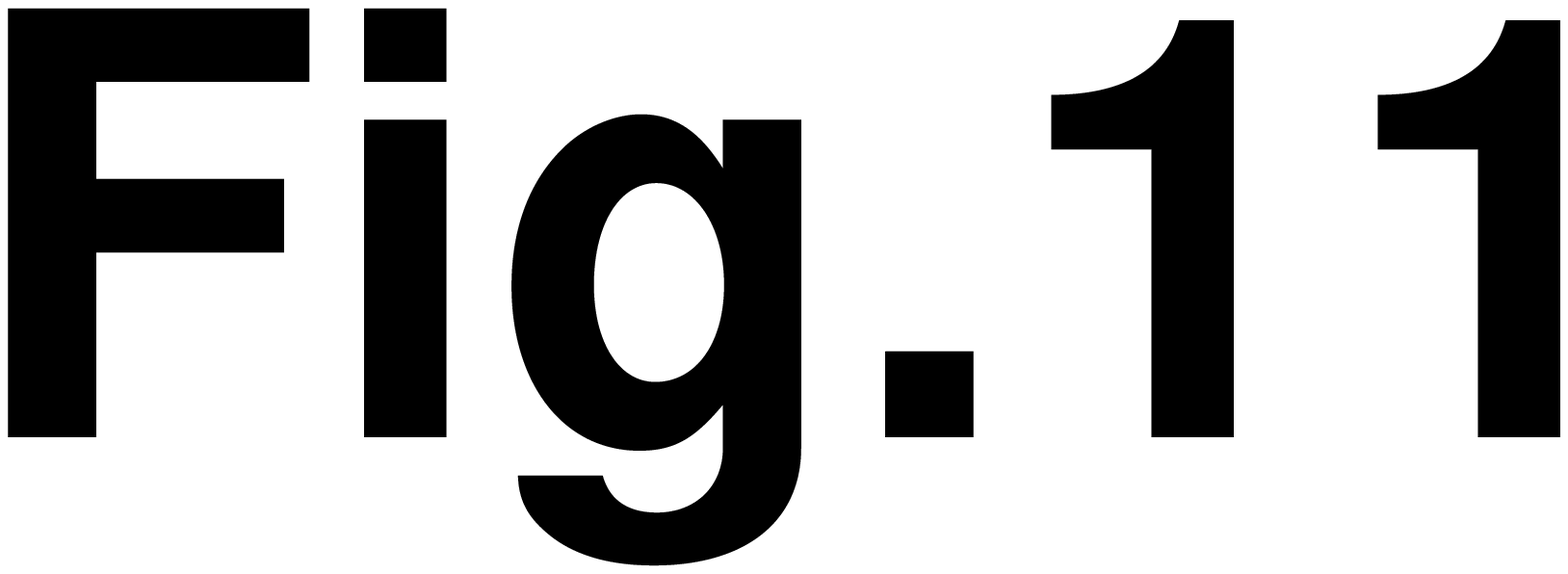,width=4cm}
\newpage
\vglue 1.60cm \hglue 2.0cm
\epsfig{file=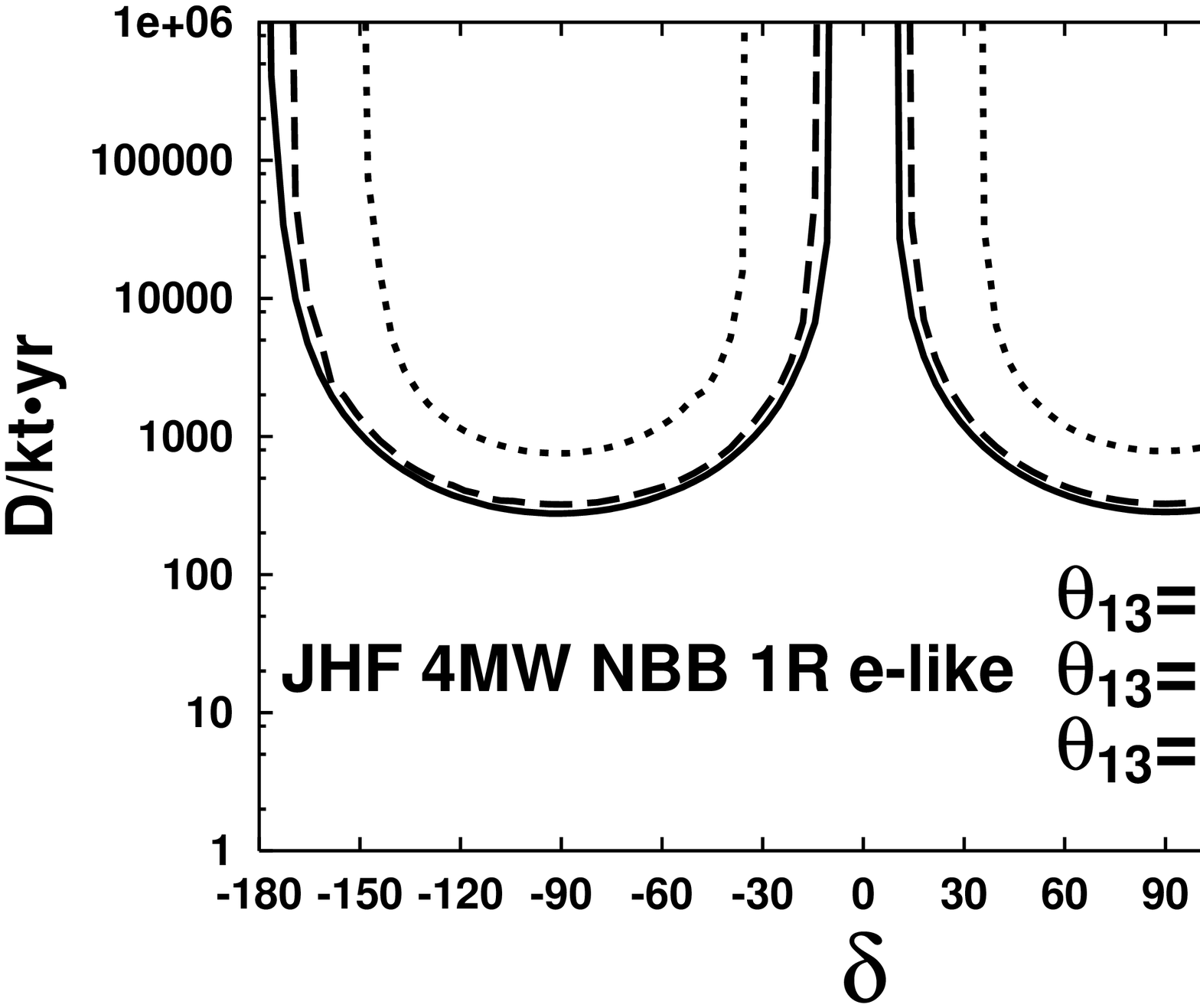,width=6cm}
\vglue -2.3cm \hglue 2.0cm
\epsfig{file=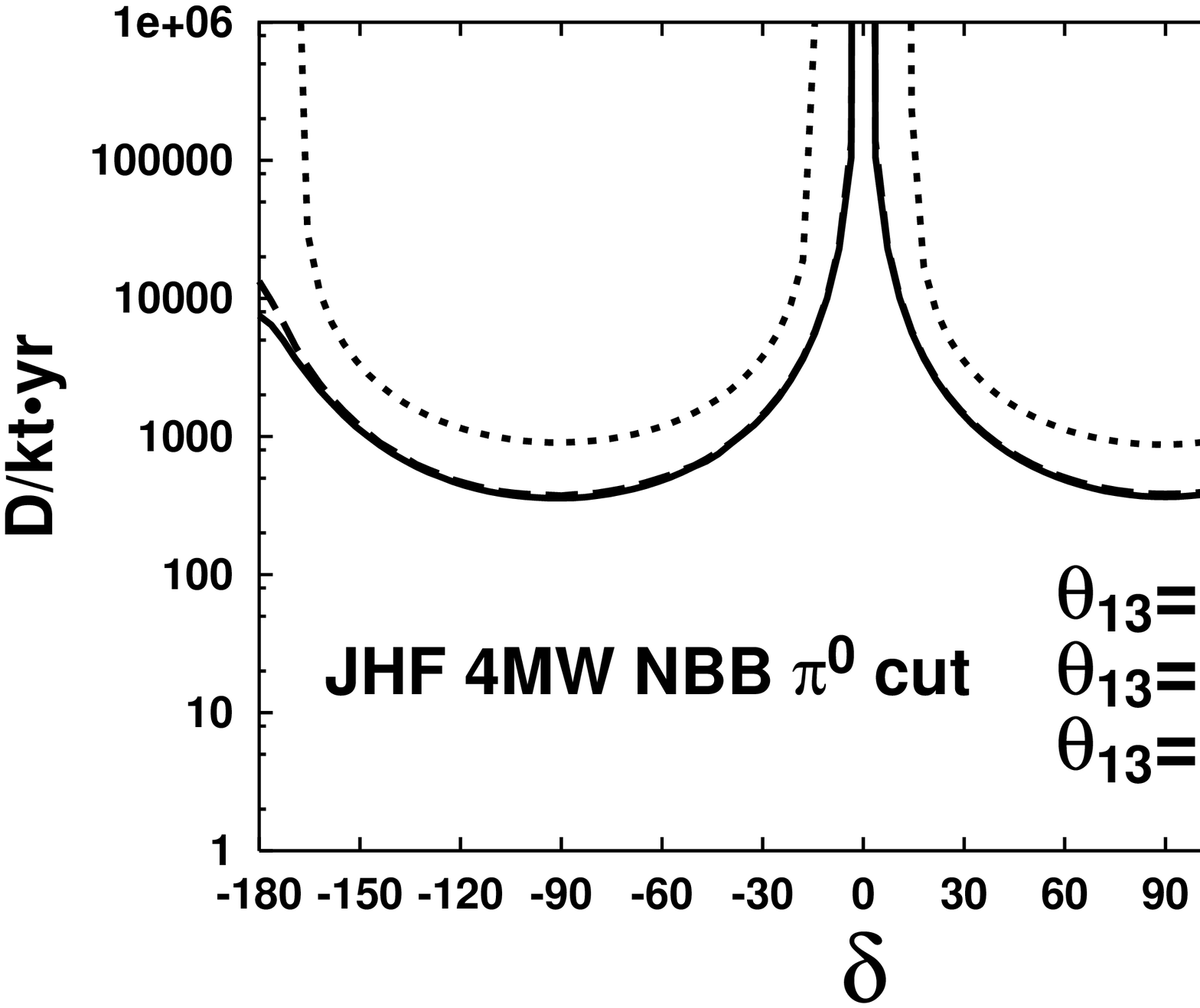,width=6cm}
\vglue -0.0cm\hglue 5.3cm
\epsfig{file=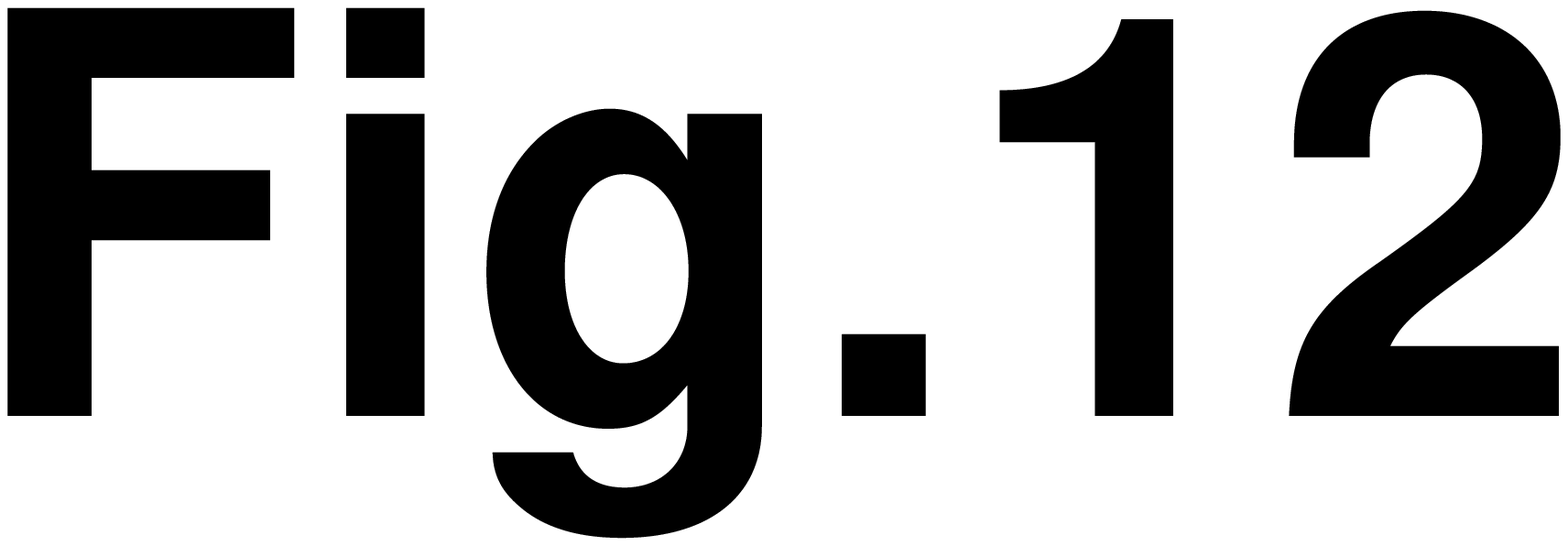,width=4cm}
\end{document}